\documentclass[11pt]{article}

\usepackage{graphicx}
\usepackage{color}
\usepackage{epstopdf}

\usepackage{amsmath}
\usepackage{amssymb}
\usepackage{latexsym}
\usepackage{bbm}

\usepackage[notlot,notlof,nottoc]{tocbibind}

\numberwithin{equation}{section}

\definecolor{wtmred}{rgb}{0.9,0.0,0}
\definecolor{wtmgreen}{rgb}{0,0.7,0}
\definecolor{wtmblue}{rgb}{0,0,0.9}

\usepackage[pdfstartview={FitH},
pdfpagemode={None},linktoc=page,colorlinks=true,pdfhighlight=/O,
linkcolor=wtmblue,citecolor=wtmred,urlcolor=wtmgreen]{hyperref}

\def\sst{\scriptscriptstyle}
\def\st{\scriptstyle}

\newcommand{\wht}{\widehat}
\newcommand{\wbr}{\overline}
\newcommand{\txfc}[2]{{\textstyle{\frac{#1}{#2}}}}

\newcommand{\txsum}{{\textstyle \sum}}
\newcommand{\txprd}{{\textstyle \prod}}

\newcommand{\com}[2]{[#1,#2]}

\newcommand{\spwd}[2]{\hspace{#1}\mbox{#2}\hspace{#1}}
\newcommand{\mvsp}[1]{\mbox{\rule{0cm}{#1}}}

\newcommand{\mc}{\mathcal}

\newcommand{\goto}{\rightarrow}

\newcommand{\be}{\begin{equation}}
\newcommand{\ee}{\end{equation}}

\newcommand{\ket}[1]{\left|#1 \right\rangle}
\newcommand{\bra}[1]{\left\langle #1 \right|}
\newcommand{\brkt}[2]{\left\langle #1\,|\, #2\right\rangle}

\newcommand{\vev}[1]{\left\langle #1 \right\rangle}
\newcommand{\amp}[2]{\left\langle\rule{0mm}{#1}\right.\!#2\!\left.\rule{0mm}{#1}\right\rangle}
\newcommand{\ampbox}[2]{\left[\rule{0mm}{#1}\right.\!#2\!\left.\rule{0mm}{#1}\right]}
\newcommand{\abs}[2]{\left|\rule{0mm}{#1}\right.\!#2\!\left.\rule{0mm}{#1}\right|}

\newcommand{\brktamp}[3]{\left\langle\rule{0mm}{#1}\right.\!#2\,|\,#3\!\left.\rule{0mm}{#1}\right\rangle}

\newcommand{\bbb}{\begin{eqnarray}}
\newcommand{\eee}{\end{eqnarray}}

\newlength{\slashwd}
\newlength{\txtwd}

\newlength{\ofst}

\begin{document}
\vspace*{2cm}

\begin{center}
{\Large\bfseries Notes on the \(SL(2,R)\) CFT}

\vspace*{1cm}

Will McElgin\footnote{\texttt{mcelgin@lsu.edu}} \\

\vspace*{1cm}
Department of Physics \& Astronomy\\
Louisiana State University \\
Baton Rouge, LA 70803, USA \\
\vspace*{1.5cm}
{\large\bfseries Abstract} \\
\vspace*{0.5cm}

\parbox{13.5cm}{This is a set of notes which reviews and addresses issues in the \(SL(2,R)\) conformal field theory, while working primarily in a basis of vertex operators of definite weight under the affine algebra. Following a review of the \(H_{3}\) coset model and the spectrum of the \(SL(2,R)\) CFT, derivations are given for the selection rules and the reduction of descendant to primary correlators involving arbitrary spectral flowed modules. Together with a description of the equivalence of correlation functions of fixed total spectral flow, this leads to an enumeration of the minimal set of three-point amplitudes required to determine the fusion rules. The corresponding fusion rules for elements of the spectrum are compared to those which arise through analytic continuation of the OPE of primary fields in the \(H_{3}\) model, and the apparent failure of the closure of the associated \(SL(2,R)\) OPE is described in detail. A discussion of the consistency of the respective fusion rules with the conjectured structure of equivalent factorizations of \(SL(2,R)\) four-point correlators is presented.}

\vspace*{3.5cm}

\end{center}

\pagebreak

\tableofcontents

\section{Introduction \label{sec_intro}}

The \(SL(2,R)\) conformal field theory has elicited great interest for more than two decades as a non-rational WZNW model incorporating a timelike direction. Much of this interest has been based on the expectation that the model may be tractable enough to allow an exact description of string propagation on the \(AdS_3\) spacetime, defined here to be the universal cover of the \(SL(2,R)\) group manifold. Physical motivation for understanding string theory on \(AdS_3\) has followed partly from the realization that black holes may be formed from \(AdS_3\) by both a coset construction, as in the \(SL(2,R)/U(1)\) CFT, and through orbifolding, as in the case of strings on BTZ black holes. More significantly, the \(AdS_3\) spacetime has served a very important role in elucidating the relationship between string theory on anti-deSitter spacetimes and conformal field theories defined on their conformal boundaries (for a review see~\cite{ADSCFT_9905}). In particular, since only \(\rm NS\) fields are required to define an exact background for the superstring, and since the corresponding boundary CFT, while possessing somewhat unfamiliar properties~\cite{SiebergWitten_9903}\cite{KutasovSeiberg_9903}\cite{GiveonKutasov_0106}, is two-dimensional, the \(SL(2,R)\) model has permitted a description of the AdS/CFT correspondence beyond the supergravity approximation that is more explicit than is available through other means. Related to these motivations has been the hope that the \(SL(2,R)\) CFT can play a role in clarifying the status of time in string theory, and in helping to resolve enduring mysteries in black hole physics.

The development of the technical description of the \(SL(2,R)\) CFT has moved at a slower pace than theoretical developments that have required its use. More progress has been made in understanding the CFT associated with \(H_{3}\,\), the Euclidean continuation of \(AdS_3\,\), which corresponds to the \(SL(2,C)/SU(2)\) coset model. Using methods which were originally applied to rational CFTs, and later used successfully in the solution of Liouville theory~\cite{Teschner_9507}\cite{Teschner_0104}\cite{Teschner_0303}, the presence of degenerate affine modules~\cite{Andreev_95} was exploited to derive the three-point function and the operator product expansion for the \(H_{3}\) CFT on the Riemann sphere~\cite{Teschner_9712}\cite{Teschner_9906}. The close relationship with Liouville theory led to a proof of the crossing symmetry of the \(H_{3}\) four-point function~\cite{Teschner_0108}, with a form of effective equivalence between the two theories subsequently demonstrated conclusively~\cite{RibaultTeschner_0502}. Significant efforts to describe string theory on \(AdS_{3}\) in the context of the AdS/CFT correspondence succeeded in uncovering the relationship between string vertex operators and operators in the boundary CFT~\cite{GiveonKutasovSeiberg_9806}\cite{deBoerOoguriRobinsTannenhauser_9812}\cite{KutasovSeiberg_9903}\cite{GiveonKutasov_0106}. However, glaring problems associated with the unitarity of the string spectrum on \(AdS_3\) remained. In particular, the requirement of unitarity for Virasoro primary states in conventional discrete affine representations of \(SL(2,R)\) imposes an apparent upper bound through the string constraints on the conformal dimensions of operators in a unitary CFT \(\mc{M}\) which appears as part of a string background \(AdS_{3}\times\mc{M}\).  These seeming pathologies were eliminated in a series of three papers, the first of which proved a no-ghost theorem for an extended spectrum which includes representations which follow from the spectral flow automorphism of the \(SL(2,R)\) current algebra~\cite{MaldacenaOoguri_0001}. That these affine representations are distinct from those arising from the representations of the global \(SL(2,R)\) algebra reflects the fact that the isometry group of \(AdS_3\) is not simply connected, but has fundamental group \(\mathbb{Z}\) which it inherits from the \(\mathbb{Z}\times\mathbb{Z}\) fundamental group of the group of isometries \((SL(2,R)\times SL(2,R))/\mathbb{Z}_2\,\) of the group manifold \(SL(2,R)\,\). This is unlike the case of the corresponding automorphism of the analogous \(SU(2)\) affine algebra, where the spectrum of states is mapped to itself under spectral flow. Further evidence that these representations give rise to a sensible string spectrum was provided through a path-integral calculation of the partition function of strings on the quotient of \(H_3\) corresponding to the Euclidean BTZ black hole~\cite{MaldacenaOoguriSon_0005}. Following this, correlation functions involving states in spectral flow sectors were investigated~\cite{MaldacenaOoguri_0111}, and many of the outstanding problems in the \(SL(2,R)\) CFT were resolved. 

Following these advances, the \(SL(2,R)\) model has been the focus of attention of numerous research efforts, including~\cite{GiribetNunez_0105}\cite{HosomichiSatoh_0105}\cite{Satoh_0109}. Prominent among these was the extension~\cite{Ribault_0507} of the work of~\cite{RibaultTeschner_0502} which presented an explicit expression for arbitrary \(n\)-point primary correlation functions\footnote{It should be noted that spectral flowed operators in the \(H_3\) CFT are non-normalizable, unlike the corresponding operators in the \(SL(2,R)\) model to which they may be related though analytic continuation.} in the \(H_3\) model with total spectral flow \(|w|\) via the computation of an amplitude in the Liouville CFT involving \(2n-2-|w|\) primary operators, of which \(n-2-|w|\) are in degenerate Virasoro modules. The relations appearing in~\cite{RibaultTeschner_0502}\cite{Ribault_0507} were further investigated~\cite{GiribetNakayama_0505}\cite{Giribet_0508}\cite{HikidaSchomerus_0706} by numerous authors, and many additional efforts~\cite{HerscovichMincesNunez_0512}\cite{MincesNunez_0701}\cite{IguriNunez_0705}\cite{IguriNunez_0908} were made using various techniques to compute correlation functions of the \(SL(2,R)\) CFT on the Riemann sphere, and to understand the \(H_3\) and \(SL(2,R)\) CFTs on both open and higher-genus closed Riemann surfaces~\cite{GiveonKutasovSchwimmer_0106}\cite{Ponsot_0204}\cite{Ribault_0512}\cite{HosomichiRibault_0610}\cite{BaronNunez_1012}. Among these investigations, building on the work of~\cite{Satoh_0109} for the unflowed sectors, was the examination in~\cite{BaronNunez_0810} of a definition of the OPE of the \(SL(2,R)\) model for generally flowed primary states via analytic continuation from the \(H_3\) model, with discrete intermediate states of the \(SL(2,R)\) CFT appearing as poles crossing the contour of integration of the spectrum of continuous states. It is important to note that an ansatz for the \(SL(2,R)\) OPE following from a presumed completeness of the spectrum requires only that the associated \(|w|\leq 1\) three-point amplitudes be derived from the \(H_3\) model. A corresponding definition of the \(SL(2,R)\) CFT thus presently lacks only a proof of crossing symmetry, though it might be hoped that such a proof would follow naturally through analytic continuation from those for the \(H_3\) or Liouville CFTs. Additionally, the factorization of string amplitudes is generally formulated\footnote{Many of the difficulties expected to be encountered in defining string amplitudes via analytic continuation in the \(\mc{V}_{j}(m)\) basis are not evident in the treatment involving the \(\Phi_{j}(x)\) basis appearing in~\cite{MaldacenaOoguri_0001}.} in terms of the analytic continuation of factorized Euclidean CFT amplitudes. The definition of the OPE introduced in~\cite{BaronNunez_0810} was motivated by the observation~\cite{Ribault_0507} that the appearance of intermediate states which are flowed elements of the \(SL(2,R)\) spectrum would seem to require the introduction of non-normalizable flowed elements in the \(H_3\) model prior to analytic continuation. However, at least in the case of the spectral flow conserving four-point amplitude, the additional flowed intermediate states appear to be superfluous since this amplitude is assumed to be computable entirely in terms of unflowed external and intermediate \(H_3\) states. Instead, the OPE coefficients involving states in spectral flowed sectors make an appearance in associated alternative factorizations of equivalent spectral flow conserving amplitudes. As suggested in~\cite{Ribault_0507}, this picture was extended in~\cite{BaronNunez_0810} to a conjectured equivalence for generally spectral flow non-conserving four-point correlation functions for alternative choices of factorization on respective \(|w|\leq 1\) intermediate \(H_3\) states. However, it is not clear that this leads to a consistent picture since the fusion rules associated with the respective factorizations do not in general agree with those derived under the assumption of the completeness of the \(SL(2,R)\,\) spectrum. As is evident in the selection rules discussed in section~\ref{sec_4}\,, for unflowed product states in the OPE involving at least one state in the continuous representations, no choice of factorization on states in a single spectral flow sector \(|w|\leq 1\) can produce all of the required flowed affine modules. Furthermore, as outlined in~\cite{BaronNunez_0810}, in choices of factorization where these fusion rules do not agree, discrete intermediate states which are outside of the \(SL(2,R)\,\) spectrum make an appearance.

The order of topics covered in these notes is as follows. In Section~\ref{sec_2} an introduction to the \(H_3\) CFT is given, with a comparison of the two- and three-point functions in a number of bases of vertex operators. The respective bases comprise the \(\Phi_{j}(x)\) basis in which the \(H_3\) model is often defined, the \(\varphi_{j}(\mu)\) basis which is useful for elucidating the relationship between the \(H_3\) and Liouville CFTs, and the basis \(\mc{V}_{j}(m)\) of definite affine weight which is predominantly employed in these notes. Following this are descriptions of both Liouville theory and the mapping presented in~\cite{RibaultTeschner_0502} which relates \(n\)-point correlation functions of primary operators in the \(H_3\) model and correlation functions of \(2n-2\) primary operators in Liouville theory. In Section~\ref{sec_3} a description of the spectrum of the \(SL(2,R)\) CFT introduced in \cite{MaldacenaOoguri_0001} is given, including those sectors which appear due to the spectral flow automorphism of the \(SL(2,R)\) current algebra, and the consequences of the condition of unitarity of the string spectrum are discussed. In Section~\ref{sec_4} the correlation functions of the \(SL(2,R)\) CFT are investigated, beginning with a detailed derivation of the selection rules~\cite{MaldacenaOoguri_0111} for flowed affine modules and associated primary states. Following this is a description of the process, which proceeds somewhat differently for the \(SL(2,R)\) model than for more conventional CFTs, of the reduction of \(n\)-point descendant correlators to those of primary states. The reduction to primary states of all three-point amplitudes permitted by the selection rules is described, and the minimal set of correlators required to compute the fusion rules involving elements of the spectrum is enumerated. The three-point function of unflowed primary fields is then analytically continued to produce a discrete two-point function which has a finite coefficient in a vertex operator normalization which imposes \(\mc{V}_{0}(0)=\Phi_{0}(x)=\mathbbm{1}\,\). This is followed by a description of the flowed primary vertex operators \(\mc{V}^{w}_{j}(m)\) in terms of parafermions and a timelike free boson which leads to a demonstration of the equivalence of spectral flow conserving correlators. This construction also leads to a finite normalization for the factor \(\omega_{j}\) which appears both in the operator equivalence \(\mc{V}^{w}_{j}(\pm j)=\omega_{j}\mc{V}^{w\pm 1}_{\tilde{\jmath}}(\mp \tilde{\jmath})\,\), where \(\tilde{\jmath}=k/2-j\,\), and in the expression for the unit spectral flow operator \(\mc{V}_{k/2}(\pm k/2)\,\) in terms of free field exponentials. Using some of the results of~\cite{Ribault_0507}, the equivalence between primary correlators\footnote{The summation considered here is over the spectral flow parameter of the primary representations, not of the corresponding affine modules. While the equivalence \(\hat{\mc{D}}^{\pm}_{w}=\hat{\mc{D}}^{\mp}_{w\pm 1}\) exists for the affine modules, with the exception of the state \(\mc{V}^{w}_{j}(\pm j)\,\), the corresponding primary representations \(\mc{D}^{\pm}_{w}\) and \(\mc{D}^{\mp}_{w\pm 1}\) are distinct.} of fixed total spectral flow is then described. The unit spectral flowed three-point function is then computed using a similar approach to that outlined in~\cite{MaldacenaOoguri_0111}, but involving a distinct basis of flowed primary fields \(\Phi_{j}^{w}(x)\,\). A finite normalization of this amplitude is provided both by requiring consistency of the identification \(\mc{V}^{w}_{j}(\pm j)=\omega_{j}\mc{V}^{w\pm 1}_{\tilde{\jmath}}(\mp \tilde{\jmath})\,\) for the discrete two-point function, and through an explicit computation involving analytic continuation of the factorization of the corresponding four-point amplitude in the \(H_{3}\) model. Section~\ref{sec_5} involves a derivation of the fusion rules for all flowed affine representations of the \(SL(2,R)\) spectrum. This is done by making use of the discrete two-point function, the spectrum of the \(SL(2,R)\) model, and the normalized three-point amplitudes, rather than through analytic continuation~\cite{BaronNunez_0810} as described in section~\ref{sec_6}. As is suggested by the results in~\cite{BaronNunez_0810}, and unlike in the treatment of the OPE of the corresponding string amplitudes given in~\cite{MaldacenaOoguri_0111}, it will be found that all continuous and discrete spectral flow sectors which are permitted by the selection rules have finite and non-vanishing OPE coefficients for some range of values of the \(j\) quantum number. Section~\ref{sec_6} contains an investigation of the factorization of the four-point function in the \(H_3\) model and the associated analytic continuation to the \(SL(2,R)\) CFT. A description of the pole structure in the complex \(j\) plane of the integral over intermediate states in the \(\Phi_{j}(x)\) basis is given. This is followed by a demonstration that the \(x\ll 1\) limit of the conformal blocks corresponds to a projection onto \(w=1\) flowed primary states. A detailed discussion is then given of the fusion rules which follow via analytic continuation of the OPE of primary states in the \(H_3\) model, with a particular emphasis on the appearance of discrete intermediate states outside of the spectrum of the \(SL(2,R)\) CFT. Also discussed is the structure of alternative factorizations~\cite{BaronNunez_0810} of the spectral flow conserving four-point correlator, which depends crucially on the equivalence of amplitudes of fixed total spectral flow. A conjectured picture of alternative factorizations for spectral flow non-conserving four-point correlators is then presented. This is followed by a conclusion in section~\ref{sec_7} which contains some remarks on the relationship between the set of apparent alternative factorizations of \(SL(2,R)\) four-point correlation functions and the failure of the closure of the related OPE coefficients on the \(SL(2,R)\) spectrum. Some further review and technical details are contained in the appendix.

\section{The \texorpdfstring{$H_{3}$}{} CFT \label{sec_2}}

The \(H_{3}\) coset CFT, like the \(SL(2,R)\) model considered below, has an affine symmetry with associated holomorphic currents \(J^{a}(z)\) which have the following OPE's
\bbb\label{sl2curope}
J^{3}(z)J^{3}(w) & \sim & -\frac{k/2}{(z-w)^2}\\
J^{3}(z)J^{\pm}(w) & \sim & \pm\frac{J^{\pm}(w)}{(z-w)}\\
J^{+}(z)J^{-}(w) & \sim & \frac{k}{(z-w)^2}\,-\,\frac{2J^{3}(w)}{(z-w)}
\eee
These are assembled into the worldsheet stress tensor via the Sugawara construction
\be\label{sugawara}
T(z)\,=\,(k-2)^{-1}\,\eta_{ab}:\!J^{a}J^{b}\!:\!(z)
\ee
The corresponding central charge is given by
\be\label{sl2cntchg}
c\,=\,\frac{3k}{k-2}\,=\,3+6b^2
\ee
where we have introduced the Liouville coupling \(b^{-2}=k-2\,\).

\subsection{The \texorpdfstring{$\Phi_{j}(x)$}{} primaries}

We consider the primary fields \(\Phi_{j}(x)\) with
worldsheet conformal dimension given by
\be
h(j)\,=\,h(1-j)\,=\,\frac{j(1-j)}{k-2}
\ee
Suppressing anti-holomorphic dependence, these fields satisfy the following OPEs
\be
J^{a}(z)\,\Phi_{j}(x|w) \,\sim\,
\frac{-D^{a}\,\Phi_{j}(x|w)}{(z-w)}
\ee
Here the generators \(D^{a}\) are given by
\be\label{phigen}
D^{+}\,=\,-\frac{\partial\,}{\partial x} \hspace{8mm} 
D^{3}\,=\,-x\frac{\partial\,}{\partial x}-j \hspace{8mm} 
D^{-}\,=\,-x^2\frac{\partial\,}{\partial x}-2jx
\ee
and satisfy the \(SL(2,R)\,\) algebra \(\com{D^{3}}{D^{\pm}}=\pm D^{\pm}\,\), \(\com{D^{+}}{D^{-}}=-2 D^{3}\,\) with the Casimir charge
\be
-D^{3}D^{3}+\txfc{1}{2}\left(D^{+}D^{-}+D^{-}D^{+}\right)\,=\,j(1-j)
\ee
The (Mobius-fixed) three-point correlation function\footnote{Here the convention \(\Phi_{j}(x)=\Phi_{j}(x|0)\) has been used, and \(u=1/z\) frame operators are given by \(\hat{\Phi}_{j}(x|u)\,\) as described in appendix~\ref{sec_D}\,.} of the \(H_3\) CFT is given by
\be\label{phi3pt}
\amp{3mm}{\hat{\Phi}_{j_3}(x_3)\Phi_{j_2}(x_2|1)\Phi_{j_1}(x_1)}=C(j_p)\,W(j_p,x_p)
\ee
where, defining \(\hat{\jmath}=\sum_{p}^{3} j_p\,\), the global \(SL(2,C)\) symmetry leads to
\be
W(j_p,x_p)\,=\,|x_{1}-x_{2}|^{2(2j_3-\hat{\jmath})}\,|x_{2}-x_{3}|^{2(2j_1-\hat{\jmath})}\,
|x_{3}-x_{1}|^{2(2j_2-\hat{\jmath})}
\ee
and the imposition of crossing symmetry on four-point correlators involving the insertion of operators in degenerate affine modules~\cite{Teschner_9712} leads to
\be\label{cfunc}
C(j_1,j_2,j_3)\,=\,-\nu^{1-\hat{\jmath}}\,\frac{b^{-2}}{2\pi^2}\,
\frac{\Gamma(1-b^2)}{\Gamma(1+b^2)}\,
\frac{G(\hat{\jmath}-1)}{G(1)}
\prod_{p=1}^{3}\frac{G(\hat{\jmath}-2j_p)}{G(2j_p-1)}
\ee
The function \(G(j)\) in (\ref{cfunc}) may be defined\footnote{Note that this definition differs from that in much of the literature in the sign of the argument.} through the \(\Upsilon_{b}(x)\) function (\ref{upsdefs1}) which appears in Liouville theory
\be\label{upsdefs}
G(j)\,=\,(k-2)^{\frac{-j(k-1-j)}{2(k-2)}}\;\Upsilon^{-1}_{b}(bj)
\ee
Defining \(\gamma(x)=\Gamma(x)/\Gamma(1-x)\,\), given the relations (\ref{upsdefs2}) for \(\Upsilon_{b}(x)\,\), it may be seen that \(G(j)\) satisfies
\be\label{gdefs1}
G(j)\,=\,G(k-1-j)
\ee
\be\label{gdefs2}
G(1-j)\,=\,(k-2)^{(2j-1)}\,\gamma(1-j)\,G(j)
\ee
\be\label{gdefs3}
G(j+1)\,=\,\gamma(1-j/(k-2))\,G(j)
\ee
Also, given the locations of the zeros (\ref{upsdefs5}) in \(\Upsilon_{b}(x)\,\), \(G(j)\) is meromorphic with poles at 
\be\label{gpoles}
j\,=\,-n-m(k-2)\spwd{5mm}{and}j\,=\,(n+1)+(m+1)(k-2)\spwd{5mm}{for}n,m\in\mathbb{Z}_{\geq 0}
\ee
The constant \(\nu\) is undetermined by the bootstrap procedure through which (\ref{cfunc}) may be derived, but will be fixed here to 
\be\label{Anrm}
\nu\,=\,\frac{\Gamma(1-b^2)}{\Gamma(1+b^2)}
\ee
as required by the relations between the \(H_3\) model and Liouville theory which appear in section~\ref{sec_lvtoh3}\,, as well as to produce the normalizations of the zero-mode quantum mechanics described in section~\ref{sec_B}\, in the \(b^2\goto 0\,\) limit.
Here \(C(j_p)\) is normalized such that, with \(\Phi_{0}(x)=\mathbbm{1}\,\), the two-point function is given by
\be\label{h32pf}
\amp{3mm}{\hat{\Phi}_{j_2}(x_2)\Phi_{j_1}(x_1)}=
\delta^2(x_2-x_1)\,\delta(s_2+s_1)+B_{j_1}\,|x_2-x_1|^{-4j_1}\,\delta(s_2-s_1)
\ee
where we have defined \(j_q=\frac{1}{2}+is_q\,\) with \(s_{q}\in\mathbb{R}\,\). This may be seen to follow from
\be\label{cfuncdlt}
C(j_1,j_2,\epsilon)\,\simeq\,B_{j_1}\,\delta(s_2-s_1)+\,\epsilon\,\pi^{-1}\delta(s_2+s_1)
\ee
as well as (\ref{phidlt}). Here \(B_{j}\) is defined in terms of
\be
\mc{R}_{j}\,=\,\mc{R}^{-1}_{1-j}\,=\,\frac{\pi B_{j}}{(2j-1)}\,=\,
\nu^{1-2j}\,\frac{\Gamma(1-b^2(2j-1))}{\Gamma(1+b^2(2j-1))}
\ee
Note that the choice (\ref{Anrm}) fixes \(\mc{R}_{0}=\lim_{k\goto\infty}\mc{R}_{j}=1\,\). It may also be seen that
\be\label{cfuncref}
\frac{C(j_1,j_2,j_3)}{C(j_1,j_2,1-j_3)}=
\mc{R}_{j_3}\,\frac{\Gamma(1-2j_3)}{\Gamma(2j_3-1)}\,
\frac{\gamma(j_{3}+j_{2}-j_{1})}{\gamma(1-j_{3}+j_{2}-j_{1})}
\ee
The primaries \(\Phi_{j}(x)\) must then satisfy the reflection relation
\be\label{phiref}
\Phi_{j}(x)\,=\,B_{j}\!\int d^2y\,|x-y|^{-4j}\,\Phi_{1-j}(y)
\ee
which is defined in the sense of analytic continuation from a 
convergent domain, and follows from the integral (\ref{louvint}). The reflection symmetry and the two-point function (\ref{h32pf}) imply that the (continuum normalizable) spectrum of the \(H_3\) model is given by operators \(\Phi_{j}(x)\) for \(j=\frac{1}{2}+is\,\) with \(s\in\mathbb{R}_{\geq 0}\,\). This is denoted below by \(j\in\mc{S}^{+}_{\sst \mc{C}}=\frac{1}{2}+i\mathbb{R}_{+}\,\). 

All descendant correlation functions in the \(H_3\) model may be readily reduced\footnote{See section \ref{reduce_sec} for a discussion of subtleties which arise in the corresponding reduction of descendant correlators in the \(SL(2,R)\) CFT.} to those of primary fields, which may in turn be computed by making use of the \(H_3\) OPE~\cite{Teschner_9906}
\be\label{H3OPEPhi}
\Phi_{j_2}(x_2|z)\Phi_{j_1}(x_1)\,=\,
\int\! d^2 x\!\int_{\mathbb{R}_{+}}\!\!ds
\amp{3mm}{\hat{\Phi}_{1-j}(x)\Phi_{j_2}(x_2|z)\Phi_{j_1}(x_1)}\left[\Phi_{j}(x)+\ldots\right]
\ee
where \(j=\frac{1}{2}+is\,\). The sum includes descendants whose coefficients are determined entirely by the affine symmetry. The reflection symmetry (\ref{phiref}) implies that the integrand in (\ref{H3OPEPhi}) is invariant under \(j\goto 1-j\,\), and thus that the \(s\) integral may be extended to the entire real axis (equivalently to \(j\in\mc{S}_{\sst \mc{C}}=\frac{1}{2}+i\mathbb{R}\,\)). The domain of validity of this expression is given by
\be\label{h3opedom}
|{\rm Re}(j_1-j_2)|<1/2\ \ ,\hspace{0.3cm}|{\rm Re}(1-j_1-j_2)|<1/2
\ee
Outside of this domain, which clearly includes \(j_{p}\in\mc{S}_{\sst\mc{C}}\,\), (\ref{H3OPEPhi}) possesses a well-defined analytic continuation to \(j_{p}\in\mathbb{C}\,\) under which poles in the integrand cross the contour of integration and produce discrete contributions. Further features of the \(H_3\) CFT in the \(\Phi_{j}(x)\) basis will be examined in section~\ref{sec_6}\, in the context of the factorization of the four-point function.

\subsection{The \texorpdfstring{$\varphi_{j}(\mu)$}{} primaries}

To examine the correspondence between the \(H_3\) and Liouville CFTs, it is very useful to introduce~\cite{RibaultTeschner_0502} the Fourier transformed fields
\be\label{phitolv}
\varphi_{j}(\mu)\,=\,-|\mu|^{2j}\,\pi^{-1}\!\!\int d^2x\,
e^{\mu x-\bar{\mu}\bar{x}}\;\Phi_{1-j}(x)
\ee
Here we have chosen the normalization \(\varphi_{0}(0)=\mc{R}^{-1}_{0}\,\mathbbm{1}\,\), where \(\mc{R}_{0}=1\) in the normalization given by (\ref{Anrm})\,.
The above fields may be seen to satisfy the reflection relation
\be\label{h3lvref}
\varphi_{j}(\mu)\,=\,R_{j}\,\varphi_{1-j}(\mu)
\ee
where, given (\ref{int1}),
\be\label{h3ref}
R_{j}\,=\,R^{-1}_{1-j}\,=\,
\mc{R}_{1-j}\,\frac{\Gamma(2j-1)}{\Gamma(1-2j)}
\ee
It may be seen that the fields \(\varphi_{j}(\mu)\) satisfy the following OPEs
\be
J^{a}(z)\,\varphi_{j}(\mu|w) \,\sim\,
\frac{-F^{a}\,\varphi_{j}(\mu|w)}{(z-w)}
\ee
Here the \(SL(2,R)\,\) generators \(F^{a}\) are given by
\be\label{fops}
F^{+}\,=\,\mu \hspace{8mm} 
F^{3}\,=\,\mu\frac{\partial\,}{\partial \mu}\hspace{8mm} 
F^{-}\,=\,\mu\frac{\partial^2\,}{\partial \mu^2}+j(1-j)/\mu
\ee
The three-point function in this basis takes the form
\be\label{mu3pf}
\amp{3mm}{\hat{\varphi}_{j_3}(\mu_3)\varphi_{j_2}(\mu_2|1)\varphi_{j_1}(\mu_1)}
\,=\,\delta^2({\textstyle \sum_{p=1}^{3}}\mu_{p})\,|\mu_3|^2\,C(1-j_p)\,W(j_p|y)
\ee
where we have defined \(y=\mu_1/(\mu_1+\mu_2)\,\) and 
\be
W(j_p|y)\,=\,-\gamma(\hat{\jmath}-1)\,
|y|^{2j_{1}}\,
|1-y|^{2j_{2}}\int d^2x\,
|x|^{2(\hat{\jmath}-2j_{1}-1)}\,|1-x|^{2(\hat{\jmath}-2j_{2}-1)}\,
|x-y|^{2(1-\hat{\jmath})}
\ee
which may be computed using (\ref{louvint}).
The corresponding two-point function is given by
\be\label{mu2pf}
\amp{3.5mm}{\hat{\varphi}_{j_2}(\mu_2)\,\varphi_{j_1}(\mu_1)} \,=\,
|\mu_1|^2\,\delta^2(\mu_2+\mu_1)\left(\delta(s_2+s_1)+R_{j_1}\delta(s_2-s_1)\right)
\ee
The OPE for the \(H_{3}\) model in the \(\varphi_{j}(\mu)\) basis takes the form
\be\label{H3OPEvarphi}
\varphi_{j_2}(\mu_2|1)\varphi_{j_1}(\mu_1)\,=\,\int_{\mathbb{R}_{+}}\!\!ds\,
C(1-j_1,1-j_2,j)\,W(j_1,j_2,1-j|y)\,\left[\,\varphi_{j}(\mu)+\ldots\right]
\ee
where \(\mu=\mu_1+\mu_2\,\). The OPE (\ref{H3OPEvarphi}) shares the domain of validity (\ref{h3opedom}) with (\ref{H3OPEPhi}).

\subsection{The \texorpdfstring{$\mc{V}_{j}(m)$}{} primaries}

We would also like to examine modes of definite affine weight
\be\label{phitoglb}
\mc{V}_{j}(m)\,=\,\frac{B_{j}}{(1-2j)}\int d^2x\,
x^{-j-m}\bar{x}^{-j-\bar{m}}\;\Phi_{1-j}(x)
\ee
where the normalization \(\mc{V}_{j}(j)=\Phi_{j}(0)/(1-2j)\,\) has been chosen given the analytic continuation of (\ref{phiref}). The currents have the following OPE's with the modes
\bbb
J^{3}(z)\,\mc{V}_{j}(m|w) & \sim & 
\frac{m\,\mc{V}_{j}(m|w)}{(z-w)}\\
J^{\pm}(z)\,\mc{V}_{j}(m|w) & \sim & 
\frac{(m\pm j)\,\mc{V}_{j}(m\pm 1|w)}{(z-w)}
\eee
The spacetime energy and angular momentum are given by \(E=m+\bar{m}\,\) and \(\ell=m-\bar{m}\in\mathbb{Z}\,\), respectively. For the \(H_{3}\) coset model \(E=-i\omega\in i\mathbb{R}\,\), while for the \(SL(2,R)\) model \(E\in\mathbb{R}\,\). From (\ref{phiref}) the modes satisfy the reflection relation
\be\label{glbref}
\mc{V}_{j}(m)\,=\,R_{j}(m)\,\mc{V}_{1-j}(m)
\ee
where, given (\ref{int2}),
\be\label{glbrefcof}
R_{j}(m)\,=\,R^{-1}_{1-j}(m)\,=\,
\mc{R}_{j}\,\frac{\Gamma(2j-1)}{\Gamma(1-2j)}\,\frac{\Gamma(1-j-m)}{\Gamma(j-m)}\,
\frac{\Gamma(1-j+\bar{m})}{\Gamma(j+\bar{m})}
\ee
These relations lead to the transforms
\be\label{phitoglb2}
\mc{V}_{j}(m)\,=\,-\mc{R}^{-1}_{j} R_{j}(m)\,\pi^{-1}\!\!\int d^2x\,
x^{j-1-m}\bar{x}^{j-1-\bar{m}}\;\Phi_{j}(x)
\ee
and, given (\ref{int3}),
\be\label{h3lvtoglb}
\mc{V}_{j}(m)\,=\,\mc{R}_{j}\,\frac{\Gamma(1-j-m)}{\Gamma(j+\bar{m})}\,\pi^{-1}
\!\!\int d^2\mu\,\mu^{m-1}\bar{\mu}^{\bar{m}-1}\,\varphi_{j}(\mu)
\ee
The three-point function in this basis takes the form
\be\label{glb3pt}
\amp{3mm}{\hat{\mc{V}}_{j_3}(m_3)\mc{V}_{j_2}(m_2|1)\mc{V}_{j_1}(m_1)}\,=\,
\pi^{-2}\,\delta^2({\textstyle \sum_{p=1}^{3} m_{p}})\,A(j_p|m_1,m_2)
\ee
where
\be\label{glb3ptcof}
A(j_p|m_1,m_2)\,=\,\pi^{2}\,C(1-j_p)\,W(j_p|m_1,m_2)\,
{\textstyle \prod_{p=1}^{3}} \frac{B_{j_p}}{(1-2j_p)}
\ee
Here \(\delta^2(m)\) is defined by (\ref{mdelta}) and,
\bbb\label{wfunc}
W(j_p|m_1,m_2) &=& \int d^2x\,d^2y\;x^{-j_{1}-m_{1}}\bar{x}^{-j_{1}-\bar{m}_{1}}\,
y^{-j_{2}-m_{2}}\bar{y}^{-j_{2}-\bar{m}_{2}} \nonumber \\
& & \hspace*{1cm}\times\;|1-x|^{2(\hat{\jmath}-2j_{2}-1)}\, 
|1-y|^{2(\hat{\jmath}-2j_{1}-1)}\,|x-y|^{2(\hat{\jmath}-2j_{3}-1)}
\eee
This may be computed~\cite{FukudaHosomichi_0105} using (\ref{glbint}). The corresponding two-point function is given by
\be\label{glb2pt}
\amp{3.5mm}{\hat{\mc{V}}_{j_2}(m_2)\,\mc{V}_{j_1}(m_1)} \,=\,\pi^{-2}\,
\delta^2(m_2+m_1)\left(\delta(s_2+s_1)+R_{j_1}(m_1)\,\delta(s_2-s_1)\right)
\ee

For the discrete primary operators \(\mc{V}_{j}(m)\) considered below with \(m=j+n\) and \(\bar{m}=j+\bar{n}\,\), where \(n,\bar{n}\in\mathbb{Z}_{\geq 0}\) and \(j\in\mathbb{R}_{+}\,\), we have
\be\label{phitohighglb}
\mc{V}_{j}(m)\,=\,\frac{\Gamma(2j)}{\Gamma(2j+n)}\,
\frac{\Gamma(2j)}{\Gamma(2j+\bar{n})}\,
\frac{\partial^{n+\bar{n}\,}}{\partial x^n\partial\bar{x}^{\bar{n}}}\,\frac{\Phi_{j}(0)}{(1-2j)}
\ee 
This operator expression, which follows from the relations (\ref{phiref}, \ref{phitoglb}), should be understood in the sense of analytic continuation of correlation functions from a convergent domain in the absence of another operator \(\Phi_{j'}(x')\) at \(x'=0\,\), and is exhibited in the form of the wavefunctions (\ref{pardef}, \ref{phifuncs}). As discussed below, these vertex operators are normalizable in the \(SL(2,R)\) model if \(1/2<j<(k-1)/2\,\). To similarly express the corresponding conjugate operator \(\mc{V}_{j}(-m)\,\), denote the \(x\) frame operator as \(\Phi^{x}_{j}(x)\) and define \(\Phi^{r}_{j}(r)\) to be the \(r=1/x\) frame operator given by
\be\label{rdef}
\Phi^{r}_{j}(r)\,=\,\left|\frac{\partial r}{\partial x}\right|^{-2j}\Phi^{x}_{j}(x)
\,=\,x^{2j}\,\bar{x}^{2j}\,\Phi^{x}_{j}(x)
\ee
Here we have taken \(\Phi_{j}\) to have a boundary conformal dimension given by \(j\). For example, suppressing the superscript on \(\Phi^{x}_{j}(x)\,\), we have from (\ref{phi3pt})
\be
\amp{3mm}{\hat{\Phi}^{r}_{j_3}(0)\Phi_{j_2}(1|1)\Phi_{j_1}(0)}\,=\,
C(j_1,j_2,j_3)
\ee
and,
\be\label{phitwo}
\amp{3mm}{\hat{\Phi}^{r}_{j_2}(0)\Phi_{j_1}(0)}\,=\,B_{j_1}\delta(s_2-s_1)
\ee
Note that the reflection relation (\ref{phiref}) is preserved by this change of frames, and that the transform (\ref{phitoglb}) takes the form
\be\label{pointoglbcnj}
\mc{V}_{j}(-m)\,=\,\frac{B_{j}}{(1-2j)}\int d^2r\,
r^{-j-m}\,\bar{r}^{-j-\bar{m}}\;\Phi^{r}_{1-j}(r)
\ee
Thus the conjugate relation to (\ref{phitohighglb}) is expressed essentially identically as
\be\label{phitohighglb2}
\mc{V}_{j}(-m)\,=\,\frac{\Gamma(2j)}{\Gamma(2j+n)}\,
\frac{\Gamma(2j)}{\Gamma(2j+\bar{n})}\,
\frac{\partial^{n+\bar{n}\,}}{\partial r^n\partial\bar{r}^{\bar{n}}}\,\frac{\Phi^{r}_{j}(0)}{(1-2j)}
\ee 

The OPE for the \(H_{3}\) model in the \(\mc{V}_{j}(m)\) basis is given by
\be\label{H3OPEglb}
\mc{V}_{j_2}(m_2|1)\mc{V}_{j_1}(m_1)\,=\,\int_{\mathbb{R}_{+}}\!\!ds
\,A(j_1,j_2,1-j|m_1,m_2)\left[\,\mc{V}_{j}(m)+\ldots\right]
\ee
where \(m=m_1+m_2\,\). In terms of the \(j_p\) quantum numbers the OPE (\ref{H3OPEglb}) shares the domain of validity (\ref{h3opedom}). In terms of \(E=m+\bar{m}\) and \(L=m-\bar{m}\,\), the domain of validity is described by
\be\label{h3glbdom}
|{\rm Re}(E)|\,<\,1+|L|
\ee
This is equivalent to
\be
{\rm Max}({\rm Re}(m),{\rm Re}(\bar{m}))>-1/2\hspace{0.8cm}
{\rm and}\hspace{0.8cm}{\rm Min}({\rm Re}(m),{\rm Re}(\bar{m}))<1/2
\ee
Clearly this includes the spectrum of the \(H_3\) model in the \(\mc{V}_{j}(m)\) basis since in this case \({\rm Re}(E)=0\) and \(j\in\mc{S}_{\sst\mc{C}}\,\). As for the \(\Phi_j(x)\) and \(\varphi_j(\mu)\) bases considered above, outside of this domain the OPE (\ref{H3OPEglb}) possesses a well-defined analytic continuation under which poles in the integrand cross the contour of integration and produce discrete contributions.

\subsection{The \texorpdfstring{$H_3$}{} model from Liouville theory\label{sec_lvtoh3}}

In order to more clearly describe the relation between Liouville theory and the \(H_3\) model, we introduce a basis of primaries\footnote{Here \(V_{a}\) is equated with the conventional Liouville operator \(V_{\alpha}\) with \(\alpha=ba\,\), and the conventional parameter \(Q\) is given by \(Q=b+b^{-1}=bq\,\) with \(q=k-1\,\).} \(V_{a}\) for Liouville theory. In the relation (\ref{lvtoh3}) below we will identify the Liouville parameter \(a\) with the \(SL(2,R)\) parameter \(j\,\) as follows
\be\label{tratoj}
a(j)\,=\,\frac{k}{2}-1+j
\ee
The central charge of the Liouville CFT is given by
\be
c_{\sst L}\,=\,1+6\,(b+b^{-1})^2\,=\,25+6\,\frac{(k-3)^2}{k-2}
\ee
The conformal dimension of the operator \(V_{a}\) is given by\footnote{For notational simplicity, in this subsection the subscript on \(h_{\sst L}(a)=h(j)+k/4\) will henceforth be dropped and \(h(a)\) will denote the conformal dimension of Liouville operators rather than those of the \(H_3\) model.}
\be
h_{\sst L}(a)\,=\,h_{\sst L}(q-a)\,=\,\frac{a(q-a)}{k-2}
\,=\,\frac{j(1-j)}{k-2}+\frac{k}{4}
\ee
where \(q=k-1\,\). The Liouville three-point function is given by
\be\label{lvcfunc}
\amp{3mm}{\hat{V}_{a_3}V_{a_2}(1)V_{a_1}\!}\,=\,C_{a_3 a_2 a_1}\,=\,
\frac{b}{2\pi}\,\left(\nu_{{\sst L}}^{b}\,b^{4}\right)^{q-\hat{a}}\,
\frac{G(\hat{a}-q)}{G(1)}
\prod_{p=1}^{3}\frac{G(\hat{a}-2a_p)}{G(2a_p)}
\ee
with \(\hat{a}=\sum_{p=1}^{3} a_p\,\). This expression is invariant under the duality transformation \(b\goto b^{-1}\,\) and \(a_{p}\goto b^2\,a_p\,\). The constant\footnote{The correspondence with convention is \(\nu_{\sst L}=(\pi\mu_{\sst L}\gamma(b^2))^{1/b}\,\) with the choices \(\nu_{\sst L}=\nu^{-1/b}\,\) and (\ref{Anrm}) leading to \(\pi\mu_{\sst L}=b^2\,\).} \(\nu_{\sst L}\) is undetermined by the Liouville bootstrap and is self-dual in Liouville theory.  However, the correspondence between Liouville theory and the \(H_3\) model fixes \(\nu_{\sst L}=\nu^{-1/b}\,\). With this identification, the Liouville primaries satisfy the reflection relation \(V_{a}=R^{\sst L}_{a}\,V_{q-a}\,\), where
\be\label{lvref}
R^{\sst L}_{a}\,=\,(R^{\sst L}_{q-a})^{-1}\,=\,
\mc{R}_{1-j}\,\frac{\Gamma(2j-1)}{\Gamma(1-2j)}
\ee
This has been expressed in terms of the \(H_3\) parameter \(j=a-\frac{1}{2}b^{-2}\,\) to demonstrate that 
\(R^{\sst L}_{a}=R_{j}\) (\ref{h3ref}). Taking \(V_{0}=\mathbbm{1}\,\), the two-point function is given by
\be\label{lv2pf}
\amp{3mm}{\hat{V}_{a_2}V_{a_1}\!}\,=\,G_{a_2 a_1}\,=\,b^{-1}\delta(s_2+s_1)+R^{\sst L}_{a_1}b^{-1}\delta(s_2-s_1)
\ee
Here we have defined \(a=q/2+is\,\), which is equivalent to \(j=1/2+is\,\),
and introduced the notation \(G_{a_2 a_1}\) to emphasize the role of the two-point function as a metric on fields. The reflection relation (\ref{lvref}) and the form of (\ref{lv2pf}) implies that the spectrum of normalizable states in Liouville may be taken to correspond to \(s\in\mathbb{R}_{+}\,\). The corresponding inverse metric then leads to the identity operator on primary fields
\be
{\delta^{a_1}}_{a_2}\,=\,\int_{\mathbb{R}_{+}}\!ds\,G^{a_1 a}\,G_{a a_2}\,=\,
\delta(s_2-s_1)+R^{\sst L}_{a_1}\delta(s_2+s_1)
\ee
Here again \(a=q/2+is\,\), and the form of the identity in the basis \(V_{a}\) is consistent with the reflection relation (\ref{lvref}). The operator product expansion in Liouville theory takes the form
\be
V_{a_2}(z)V_{a_1}\,=\,\int_{\mathbb{R}_{+}}\!ds_3\,{C^{a_3}}_{a_2 a_1}\,|z|^{-2(\hat{h}-2h_3)}\left[V_{a_3}+\ldots\right]
\ee
where \(h_p=h(a_{p})\,\), and \(\hat{h}=\sum_{p=1}^{3} h_p\,\). The sum includes Virasoro descendant states which are determined by the conformal symmetry. It may be seen that
\be
{C^{a_3}}_{a_2 a_1}\,=\,
\int_{\mathbb{R}_{+}}\!ds\,G^{a_3 a}\,C_{a a_2 a_1}\,=\,
R^{\sst L}_{q-a_3}C_{a_3 a_2 a_1}
\ee

Building on the work of Stoyanovsky, which examined the relationship between the Knizhnik-Zamolodchikov equation of the \(H_{3}\) CFT and the Belavin-Polyakov-Zamolodchikov differential equation for degenerate representations in Liouville theory, Teschner and Ribault~\cite{RibaultTeschner_0502} constructed the following map between \(H_{3}\) and Liouville correlators
\bbb\label{lvtoh3}
\lefteqn{\amp{3mm}{\varphi_{j_n}(\mu_n|z_n)\ldots\varphi_{j_1}(\mu_1|z_1)}
}\hspace{1cm}\nonumber\\[0.1cm]
 & = & b\,\delta^2({\textstyle \sum_{p=1}^{n}}\mu_{p})\,\left|\Theta_{n}(z_r,y_s)\right|^2
\,\amp{3mm}{V_{a_n}(z_n)\ldots V_{a_1}(z_1)\,
V_{a_{-}}(y_{n-2})\ldots V_{a_{-}}(y_{1})}
\eee
where \(a_{-}=-\frac{1}{2}b^{-2}\,\), and \(a_{p}=a(j_p)=j_p-a_{-}\,\) as defined by (\ref{tratoj}). The respective variables are related through
\be\label{stoyrel}
\sum_{p=1}^{n}\frac{\mu_{p}}{t-z_{p}}\,=\,
u\,\frac{\prod_{p=1}^{n-2}(t-y_p)}{\prod_{p=1}^{n}(t-z_p)}
\ee
which implies \(u=\sum_{p=1}^{n}\mu_{p}z_{p}\,\). We also have
\be
\Theta_{n}(z_r,y_s)\,=\,u\,\prod_{r<s\leq n} (z_{r}-z_{s})^{-a_{-}}\ \prod_{r<s\leq n-2} (y_{r}-y_{s})^{-a_{-}}\ 
\prod_{r\leq n}\,\prod_{s\leq n-2} (z_{r}-y_{s})^{a_{-}}
\ee
In particular, the relation between the two-point functions (\ref{mu2pf},\,\ref{lv2pf}) follows from \(R^{\sst L}_a=R_j\,\). Furthermore, the Mobius-fixed three point function (\ref{mu3pf}) may be computed as
\be
\amp{3mm}{\hat{\varphi}_{j_3}(\mu_3)
\varphi_{j_2}(\mu_2|1)\varphi_{j_1}(\mu_1)}\,=\,
b\,|\mu_3|^2\delta^2({\textstyle \sum_{p=1}^{3}}\mu_{p})\,|y(1-y)|^{2a_{-}}
\amp{3mm}{\hat{V}_{a_3}V_{a_2}(1)V_{a_{-}}(y)V_{a_1}}
\ee
where (\ref{stoyrel}) implies \(y=\mu_1/(\mu_1+\mu_2)\,\). This requires
\be\label{lv4ptdeg}
\amp{3mm}{\hat{V}_{a_3}V_{a_2}(1)V_{a_{-}}(y)V_{a_1}}\,=\,b^{-1}C(1-j_p)\,|y(1-y)|^{-2a_{-}}\,W(j_p|y)
\ee
In general the Liouville four-point function may be factorized as 
\be\label{lv4ptfac}
\amp{3mm}{\hat{V}_{a_4}V_{a_3}(1)V_{a_{2}}(y)V_{a_1}}\,=\,\int_{\mathbb{R}_{+}} ds\,{C_{a_4a_3}}^{a}\,C_{a a_2 a_1}
\abs{3mm}{\mc{F}^{a}_{a_4a _3 a_2 a_1}(y)}^2
\ee
where the conformal block \(\mc{F}^{a}_{a_4a _3 a_2 a_1}(y)\) is normalized as
\be
\mc{F}^{a}_{a_4a _3 a_2 a_1}(y)\,=\,y^{h(a)-h_2-h_1}\,(1+\mc{O}(y))
\ee
Degenerate Virasoro representations in Liouville theory appear for
\be
a^{\pm}_{n,m}\,=\,q/2\pm (n+b^{-2}m)/2\,=\,q-a^{\mp}_{n,m}
\ee
where \(n,m\in\mathbb{Z}_{>0}\,\). These representations correspond to the degenerate affine representations of \(SL(2,\mathbb{R})\) given by (\ref{refdeg}). The null descendant of \(V_{a_{-}}\,\), where \(a_{-}=a^{-}_{1,2}=-\frac{1}{2}b^{-2}\,\), satisfies
\be
\left(L_{-2}+b^2L^{2}_{-1}\right)\cdot V_{a_{-}}\,=\,0
\ee
Making use of the conformal Ward identities we find the hypergeometric equation
\be\label{hypeq1}
\left(b^2\frac{\partial^2\,}{\partial y^2}+\left(\frac{1}{1-y}-\frac{1}{y}\right)\frac{\partial\,}{\partial y}+\frac{h_1}{y^2}+\frac{h_2}{(1-y)^2}+\frac{h_1+h_2-h_3+h_{-}}{y(1-y)}\right)\mc{F}^{a}_{a_3 a_2 a_{-} a_1}(y)\,=\,0
\ee
where \(h_{-}=h(a_{-})\,\). The corresponding two solutions imply that for the amplitude (\ref{lv4ptdeg}) the integral (\ref{lv4ptfac}) may be written as the sum of two terms as follows
\be
\amp{3mm}{\hat{V}_{a_3}V_{a_2}(1)V_{a_{-}}(y)V_{a_1}}\,=\,
A_{+}\abs{3mm}{\mc{F}^{a_1+a_{-}}_{a_3 a _2 a_{-} a_1}(y)}^2\,+\,
A_{-}\abs{3mm}{\mc{F}^{a_1-a_{-}}_{a_3 a _2 a_{-} a_1}(y)}^2
\ee
It may be seen that (\ref{hypeq1}) has the solutions
\bbb
\mc{F}^{a_1\pm a_{-}}_{a_3 a _2 a_{-} a_1}(y) & = & y^{h(a_{1}\pm a_{-})-h_{-}-h_{1}}(1-y)^{h(a_{2}\pm a_{-})-h_{-}-h_{2}}\nonumber \\ & & \times\ {_{2}{\rm F}_{1}}(\pm\hat{\jmath}+\txfc{1}{2}\mp\txfc{3}{2},\pm(\hat{\jmath}-2j_{3})+\txfc{1}{2}\mp\txfc{1}{2};1\pm (2j_{1}-1);y)
\eee
where \(\hat{\jmath}=\sum_{p=1}^{3} j_{p}=\hat{a}+3a_{-}\,\). Using the hypergeometric identity (\ref{hypid1}), we find that (\ref{lv4ptdeg}) requires
\be
A_{+}\,=\,-\pi\,b^{-1}\,C(1-j_p)\,\gamma(1-2j_{1})\gamma(\hat{\jmath}-1)\gamma(\hat{\jmath}-2j_2)\gamma(\hat{\jmath}-2j_3)
\ee
and 
\be
A_{-}\,=\,-\pi\,b^{-1}\,C(1-j_p)\,\gamma(2j_{1}-1)\gamma(\hat{\jmath}-2j_1)
\ee
Making use of the normalization (\ref{Anrm}), a careful treatment of the countour integral in (\ref{lv4ptfac}) demonstrates that this is the correct result. In particular, by recognizing that the integrand is invariant under \(a\goto q-a\,\), the contour may be extended to the entire real \(s\) axis. Under the analytic continuation of \(a_{-}\) from \(q/2+i\mathbb{R}_{+}\) to \(a_{-}=-\txfc{1}{2}b^{-2}\,\), it may be shown that the only non-vanishing contributions to (\ref{lv4ptfac}) lead to
\be
A_{\pm}\,=\,\mp\oint_{a_{1}\pm a_{-}}\! ds\,{C_{a_3a_2}}^{a}\,C_{a a_{-} a_1}
\ee
Here the respective (counter-clockwise) contours encircle poles\footnote{In addition there are identical contributions from reflected poles at \(q-a=q-a_1\mp a_{-}\) which cross the \(s\) axis and cancel the factor of \(1/2\) from the extension of the contour of integration.} at \(a=q/2+is=a_{1}\pm a_{-}\) which cross the real \(s\) axis. This is consistent with the operator product expansion
\be
V_{a_{-}}(z)V_{a_1}\,=\,\sum_{\sigma=\pm 1} B_{\sigma}\,
|z|^{2(h(a_{1}+\sigma a_{-})-h_{-}-h_{1})}\,
\left[V_{a_{1}+\sigma a_{-}}+\ldots\right]
\ee
where \(B_{+}=1\) and
\be
B_{-}\,=\,b^{-4}\,\nu^{-b^{-2}}\,\gamma(2a_{1}-q)/\gamma(2a_{1})
\ee
As has been shown in~\cite{RibaultTeschner_0502}, having treated the \(H_3\) two-point and three-point functions, the formula (\ref{lvtoh3}) may then be proven through induction.

\section{The spectrum of the \texorpdfstring{$SL(2,R)$}{} CFT \label{sec_3}}

The holomorphic worldsheet\footnote{In what follows we will primarily work with holomorphic quantities on a Euclidean worldsheet which is defined through the mapping from the Lorenztian cylinder (\(w^{0},w^{1}\)) to the Riemann sphere (\(z,\bar{z}\)) given by \(z=e^{w^{2}-iw^{1}}\) with the continuation \(w^{0}\goto -iw^{2}\,\). In the notation for the related Killing vector fields in appendix \ref{sec_A}, we will also define \(J^{a}_{+}\equiv J^{a}\) and \(J^{a}_{-}\equiv \bar{J}^{a}\,\). However, this notation invites confusion since, unlike the case of the \(H_3\) model where \((J^{\pm}_{+})^{*}=J^{\pm}_{-}\) and \((J^{3}_{+})^{*}=J^{3}_{-}\,\), for the \(SL(2,R)\) model \(J^{a}_{+}\) and \(J^{a}_{-}\) are not related through complex conjugation.} currents of the \(SL(2,R)\) model have the mode expansions
\be
J^{a}(z)\,=\,\sum_{n\in\mathbb{Z}}\,J^{a}_{n}\,z^{-n-1}
\ee
The OPEs (\ref{sl2curope}) of the currents lead to the associated mode algebra
\bbb\label{modealg}
&\com{J^{3}_{n}}{J^{3}_{m}}\!\!&=\,-\txfc{1}{2}\,k n\,\delta_{n+m} \nonumber\\
& \com{J^{3}_{n}}{J^{\pm}_{m}}\!\!&=\,\pm J^{\pm}_{n+m} \nonumber\\
& \com{J^{+}_{n}}{J^{-}_{m}}\!\!&=\,-2J^{3}_{n+m}+k n\,\delta_{n+m}
\eee
The stress tensor (\ref{sugawara}) has the following expansion in Virasoro generators
\be
T(z)\,=\,\sum_{n\in\mathbb{Z}}\,L_{n}\,z^{-n-2}
\ee
In terms of the modes, for \(p=n/2\in\mathbb{Z}\,\) the Virasoro generators have the form 
\bbb\label{viramodes1}
(k-2)\,L_{n} & = & 
\txfc{1}{2}\left(-2J^{3}_{p}J^{3}_{p}+J^{+}_{p}J^{-}_{p}+J^{-}_{p}J^{+}_{p}\right)\nonumber\\[2mm] 
& & + \ {\textstyle\sum_{m=1}^{\infty}}
\left(-2J^{3}_{p-m}J^{3}_{p+m}+J^{+}_{p-m}J^{-}_{p+m}+J^{-}_{p-m}J^{+}_{p+m}\right)
\eee
And for \(p=(n-1)/2\in\mathbb{Z}\,\) they have the form
\be\label{viramodes2}
(k-2)\,L_{n}\,=\,{\textstyle\sum_{m=0}^{\infty}}
\left(-2J^{3}_{p-m}J^{3}_{p+1+m}+J^{+}_{p-m}J^{-}_{p+1+m}+J^{-}_{p-m}J^{+}_{p+1+m}\right)
\ee
Given the mode algebra (\ref{modealg}), the Virasoro generators satisfy
\be\label{vircaalg}
\com{L_{n}}{J^{a}_{m}}\,=\,-mJ^{a}_{m+n}
\ee
and the Virasoro algebra
\be
\com{L_{n}}{L_{m}}\,=\,(n-m)L_{n+m}+\frac{c}{12}\,n(n^2-1)\,\delta_{n+m}
\ee
with central charge \(c=3k/(k-2)\,\). 

\subsection{The unflowed spectrum}

To determine the spectrum of the \(SL(2,R)\) CFT we first examine primary vertex operators \(\mc{V}_{j}(m)\) of definite weight under the global \(SL(2,R)\) algebra generated by \(J^{a}_{0}\,\), out of which affine modules are built by application of the modes of positive conformal weight \(J^{a}_{-n}\,\) for \(n>0\,\). To contribute to a basis of physical string states, these primary operators, unlike their affine descendants, must lie in unitary representations. The unitary representations of the \(SL(2,R)\) algebra are described in detail in appendix \ref{sec_B}. Elements of the spectrum must also be normalizable under the norm derived from the associated two-point function, which, like all correlation functions of the \(SL(2,R)\) CFT, we take to be inherited from the \(H_3\) model under analytic continuation. This norm, which is the CFT analog of the quantum mechanical \(\mc{L}^2\) norm given by (\ref{l2nrm}), is described in subsection \ref{subsec_ufcf}\,. As in the zero-mode quantum mechanics, normalizability places restrictions on the range of unitary representations which appear in the \(SL(2,R)\) spectrum. We will find below that the associated affine representations must be extended to include representations which arise due to the spectral flow automorphism of the \(SL(2,R)\) current algebra. The unflowed spectrum of primary states consists\footnote{In general all primary and affine representations of the \(SL(2,R)\) CFT appearing in this paper are diagonal products of holomorphic and anti-holomorphic copies. This is suppressed for notational simplicity with , for example, \(\mc{C}_{0}\otimes\mc{C}_{0}\,\) simply denoted as \(\mc{C}_{0}\,\). Also, in anticipation of the introduction of spectral flow, the unflowed representations are denoted by the quantum number \(w=0\,\).} of the direct sum \(\mc{H}_{0}=\mc{D}^{+}_{0}\!\oplus\mc{D}^{-}_{0}\!\oplus\mc{C}_{0}\,\). The continuous representations \(\mc{C}_{0}\) comprise the irreducible representations \(\mc{C}_{j}^{\alpha}(0)\,\) as 
\be\label{contintspec}
\mc{C}_{0}\,=\,\int_{\sst -\frac{1}{2}}^{\sst \frac{1}{2}}\!d\alpha
\int_{\mc{S}^{+}_{\sst \mc{C}}}\!dj\;\mc{C}_{j}^{\alpha}(0)
\ee
where
\be\label{contspec}
\mc{S}^{+}_{\sst \mc{C}}\,=\,1/2+i\,\mathbb{R}_{\geq 0}
\ee
Suppressing anti-holomorphic dependence, for \(\mc{V}_{j}(m)\in\mc{C}_{j}^{\alpha}(0)\) we have\footnote{The range \(-1/2\leq\alpha<1/2\) has been chosen here, rather than the conventional range \(0\leq\alpha<1\,\), to facilitate discussion of the spectral flowed current algebra representations treated below. Note that \(\mc{C}_{j}^{-\sst 1/2}=\mc{C}_{j}^{\sst 1/2}\).} \(m=\alpha+p\) with \(p\in\mathbb{Z}\,\). The corresponding affine modules are denoted by \(\hat{\mc{C}}_{0}\) and \(\hat{\mc{C}}_{j}^{\alpha}(0)\,\). The discrete representations \(\mc{D}^{\pm}_{0}\) comprise the irreducible representations \(\mc{D}_{j}^{\pm}(0)\,\) as 
\be\label{discintspec}
\mc{D}^{\pm}_{0}\,=\,\int_{\mc{S}_{\sst \mc{D}}}\!dj\;\mc{D}_{j}^{\pm}(0)
\ee
where
\be\label{discspec}
\mc{S}_{\sst \mc{D}}\,=\,(1/2\,,(k-1)/2)\subset\mathbb{R}
\ee
For \(\mc{V}_{j}(m)\in\mc{D}_{j}^{\pm}(0)\) we have \(m=\pm(j+q)\) with \(q\in\mathbb{Z}_{\geq 0}\,\). The range appearing in (\ref{discspec}), which will be seen below to be consistent with the complete \(SL(2,R)\) spectrum including flowed representations, is a restriction from the range \(0<j<k/2\) required by unitarity. As in the quantum mechanical picture described in appendix \ref{sec_B}, this is a consequence of the requirement of normalizability under the two-point function (\ref{sl2pt}). The corresponding affine modules are denoted by \(\hat{\mc{D}}^{\pm}_{0}\) and \(\hat{\mc{D}}_{j}^{\pm}(0)\,\).

\subsection{Spectral flow}

Absent an extension of the unflowed spectrum, the restriction (\ref{discspec}) associated with the discrete representations \(\mc{D}_{j}^{\pm}(0)\) places highly unphysical constraints on the spectrum of physical string vertex operators. In particular, the relevant string constraint (\ref{wzerophys}) on unflowed descendant states in \(\hat{\mc{D}}_{j}^{\pm}(0)\) places an upper bound on both the affine \(SL(2,R)\,\) level number and on the conformal dimensions of operators in a unitary CFT \(\mc{M}\) which appears as part of a string background \(AdS_{3}\times\mc{M}\). The solution to this problem was found in~\cite{MaldacenaOoguri_0111}, where use was made of the fact that the affine algebra (\ref{modealg}) is preserved by the spectral flow symmetry defined by
\be\label{flowmodes}
J^{3}_{n}(w)\,=\,J^{3}_{n}-\txfc{1}{2}kw\delta_{n}\spwd{1cm}{and}
J^{\pm}_{n}(w)\,=\,J^{\pm}_{n\pm w}
\ee
where \(w\in\mathbb{Z}\,\). From (\ref{viramodes1}, \ref{viramodes2}), the corresponding Virasoro generators are given by
\be\label{virflow}
L_{n}(w)\,=\,L_{n}+w J^{3}_{n}-\txfc{1}{4}kw^2\delta_{n}
\ee
The spectrum \(\,\hat{\mc{H}}_{\sst SL(2,R)}=\sum_{w\in\mathbb{Z}}\hat{\mc{H}}_{w}\) of the \(SL(2,R)\) CFT  is parameterized by the spectral flow integer \(w\in\mathbb{Z}\,\). The affine module \(\hat{\mc{H}}_{w}\) comprises the descendants\footnote{The corresponding affine representations are denoted by \(\hat{\mc{C}}_{j}^{\alpha}(w)\,\), \(\hat{\mc{C}}_{w}\,\), and similarly for the discrete representations. For reasons of notational clarity, the notation \(\hat{\mc{C}}(w)=\hat{\mc{C}}_{w}\,\), \(\mc{C}(w)=\mc{C}_{w}\,\), and etc. will sometimes be used.} of the spectrum of flowed primaries given by
\be\label{sl2space}
\mc{H}_{w}\,=\,\mc{D}^{+}_{w}\oplus\mc{D}^{-}_{w}\oplus\mc{C}_{w}
\ee
Here, 
\be \label{flwspec}
\mc{D}^{\pm}_{w}\,=\,\int_{\mc{S}_{\sst \mc{D}}}\!dj\;\mc{D}_{j}^{\pm}(w)\spwd{1.5cm}{and}
\mc{C}_{w}\,=\,\int_{\sst -\frac{1}{2}}^{\sst \frac{1}{2}}\!d\alpha
\int_{\mc{S}^{+}_{\sst \mc{C}}}\!dj\;\mc{C}_{j}^{\alpha}(w)
\ee
where \(\mc{S}^{+}_{\sst \mc{C}}\) and \(\mc{S}_{\sst \mc{D}}\) are given by (\ref{contspec}) and (\ref{discspec}), respectively. More precisely, \(\hat{\mc{H}}_{w}\) is given by the action of the flowed raising operators \(J^{a}_{-n}(w)\) with \(n>0\) on the vertex operators\footnote{As for the \(H_{3}\) CFT, the anti-holomorphic charge \(\bar{m}\) is suppressed here. The behavior of the anti-holomorphic sector follows straightforwardly since on \(AdS_{3}\), which is the universal cover  of the \(SL(2,R)\) group manifold, the spectral flow parameter satisfies \(w=\bar{w}\,\).} \(\mc{V}_{j}^{w}(m)\in\mc{H}_{w}\,\) which satisfy
\be\label{prim1}
J^{a}_{n}(w)\cdot\mc{V}_{j}^{w}(m)\,=\,0
\ee
for \(n>0\,\), and
\be\label{prim2}
J^{3}_{0}(w)\cdot\mc{V}_{j}^{w}(m)\,=\,m\,\mc{V}_{j}^{w}(m)
\ee
Taking \(C(w)\) to be the Casimir operator built from the flowed zero modes,
\be
C(w)\,=\,-J^{3}_{0}J^{3}_{0}+kwJ^{3}_{0}-\txfc{1}{4}k^2w^2+
\txfc{1}{2}\left(J^{+}_{+w}J^{-}_{-w}+J^{-}_{-w}J^{+}_{+w}\right)
\ee
for the primaries \(\mc{V}_{j}^{w}(m)\,\) we have
\be\label{flwcsm}
C(w)\cdot\mc{V}_{j}^{w}(m)\,=\,(k-2)\,L_{0}(w)\cdot\mc{V}_{j}^{w}(m)\,=\,j(1-j)\,\mc{V}_{j}^{w}(m)
\ee
The most general state in \(\hat{\mc{H}}_{w}\,\) is given by a superposition of eigenstates of \(J^{3}_{0}\) and \(L_{0}\) with eigenvalues \(M\) and \(H\,\), respectively. The corresponding eigenstates will be denoted in a condensed notation by \(\ket{j,M,N,w}\). Using (\ref{virflow}) the eigenvalues may be seen to be related by 
\be\label{Hgen}
H\,=\,\frac{j(1-j)}{k-2}+N-wM+\txfc{1}{4}kw^2
\ee
where \(N=L_{0}(w)-C(w)/(k-2)\) is the level of the state in \(\hat{\mc{H}}_{w}\,\). 

\begin{figure}[ht]
\begin{center}
\begin{picture}(200,160)
\put(96,156){\(L_{0}\)}
\put(205,119){\(J^{3}_{0}\)}
\includegraphics{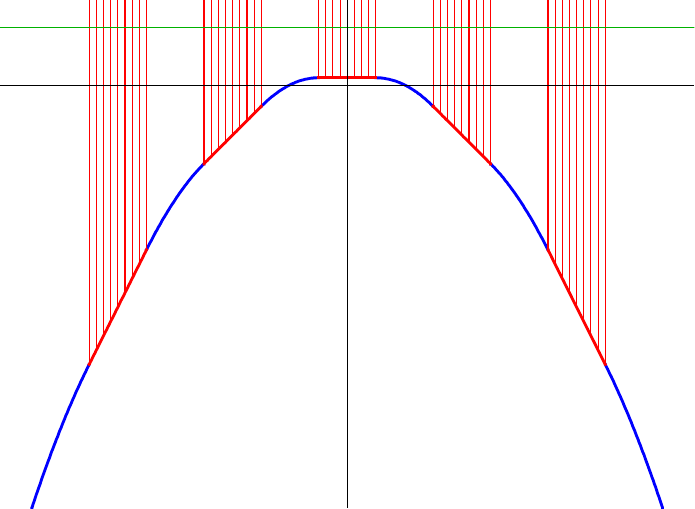}
\end{picture}
\caption{The figure above shows the \(J^{3}_{0}\) and \(L_{0}\) charges for representative primary states \(\mc{V}_{j}^{w}(\alpha)\in\mc{C}^{\alpha}_{j}(w)\,\) and \(\mc{V}^{w}_{j}(\pm j)\in\mc{D}^{\pm}_{j}(w)\,\), where the choice \(k=4\) has been made to emphasize features in the figure. The charges of the primary states \(\mc{V}_{j}^{w}(\alpha)\)  appear in \textcolor{wtmred}{red} for \(-2\leq w\leq 2\,\) and \(j\in\mc{S}^{+}_{\sst \mc{C}}\) with \(-1/2\leq\alpha<1/2\,\). Among these are primaries \(\mc{V}_{\sst 1/2}^{w}(\alpha)\) of lowest \(L_{0}\) for a given \(J^{3}_{0}\,\), with charges which appear above as (\textcolor{wtmred}{red}) line segments given by (\ref{CM}, \ref{CH}) with \(s=N=P=0\,\). The charges of the primaries \(\mc{V}_{j}^{w}(\pm j)\) appear in \textcolor{wtmblue}{blue} for \(0\leq \pm w\leq 2\,\) and \(j\in\mc{S}_{\sst \mc{D}}\) as segments of the parabolas given by (\ref{DM}, \ref{DH}) with \(N=Q=0\). Also shown in \textcolor{wtmgreen}{green} is the \(L_0=1\) line below which physical string states lie.}
\label{fig1}
\end{center}
\end{figure}

\begin{figure}[ht]
\begin{center}
\begin{picture}(150,120)
\includegraphics{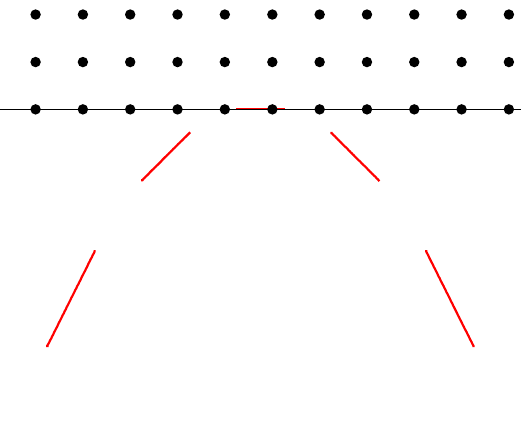}
\put(-87,78){\(\hat{\mc{C}}_{\sst 1/2}^{\alpha}(0)\)}
\end{picture}
\begin{picture}(150,110)
\includegraphics{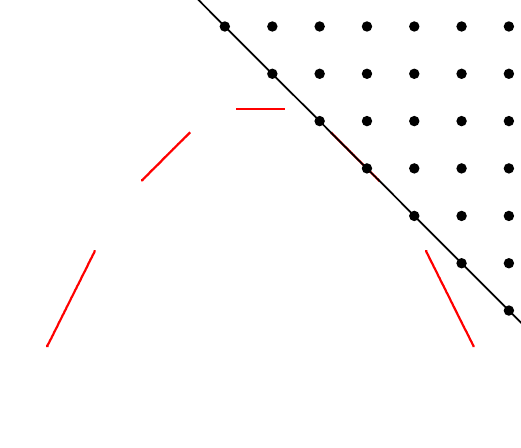}
\put(-80,70){\(\hat{\mc{C}}_{\sst 1/2}^{\alpha}(1)\)}
\end{picture}
\begin{picture}(150,110)
\includegraphics{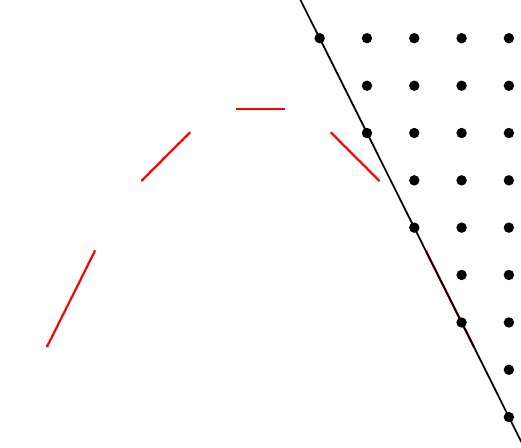}
\put(-55,30){\(\hat{\mc{C}}_{\sst 1/2}^{\alpha}(2)\)}
\end{picture}
\caption{The figure above shows the \(J^{3}_{0}\) and  \(L_{0}\) charges for the continuous representations \(\hat{\mc{C}}_{j}^{\alpha}(w)\) for \(j=1/2\,\), \(\alpha=1/4\,\), and \(k=4\) for \(w=0\) through \(w=2\,\). These are examples of representations built on the primaries \(\mc{V}_{\sst 1/2}^{w}(\alpha)\in\mc{C}^{\alpha}_{\sst 1/2}(w)\) of lowest \(L_{0}\) for a given \(J_{0}^{3}\) charge, with charges which are given by the \textcolor{wtmred}{red} line segments in the figure. Note that the lines of primary states \(\mc{C}_{j}^{\alpha}(w)\) have slope \(-w\,\), and that \(\hat{\mc{C}}_{j}^{\alpha}(w)\) is \(J_{0}^{3}\) charge conjugate to \(\hat{\mc{C}}_{j}^{-\alpha}(-w)\,\).}
\label{fig2}
\end{center}
\end{figure}

Through application of \(J^{a}_{-n}(w)\) with \(n\geq 0\) all states in \(\hat{\mc{C}}^{\alpha}_{j}(w)\) can be generated from the state \(\mc{V}_{j}^{w}(\alpha)\,\). For states in \(\hat{\mc{C}}^{\alpha}_{j}(w)\)
\be\label{CM}
M\,=\,\alpha+\txfc{1}{2}kw+P
\ee
where \(P\in\mathbb{Z}\) and, taking \(j=1/2+is\in\mc{S}^{+}_{\sst\mc{C}}\,\),
\be\label{CH}
H\,=\,\frac{(1/4+s^2)}{(k-2)}-w\alpha-\txfc{1}{4}kw^2+N-wP
\ee
The corresponding primary (\(N=0\)) states \(\mc{V}^{w}_{j}(m)\in\mc{C}^{\alpha}_{j}(w)\,\) have \(m=\alpha+P\,\).  Of particular interest are primary states \(\mc{V}_{\sst 1/2}^{w}(\alpha)\in\mc{C}^{\alpha}_{\sst 1/2}(w)\) with \(s=0\) and \(P=0\,\). As shown in Figure~\ref{fig1}, these are states with eigenvalues of lowest \(H\) for a given \(M\) among all states in the \(SL(2,R)\) CFT. For a given \(w\,\), for \(-1/2\leq \alpha<1/2\) the eigenvalues of these states have the ranges \(\Delta M=1\) and \(\Delta H=-w\,\). Figure~\ref{fig2} shows the \((M,H)\) eigenvalues for a number of spectral flow sectors of \(\hat{\mc{C}}^{\alpha}_{j}(w)\) for fixed \(j\) and \(\alpha\,\). It may be seen from (\ref{CM}, \ref{CH}) that \(J_{0}^{3}\) charge conjugation, \((M,H)\goto (-M,H)\,\), acts as \(j\goto j\,\), \(w\goto -w\,\), \(\alpha\goto -\alpha\,\), \(N\goto N\,\), and  \(P\goto -P\,\). Thus \(\hat{\mc{C}}_{j}^{\alpha}(w)\) and \(\hat{\mc{C}}_{j}^{-\alpha}(-w)\,\) are \(J_{0}^{3}\) charge conjugate representations.

\begin{figure}[ht]
\begin{center}
\begin{picture}(150,130)
\includegraphics{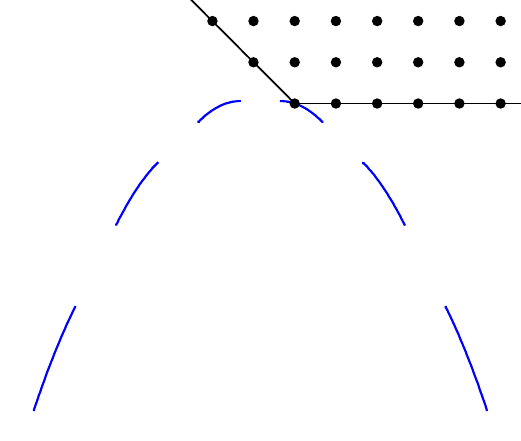}
\put(-84,78){\(\hat{\mc{D}}_{j}^{+}(0)\)}
\end{picture}
\begin{picture}(150,130)
\includegraphics{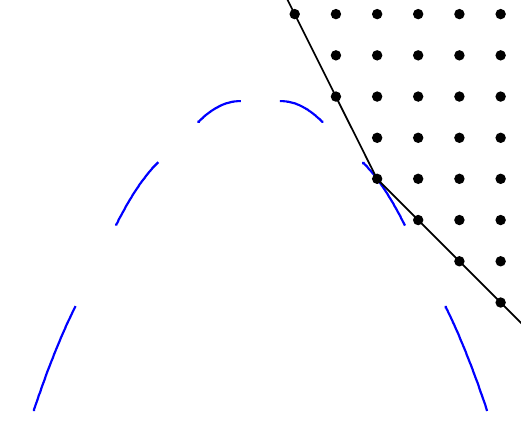}
\put(-76,66){\(\hat{\mc{D}}_{j}^{+}(1)\)}
\end{picture}
\begin{picture}(150,130)
\includegraphics{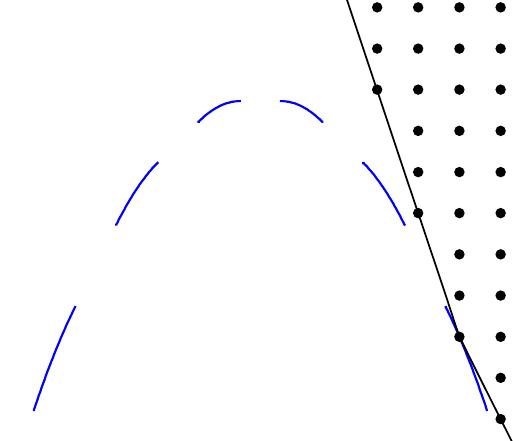}
\put(-54,25){\(\hat{\mc{D}}_{j}^{+}(2)\)}
\end{picture}
\caption{The figure above shows the \(J^{3}_{0}\) and  \(L_{0}\) charges for the discrete representations \(\hat{\mc{D}}^{+}_{j}(w)\) for \(2j-1=(k-2)/3\) and \(k=4\) for \(w=0\) through \(w=2\,\).  These are examples of representations built on the primaries \(\mc{V}_{j}^{w}(\pm j)\in\mc{D}^{\pm}_{j}(w)\) of lowest \(L_{0}\) for a given \(J_{0}^{3}\) charge, with charges which are shown in the figure as segments of \textcolor{wtmblue}{blue} parabolas. The lines of primary states \(\mc{D}^{\pm}_{j}(w)\) have slope \(-w\) and are tangent to the respective parabolic segments at \(j=1/2\,\), while the lines of descendant states (\ref{weylprim}), which are identified with the primaries \(\mc{D}^{\mp}_{\tilde{\jmath}}(w\pm 1)\,\) for \(\tilde{\jmath}=k/2-j\,\), have slope \(-w\mp 1\) and are tangent to the respective parabolic segment at \(j=(k-1)/2\,\). Note that the affine representation \(\hat{\mc{D}}^{\pm}_{j}(w)=\hat{\mc{D}}^{\mp}_{\tilde{\jmath}}(w\pm1)\) is the \(J_{0}^{3}\) charge conjugate of \(\hat{\mc{D}}^{\mp}_{j}(-w)\,\).}
\label{fig3}
\end{center}
\end{figure}

Through application of \(J^{a}_{-n}(w)\) with \(n\geq 0\) all states in \(\hat{\mc{D}}^{\pm}_{j}(w)\) can be generated from the primary state \(\mc{V}_{j}^{w}(\pm j)\,\) which satisfies \(J^{\mp}_{0}(w)\mc{V}_{j}^{w}(\pm j)=0\,\). For states in \(\hat{\mc{D}}^{\pm}_{j}(\pm w)\,\)
\be\label{DM}
\pm M\,=\,j+\txfc{1}{2}kw+Q-N
\ee
where \(Q\in\mathbb{Z}_{\geq 0}\) and, for \(j\in\mc{S}_{\sst\mc{D}}\,\),
\be\label{DH}
H\,=\,\frac{j(1-j)}{k-2}-wj-\txfc{1}{4}kw^2+(1+w)N-wQ
\ee
The corresponding primary states \(\mc{V}^{w}_{j}(m)\in\mc{D}^{\pm}_{j}(w)\,\) have \(m=\pm(j+Q)\,\). As shown in Figure~\ref{fig1}, together with the states \(\mc{V}_{\sst 1/2}^{w}(\alpha)\in\mc{C}^{\alpha}_{\sst 1/2}(w)\,\), the states \(\mc{V}_{j}^{w}(\pm j)\in\mc{D}^{\pm}_{j}(w)\,\) trace out a curve in the \((M,H)\) plane below which no states in the \(SL(2,R)\) CFT appear. For fixed \(w\) and \(j\in\mc{S}_{\sst \mc{D}}\,\) for \(\mc{D}^{\pm}_{j}(\pm w)\) the eigenvalues of these states have the ranges \(\Delta M=\pm(k-2)/2\,\) and \(\Delta H=-(w+\txfc{1}{2})(k-2)/2\,\). The fact that the curve in figure~\ref{fig1} can clearly be continuously mapped to the real line relies on the range of \(\mc{S}_{\sst \mc{D}}\,\)(\ref{discspec}). Figure~\ref{fig3} shows the \((M,H)\) eigenvalues for a number of spectral flow sectors of \(\hat{\mc{D}}^{\pm}_{j}(w)\,\) for fixed \(j\,\). It is clear from  from (\ref{DM}, \ref{DH}) that \(\hat{\mc{D}}^{+}_{j}(w)\) and \(\hat{\mc{D}}^{-}_{j}(-w)\,\) are \(J_{0}^{3}\) charge conjugate representations.

The states with \(Q=0\) in \(\hat{\mc{D}}^{\pm}_{j}(w)\,\), which are annihilated by \(J^{\mp}_{0}(w)\,\), are given by 
\be\label{weylprim}
\left(J^{\mp}_{-1}(w)\right)^{N}\mc{V}_{j}^{w}(\pm j)
\ee
It may be seen from (\ref{DM},\,\ref{DH}) that the states (\ref{weylprim}) are physically equivalent to 
\be\label{weylconj}
\left(J^{\mp}_{0}(w\pm 1)\right)^{N}\mc{V}_{\tilde{\jmath}}^{w\pm 1}(\mp \tilde{\jmath})
\in\mc{D}^{\mp}_{\tilde{\jmath}}(w\pm 1)
\ee 
where \(\tilde{\jmath}=k/2-j\,\). For \(N=0\) the corresponding primaries  are identified by the relation
\be\label{weylprimeq}
\mc{V}_{j}^{w}(\pm j)\,=\,\omega_{j}\mc{V}_{\tilde{\jmath}}^{w\pm 1}(\mp\tilde{\jmath})
\ee
where the factor \(\omega_{j}\) is calculated in section~\ref{sfcons_sec} below. The fact that each of these operators are terminating states under the global algebra in their respective spectral flow sectors is a consequence of the other being primary under the corresponding current algebra generators. This results from
\be
J_{1}^{\pm}(w)\,=\,J_{0}^{\pm}(w\pm 1)\,=\,J_{1\pm w}^{\pm}
\ee
Thus the affine representations are identified as
\be\label{affinerepid}
\hat{\mc{D}}^{\pm}_{j}(w)\,=\,\hat{\mc{D}}^{\mp}_{\tilde{\jmath}}(w\pm 1)
\ee
However, except for the identification (\ref{weylprimeq}), the corresponding spaces of primaries \(\mc{D}^{\pm}_{j}(w)\,\) and \(\,\mc{D}^{\mp}_{\tilde{\jmath}}(w\pm 1)\) are not identified, but are related as described by (\ref{weylprim},\,\ref{weylconj},\,\ref{weylprimeq}). In what follows the convention will be to mainly choose \(\hat{\mc{D}}^{\pm}_{w}=\hat{\mc{D}}^{\pm}(w)\) for \(\pm w\geq 0\,\).

\subsection{Physical string states}

The single string Hilbert space is the BRST cohomology of a tensor product of the worldsheet ghost state space with a matter state space given by \(\hat{\mc{H}}_{\sst SL(2,R)}\otimes\hat{\mc{H}}_{\sst \mc{M}}\,\), where \(\mc{M}\) denotes the target space for a unitary CFT with central charge
\be\label{mcchg}
c_{\sst \mc{M}}\,=\,26-\frac{3k}{k-2}
\ee
Taking \(\mc{L}_{n}\) to be the Virasoro generators for the \(\mc{M}\) CFT, and \(\ket{h}\) to be a corresponding state with conformal dimension \(h\geq 0\,\), the physical state conditions 
\be\label{phys}
(L_{n}+\mc{L}_{n}-\delta_{n})\ket{j,M,N,w}\!\ket{h}\,=\,0\spwd{1cm}{for}n\geq 0
\ee
lead to
\be\label{MS}
\frac{j(1-j)}{k-2}+N-wM+\txfc{1}{4}kw^2+h-1\,=\,0\spwd{1cm}{for}n=0
\ee
and
\be
(L_{n}(w)-w\,J^{3}_{n}+\mc{L}_{n})\ket{j,M,N,w}\!\ket{h}\,=\,0\spwd{1cm}{for}n>0
\ee
where the expression (\ref{Hgen}) for \(H=1-h\) has been used. For \(\hat{\mc{C}}_{j}^{\alpha}(w)\,\), making use of the inequality \(-\txfc{1}{2}\leq \alpha<\txfc{1}{2}\,\), (\ref{MS}) and (\ref{CM}) lead to the constraint
\be
-\frac{w}{2}\leq
\frac{(1/4+s^2)}{k-2}+N-Pw-\txfc{1}{4}kw^2+h-1
<\frac{w}{2}
\ee
For \(\hat{\mc{D}}^{\pm}_{j}(\pm w)\,\), given the inequality \(1/2<j<(k-2)/2\,\), (\ref{MS}) and (\ref{DM}) allow a solution for \(j\) 
\be
2j-1\,=\,-w(k-2)+\left[1+4(k-2)\left(h-1+(1+w)N-wQ-w(w+1)/2\right)\right]^{1/2}
\ee
which leads to the constraint
\be
\txfc{1}{4}kw^2+\txfc{1}{2}w<
\frac{1}{4(k-2)}+h-1+(1+w)N-wQ
<\txfc{1}{4}k(w+1)^2-\txfc{1}{2}(w+1)
\ee
For \(H=1-h<\txfc{1}{4}(k-2)^{-1}\,\) and \(w\geq 0\) it may be seen from Figure~\ref{fig1} that the states \(\mc{V}_{\sst 1/2}^{w}(\alpha)\in\mc{C}^{\alpha}_{\sst 1/2}(w)\) and \(\mc{V}_{j}^{w}(j)\in\mc{D}^{+}_{j}(w)\,\), which are states\footnote{The restriction \(w\geq 0\) is imposed here (and \(\alpha\geq0\) for \(\hat{\mc{C}}_{j}^{\alpha}(0)\)) because of the identification of the representations under \(J_{0}^{3}\) conjugation which has been described above.} of lowest \(H\) for a given \(M\), are also states of lowest \(M\) for a given \(H\,\), and thus may be considered ground states of the \(SL(2,R)\) CFT for a fixed conformal weight \(h\) of the \(\mc{M}\) CFT. The picture is more complicated in the gap given by \(\,1>H>\txfc{1}{4}(k-2)^{-1}\,\), where a discrete set of ground states appears. Note that this gap is present for all \(k\) satisfying (\ref{mcchg}) since \(c_{\sst \mc{M}}>0\) implies \(k>52/23>9/4\,\). Thus for \(\hat{\mc{C}}_{j}^{\alpha}(0)\) a tachyon (\(N=0\)) is always present in the physical spectrum, and from (\ref{CM}) there are states in \(\hat{\mc{C}}_{j}^{\alpha}(0)\) with \(M\leq 0\) for \(\alpha\geq 0\,\). While the structure of the ground states will not be elucidated in detail here, it will be shown that for all other representations with \(w\geq 0\) we have \(M\geq 0\,\), with a unique state \((J^{-}_{-1}\bar{J}^{-}_{-1})\cdot\mc{V}_{1}(1)\in\hat{\mc{D}}^{+}_{1}(0)\,\) which has \(M=\bar{M}=0\,\).

For \(\hat{\mc{D}}^{+}_{j}(w)\,\) with \(w\geq 0\,\) the physical state of lowest \(M\) will appear for \(Q=0\) and \(H\leq 1\) in (\ref{DM}, \ref{DH}), which correspond to states in \(\mc{D}^{-}_{\tilde{\jmath}}(w+1)\,\) with \(\tilde{\jmath}=k/2-j\,\). Given \(Q=0\), for \(N=0\) we have \(M=j+\txfc{1}{2}kw>1/2\,\), and for \(N>0\,\) equations (\ref{DM}) and (\ref{DH}) may be solved to give
\be\label{Mequal}
M\,=\,\frac{1}{(w+1)}\left(h+\frac{(j-1)(k-2-j)}{(k-2)}+\txfc{1}{4}kw(w+2)\right)
\ee
This is minimized for \(h=0\) and \(j=1/2\,\), producing the following inequality 
\be
M\,\geq\,\frac{1}{4(w+1)}\left(\frac{1}{(k-2)}-2+kw(w+2)\right)
\ee
It may be seen that for \(w>0\) the right hand side is strictly positive. For \(\hat{\mc{D}}^{\pm}_{j}(0)\,\) (\ref{DH}) reads
\be\label{wzerophys}
N\,=\,1-h-\frac{j(1-j)}{(k-2)}
\ee
For \(N>0\) and \(j\in\mc{S}_{\sst\mc{D}}\,\), this has solutions for \(j\geq 1\) and \(k\geq 3\,\). From the form of (\ref{Mequal},\,\ref{wzerophys}), this implies that the state of lowest \(M\) satisfies \(w=0\,\), \(j=1\,\), \(h=0\,\) and \(N=1\,\), leading to the eigenvalues \(M=0\,\) and \(H=1\,\). Thus, the unique physical state with \(E=M+\bar{M}=0\) is given by
\be
\ket{j=1,M=0,N=1,w=0}\,=\,J^{-}_{-1}\bar{J}^{-}_{-1}\ket{\mc{V}_{1}(1)}
\ee
This state is associated with the zero-mode of the dilaton on \(AdS_{3}\,\), and may be integrated to form the identity operator\footnote{Note that the state \(\mc{V}_{1}(1)\,\) has \(H=0\) but \(M=1\,\), and thus is not the identity operator of the \(SL(2,R)\) CFT, which arises through analytic continuation via \(\lim_{j\goto 0}\mc{V}_{j}(j)=\mathbbm{1}\,\).} of the spacetime boundary CFT
\be
I\,=\,\frac{1}{k^2}\int d^2z\,(J^{-}_{-1}\bar{J}^{-}_{-1})\cdot\mc{V}_{1}(1|z)
\ee
The operator \(I\) may be seen to be invariant under the global algebra. This results from \((J_{0}^{3}J^{-}_{-1})\cdot\mc{V}_{1}(1)=0\,\), \((J_{0}^{-}J^{-}_{-1})\cdot\mc{V}_{1}(1)=0\,\), and 
\be
(J_{0}^{+}J^{-}_{-1})\cdot\mc{V}_{1}(1|z)\,=\,(k-2)\,L_{-1}\cdot\mc{V}_{1}(1|z)\,=\,\
(k-2)\,\frac{\partial\,}{\partial z}\,\mc{V}_{1}(1|z)
\ee
which follows from the Knizhnik-Zamolochikov equation (\ref{sugmnsone}) and the affine algebra (\ref{modealg}). The total derivative requires the treatment of contact terms, which may be shown to vanish.

For \(\hat{\mc{C}}_{j}^{\alpha}(w)\) with \(w>0\) it may be seen that for \(h\geq0\) we have the inequality 
\be
M\,=\,\frac{1}{w}\left(h-1+\frac{(1/4+s^2)}{(k-2)}+N+\txfc{1}{4}kw^2\right)\,\geq\,
\frac{1}{4w}\left(\frac{1}{(k-2)}-4+kw^2\right)\,\geq\,0
\ee
Note that for \(w=1\) and \(k=3\) the bound on the right is saturated. In the notation introduced above (\ref{CM}, \ref{CH}), this corresponds to the single state given by \(s=0\,\), \(N=0\,\), \(h=0\,\), \(\alpha=-1/2\) and \(P=-1\,\). Using \(\hat{\mc{D}}^{+}_{j}(w)=\hat{\mc{D}}^{-}_{\tilde{\jmath}}(w+1)\) with \(\tilde{\jmath}=k/2-j\,\), the significance of this state may be seen as follows. The current algebra representation \(\hat{\mc{C}}_{\sst 1/2}^{-\sst 1/2}(w)\) may be seen to be equivalent~\cite{MaldacenaOoguri_0001} to \(\hat{\mc{D}}^{+}_{\sst 1/2}(w)\oplus\hat{\mc{D}}^{-}_{\sst 1/2}(w)\,\). Thus for \(w=1\)
\be
\hat{\mc{C}}_{\sst 1/2}^{-\sst 1/2}(1)\,=\,\hat{\mc{D}}^{+}_{\sst 1/2}(1)\oplus\hat{\mc{D}}^{-}_{\sst 1/2}(1)\,=\,\hat{\mc{D}}^{+}_{\sst 1/2}(1)\oplus\hat{\mc{D}}^{+}_{\sst (k-1)/2}(0)
\ee
and for \(k=3\) the \(M=0\) state is contained in \(\hat{\mc{D}}^{+}_{\sst (k-1)/2}(0)=\hat{\mc{D}}^{+}_{1}(0)\,\).

The consistency of the structure of the spectrum of the \(SL(2,R)\) CFT depends upon the precise bound \(j\in\mc{S}_{\sst\mc{D}}\) (\ref{discspec}) for \(\hat{\mc{D}}^{\pm}_{j}(w)\,\). It is helpful to briefly discuss how this bound arises. As discussed in section~\ref{sec_B} below, unitary representations \(\mc{D}^{\pm}_{j}\,\) of the global \(SL(2,R)\) algebra \(J^{a}_{0}\,\) appear for \(j>0\,\). Furthermore, as discussed in subsection~\ref{sfequiv_sec}, the (Mobius fixed) two-point function is independent of the spectral flow sector. Thus the identification (\ref{weylprimeq}) implies that if \(\mc{D}^{\pm}_{j}(w)\) is a unitary representation, then \(\mc{D}^{\mp}_{\tilde{\jmath}}(w\pm 1)\) for \(\tilde{\jmath}=k/2-j\) is also unitary. This suggests the unitarity bound \(0<j<k/2\,\) for the respective affine representations in (\ref{affinerepid}) which may appear in string amplitudes. It may be shown that all states which satisfy the physical state conditions (\ref{phys}) can be expressed as the sum of a BRST exact state and a state\footnote{More precisely, the relevant cohomology may be described by states which are annihilated by the lowering operators \(J^{3}_{>0}(w)\,\), without necessarily satisfying the \(J^{3}_{0}+\bar{J}^{3}_{0}=0\,\) condition imposed in the \(SL(2,R)/U(1)\,\) model.} in the \(SL(2,R)/U(1)\) coset CFT. Thus unitarity of the string spectrum requires the restriction to those representations of the \(SL(2,R)\) current algebra which are built by applying \(J^{3}_{<0}(w)\) to unitary representations of the coset \(SL(2,R)/U(1)\,\). It turns out that the condition \(0<j<k/2\,\) is sufficient to ensure unitarity of representations of \(SL(2,R)/U(1)\,\). As an example, the coset state \(J^{-}_{-1}\cdot\mc{V}_{j}(j)\in\hat{\mc{D}}_{j}^{+}(0)\,\) has norm
\be
\bra{\mc{V}_{j}(j)}J^{+}_{+1}J^{-}_{-1}\ket{\mc{V}_{j}(j)}
\,=\,(k-2j)\brkt{\mc{V}_{j}(j)}{\mc{V}_{j}(j)}
\ee
Note that this is also a Virasoro primary state in the flowed primary representation \(\mc{D}^{-}_{\tilde{\jmath}}(1)\,\), where \(\tilde{\jmath}=k/2-j\,\), and so must have positive norm to appear in the spring spectrum. While this imposes \(j<k/2\,\) (or \(\tilde{\jmath}>0\)), in the quantum mechanical picture described in section~\ref{sec_B} normalizability in the \(\mc{L}^2\) norm (\ref{l2nrm}) requires \(j>1/2\,\). This suggest the bound \(1/2<j<(k-1)/2\,\) for the discrete affine representations \(\hat{\mc{D}}^{\pm}_{j}(w)\) which appear in the spectrum of the \(SL(2,R)\) CFT. This unitarity bound may be seen to follow directly from the form of the two-point function (\ref{sl2pt}), which is the CFT analogue of the quantum mechanical \(\mc{L}^2\) norm.

\section{\texorpdfstring{$SL(2,R)$}{} correlation functions \label{sec_4}}

We now consider the computation of correlation functions of the \(SL(2,R)\) CFT for vertex operators in the spectrum of definite affine weight described in the previous section. This spectrum consists\footnote{Here we introduce the notation \(\hat{\mc{C}}=\sum_{w\in\mathbb{Z}}\hat{\mc{C}}_{w}\,\), \(\mc{C}=\sum_{w\in\mathbb{Z}}\mc{C}_{w}\,\), and etc.} of the flowed affine modules 
\be
\hat{\mc{H}}_{\sst SL(2,R)}\,=\,\hat{\mc{C}}\oplus\hat{\mc{D}}^{+}\oplus\hat{\mc{D}}^{-}\,=\,
\sum_{w\in\mathbb{Z}}\hat{\mc{C}}_{w}\,\oplus\,
\sum_{w=0}^{\infty}\hat{\mc{D}}^{+}_{w}\oplus\hat{\mc{D}}^{-}_{-w}
\ee
where the identification \(\hat{\mc{D}}^{\pm}_{w}=\hat{\mc{D}}^{\mp}_{w\pm 1}\) (\ref{affinerepid}) has been made. The corresponding Virasoro primary states are 
\(\mc{C}_{w}\subset\hat{\mc{C}}_{w}\,\) and \(\mc{D}^{\pm}_{w}\oplus\mc{D}^{\mp}_{w\pm 1}\subset\hat{\mc{D}}^{\pm}_{w}\,\). In what follows it will also be helpful to define 
\be
\mc{B}^{\pm}_{w}=\mc{B}^{\mp}_{w\pm 1}=\mc{D}^{\pm}_{w}\cap\mc{D}^{\mp}_{w\pm 1}
\ee
which consist of the terminating states \(\mc{V}^{w}_{j}(\pm j)\in\mc{B}_{j}^{\pm}(w)\,\). As for the flowed affine modules \(\hat{\mc{D}}^{\pm}_{w}\,\), the convention will be to generally take \(\mc{B}^{\pm}_{w}\) for \(\pm w\geq 0\,\) to avoid redundancy due to (\ref{weylprimeq}). Similarly, to lighten the notation, general descendant operators will be denoted by \(\varphi\,\), or more specifically by \(\varphi^{c}\in\hat{\mc{C}}\,\) and \(\varphi^{\pm}\in\hat{\mc{D}}^{\pm}\,\). Also, affine primaries will be denoted by \(\phi\,\), or more specifically by \(\phi^{c}\in\mc{C}\,\) and \(\phi^{\pm}\in\mc{D}^{\pm}\,\). Finally, elements of \(\mc{B}^{\pm}\) will be denoted by \(\omega^\pm\,\).

A reasonably complete description of the \(SL(2,R)\) CFT on closed genus zero Riemann surfaces would include a demonstration that all descendant correlation functions 
\be\label{deccor}
\amp{3mm}{{\textstyle\prod}_{p}^{n}\,\varphi_{p}(z_{p})}
\ee
are in principle computable. Also required would be a demonstration that the associated operator product expansion gives rise to a well-defined CFT in terms of a generalization to non-rational and non-unitary theories of the conditions described in~\cite{MooreSeibergCFT_89}. These conditions include the requirement of crossing symmetry of correlation function, or equivalently the associativity of the OPE. In a conventional WZNW model, where primary operators are annihilated by affine modes \(J^{a}_{{\sst >}0}\) of negative conformal weight, an analysis would begin with the observation that \(n\)-point correlation functions of descendant fields may be straightforwardly reduced to \(n\)-point correlation functions of corresponding primary fields. This also applies to correlators of unflowed primaries in the \(SL(2,R)\) CFT. Considering the unflowed primaries \(\mc{V}_{j_p}(m_p|z_p)\,\),  this relies on the fact that the contours of modes \(J^{a}_{{\sst <}0}\) of positive conformal weight in the \(n\)-point correlator
\be
\amp{3mm}{J^{a}_{{\sst <}0}\cdot\mc{V}_{j_n}(m_n|z_n)\,
{\textstyle\prod}_{p}^{n-1}\mc{V}_{j_p}(m_p|z_p)}
\ee
can be pulled back both to annihilate the identity at infinity as well as onto the other primaries to produce a linear combination of primary correlators. This process of reduction of descendant correlators does not proceed as simply for descendants in spectral flow sectors. For example, assuming that \(\pm w_n>0\,\) and \(q>0\,\), consider the correlator 
\be
\amp{3mm}{J^{\pm}_{-q}(w_n)\cdot\mc{V}^{w_n}_{j_n}(m_n|z_n)\,
{\textstyle\prod}_{p}^{n-1}\mc{V}^{w_p}_{j_p}(m_p|z_p)}
\ee
For \(q<\pm w_n\,\), the mode operator \(J^{\pm}_{-q}(w_n)=J^{\pm}_{\pm w_n-q}\) is of negative conformal weight, and does not annihilate the identity at infinity when the associated contour is pulled back. Furthermore, when the contour is pulled onto the respective spectral flowed primaries, descendant operators are in general produced. Thus, this somewhat naive argument merely reduces correlation functions of spectral flowed descendants to those involving only mode operators of non-positive conformal weight of the following form
\be\label{nonposform}
{\textstyle\prod}_{q=1}^{\pm w}\left(J^{\pm}_{-q}(w)\right)^{s^{\pm}_{q}}\cdot\mc{V}^{w}_{j}(m)
\ee
Here we have chosen \(\pm w>0\,\) and introduced \(s^{\pm}_{q}\in\mathbb{Z}_{\geq 0}\,\). The process of reduction to spectral flowed (and thus Virasoro) primaries is somewhat more involved, and will be discussed in detail in section~\ref{reduce_sec}\,. It is helpful in this context to introduce the spectral flow selection rules~\cite{MaldacenaOoguri_0111}\,, the derivation of which is described in section~\ref{select_sec}\,, imposed by the affine algebra on the spectral flowed modules.

A fact that will prove useful in the following section is that a general \(n\)-point descendant correlator (\ref{deccor}) 
can be simply reduced to sums of correlators of the following form
\be
\amp{3mm}{\varphi_{n}(z_n)\prod_{p}^{n-1}\phi_{p}(z_{p})}
\ee
where repeated application of operators of the form
\be
\int_{\infty} dz\,J^{\pm}(z)\prod_{p}^{n}(z-z_p)^{\pm w_{p}+r_p}\,=\,0
\ee
with \(r_{p}\in\mathbb{Z}\,\), has been used to reduce the \(n-1\) vertex operators to flowed affine primaries \(\phi_{p}\,\).  These operators, which encircle all of the fields, allow flowed affine raising modes of negative conformal dimension to be pulled away from \(n-1\) of the fields and applied to the field \(\varphi_{n}\,\). The operator annihilates the identity at infinity if \(\sum_{p}^{n}(\pm w_{p}+r_p)\leq 0\,\), a choice that can always be made by applying a mode of sufficiently negative \(r_n\) to \(\varphi_{n}\,\). More precisely, defining \(n_{c}\) to be the number of insertions of elements of \(\hat{\mc{C}}\), and \(n_{\pm}\) to be the number of insertions of elements of \(\hat{\mc{D}}^{\pm}\), we will use the conjugation symmetry to set \(n_+\geq n_-\,\),  and reduce the most general correlator as follows. If \(n_c>0\) we will choose the descendant to be \(\varphi^{c}_{n_c}\in\hat{\mc{C}}\) and reduce to correlators of the form
\be\label{contcf}
\amp{3.5mm}{\hat{\varphi}^{c}_{n_{c}}\!\prod_{p}^{n_{c}-1}\!\phi^{c}_{p}(z^{c}_{p})
\prod_{a}^{n_{+}}\omega^{+}_{a}(z^{+}_{a})\prod_{b}^{n_{-}}\,\omega^{-}_{b}(z^{-}_{b})}
\ee
Here a Mobius transformation has been used to place the field \(\varphi^{c}_{n}\) at infinity, and the primary fields \(\phi^{c}_{p}\) have been reduced to spinless fields 
\be
H_{p}-\bar{H}_{p}=-w_{p}(M_{p}-\bar{M}_{p})\,=\,0
\ee
where \(M_{p}=m_{p}+\txfc{1}{2}kw_p\,\) is the global \(J_{0}^{3}\) eigenvalue, to simplify further analysis. Since \(\omega^{\pm}_{a}\,\) are also spinless primaries, global \(J^{3}_{0}\) conservation implies  \(M_{n_{c}}=\bar{M}_{n_{c}}\,\) for \(\varphi^{c}_{n_{c}}\,\). If \(n_c=0\,\) we will choose the descendant to be \(\varphi^{+}_{n_+}\in\hat{\mc{D}}^{+}\) and reduce to correlators of the form
\be\label{disccf}
\amp{3.5mm}{\hat{\varphi}^{+}_{n_{+}}\!\!
\prod_{a}^{n_{+}-1}\omega^{+}_{a}(z^{+}_{a})\prod_{b}^{n_{-}}\,\omega^{-}_{b}(z^{-}_{b})}
\ee
Again, here global \(J^{3}_{0}\) conservation implies  \(M_{n_{+}}=\bar{M}_{n_{+}}\) for \(\varphi^{+}_{n_{+}}\,\), which has been pushed to infinity.

\subsection{Derivation of spectral flow selection rules\label{select_sec}}

Spectral flow selection rules~\cite{MaldacenaOoguri_0111} apply to all correlators, with amplitudes vanishing if the following bounds are violated. If \(n_{c}>0\,\) for correlators of the form (\ref{contcf}) we have 
\be\label{sfslccont}
-(n_{c}+n_{+}-2)\,\leq\,\sum_{p}^{n}w_{p}\,\leq\,(n_{c}+n_{-}-2)
\ee
If \(n_{c}=0\,\) for correlators of the form (\ref{disccf}) we have
\be\label{sfslcdisc}
-(n_{+}-1)\,\leq\,\sum_{p}^{n}w_{p}\,\leq\,(n_{-}-1)
\ee
Note that both of these inequalities are invariant under \(J^{3}_{0}\) charge conjugation, which takes \(w_{p}\goto -w_{p}\,\) and \(n_{+}\leftrightarrow n_{-}\,\). They are also invariant under a relabeling using \(\hat{\mc{D}}^{\pm}_{w}=\hat{\mc{D}}^{\mp}_{w\pm 1}\,\) which preserves \(n_{+}+n_{-}\,\). It is important to recognise that these selection rules apply to the respective current algebra modules, \(\hat{\mc{C}}_{w}\,\) and \(\hat{\mc{D}}^{\pm}_{w}\,\), and that more restrictive selection rules could apply to states (for example to sets of flowed primaries) within these modules.

The derivation of selection rules which appears below requires that the two-point function \(\vev{\varphi_{2}(z_2)\varphi_{1}(z_1)}\,\) of general flowed descendants of definite affine weight vanishes unless the vertex operators are in \(J_{0}^{3}\) conjugate modules related by \(\hat{\mc{C}}_{j}^{\alpha}(w)\goto\hat{\mc{C}}_{j}^{-\alpha}(-w)\,\), and  \(\hat{\mc{D}}_{j}^{\pm}(w)\goto\hat{\mc{D}}_{j}^{\mp}(-w)\,\). Using Mobius invariance (\ref{twoptdesc}) to produce linear combinations of descendant correlators of the form \(\vev{\hat{\varphi}^{w_2}_{2}\varphi^{w_1}_{1}}=\vev{\hat{\varphi}^{w_1}_{1}\varphi^{w_2}_{2}}\,\), with \(M_2=-M_1\) and \(H_2=H_1\,\), this may be seen as follows. Pulling all flowed raising operators away from \(\varphi^{w_1}_{1}\) produces a correlator of the form \(\vev{\hat{\varphi}^{w_2}_{2}\phi^{w_1}_{1}}\,\) where \(\phi^{w_1}_{1}\) is a primary field. If \(\phi^{w_1}_{1}\in \mc{D}_{j_1}^{\pm}(w_1)\) then the correlator can be further reduced to the form \(\vev{\hat{\varphi}^{w_2}_{2}\omega^{w_1}_{1}}\,\) where \(\omega^{w_1}_{1}\in\mc{B}_{j_1}^{\pm}(w_1)\,\). In this case a non-zero correlator only arises if \(\varphi^{w_2}_{2}\in \mc{B}_{j_1}^{\mp}(-w_1)\subset\mc{D}_{j_1}^{\mp}(-w_1)\,\) since, as may be seen from figure~\ref{fig1}\,, the states \(\mc{V}^{w}_{j}(\pm j)\in\mc{B}_{j}^{\pm}(w)\) have unique \(L_0\) and \(J_{0}^{3}\) eigenvalues. Alternatively, if \(\phi^{w_1}_{1}\in \mc{C}_{j_1}^{\alpha_1}(w_1)\) then \(J^{3}_{0}\) conjugation symmetry allows the choice \(w_1 \geq |w_2|\geq -w_2\) with \(\varphi^{w_2}_{2}\in \hat{\mc{C}}_{j_2}^{\alpha_2}(w_2)\). In this case the zero mode \(J^{-}_{0}(w_1)\) can be extracted from \(\phi^{w_1}_{1}\) to produce a correlator of the form 
\be
\amp{3.5mm}{\hat{\varphi}^{w_2}_{2}J^{-}_{0}(w_1)\phi^{w_1}_{1}}=
-\amp{3.5mm}{\hat{\phi}^{w_1}_{1}J^{-}_{w_1+w_2}(w_2)\varphi^{w_2}_{2}}
\ee
Carrying this out repeatedly will lead to a vanishing correlator unless \(w_1+w_2=0\,\). Furthermore, in this case any flowed raising mode \((n>0)\) may be extracted from \(\varphi^{w_2}_{2}\) to produce
\be
\amp{3.5mm}{\hat{\phi}^{w_1}_{1}J^{a}_{-n}(w_2)\varphi^{w_2}_{2}}=
-\amp{3.5mm}{\hat{\varphi}^{w_2}_{2}J^{a}_{n}(w_1)\phi^{w_1}_{1}}=0
\ee
Thus \(\varphi^{w_2}_{2}\) is a flowed affine primary state \(\varphi^{w_2}_{2}=\phi^{-w_1}_{2}\in\mc{C}_{j_2}^{\alpha_2}(-w_1)\,\). From the spectrum described above, \(\phi^{-w_1}_{2}\) must have eigenvalues \(j_2=j_1\) and \(\alpha_2=-\alpha_1\,\), a fact that may also be seen from (\ref{sfcons2pf}), which imposes \(\amp{3.5mm}{\hat{\phi}^{-w}_{2}\phi^{w}_{1}}=
\amp{3.5mm}{\hat{\phi}_{2}\,\phi_{1}}\,\) for spectral flow conserving two-point primary correlators.

To demonstrate the inequality (\ref{sfslccont}) we make use of the orthogonality of the respective flowed modules to write
\be
\amp{3.5mm}{\hat{\varphi}^{c}_{n_{c}}\!\prod_{p}^{n_{c}-1}\!\phi^{c}_{p}(z^{c}_{p})
\prod_{a}^{n_{+}}\omega^{+}_{a}(z^{+}_{a})\prod_{b}^{n_{-}}\,\omega^{-}_{b}(z^{-}_{b})}\,=\,
\amp{3.5mm}{\hat{\varphi}^{c}_{n_{c}}\Psi^{c}_{n_{c}}}
\ee
where \(\Psi^{c}_{n_{c}}\in\hat{\mc{C}}(-w^{c}_{n_c})\,\) is taken to arise from the operator product expansion. The state \(\Psi^{c}_{n_{c}}\) is annihilated by 
\be
\int_{\infty} dz\,J^{\pm}(z)\prod_{p}^{n_{c}-1}\!(z-z^{c}_{p})^{\pm w^{c}_{p}+1}
\prod_{a}^{n_{\pm}}(z-z^{\pm}_{a})^{\pm w^{\pm}_{a}+1}
\prod_{b}^{n_{\mp}}(z-z^{\mp}_{b})^{\pm w^{\mp}_{b}}
\ee
where the contour encircles all of the \(n-1\) primary operators. Pulling the contour back onto \(\varphi^{c}_{n_{c}}\) and using the Mobius symmetry leads to
\be\label{contslct0}
0\,=\,-\sum_{q=0}^{\infty}c_{q}\amp{3.5mm}{\hat{\Psi}^{c}_{n_{c}}
J^{\pm}_{\lambda_{\pm}+1+q}(w^{c}_{n_{c}})\,\varphi^{c}_{n_{c}}}
\,=\,\sum_{q=0}^{\infty}c_{q}\amp{3.5mm}{\hat{\varphi}^{c}_{n_{c}}
J^{\pm}_{-\lambda_{\pm}-1-q}(-w^{c}_{n_{c}})\,\Psi^{c}_{n_{c}}}
\ee
where \(c_q\) is a function of the operator insertion points and 
\be
\lambda_{\pm}\,=\,\mp\sum_{p}w_{p}-n_{c}-n_{\pm}
\ee
Since the affine modules are non-degenerate, if \(\Psi^{c}_{n_{c}}\neq 0\,\) and \(\lambda_{\pm}+1\geq 0\,\), then 
\be
\sum_{q=0}^{\infty}c_{q}\,J^{\pm}_{-\lambda_{\pm}-1-q}(-w^{c}_{n_{c}})\,\Psi^{c}_{n_{c}}\neq 0
\ee
Thus for \(\,\mp\sum_{p}w_{p}\geq (n_{c}+n_{\pm}-1)\,\) the vanishing of (\ref{contslct0}) implies \(\Psi^{c}_{n_{c}}=0\,\), which leads to the selection rules (\ref{sfslccont}). 

For the inequality (\ref{sfslcdisc}) we write
\be
\amp{3.5mm}{\hat{\varphi}^{+}_{n_{+}}\prod_{p}^{n_{c}}\phi^{c}_{p}(z^{c}_{p})\!
\prod_{a}^{n_{+}-1}\!\omega^{+}_{a}(z^{+}_{a})\prod_{b}^{n_{-}}\,\omega^{-}_{b}(z^{-}_{b})}\,=\,
\amp{3.5mm}{\hat{\varphi}^{+}_{n_{+}}\Psi^{+}_{n_{+}}}
\ee
where \(\Psi^{+}_{n_{+}}\in\hat{\mc{D}}^{-}(-w^{+}_{n_+})\,\) is in the module conjugate to that of \(\varphi^{+}_{n_{+}}\in\hat{\mc{D}}^{+}(w^{+}_{n_+})\,\). Primary fields \(\phi^{c}_{p}\in\mc{C}(w^{c}_{p})\) have been inserted for comparison with the inequality (\ref{sfslccont}). The state \(\Psi^{+}_{n_{+}}\) is annihilated by the following operator which encircles the primary fields
\be
\int_{\infty} dz\,J^{\pm}(z)\prod_{p}^{n_{c}}(z-z^{c}_{p})^{\pm w^{c}_{p}+1}
\prod_{a}^{n_{+}-1}\!(z-z^{+}_{a})^{\pm w^{+}_{a}+\theta_{\pm}}
\prod_{b}^{n_{-}}(z-z^{-}_{b})^{\pm w^{-}_{b}+\theta_{\mp}}
\ee
where \(\theta_{+}=1\,\) and \(\theta_{-}=0\,\). Pulling the contour back onto \(\varphi^{+}_{n_{+}}\) leads to
\be\label{discslct0}
0\,=\,-\sum_{q=0}^{\infty}d_{q}\amp{3.5mm}{\hat{\Psi}^{+}_{n_{+}}
J^{\pm}_{\lambda_{\pm}+\theta_{\pm}+q}(w^{+}_{n_{+}})\,\varphi^{+}_{n_{+}}}
\,=\,\sum_{q=0}^{\infty}d_{q}\amp{3.5mm}{\hat{\varphi}^{+}_{n_{+}}
J^{\pm}_{-\lambda_{\pm}-\theta_{\pm}-q}(-w^{+}_{n_{+}})\,\Psi^{+}_{n_{+}}}
\ee
where, again, \(d_q\) is a function of the operator insertion points. Since \(\hat{\mc{D}}^{-}(-w^{+}_{n_+})\,\) is non-degenerate, and is preserved by \(J^{\pm}_{-\theta_{\pm}}(-w^{+}_{n_{+}})\) and modes of higher conformal weight, if \(\Psi^{+}_{n_{+}}\neq 0\,\) and \(\lambda_{\pm}\geq 0\,\), then
\be
\sum_{q=0}^{\infty}d_{q}\,J^{\pm}_{-\lambda_{\pm}-\theta_{\pm}-q}(-w^{+}_{n_{+}})\,\Psi^{+}_{n_{+}}\neq 0
\ee
Thus for \(\,\mp\sum_{p}w_{p}\geq (n_{c}+n_{\pm})\,\) (\ref{discslct0}) implies \(\Psi^{+}_{n_{+}}=0\,\), which leads to the selection rule
\be
-(n_{c}+n_{+}-1)\,\leq\,\sum_{p}^{n}w_{p}\,\leq\,(n_{c}+n_{-}-1)
\ee
In general this is a less restrictive condition than (\ref{sfslccont}), but in the case that \(n_c=0\) it gives the selection rule (\ref{sfslcdisc}).

In the treatment of the fusion rules the focus will be on the flowed primaries \(\mc{C}_{w}\) and \(\mc{D}^{\pm}_{w}\oplus\mc{D}^{\mp}_{w\pm 1}\) associated with the modules \(\hat{\mc{C}}_{w}\) and \(\hat{\mc{D}}^{\pm}_{w}=\hat{\mc{D}}^{\mp}_{w\pm 1}\,\). The selection rules which apply to a particular correlation function involving flowed primaries in the discrete modules \(\hat{\mc{D}}^{\pm}\) will in general be more restrictive than those which apply to the modules themselves. In the analysis below, elements \(\omega^{\pm}\in\mc{B}^{\pm}_{w}=\mc{D}^{\pm}_{w}\cap\mc{D}^{\mp}_{w\pm 1}\,\), which are terminating states of the discrete representations \(\mc{D}^{\pm}_{w}\,\), will be distinguished from other flowed primaries 
\(\phi\in\mc{P}_{w}\,\) in the \(w\) spectral flow sector. This distinction will be expressed here as
\be\label{defprime}
\mc{P}_{w}=\mc{C}_{w}\oplus\mc{D}^{+}_{w}\oplus\mc{D}^{-}_{w}\ominus\mc{B}^{+}_{w}\ominus\mc{B}^{-}_{w}
\ee
Note again that, while \(\mc{B}^{\pm}_{w}=\mc{B}^{\mp}_{w\pm 1}\,\), the states in \(\mc{D}^{\pm}_{w}\,\) and \(\mc{D}^{\mp}_{w\pm 1}\,\) are otherwise distinct. For the sake of clarity, in this expression we have deviated from the convention that \(\mc{B}^{\pm}_{w}\) will be chosen for \(\pm w\geq 0\,\). Now consider an \(n\)-point primary correlator with \(N\) elements \(\phi_{p}\in\mc{P}\,\) and \(N_{\pm}\) elements \(\omega^{\pm}_{a}\in\mc{B}^{\pm}\,\)
\be\label{primecor}
\amp{3mm}{\prod_{p}^{N}\phi_{p}(z_{p})\prod_{a}^{N_{+}}
\omega^{+}_{a}(z^{+}_{a})\prod_{b}^{N_{-}}\omega^{-}_{a}(z^{-}_{b})}
\ee
For \(N>0\,\), a correlator of the form (\ref{primecor}) can always be produced by applying the operator
\be
\int_{\infty} dz\,J^{\pm}(z)\,(z-z_{\sst N})^{\pm w_{\sst N}}\prod_{p}^{N-1}\!(z-z_{p})^{\pm w_{p}+1}
\prod_{a}^{N_{\pm}}(z-z^{\pm}_{a})^{\pm w^{\pm}_{a}+1}
\prod_{b}^{N_{\mp}}(z-z^{\mp}_{b})^{\pm w^{\mp}_{b}}
\ee
to the same correlator with \(\phi_{\sst N}=J^{\pm}_{0}(w_{\sst N})\tilde{\phi}^{\pm}_{\sst N}\) 
replaced by \(\tilde{\phi}^{\pm}_{\sst N}\in\mc{P}(w_{\sst N})\oplus\mc{B}^{\pm}(w_{\sst N})\,\). The conditions for this operator to vanish when pulled back onto the identity at infinity lead to the selection rule
\be\label{primeslc}
-(N+N_{+}-2)\,\leq\,\sum_{p}^{n}w_{p}\,\leq\,(N+N_{-}-2)
\ee
where \(n=N+N_{+}+N_{-}=n_{c}+n_{+}+n_{-}\,\). A similar construction can be applied for one of the states \(\omega^{\pm}_{a}\in\mc{B}^{\pm}\,\), but the corresponding selection rule is less restrictive than (\ref{primeslc}) except in the case \(N=0\,\), where it reproduces the selection rule (\ref{sfslcdisc}) with \(n_{\pm}\) replaced with \(N_{\pm}\,\).

\subsection{Reduction to flowed primary correlators\label{reduce_sec}}

Consider the correlator (\ref{contcf}) with the descendant \(\varphi^{c}_{n_{c}}\) Mobius transformed away from infinity
\be\label{contcf2}
\amp{3.5mm}{\varphi^{c}_{n_{c}}(z^{c}_{n_{c}})\!\prod_{p}^{n_{c}-1}\!\phi^{c}_{p}(z^{c}_{p})
\prod_{a}^{n_{+}}\omega^{+}_{a}(z^{+}_{a})\prod_{b}^{n_{-}}\,\omega^{-}_{b}(z^{-}_{b})}
\ee
Under the condition that it annihilates the identity at infinity, the following operator 
\be
\int_{\infty} dz\,J^{\pm}(z)\,(z-z^{c}_{n_{c}})^{\pm w^{c}_{n_{c}}-q}\prod_{p}^{n_{c}-1}\!(z-z^{c}_{p})^{\pm w^{c}_{p}}
\prod_{a}^{n_{\pm}}(z-z^{\pm}_{a})^{\pm w^{\pm}_{a}+1}
\prod_{b}^{n_{\mp}}(z-z^{\mp}_{b})^{\pm w^{\mp}_{b}}
\ee
for \(q\geq 1\) both preserves the form of the correlator as involving an element of 
\be
\hat{\mc{C}}\otimes\left(\mc{C}\right)^{n_c-1}\otimes\left(\mc{B}^{+}\right)^{n_+}\otimes\left(\mc{B}^{-}\right)^{n_-}
\ee
and permits the reduction of the flowed affine level of \(\varphi^{c}_{n_{c}}\,\). That is, the corresponding 
mode operators \(J^{\pm}_{-q}(w^{c}_{n_{c}})\,\) can be extracted from \(\varphi^{c}_{n_{c}}\,\) under the condition
\be\label{redcond}
\mp\sum_{p}^{n}w_{p}+q-n_{\pm}\geq 0
\ee
Thus, choosing \(q=1\,\) and imposing the selection rule (\ref{sfslccont}), 
it is apparent that an \(n\)-point correlator of the form (\ref{contcf2}) can be directly reduced to flowed primaries in 
\(\left(\mc{C}\right)^{n_c}\otimes\left(\mc{B}^{+}\right)^{n_+}\otimes(\mc{B}^{-})^{n_-}\) if it satisfies
\be
1-n_{\pm}\geq n_c+n_{\mp}-2
\ee
or \(n=n_c+n_{+}+n_{-}\leq 3\,\). Note that while a given \(n>3\,\) correlation function of the form (\ref{contcf2}) may satisfy both (\ref{redcond}) for \(q\geq 1\,\) and the selection rule (\ref{sfslccont}), this is always true for \(n\leq 3\,\).

We now take \(n_c=0\,\) and examine the reduction of correlators of the form (\ref{disccf}) with the descendant \(\varphi^{+}_{n_{+}}\) Mobius transformed away from infinity
\be\label{disccf2}
\amp{3.5mm}{\varphi^{+}_{n_{+}}(z^{+}_{n_{+}})\!
\prod_{a}^{n_{+}-1}\omega^{+}_{a}(z^{+}_{a})\prod_{b}^{n_{-}}\,\omega^{-}_{b}(z^{-}_{b})}
\ee
This involves an element of \(\hat{\mc{D}}^{+}\otimes\left(\mc{B}^{+}\right)^{n_{+}-1}\otimes\left(\mc{B}^{-}\right)^{n_-}\,\) and is reduced using operators of the form
\be
\int_{\infty} dz\,J^{\pm}(z)\,(z-z^{+}_{n_{+}})^{\pm w^{+}_{n_{+}}-q}\,(z-z^{-}_{n_{-}})^{\pm w^{-}_{n_{-}}}
\prod_{a}^{n_{\pm}-1}(z-z^{\pm}_{a})^{\pm w^{\pm}_{a}+1}
\prod_{b}^{n_{\mp}-1}(z-z^{\mp}_{b})^{\pm w^{\mp}_{b}}
\ee
for \(q\geq 1\,\). These operators preserve the form of the correlator as involving an element of
\be
\hat{\mc{D}}^{+}\otimes\mc{D}^{-}\otimes\left(\mc{B}^{+}\right)^{n_{+}-1}\otimes\left(\mc{B}^{-}\right)^{n_{-}-1}
\ee
Here it is helpful to recognize that the selection rule (\ref{sfslcdisc}), along with the choices \(\pm w_{a}^\pm\geq 0\,\) and \(n_{+}\geq n_{-}\,\), imply that \(n_{-}>0\,\). Thus, the mode operators \(J^{\pm}_{-q}(w^{+}_{n_{+}})\,\) can be extracted from \(\varphi^{+}_{n_{+}}\,\) under the condition
\be\label{redconddisc}
\mp\sum_{p}^{n}w_{p}+q-n_{\pm}+1\geq 0
\ee
Choosing \(q=1\,\) and imposing the selection rule (\ref{sfslcdisc}), 
it is apparent that an \(n\)-point correlator of the form (\ref{disccf2}) can be directly reduced to flowed primaries in 
\(\mc{D}^{+}\otimes\mc{D}^{-}\otimes\left(\mc{B}^{+}\right)^{n_{+}-1}\otimes(\mc{B}^{-})^{n_{-}-1}\) if it satisfies
\be
2-n_{\pm}\geq n_{\mp}-1
\ee
or \(n=n_{+}+n_{-}\leq 3\,\).

An example of a generally non-zero correlation function which fails 
to satisfy (\ref{redcond}) for \(q\geq 1\,\) is the following correlator involving an element of 
\(\hat{\mc{C}}(2)\otimes\mc{C}(0)^{3}\)
\be\label{4ptexmp}
\amp{3.5mm}{\phi_{4}\,\phi_{3}\,\phi_{2}\,J^{+}_{-1}(2)\cdot\phi_{1}}
\ee
where \(w_4=w_3=w_2=0\,\) and \(w_1=2\,\). Despite this seeming obstruction, it may be shown that all flowed descendant correlators can be reduced to corresponding primary correlators by making use of the Sugawara construction of the stress tensor. In the case of (\ref{4ptexmp}), taking \(\phi_{1}=J^{-}_{0}(2)\cdot\tilde{\phi}_{1}\,\) for the flowed primary \(\tilde{\phi}_{1}\,\), we have\footnote{Note that in the case of the relevant corresponding discrete flowed primary states \(\mc{D}^{+}(2)\) the state \(\tilde{\phi}_{1}\) would also exist.} 
\be
J^{+}_{-1}(2)\cdot\phi_{1}\,=\,\left((k-2)L_{-1}(2)+2J^{3}_{-1}(2)J^{3}_{0}(2)-
J^{-}_{-1}(2)J^{+}_{0}(2)\right)\cdot\tilde{\phi}_{1}
\ee
Each of the terms on the right side of this equations involve an operator of positive conformal weight which acts on a flowed primary field and which can be pulled back to produce correlators involving only flowed primary fields. Although this is essentially the simplest such example that could be considered, a careful examination of the most general flowed descendant correlator shows that a similar, though potentially much more cumbersome, use of the Sugawara form of the stress tensor will always permit a reduction to correlators involving only flowed primary fields. In the following subsections the further simplification and computation of the three-point primary correlation functions will be considered, and the fusion rules for spectral elements will be examined.

\subsection{Selection rules and reduction of three-point correlators\label{subsec3ptred}}

The selection rules (\ref{sfslccont}\,,\,\ref{sfslcdisc}) imply the following non-vanishing three-point 
correlators for the flowed affine modules
\be\label{3ptslchat}
\begin{array}{cl}
\hat{\mc{C}}_{w_3}\otimes\hat{\mc{C}}_{w_2}\otimes\hat{\mc{C}}_{w_1}&\hspace{0.4cm}\hat{w}=-1,0,1\\[0.2cm]
\hat{\mc{C}}_{w_3}\otimes\hat{\mc{C}}_{w_2}\otimes\hat{\mc{D}}^{\pm}_{w_1}&\hspace{0.4cm}\hat{w}=0,\mp 1\\[0.2cm]
\hat{\mc{C}}_{w_3}\otimes\hat{\mc{D}}^{\pm}_{w_2}\otimes\hat{\mc{D}}^{\pm}_{w_1}&\hspace{0.4cm}\hat{w}=\mp 1\\[0.2cm]
\hat{\mc{C}}_{w_3}\otimes\hat{\mc{D}}^{+}_{w_2}\otimes\hat{\mc{D}}^{-}_{w_1}&\hspace{0.4cm}\hat{w}=0\\[0.2cm]
\hat{\mc{D}}^{\pm}_{w_3}\otimes\hat{\mc{D}}^{\pm}_{w_2}\otimes\hat{\mc{D}}^{\mp}_{w_1}&\hspace{0.4cm}\hat{w}=0,\mp 1
\end{array}
\ee
Here we have defined \(\hat{w}=\sum_{p}^{3}w_{p}\) and made the choice \(\pm w\geq 0\) for \(\hat{\mc{D}}^{\pm}_{w}\,\).  Defining \(w_3=w_1+w_2\,\), the selection rules (\ref{sfslccont}\,,\,\ref{sfslcdisc}) then lead to the following fusion rules\footnote{The continuous representation \(\hat{\mc{C}}_{w_3}\) should appear with multiplicity two in the fusion rules for \(\hat{\mc{C}}_{w_2}\times\hat{\mc{C}}_{w_1}\,\). However, this factor will be suppressed until the discussion appearing in section \ref{fussum}\,.} for the spectral flowed sectors
\bbb\label{opeslct1}
\hat{\mc{D}}^{-}_{w_2}\times\hat{\mc{D}}^{+}_{w_1}&\goto& 
\hat{\mc{D}}^{+}_{w_3}\oplus\hat{\mc{D}}^{-}_{w_3}\oplus\hat{\mc{C}}_{w_3}\\[0.2cm]
\label{opeslct2}
\hat{\mc{C}}_{w_2}\times\hat{\mc{D}}^{\pm}_{w_1}&\goto&
\hat{\mc{D}}^{\pm}_{w_3}\oplus\hat{\mc{C}}_{w_3}\oplus\hat{\mc{C}}_{w_3\pm 1}\\[0.2cm]
\label{opeslct3}
\hat{\mc{C}}_{w_2}\times\hat{\mc{C}}_{w_1}&\goto&
\hat{\mc{C}}_{w_3}\oplus\hat{\mc{C}}_{w_3+1}\oplus\hat{\mc{C}}_{w_3-1}\oplus
\hat{\mc{D}}^{+}_{w_3}\oplus\hat{\mc{D}}^{-}_{w_3}
\eee
Here use has been made of \(\hat{\mc{D}}^{\pm}_{w}=\hat{\mc{D}}^{\mp}_{w\pm 1}\,\), under which (\ref{opeslct1}) is equivalent to
\be\label{opeslct4}
\hat{\mc{D}}^{\pm}_{w_2}\times\hat{\mc{D}}^{\pm}_{w_1} \,\goto\,
\hat{\mc{D}}^{\pm}_{w_3}\oplus\hat{\mc{D}}^{\pm}_{w_3\pm 1}
\oplus\hat{\mc{C}}_{w_3\pm 1}
\ee
The specific dependence of the fusion rules on the \(j\) quantum numbers through the computation of non-vanishing OPE coefficients of spectral elements will be considered below.

Due to the further restrictions imposed by the selection rule (\ref{primeslc}), the structure of the fusion rules involving the flowed primary states \(\mc{D}^{\pm}_{w}\oplus\mc{D}^{\mp}_{w\pm 1}\subset\hat{\mc{D}}^{\pm}_{w}\,\) is somewhat more intricate than that which follows directly from (\ref{opeslct1}) for the corresponding affine modules. It is those flowed primaries which are in the intersection of the corresponding sets defined by (\ref{sfslccont}\,,\,\ref{sfslcdisc}\,,\,\ref{primeslc}) that appear in the fusion rules. More precisely, three-point amplitudes for primary states which are elements of the affine modules (\ref{3ptslchat}) are generally non-vanishing except in the cases
\be\label{3ptslcexc}
\begin{array}{cl}
\mc{D}^{\pm}_{w_3}\otimes\mc{D}^{\pm}_{w_2}
\otimes\mc{D}^{\pm}_{w_1\mp 1}&\hspace{0.4cm}\hat{w}=\mp 1\\[0.2cm]
\mc{D}^{\mp}_{w_3\pm 1}\otimes\mc{D}^{\mp}_{w_2\pm 1}
\otimes\mc{D}^{\mp}_{w_1}&\hspace{0.4cm}\hat{w}=0
\end{array}
\ee
which vanish\footnote{However, the correlator associated with \(\mc{B}^{\pm}_{w_3}\otimes\mc{B}^{\pm}_{w_2}
\otimes\mc{B}^{\mp}_{w_1}\,\), which is common to all sets of primary correlators in \(\hat{\mc{D}}^{\pm}_{w_3}\otimes\hat{\mc{D}}^{\pm}_{w_2}\otimes\hat{\mc{D}}^{\mp}_{w_1}\,\), is not constrained to vanish.} by the selection rules (\ref{primeslc}). Here again \(\pm w\geq 0\) for \(\hat{\mc{D}}^{\pm}_{w}\,\) has been imposed for the corresponding affine modules. Note that \(\hat{w}=\sum_{p}^{3}w_{p}\) is the total spectral flow of the affine modules (\ref{3ptslchat}), not the total spectral flow in the respective correlators of primary states. If we denote the total spectral flow in a given correlator of primary states by \(\tilde{w}\,\), then the non-vanishing three-point primary correlators satisfy \(|\tilde{w}|\leq 1\,\), while the excluded correlators (\ref{3ptslcexc}) satisfy \(|\tilde{w}|=2\,\). As discussed below, it is the \(|\tilde{w}|\leq 1\,\) three-point primary correlators, including elements of \(\mc{D}^{\pm}\), which are ultimately defined via analytic continuation of \(H_3\) correlators. 

Given the form of the non-vanishing three-point primary correlators, the fusion rules for flowed primaries in the cases 
(\ref{opeslct2}\,,\,\ref{opeslct3}) follow without further restriction
\bbb\label{opeslctprm2}
\mc{C}_{w_2}\times\mc{D}^{\pm}_{w_1}&\goto& 
\mc{D}^{\pm}_{w_3}\oplus\mc{D}^{\mp}_{w_3\pm 1}\oplus\mc{C}_{w_3}\oplus\mc{C}_{w_3\pm 1}\,+\,\ldots\\[0.2cm]
\label{opeslctprm3}
\mc{C}_{w_2}\times\mc{C}_{w_1}&\goto&
\mc{C}_{w_3}\oplus\mc{C}_{w_3+1}\oplus\mc{C}_{w_3-1}
\oplus\,\mc{D}^{+}_{w_3}\oplus\mc{D}^{-}_{w_3+1}\oplus
\mc{D}^{-}_{w_3}\oplus\mc{D}^{+}_{w_3-1}\,+\,\ldots
\eee
where \(w_3=w_1+w_2\,\), and flowed descendants have been suppressed on the right side of these equations. The fusion rules for the flowed primaries corresponding to (\ref{opeslct1}) and (\ref{opeslct4}) are not equivalent since \(\mc{D}^{\pm}_{w}\) and \(\mc{D}^{\mp}_{w\pm 1}\) are distinct sets except for \(\mc{B}^{\pm}_{w}=\mc{B}^{\mp}_{w\pm 1}\,\). In the case (\ref{opeslct1}) there are no further restrictions
\be\label{opeslctprm1}
\mc{D}^{-}_{w_2}\times\mc{D}^{+}_{w_1}\,\goto\,
\mc{D}^{+}_{w_3}\oplus\mc{D}^{-}_{w_3+1}\oplus\mc{D}^{-}_{w_3}\oplus\mc{D}^{+}_{w_3-1}\oplus\mc{C}_{w_3}\,+\,\ldots
\ee
However, the case (\ref{opeslct4}) involves further restrictions due to (\ref{primeslc})
\be\label{opeslctprm4}
\mc{D}^{\pm}_{w_2}\times\mc{D}^{\pm}_{w_1}\,\goto\,
\mc{D}^{\pm}_{w_3}\oplus\mc{D}^{\mp}_{w_3\pm 1}\oplus\mc{D}^{\pm}_{w_3\pm 1}\oplus\mc{C}_{w_3\pm 1}\,+\,\ldots
\ee
In particular, a contribution \(\mc{D}^{\mp}_{w_3\pm 2}\,\) associated with the vanishing three-point correlators (\ref{3ptslcexc}) does not appear. The computation of the three-point flowed primary correlators is described below, along with a reduced set of OPE coefficients required to establish the \(j\) dependence of the fusion rules for the affine modules. Given the condition \(|\tilde{w}|\leq 1\,\), the further enforcement of the selection rules may be seen to follow from the form of the spectral flow conserving (\ref{wdisc}) and non-conserving (\ref{onesf}) three-point correlators involving discrete states. It may be also be shown that those primary representation which appear in the fusion rules (\ref{opeslctprm2}\,,\,\ref{opeslctprm3}\,,\,\ref{opeslctprm1}\,,\,\ref{opeslctprm4}) have generally non-vanishing OPE coefficients. That is, in terms of the spectral flow sectors of the primary states, the selection rules entirely determine the fusion rules.

As shown above, making use of the conjugation symmetry and defining \(\hat{w}=\sum_{p}^{n}w_{p}\,\), 
for \(n=3\) all correlation functions can be reduced to the flowed primary representations
\be
\begin{array}{cl}
\mc{C}_{w_{3}}\otimes\mc{C}_{w_{2}}\otimes\mc{C}_{w_{1}}&\hspace{0.3cm}\hat{w}=-1,0\\[0.1cm]
\mc{C}_{w_{3}}\otimes\mc{C}_{w_{2}}\otimes\mc{B}^{+}_{w_{1}}&\hspace{0.3cm}\hat{w}=-1,0\\[0.1cm]
\mc{C}_{w_{3}}\otimes\mc{B}^{+}_{w_{2}}\otimes\mc{B}^{+}_{w_{1}}&\hspace{0.3cm}\hat{w}=-1\\[0.1cm]
\mc{C}_{w_{3}}\otimes\mc{B}^{+}_{w_{2}}\otimes\mc{B}^{-}_{w_{1}}&\hspace{0.3cm}\hat{w}=0\\[0.1cm]
\mc{D}^{+}_{w_{3}}\otimes\mc{B}^{+}_{w_{2}}\otimes\mc{D}^{-}_{w_{1}}&\hspace{0.3cm}\hat{w}=-1,0
\end{array}
\ee
Here we are using the convention \(\pm w\geq 0\,\) for \(\mc{B}^{\pm}_{w}\,\) and, in the notation introduced above, have taken \(n_{+}\geq n_{-}\) with \(n_{-}>0\) for \(n_c=0\,\). In the case 
\(\mc{D}^{+}_{w_{3}}\otimes\mc{B}^{+}_{w_{2}}\otimes\mc{D}^{-}_{w_{1}}\) for \(\hat{w}=0\,\) the operator 
\be
\int_{\infty} dz\,J^{-}(z)\,(z-z_{3})^{-w_{3}}\,(z-z_{2})^{-w_{2}}\,(z-z_{1})^{-w_{1}}
\ee
permits reduction to \(\,\mc{D}^{+}_{w_{3}}\otimes\mc{B}^{+}_{w_{2}}\otimes\mc{B}^{-}_{w_{1}}\,\). Alternatively,
the operator
\be
\int_{\infty} dz\,J^{+}(z)\,(z-z_{3})^{w_{3}}\,(z-z_{2})^{w_{2}+1}\,(z-z_{1})^{w_{1}-1}
\ee
then provides the choice of reduction to \(\,\mc{B}^{+}_{w_{3}}\otimes\mc{B}^{+}_{w_{2}}\otimes\mc{D}^{+}_{w_{1}-1}\,\).
Similarly, in the case \(\hat{w}=-1\,\) the operator
\be
\int_{\infty} dz\,J^{+}(z)\,(z-z_{3})^{w_{3}}\,(z-z_{2})^{w_{2}+1}\,(z-z_{1})^{w_{1}}
\ee
permits reduction to \(\,\mc{B}^{+}_{w_{3}}\otimes\mc{B}^{+}_{w_{2}}\otimes\mc{D}^{-}_{w_{1}}\,\). The operator
\be
\int_{\infty} dz\,J^{-}(z)\,(z-z_{3})^{-w_{3}-1}\,(z-z_{2})^{-w_{2}}\,(z-z_{1})^{-w_{1}}
\ee
then provides the choice of reduction to \(\,\mc{D}^{-}_{w_{3}+1}\otimes\mc{B}^{+}_{w_{2}}\otimes\mc{B}^{-}_{w_{1}}\,\). These primary correlation functions can be written in terms of a smaller subset of correlators when taking into account the equivalence between those of equal total spectral flow. This equivalence will be discussed in more detail in \ref{sfequiv_sec} but ultimately permits all three-point correlators of spectral elements to be written in terms of
\be
\begin{array}{c}\label{redreps}
\mc{C}_{0}\otimes\mc{C}_{0}\otimes\mc{C}_{0}\\[0.1cm]
\mc{C}_{0}\otimes\mc{C}_{0}\otimes\mc{C}_{1}\\[0.1cm]
\mc{C}_{0}\otimes\mc{C}_{0}\otimes\mc{B}^{+}_{0}\\[0.1cm]
\mc{C}_{0}\otimes\mc{B}^{+}_{0}\otimes\mc{B}^{-}_{0}\\[0.1cm]
\mc{D}^{+}_{0}\otimes\mc{B}^{+}_{0}\otimes\mc{B}^{-}_{0}
\end{array}
\ee
Here free use has been made of the \(J_{0}^{3}\) conjugation symmetry. Note that the only spectral flow non-conserving amplitude required involves the continuous flowed representation \(\mc{C}_{1}\,\). Alternatively, the identification \(\mc{B}^{\pm}_{w}=\mc{B}^{\mp}_{w\pm 1}\) allows for the computation of all correlators in terms of either total spectral flow \(w=1\) three-point primary correlators or unflowed correlators of the form \(\mc{C}_{0}\otimes\mc{C}_{0}\otimes\mc{C}_{0}\,\).

\subsection{Unflowed three-point primary correlators\label{subsec_ufcf}}

We now consider the two-point and three-point correlation functions of primary states of the unflowed\footnote{As in appendix \ref{sec_B}, in this section we omit the spectral flow quantum number from the \(w=0\) primary representations. Thus, for example, \(\mc{C}(0)=\mc{C}_{0}\) is written here as \(\mc{C}\,\).} representations of the \(SL(2,R)\) CFT. As discussed above, the unflowed spectrum of primaries is given by \(\mc{D}^{+}\oplus\mc{D}^{-}\oplus\mc{C}\,\) with \(J_{0}^{3}\) charge conjugation implemented as \(\mc{C}_{j}^{\alpha}\goto\mc{C}_{j}^{-\alpha}\,\), and  \(\mc{D}_{j}^{\pm}\goto\mc{D}_{j}^{\mp}\,\). In general, the corresponding three-point functions, as well those which involve unflowed non-normalizable operators, arise through analytic continuation\footnote{Here \(\delta^2(m)\) takes the form (\ref{mdelta}) with \(\delta(\omega)\goto\delta(E)\) where \(E=-i\omega\in\mathbb{R}\,\).} of the expression (\ref{glb3pt}). The two-point function is given by
\be\label{gensl2pt}
\amp{3.5mm}{\hat{\mc{V}}_{j_2}(m_2)\mc{V}_{j_1}(m_1)}\,=\,
\lim_{\epsilon\goto 0}\,\amp{3.5mm}{\hat{\mc{V}}_{j_2}(m_2)
\mc{V}_{\epsilon}(\epsilon|1)\mc{V}_{j_1}(m_1)}
\ee
This amplitude will only be finite for elements of the spectrum, and for two vertex operators in \(\mc{C}\) it takes the form (\ref{glb2pt}). Given (\ref{h3lvtoglb}), this may be seen to agree with the quantum mechanical result (\ref{qmpoin2pt}) in the limit \(k\goto\infty\,\).

The three-point functions involving tensor products of the form \(\mc{C}\otimes\mc{C}\otimes\mc{C}\) do not generally simplify considerably from the expression (\ref{glb3pt}), and thus we concentrate on three-point functions in which at least one primary is in \(\mc{D}^{+}\oplus\mc{D}^{-}\). That is, we consider\footnote{The reason \(\mc{D}_{j_1}^{-}\) is preferred here and below, rather than \(\mc{D}_{j_1}^{+}\,\), is that the form of the expression (\ref{glbint}) is somewhat more tractable in this case. Hypergeometric identities would lead to a form in which \(\mc{D}_{j_1}^{+}\,\) would be preferred.} (\ref{glb3pt},\,\ref{gensl2pt}) with \(\mc{V}_{j_{1}}(m_1)\in\mc{D}_{j_{1}}^{-}\,\), where \(j_{1}\in\mc{S}_{\sst\mc{D}}\,\) and \(m_{1}=-j_{1}-n_{1}\,\) with \(n_{1}\in\mathbb{Z}_{\geq 0}\,\). The function \(W(j_i,m_i)\) in (\ref{glb3pt}) then becomes
\be\label{wdisc}
W(j_i,m_i)\,=\,-\pi^2 F(j_{i},m_{2},\bar{m}_{2})\,
\wht{W}(j_{i},m_{2},n_{1})\,\wht{W}(j_{i},\bar{m}_{2},\bar{n}_{1})
\ee
where,
\be\label{ffunc}
F(j_{i},m_{2},\bar{m}_{2})\,=\,
\frac{1}{2}\;\frac{s(\hat{\jmath})\,s(\hat{\jmath}-2j_{2})}
{s(2j_{1})\,s(\hat{\jmath}-2j_{3})}
\left(\frac{s(j_{2}+m_{2})}{s(j_{3}+j_{1}-m_{2})}+
\frac{s(j_{2}+\bar{m}_{2})}{s(j_{3}+j_{1}-\bar{m}_{2})}\right)
\ee
and, using the identity (\ref{gid1}),
\bbb\label{whatdisc}
\wht{W}(j_{i},m,n)&=&\frac{\Gamma(\hat{\jmath}-1)\,\Gamma(\hat{\jmath}-2j_{2})}
{\Gamma(2j_{1})\,\Gamma(1-\hat{\jmath}+2j_{3})}\,\frac{\Gamma(1-j_{3}+j_{1}-m+n)\,\Gamma(1-j_{2}-m)}{\Gamma(1-j_{3}+j_{1}-m)\,\Gamma(j_{3}+j_{1}-m+n)}\nonumber \\[0.2cm]
& & \times\,{_{3}{\rm F}_{2}}(1+2j_{1}-\hat{\jmath},\hat{\jmath}-2j_{3},-n\,;2j_{1},1-j_{3}+j_{1}-m;1)
\eee
Here \(s(x)=\sin(\pi x)\,\), and \(\hat{\jmath}=j_{1}+j_{2}+j_{3}\,\). Note that, since \(n_{1}\) and \(\bar{n}_{1}\) are non-negative integers, it is clear from the expression (\ref{F32sum}) that the hypergeometric function in (\ref{whatdisc}) is a finite sum. 

From the form of the functions \(W(j_i,m_i)\) in (\ref{wdisc}) and \(C(j_1,j_2,j_3)\) in (\ref{cfunc}) it may be seen that the two-point function of one primary in \(\mc{C}\) and another in \(\mc{D}^{+}\oplus\mc{D}^{-}\,\) vanishes. Of course, due to \(J_0^3\) conservation, this is also true  of two operators both of which are in either \(\mc{D}^{-}\) or \(\mc{D}^{+}\,\). Furthermore, using the \(SL(2,R)\) algebra, the two-point function for two primaries in \(\mc{D}_{j}^{-}\) and \(\mc{D}_{j'}^{+}\,\) with \(j,j'\in\mc{S}_{\sst \mc{D}}\) is given by
\be\label{sl2pt}
\amp{3.5mm}{\hat{\mc{V}}_{j}(-m)\,\mc{V}_{j'}(m')}\,=\,D_{j}(n)\,
\delta_{nn'}\,\delta_{\bar{n}\bar{n}'}\amp{3.5mm}{\hat{\mc{V}}_{j}(-j)\,\mc{V}_{j'}(j')}
\ee
where \(m=j+n\,\) for \(n\in\mathbb{Z}_{\geq 0}\,\), and
\be\label{dfnc}
D_{j}(n)\,=\,
\frac{\Gamma(2j)\,\Gamma(1+n)}{\Gamma(2j+n)}\,
\frac{\Gamma(2j)\,\Gamma(1+\bar{n})}{\Gamma(2j+\bar{n})}
\ee
The two-point function for the highest/lowest weight discrete states in (\ref{sl2pt}) may be computed 
by taking \(m_1=-j_1\,\), \(m_2=j_2\,\) and \(m_3=j_3=j_1-j_2\,\) in (\ref{glb3pt}). This yields
\be
W(j_i,m_i)\,=\,-\pi^2\,(1-2j_1)^{-2}
\ee
Now, taking \(j_2=\epsilon\goto 0\,\), and making use of
\be
\lim_{\epsilon\goto 0}\,C(1-j_1,1-\epsilon,1-j_1+\epsilon)\,=\,\mc{R}_{1}\mc{R}_{1-j_1}\,(2j_1-1)/(2\pi^2)
\ee
we find the result
\be\label{dsc2pf}
\amp{3.5mm}{\hat{\mc{V}}_{j}(-j)\,\mc{V}_{j'}(j')}\,=\,B_{j}\,\frac{\delta(j-j')}{(2j-1)^2}
\ee
which is non-zero and finite for \(j\in\mc{S}_{\sst\mc{D}}\,\)(\ref{discspec}). 
Note that (\ref{dsc2pf}), which may be seen to be formally equivalent to (\ref{phitwo}), has been derived from the limit of the three-point function (\ref{glb3pt}) rather than from analytic continuation of the two-point function (\ref{glb2pt}). Unlike the cases of (\ref{glb2pt},\,\ref{phitwo}), the term \(\delta(j-j')\) in (\ref{dsc2pf}) is not contributed by the function \(C(j_1,j_2,j_3)\) in (\ref{cfunc},\,\ref{cfuncdlt}) but follows from \(\delta(E-E')\,\) in (\ref{glb3pt}). The normalization of the vertex operators associated with the transform (\ref{phitoglb}), which fixes \(\Phi_{0}(0)=\mc{V}_{0}(0)=\mathbbm{1}\,\), avoids the introduction of a divergent factor in defining the discrete two-point function. In the choice of normalization in much of the literature, denoted here by \(\tilde{\mc{V}}_{j}(m)\,\), the operators of definite affine weight do not include the factor in front of the integral in (\ref{phitoglb2}). Thus, up to a factor which is finite for spectral elements, they are the \(j\goto 1-j\) reflection of those appearing here
\be
\mc{V}_{j}(m)\,=\,\frac{B_{j}}{(1-2j)}\,\tilde{\mc{V}}_{1-j}(m)\,=\,-\pi^{-1}\,\mc{R}_{1-j}\,R_{j}(m)\;\tilde{\mc{V}}_{j}(m)
\ee
For example, the associated lowest weight operators\footnote{Thus, for instance, the spectral element \(\mc{V}_{1}(1)\) and the identity \(\mc{V}_{0}(0)=\mathbbm{1}\) are finite operators of zero conformal weight which are normalizable and non-normalizable, respectively. The corresponding reflected operators \(\mc{V}_{0}(1)\) and \(\mc{V}_{1}(0)\) have singular vertex operator normalizations. Reflection-symmetric operators \(\mc{W}_{j}(m)=\mc{W}_{1-j}(m)\) are introduced below which avoid these singularities.} are related by the limit
\be
\mc{V}_{j}(j)\,=\,\pi^{-1}\,(1-2j)^{-1}\lim_{\epsilon\goto 0}\,\tilde{\mc{V}}_{j}(j+\epsilon)/\Gamma(-\epsilon)
\ee
From (\ref{phitoglb2}) we find the formal result
\be\label{vconf}
\tilde{\mc{V}}_{0}(0)\,=\,\mathbbm{1}\int d^2 x/|x|^2\,=\,\mathbbm{1}\,V_{\rm conf}
\ee
Here \(V_{\rm conf}\) is the (divergent) volume of the global conformal subgroup\footnote{It should be mentioned that the divergent factor \(V_{\rm conf}\) is required~\cite{MaldacenaOoguri_0111} to define the boundary two-point function (equivalently to give meaning to the string two-point function) in the AdS/CFT correspondence. However, there is no apparent need for its inclusion in the \(SL(2,R)\) CFT.} which leaves two points fixed on the Riemann sphere, and may be taken to be \(\pi\,\lim_{\epsilon\goto 0}\Gamma(-\epsilon)\) under analytic continuation.

The two-point function (\ref{glb2pt}) may be seen to be continuum normalizable with positive norm for \(j\in\mc{S}^{+}_{\sst \mc{C}}\,\), while (\ref{sl2pt}) may be seen to be continuum normalizable with positive norm for \(j\in\mc{S}_{\sst \mc{D}}\,\). The associated norm is given by 
\be
\brkt{\mc{V}_{j}(m)}{\mc{V}_{j'}(m')}\,=\,\amp{3.5mm}{\hat{\wbr{\mc{V}}}_{j}(m)\,\mc{V}_{j'}(m')}
\ee
The complex conjugation map \(\mc{V}_{j}(m)\goto\wbr{\mc{V}}_{j}(m)\) acts as \(\wbr{\mc{V}}_{j}(m)=\mc{V}_{1-j}(-m)\) for \(j=1-\bar{\jmath}\in\mc{S}^{+}_{\sst \mc{C}}\,\), and \(\wbr{\mc{V}}_{j}(m)=\mc{V}_{j}(-m)\) for \(j=\bar{\jmath}\in\mc{S}_{\sst\mc{D}}\,\). As for the quantum mechanical expression (\ref{l2vnrm}), (\ref{dsc2pf}) has no reflection term (\(j\goto 1-j\)) since the relation (\ref{glbref}) degenerates (\(R_{j}(m)=0\)) for elements of \(\mc{D}^{\pm}_{j}\,\). Note that while the operator \(\mc{V}_{j}(\pm(1-j+n))\) for \(j\in\mc{S}_{\sst \mc{D}}\,\) and \(n\in\mathbb{Z}_{\geq 0}\,\) is non-normalizable in the \(SL(2,R)\) CFT, the operator \(\mc{V}_{1-j}(\pm(j+n))\) merely suffers from a singular vertex operator normalization. As in (\ref{wref}), it is consequently helpful to introduce the vertex operators\be\label{vowref}
\mc{W}_{j}(m)\,=\,\mc{W}_{1-j}(m)\,=\,\Omega_{1-j}(m)\mc{V}_{j}(m)
\ee
where,
\be\label{vowomg}
\Omega_{j}(m)\,=\,R_{j}(m)\,\Omega_{1-j}(m)\,=\,\nu^{1-j}\,
\frac{\Gamma(1-b^2(2j-1))\,\Gamma(2j-1)}
{\Gamma(1-b^2)\,\Gamma(j-m)\,\Gamma(j+\bar{m})}
\ee
The choice here has been made so that \(\lim_{j\goto 0}\mc{W}_{j}(j)=\mathbbm{1}\,\). It may be shown that for both \(\mc{C}^{\alpha}_{j}\) and \(\mc{D}^{\pm}_{j}\) the relationship between the norm and the two-point function is given by
\be\label{wnrm}
\brkt{\mc{W}_{j}(m)}{\mc{W}_{j'}(m')}\,=\,
(\Omega_{1-j}(m))^2\amp{3.5mm}{\hat{\mc{V}}_{j}(-m)\,
\mc{V}_{j'}(m')}\,=\,\frac{\Omega_{1-j}(m)}{\Omega_{1-j}(-m)}
\amp{3.5mm}{\hat{\mc{W}}_{j}(-m)\,\mc{W}_{j'}(m')}
\ee
For elements of \(\mc{C}^{\alpha}_{j}\) with \(j=1/2+is\,\) and \(m=\alpha+p\,\) for \(p\in\mathbb{Z}\,\) and \(\alpha\in(-\txfc{1}{2},\txfc{1}{2})\,\) we compute
\bbb\label{contw2pf}
\lefteqn{\amp{3.5mm}{\hat{\mc{W}}_{j}(-m)\,\mc{W}_{j'}(m')}\,=\,
(-1)^{p+\bar{p}}\,\left|\frac{\Gamma(j+\alpha+\bar{p})}{\Gamma(j+\alpha+p)}\right|^2
\brkt{\mc{W}_{j}(m)}{\mc{W}_{j'}(m')}}\nonumber\\[0.1cm]
&=&\nu\,\frac{\Gamma(1+b^2)}{\Gamma(1-b^2)}\,\frac{\pi^{-1}\sin(\pi b^2)}{\sinh(2\pi sb^2)}\,
\frac{\cosh(2\pi s)+\cos(2\pi\alpha)}{\sinh(2\pi s)}\,(-1)^{p+\bar{p}}\,\delta_{pp'}
\delta_{\bar{p}\bar{p}'}\delta(\alpha-\alpha')\delta(s-s')\hspace{0.5cm}
\eee
Here we have fixed \(j,j'\in\mc{S}^{+}_{\sst \mc{C}}\,\) since reflected values follow from \(\mc{W}_{j}(m)=\mc{W}_{1-j}(m)\,\). For elements of \(\mc{D}^{-}_{j}\) and \(\mc{D}^{+}_{j'}\) with \(m=j+n\,\) for \(n\in\mathbb{Z}_{\geq 0}\,\) the two-point function is given by
\bbb\label{discw2pf}
\amp{3.5mm}{\hat{\mc{W}}_{j}(-m)\,\mc{W}_{j'}(m')}&=&(-1)^{n+\bar{n}}
\,\frac{\Gamma(1+\bar{n})\,\Gamma(2j+\bar{n})}{\Gamma(1+n)\,\Gamma(2j+n)}
\brkt{\mc{W}_{j}(m)}{\mc{W}_{j'}(m')}\nonumber\\[0.1cm]
&=&\nu\,\frac{\Gamma(1+b^2)}{\Gamma(1-b^2)}\,\frac{\pi^{-1}\sin(\pi b^2)}{\sin(\pi b^2(2j-1))}
\,(-1)^{n+\bar{n}}\,\delta_{nn'}\,
\delta_{\bar{n}\bar{n}'}\,\delta(j-j')\hspace{0.5cm}
\eee
Note that the norm (\ref{wnrm}) diverges at the boundaries of \(\mc{S}_{\sst \mc{D}}\,\), and agrees precisely with (\ref{qmwnrm}) in the limit \(k\goto\infty\,\) with \(\nu\) given by (\ref{Anrm}). Again, here we have fixed \(j,j'\in\mc{S}_{\sst \mc{D}}\,\), making use of the symmetry under the reflection (\ref{vowref}). Thus, while the basis element \(\mc{V}_{1-j}(\pm(j+n))\) degenerates because \(R_{j}(\pm(j+n))=0\,\), the operator \(\mc{W}_{j}(\pm(j+n))=\mc{W}_{1-j}(\pm(j+n))\in\mc{D}^{\pm}_{j}\,\) is non-degenerate. However, care must be taken since for \(j\in\mc{S}_{\sst \mc{D}}\) the operator \(\mc{W}_{j}(\pm(1-j+n))\) is not normalizable in the \(SL(2,R)\) CFT. In this regard it should be noted that while (\ref{contw2pf}) is invariant under \(j\goto 1-j\), this is of course not true of (\ref{discw2pf}).

\subsection{Parafermions and spectral flow conserving primary correlators\label{sfcons_sec}}

As discussed below, it may be shown that all correlation functions of primary states of definite affine weight which conserve spectral flow, that is those which satisfy \(\sum_{i=1}^{N}w_{i}=0\) for an \(N\)-point function involving \(\mc{V}_{j_{i}}^{w_{i}}(m_{i})\,\), have the same form as for the corresponding unflowed correlators, except for \(z_{i}\)-dependent factors which depend on the spectral flowed dimensions of the vertex operators. Some insight into the relationship between unflowed correlators and spectral flow conserving correlators may be gained by expressing the primary fields \(\mc{V}_{j}^{w}(m)\) in terms of parafermions \(\Psi_{j}(m)\) as 
\be\label{paradef}
\mc{V}_{j}^{w}(m)\,=\,\Psi_{j}(m)\,
e^{i\sqrt{2/k}\,(m+\frac{1}{2}kw)\phi}
\ee
where the anti-holomorphic dependence has been suppressed. The unflowed holomorphic currents are expressed as 
\be
J^3\,=\,-i\sqrt{k/2}\,\partial\phi\hspace{2cm}J^{\pm}\,=\,\psi^{\pm}\,e^{\pm i\sqrt{2/k}\,\phi}
\ee
where \(\psi^{\pm}\) are the parafermions of the adjoint representation, and the Hermitian field \(\phi\) satisfies the free OPE \(\,\phi(w)\phi(z)\sim \ln(z-w)\,\), with a sign associated with a timelike boson. Denoting the \(J^{3}_{0}\) charge by \(M\) and recognizing that
\be
M(\mc{V}_{j}^{w}(m))\,=\,
M(\,e^{i\sqrt{2/k}\,(m+\frac{1}{2}kw)\phi}\,)\,=\,m+\txfc{1}{2}kw
\ee
it follows that \(M(\Psi_{j}(m))=0\,\). Furthermore, denoting the \(L_{0}\) eigenvalue by \(H\,\), it may be shown that
\be
H(\,e^{i\sqrt{2/k}\,(m+\frac{1}{2}kw)\phi}\,)\,=\,-(m+\txfc{1}{2}kw)^2/k
\ee
Thus, since
\be
H(\mc{V}_{j}^{w}(m))\,=\,\frac{j(1-j)}{k-2}\,-mw\,-\txfc{1}{4}kw^2
\ee
it follows that
\be
H(\Psi_{j}(m))\,=\,\frac{j(1-j)}{k-2}\,+\,m^2/k
\ee
It is clear that all dependence on the spectral flow sector is contained in the free boson exponential in (\ref{paradef}). More generally, the parafermions describe the \(SL(2,R)/U(1)\) coset model, and all dependence on the timelike direction is contained in the free-boson factors. As mentioned above, there is the following physical equivalence between primary vertex operators
\be\label{weylprimeq2}
\mc{V}_{j}^{w}(\pm j)\,=\,\omega_{j}\mc{V}_{\tilde{\jmath}}^{w\pm 1}(\mp\tilde{\jmath})
\ee
with \(\tilde{\jmath}=k/2-j\,\). Since the arguments of the free boson exponentials in (\ref{weylprimeq2}) are identical, the respective parafermions are related by
\be\label{paraeq}
\Psi_{j}(\pm j)\,=\,\omega_{j}\Psi_{\tilde{\jmath}}(\mp\tilde{\jmath})
\ee 
The normalization parameter \(\omega_{j}\) may be determined from the discrete unflowed two-point function (\ref{sl2pt}). Since the Mobius fixed two-point function is independent of the spectral flow sector,
\be
\amp{3.2mm}{\hat{\mc{V}}_{j}^{-w}(-j)\,\mc{V}_{j'}^{w}(j')}\,=\,
\amp{3.2mm}{\hat{\mc{V}}_{j}(-j)\,\mc{V}_{j'}(j')}\,=\,B_{j}\,\frac{\delta(j-j')}{(2j-1)^2}
\ee
Considering (\ref{weylprimeq2}) this may be written as  
\be
\amp{3.2mm}{\hat{\mc{V}}_{j}^{-w}(-j)\,\mc{V}_{j'}^{w}(j')}\,=\,
\omega_{j}^2\amp{3.2mm}{\hat{\mc{V}}^{-w-1}_{\tilde{\jmath}}(\tilde{\jmath}\,)\,
\mc{V}^{w+1}_{\tilde{\jmath}\,'}(-\tilde{\jmath}\,')}\,=\,
\omega_{j}^2\,B_{\tilde{\jmath}}\,\frac{\delta(\tilde{\jmath}-\tilde{\jmath}')}{(2\tilde{\jmath}-1)^2}
\ee
Since \(\delta(j-j')=\delta(\tilde{\jmath}-\tilde{\jmath}\,')\,\), using \(B_{j}B_{\tilde{\jmath}}=\pi^{-2}b^{-4}\nu^{2-k}\,\), it follows that 
\be
\omega_{j}\,=\,\omega^{-1}_{\tilde{\jmath}}\,=\,
\nu^{k/2-1}\,b^2\pi B_{j}\,\frac{(2\tilde{\jmath}-1)}{(2j-1)}\,=\,
\frac{\nu^{\tilde{\jmath}}\,\Gamma(1+b^2(2\tilde{\jmath}-1))}
{\nu^{j}\,\Gamma(1+b^2(2j-1))}
\ee
where the choice of sign has been made so that for \(j=\tilde{\jmath}=k/4\) we have \(\Psi_{j}(j)=\Psi_{j}(-j)\,\). The relation (\ref{paraeq}) implies
\be
\mc{V}_{0}(0)\,=\,\Psi_{0}(0)
\,=\,\omega_{0}\,\Psi_{k/2}(\pm \txfc{1}{2}k)\,=\,\mathbbm{1}
\ee
and leads to the definition of the one-unit spectral flow operator\footnote{As in the case of (\ref{dsc2pf}), the normalization associated with the transform (\ref{phitoglb}) avoids the introduction~\cite{MaldacenaOoguri_0111} of a divergent factor in the definition of the one-unit spectral flow operator.}
\be\label{onesfop}
\omega_{0}\,\mc{V}_{k/2}(\pm \txfc{1}{2}k)\,=\,e^{\pm i\sqrt{k/2}\,\phi}
\ee
which is a non-normalizable vertex operator arising via analytic continuation from \(\mc{D}^{\pm}_{0}\,\). Defining \(M=m+\frac{1}{2}kw\,\), the operator (\ref{onesfop}) may be seen to act as\footnote{For \(w=0\) this definition is essentially equivalent, including the absence of a divergent factor, to that given in terms of the \(\Phi_{j}(x)\) operators in appendix~E of~\cite{MaldacenaOoguri_0111}. However, the operators are transposed in the definition given here as required for consistency with the equivalences appearing in subsection~\ref{sfequiv_sec}\,. In particular, the above definition leads to an expression for the \(w=1\) three-point amplitude (\ref{onesf}) which respects Mobius invariance.}
\be\label{oneunitsf}
\mc{V}^{w\pm 1}_{j}(m)\,=\,\omega_{0}\,\lim_{z\goto 0}\,z^{\pm M}\,\bar{z}^{\pm \bar{M}}\,
\mc{V}^{w}_{j}(m|z)\,\mc{V}_{k/2}(\pm \txfc{1}{2} k)
\ee
For the corresponding reflection symmetric operators \(\mc{W}_{j}^{w}(m)\) (\ref{vowref}) it may be shown that the choice (\ref{vowomg}) leads to
\be
\mc{W}_{j}^{w}(\pm j)\,=\,\mc{W}_{\tilde{\jmath}}^{w\pm 1}(\mp\tilde{\jmath})
\ee
and 
\be
\mc{W}_{k/2}(\pm \txfc{1}{2}k)\,=\,e^{\pm i\sqrt{k/2}\,\phi}
\ee

Consider an unflowed \(N\)-point correlator with 
\(\mc{V}_{j_{i}}(m_{i})\in\mc{D}^{+}_{0}\!\oplus\mc{D}^{-}_{0}\!\oplus\mc{C}_{0}\)
\be\label{unflwpar}
\amp{3.2mm}{{\textstyle \prod_{i=1}^{N}}\mc{V}_{j_{i}}(m_{i})}\,=\,
\amp{3.2mm}{{\textstyle \prod_{i=1}^{N}}\Psi_{j_{i}}(m_{i})}\,
\amp{3.2mm}{{\textstyle \prod_{i=1}^{N}}e^{i\,m_i\sqrt{2/k}\,\phi}}
\ee
The free boson correlator requires \(\sum_{i=1}^{N}m_{i}=0\,\). Its computation, given knowledge of the unflowed correlator, leads to an expression for the parafermion correlator. This in turn allows for the computation of the corresponding flowed correlator
\be\label{flwcnsvpar}
\amp{3.2mm}{{\textstyle \prod_{i=1}^{N}}\mc{V}^{w_i}_{j_{i}}(m_{i})}\,=\,
\amp{3.2mm}{{\textstyle \prod_{i=1}^{N}}\Psi_{j_{i}}(m_{i})}\,
\amp{3.2mm}{{\textstyle \prod_{i=1}^{N}}e^{i\,(m_i+\frac{1}{2}kw_{i})\sqrt{2/k}\,\phi}}
\ee
under the condition \(\sum_{i=1}^{N}w_{i}=0\,\) of spectral flow conservation, which is required for a non-vanishing result for the free boson correlator in (\ref{flwcnsvpar}). The treatment of amplitudes which do not conserve spectral flow is more subtle and, as described below, may be derived from the above definition of the spectral flow operator, or from the relation of the \(H_3\) correlators to those in Liouville theory. In both the spectral flow conserving and non-conserving cases, since the discussion here is aimed at the computation of the lowest order term in the OPE, only the two-point and three-point amplitudes appearing in the selection rules (\ref{opeslct1}\,-\,\ref{opeslct3}) will be considered.

\subsection{Equivalence of primary correlators with fixed total spectral flow \label{sfequiv_sec}}

We first consider spectral flow conserving correlation functions of flowed affine primary operators \(\phi^{w}\) which satisfy \(L_{0}(w)\cdot\phi^{w}=h\,\phi^{w}\,\) and \(J^{a}_{0}(w)\cdot\phi^{w}=-t^{a}\phi^{w}\,\) for some set of \(SL(2,R)\) generators \(t^{a}\,\). These satisfy \(L_{1}\cdot\phi^{w}=(L_{1}(w)-wJ_{1}^{3})\cdot\phi^{w}=0\,\) but are not in general of definite global conformal weight\footnote{However, to ensure mutual locality, the operator \((-1)^{L_{0}-\bar{L}_{0}}\,\) acts as an involution on all fields. This fact impacts many of the relations given in this section which involve fields of non-zero worldsheet spin.} under \(L_{0}\,\). As discussed in~\cite{Ribault_0507}\,, 
the identity for the flowed Virasoro generator \(L_{-1}(w)\)
\be\label{kz2pf}
\left(b^{-2}L_{-1}+(wb^{-2}-2t^3)J^{3}_{-1}+t^{-}J^{+}_{w-1}+t^{+}J^{-}_{-w-1}\right)\cdot\phi^{w}\,=\,0
\ee
which follows from the Sugawara construction of the stress tensor, leads to a twisted form of the Knizhnik-Zamolodchikov equation for spectral flow conserving correlation functions of flowed affine primary operators. These correlators are related to  unflowed correlators via a simple twist 
\be\label{sfcons}
\amp{3.5mm}{\txprd_{p}^{n}\,\phi_{p}^{w_p}(z_p)}\,=\,\kappa
\amp{3.5mm}{\txprd_{p}^{n}\,\phi_{p}(z_p)}
\ee
where, with \(\sum_{p}^{n}w_p=0\,\) and \(\sum_{p}^{n}t_{p}^{3}=0\,\), suppressing anti-holomorphic dependence
\be
\kappa\,=\,\prod_{p}^{n}\,\rho_{p}^{t^3_p-\frac{1}{4}kw_p}
\ee
and
\be
\rho_{p}\,=\,\prod_{q\neq p}^{n}\,(z_{p}-z_q)^{w_q}
\ee
Note that \(\kappa\) is symmetric under operator exchange as required by crossing symmetry. 

Given (\ref{twoptdesc}), for the two-point function we have 
\be
\amp{3.5mm}{\phi^{-w}_{2}(z_2)\phi^{w}_{1}(z_1)}\,=\,
(-z_{21}^{-2})^{h_1+wt^{3}_{1}-\frac{1}{4}kw^2}\amp{3.5mm}{\hat{\phi}^{-w}_{2}\phi^{w}_{1}}
\ee
where \(h_1=h_2\,\) due to global \(J_{0}^{3}\) conservation. Then (\ref{sfcons}) leads to
\be\label{sfcons2pf}
\amp{3.5mm}{\hat{\phi}^{-w}_{2}\phi^{w}_{1}}\,=\,\amp{3.5mm}{\hat{\phi}_{2}\,\phi_{1}\!}
\ee
The spectral flow conserving two-point function in the \(\mc{V}_{j}^{w}(m)\) basis is given by
\be
\amp{3.5mm}{\mc{V}^{-w}_{j_2}(m_2|z_2)\,\mc{V}^{w}_{j_1}(m_1|z_1)}\,=\,
(-z_{21}^{-2})^{H_{1}}\amp{3.5mm}{\hat{\mc{V}}_{j_2}(m_2)\,\mc{V}_{j_1}(m_1)}
\ee
Here \(H_1=H_2=h_1-m_1 w-\frac{1}{4}kw^2\,\) with \(h_1=h_2=j_1(1-j_1)/(k-2)\,\). In the \(\varphi_{j}^{w}(\mu)\) basis we have
\be
\amp{3.5mm}{\varphi^{-w}_{j_2}(\mu_2|z_2)\,\varphi^{w}_{j_1}(\mu_1|z_1)} \,=\,
(-z_{21}^{-2})^{h_1-\frac{1}{4}kw^2}\amp{3.5mm}{\hat{\varphi}_{j_2}(\mu_{2w})\,\varphi_{j_1}(\mu_1)}
\ee
where \(\mu_{2w}=(-z_{21}^{-2})^{-w} \mu_{2}\,\). Note the non-trivial worldsheet coordinate dependence, which reflects the fact that these fields do not have have definite \(L_0\) weight. Here we have used
\be
\alpha^{L_0}\,\varphi^{w}_{j}(\mu)\,=\,\alpha^{h-\frac{1}{4}kw^2}\,\varphi^{w}_{j}(\alpha^w\mu)
\ee
In the \(\Phi_{j}^{w}(x)\) basis\footnote{Note that this basis is different than that which appears in~\cite{MaldacenaOoguri_0111}, where the \(x\) coordinate is associated with the global generators \(J^{a}_{0}\) acting as in (\ref{phigen}), and the corresponding flowed fields are in discrete global representations. Here \(\Phi_{j}^{w}(x)\,\), which are both Virasoro and flowed affine primaries, are Fourier transforms of the fields \(\varphi_{j}^{w}(\mu)\,\) which appear in~\cite{Ribault_0507}, and are acted on by the flowed generators \(J^{a}_{0}(w)\,\) as in (\ref{phigen}). In particular, the fields \(\mc{V}^{w}_{j}(m)\) are the modes of \(\Phi_{j}^{w}(x)\,\) given by the transform (\ref{phitoglb2}).} the two-point function takes the form
\be
\amp{3.5mm}{\Phi^{-w}_{j_2}(x_2|z_2)\,\Phi^{w}_{j_1}(x_1|z_1)}\,=\,
(-z_{21}^{-2})^{h_1-wj_1-\frac{1}{4}kw^2}
\amp{3.5mm}{\hat{\Phi}_{j_2}(x_2)\,\Phi_{j_1}(x_{1w})}
\ee
where \(x_{1w}=(-z_{21}^{-2})^{-w} x_{1}\,\).

For flowed primary fields the three-point function takes the form
\be
\amp{3.5mm}{{\textstyle \prod}_{p=1}^{3}\phi_{p}^{w_p}(z_p)}\,=\,(-1)^{H_3}
\prod_{p>q}z_{pq}^{H_{pq}}\,
\amp{3.5mm}{\hat{\phi}_{3}^{w_3}\phi_{2}^{w_2}(1)\,\phi_{1}^{w_1}\!}
\ee
where \(H_p=h_p+w_p t^{3}_{p}-\frac{k}{4}w_p^2\,\), \(h_p=j_p(1-j_p)/(k-2)\,\), \(H_{pq}=\hat{H}-2H_p-2H_q\,\) and \(\hat{H}=\sum_{p}H_{p}\,\).
For the spectral flow conserving case, using \(h_p=\bar{h}_p\,\) and \(w_p=\bar{w}_p\,\), (\ref{sfcons}) leads to
\be
\amp{3.5mm}{\hat{\phi}_{3}^{w_3}\phi_{2}^{w_2}(1)\,\phi_{1}^{w_1}\!}\,=\,
(-1)^{w_2t_{1}^{3}}\amp{3.5mm}{\hat{\phi}_{3}\,\phi_{2}(1)\,\phi_{1}\!}
\ee
For the flowed primaries \(\mc{V}_{j}^{w}(m)\) of definite global conformal weight this is expressed as
\be
\amp{3.5mm}{\prod_{p}^{3}\mc{V}_{j_p}^{w_p}(m_p|z_p)}\,=\,
\prod_{p}^{3}\rho_{p}^{-m_p}\amp{3.5mm}{\hat{\mc{V}}_{j_3}(m_3)
\mc{V}_{j_2}(m_{2}|1)
\mc{V}_{j_1}(m_{1})}
\prod_{p>q}z_{pq}^{h_{pq}-\frac{k}{2}w_{p}w_{q}}
\ee 
while for the flowed primaries \(\varphi_{j}^{w}(\mu|z)\,\),
\be
\amp{3.5mm}{\prod_{p}^{3}\varphi_{j_p}^{w_p}(\mu_p|z_p)}\,=\,
\amp{3.5mm}{\hat{\varphi}_{j_3}(\rho_3\mu_3)
\varphi_{j_2}(\rho_2\mu_{2}|1)
\varphi_{j_1}(\rho_1\mu_{1})}
\prod_{p>q}z_{pq}^{h_{pq}-\frac{k}{2}w_{p}w_{q}}
\ee
and, using \(j_{p}=\bar{\jmath}_{p}\,\), for the flowed primaries \(\Phi_{j}^{w}(x|z)\,\) we have
\be
\amp{3.5mm}{\prod_{p}^{3}\Phi_{j_p}^{w_p}(x_p|z_p)}\,=\,\amp{3.5mm}{\hat{\Phi}_{j_3}(x_3/\rho_3)
\Phi_{j_2}(x_{2}/\rho_2|1)
\Phi_{j_1}(x_{1}/\rho_1)}
\prod_{p>q}z_{pq}^{h_{pq}-\frac{k}{2}w_{p}w_{q}-w_q j_p-w_p j_q}
\ee

Spectral flow non-conserving primary correlators 
\be
\amp{3mm}{{\textstyle \prod}_{p=1}^{n}\,\phi_{p}^{w_p}(z_p)}
\ee
are also discussed in~\cite{Ribault_0507}. Denoting \(\hat{w}=\sum_p w_p\) to be the total spectral flow\footnote{This notation has been recycled from the discussion in \ref{subsec3ptred} above, where \(\hat{w}\) referred to the total spectral flow assigned to the corresponding flowed affine modules, while that of the primary fields, which could differ in the case of the appearance of discrete states, was denoted by \(\tilde{w}\,\).} of the primary fields in a particular correlation function, consider the operators
\be\label{gamops}
\Gamma^{\pm}_{\alpha}\,=\,
\int_{\infty}dz\,J^{\pm}(z)\,z^{\alpha}\,{\textstyle \prod}_{p=1}^{n}(z-z_{p})^{\pm w_p}
\ee
It may be seen that the following sets vanish for \(\mp\hat{w}\geq 0\)
\be
\{\Gamma^{\pm}_{0},\ldots,\Gamma^{\pm}_{\mp\hat{w}}\}
\ee
This leads to two independent relations between spectral flowed affine primary fields for \(|\hat{w}|=0\,\), and \(|\hat{w}|+1\,\) relations for \(|\hat{w}|>0\,\). For \(|\hat{w}|>0\,\), the disappearance of \(|\hat{w}|-1\) linearly independent equations associated with the spectral flowed KZ equations is compensated for by the extra relations provided by the vanishing operators (\ref{gamops}), allowing for the computation of all correlation functions of primary fields in all spectral flow sectors permitted by the selection rules.

In the case of the three-point function, it may be seen that the selection rules for flowed primaries impose \(-1\leq \hat{w}\leq 1\,\), where  \(\hat{w}=\sum_{p=1}^{3}w_{p}\,\). Suppressing anti-holomorphic dependence and using the Mobius symmetry,
\be
\amp{3mm}{{\textstyle \prod}_{p=1}^{3}\mc{V}^{w_p}_{j_p}(m_p|z_p)}\,=\,
\kappa(w_p)\,G(\hat{w})\,=\,
\amp{3mm}{\hat{\mc{V}}^{w_3}_{j_3}(m_3)\mc{V}^{w_2}_{j_2}(m_2|1)\mc{V}^{w_1}_{j_1}(m_1)}
(-1)^{H_3}\prod_{p>q}z_{pq}^{H_{pq}}\,
\ee
Here \(G(\hat{w})=G(\hat{w}|j_p,m_p,z_p)\) is  independent of \(w_p\) for fixed \(\hat{w}\,\). Note that
\be
\kappa(w_p)\,=\,\prod_{p\neq q}(z_{pq})^{-w_q(m_p+\frac{k}{4}w_p)}
\ee
and thus \(G(\hat{w})\,\), are symmetric functions under operator exchange. This leads to
\be
G(\hat{w})\,=\,\amp{3mm}{\hat{\mc{V}}^{w_3}_{j_3}(m_3)\mc{V}^{w_2}_{j_2}(m_2|1)\mc{V}^{w_1}_{j_1}(m_1)}
(-1)^{m_1w_2}\prod_{p>q}z_{pq}^{h_{pq}+\hat{w}(m_p+m_q+\frac{k}{4}\hat{w})}
\ee
This implies that for two Mobius-fixed three point functions of equal \(\hat{w}\)
\be
(-1)^{m_1w_2}\amp{3mm}{\hat{\mc{V}}^{w_3}_{j_3}(m_3)\mc{V}^{w_2}_{j_2}(m_2|1)\mc{V}^{w_1}_{j_1}(m_1)}
\,=\,(-1)^{m_1w'_2}\amp{3mm}{\hat{\mc{V}}^{w'_3}_{j_3}(m_3)\mc{V}^{w'_2}_{j_2}(m_2|1)\mc{V}^{w'_1}_{j_1}(m_1)}
\ee
Thus,
\be\label{3pfequiv}
\amp{3mm}{\hat{\mc{V}}^{w_3}_{j_3}(m_3)\mc{V}^{w_2}_{j_2}(m_2|1)\mc{V}^{w_1}_{j_1}(m_1)}
\,=\,(-1)^{m_1w_2}\amp{3mm}{\hat{\mc{V}}_{j_3}(m_3)\mc{V}_{j_2}(m_2|1)\mc{V}^{\hat{w}}_{j_1}(m_1)}
\ee
Here only holomorphic dependence appears so that the factor \((-1)^{m_1 w_2}\) denotes the sign \((-1)^{\ell_1 w_2}\,\), where \(\ell_1=m_1-\bar{m}_1\in\mathbb{Z}\,\).

\subsection{Spectral flow non-conserving three-point primary correlator \label{sfnoncns_sec}}

The relation (\ref{oneunitsf}) may be used to determine the \(\sum_{i=1}^{3}w_{i}=\pm 1\) spectral flow non-conserving three-point function. Along with the spectral flow conserving amplitudes considered above, these determine the OPE coefficients for primary fields in the spectrum of the \(SL(2,R)\) CFT. We would like to compute
\be
\amp{3mm}{\hat{\mc{V}}_{j_3}(m_3)\mc{V}_{j_2}(m_2|1)\mc{V}^{\pm 1}_{j_1}(m_1)}\,=\,
\amp{3mm}{\hat{\mc{V}}_{j_3}(-m_3)\mc{V}_{j_2}(-m_2|1)\mc{V}^{\mp 1}_{j_1}(-m_1)}
\ee
where we have made use of the \(J^{3}_{0}\) charge conjugation symmetry. Thus, suppressing anti-holomorphic dependence, it suffices to compute
\bbb\label{sfoneamp}
\lefteqn{
\amp{3mm}{\hat{\mc{V}}_{j_3}(m_3)\mc{V}_{j_2}(m_2|1)\mc{V}^{+1}_{j_1}(m_1)}\,=\,}\hspace{3cm} \nonumber\\[0.2cm]
&&\omega_{0}\,\lim_{z\goto 0}\,z^{m_{1}}\,\bar{z}^{\bar{m}_{1}}
\amp{3mm}{\hat{\mc{V}}_{j_3}(m_3)\mc{V}_{j_2}(m_2|1)\mc{V}_{j_1}(m_1|z)\mc{V}_{k/2}(k/2)}
\eee
It is perhaps simplest to perform this computation in the \(\Phi^{w}_{j}(x)\) basis introduced in subsection~\ref{sfequiv_sec}\,. The computation equivalent to (\ref{sfoneamp}) may be shown to be
\bbb\label{sfoneampphi}
\amp{3mm}{\hat{\Phi}_{j_3}(x_3)\Phi_{j_2}(x_2|1)\Phi^{+1}_{j_1}(x_1)}&=&
\frac{\omega_{0}}{(1-k)}\lim_{z\goto 0}\,
z^{j_1}\amp{3mm}{\hat{\Phi}_{j_3}(x_3)\Phi_{j_2}(x_2|1)\Phi_{j_1}(x_1 z|z)\,\Phi_{k/2}(0)}\hspace{1cm}
\eee
where the identification \(\mc{V}_{j}(j)=\Phi_{j}(0)/(1-2j)\,\) has been used. Making use of
\bbb
\lefteqn{\amp{3mm}{\hat{\Phi}_{j_4}(x_4)\Phi_{j_3}(x_3|1)\Phi_{j_2}(x_2|z)\,\Phi_{j_1}(x_1)}\,=\,
x_{42}^{-2 j_2}\,x_{41}^{j_2+j_3-j_1-j_4}\,x_{43}^{j_1+j_2-j_3-j_4}\,x_{31}^{j_4-j_3-j_2-j_1}}
\hspace{3cm}\nonumber\\[0.2cm]&&\hspace{2cm}
\amp{3mm}{\hat{\Phi}^{r}_{j_4}(0)\Phi_{j_3}(1|1)\Phi_{j_2}((x_{21}x_{43})/(x_{42}x_{31})|z)\,\Phi_{j_1}(0)}\hspace{2cm}
\eee
and, with \(z_1=-z_2/(1-z_2)\,\) and \(x_1=-x_2/(1-x_2)\,\),
\bbb
\lefteqn{\amp{3mm}{\hat{\Phi}^{r}_{j_4}(0)\Phi_{j_3}(1|1)\Phi_{j_2}(x_2|z_2)\,\Phi_{j_1}(0)}\,=\,}
\hspace{1cm} \nonumber\\[0.2cm]&&(1-z_2)^{h_4-h_1-h_2-h_3}\,(1-x_2)^{j_4-j_1-j_2-j_3}
\amp{3mm}{\hat{\Phi}^{r}_{j_4}(0)\Phi_{j_3}(1|1)\Phi_{j_1}(x_1|z_1)\,\Phi_{j_2}(0)}\hspace{2cm}
\eee
the amplitude (\ref{sfoneampphi}) may be computed \cite{MaldacenaOoguri_0111} from
\be\label{sfoneampx}
\amp{3mm}{\hat{\Phi}^{r}_{j_3}(0)\Phi_{j_2}(1|1)\Phi_{k/2}(x|z)\Phi_{j_1}(0)}\,=\,
F(j_1,j_2,j_3)\,|\mc{F}(x|z)|^2
\ee
where we have anticipated, as shown explicitly below, a factorization of the four point function involving a single conformal block. Due to the presence of the degenerate field \(\Phi_{k/2}(x|z)\) the amplitude (\ref{sfoneampx}) satisfies the null state decoupling equation (\ref{nullphi}) in the form 
\be\label{nulldiffeq}
\left(\frac{\partial\,}{\partial x}+
\frac{j_{1}(x^2-2x+z)+j_{2}(x^2-z)-(j_{3}-k/2)(x^2-2xz+z)}{x(1-x)(z-x)}\right)\mc{F}(x|z)\,=\,0
\ee
In addition (\ref{sfoneampx}) satisfies the KZ equation with one null state (\ref{kzphi2}) in the form
\be\label{kzdiffeq}
\left(\frac{\partial\,}{\partial z}+\frac{x(1-x)}{z(1-z)}\frac{\partial\,}{\partial x}+
\frac{j_{1}(1-x)-j_{2}\,x-(j_{3}-k/2)(z-x)}{z(1-z)}\right)\mc{F}(x|z)\,=\,0
\ee
These may be combined to yield 
\be\label{combdiffeq}
\left(\frac{\partial\,}{\partial z}+
\frac{j_{1}x(1-z)+j_{2}(1-x)z+(j_{3}-k/2)z(1-z)}{z(1-z)(z-x)}\right)\mc{F}(x|z)\,=\,0
\ee
The solution to (\ref{nulldiffeq}) and (\ref{combdiffeq}) may be seen to be
\be\label{sfonecnfblk}
\mc{F}(x|z)\,=\,z^{j_1}(1-z)^{j_2}(z-x)^{k/2-j_1-j_2-j_3}x^{j_2-j_1+j_3-k/2}(1-x)^{j_1-j_2+j_3-k/2}
\ee
up to a constant which is incorporated into \(F(j_1,j_2,j_3)\,\). This leads to the following (worldsheet Mobius fixed) form of the \(w=1\) three-point function in the \(\Phi^{w}_{j}(x)\) basis
\bbb
\lefteqn{\amp{3mm}{\hat{\Phi}_{j_3}(x_3)\Phi_{j_2}(x_2|1)\Phi^{+1}_{j_1}(x_1)}\,=\,}
\hspace{2cm}\nonumber\\[0.2cm]&&\frac{\omega_{0}}{(1-k)}\,F(j_1,j_2,j_3)\ 
x_1^{-2 j_1}\,x_2^{-2 j_2}\,x_3^{-2j_3}\,(x_3^{-1}+x_1^{-1}-x_2^{-1})^{k/2-j_1-j_2-j_3}\hspace{2cm}
\eee
Finally, from the transform (\ref{phitoglb2}) and the integral (\ref{int2}), the \(w=\pm 1\) three-point function in the \(\mc{V}^{w}_{j}(m)\) basis is given by
\bbb\label{onesf}
\lefteqn{\amp{3mm}{\hat{\mc{V}}_{j_3}(m_3)\mc{V}_{j_2}(m_2|1)\mc{V}^{\pm 1}_{j_1}(m_1)}
\,=\,\frac{\pi\omega_{0}}{(k-1)}\,\frac{F(j_1,j_2,j_3)}{\gamma(\hat{\jmath}-k/2)}\nonumber}\\[0.1cm]& & \hspace{2.5cm}\times\ 
\pi^{-2}\,\delta^2(\hat{m}\pm k/2)\,(-1)^{m_2-\bar{m}_2}\;{\textstyle \prod}_{i=1}^{3}\;
\frac{\Gamma(2j_{i}-1)}{\Gamma(1-2j_{i})}\frac{\Gamma(1-j_{i}\pm\bar{m}_{i})}{\Gamma(j_{i}\mp m_{i})}
\eee
Here the definitions \(\hat{m}=\sum_{i=1}^3 m_i\) and \(\hat{\jmath}=\sum_{i=1}^3 j_i\,\) have been used. Note that this amplitude is not symmetric under \((j_2,m_2)\leftrightarrow(j_3,m_3)\) due to the factor \((-1)^{m_2-\bar{m}_2}\) which enforces Mobius invariance given \(H_{1}-\bar{H}_{1}=\mp(m_1-\bar{m}_1)\,\). This follows from the general result for Virasoro primaries of definite conformal weight
\be\label{spinids}
\amp{3mm}{\hat{V}_{3}V_{2}(1)V_{1}}\,=\,(-1)^{s_2}\amp{3mm}{\hat{V}_{1}V_{2}(1)V_{3}}
\,=\,(-1)^{s_1+s_2+s_3}\amp{3mm}{\hat{V}_{3}V_{1}(1)V_{2}}
\,=\,(-1)^{s_1}\amp{3mm}{\hat{V}_{2}V_{3}(1)V_{1}}
\ee
where we have taken \(s_{j}=H_{j}-\bar{H}_{j}\in\mathbb{Z}\,\). 

To determine the factor \(F(j_1,j_2,j_3)\) the relation between discrete states (\ref{weylprimeq2}) may be used to construct
\be
\amp{3mm}{\hat{\mc{V}}_{j_3}(m_3)\mc{V}_{j_2}(m_2|1)\mc{V}_{j_1}(-j_1)}\,=\,\omega_{j_1}
\amp{3mm}{\hat{\mc{V}}_{j_3}(m_3)\mc{V}_{j_2}(m_2|1)\mc{V}^{-1}_{\tilde{\jmath}_1}(\tilde{\jmath}_1)}
\ee
where \(\tilde{\jmath}_p=k/2-j_p\,\). A comparison of (\ref{onesf}) with (\ref{glb3pt}) using (\ref{wdisc}) yields
\be\label{sffunc}
F(j_1,j_2,j_3)\,=\,\frac{B_{j_1}}{B_{0}}\,C(\tilde{\jmath}_1,j_2,j_3)\,=\,
\frac{B_{j_2}B_{j_3}}{B_{0}B_{\tilde{\jmath}_1}}\,C(\tilde{\jmath}_1,\tilde{\jmath}_2,\tilde{\jmath}_3)
\ee
where the identity 
\be\label{sffunc2}
B_{j_1} C(\tilde{\jmath}_1,j_2,j_3)\,=\,
B_{j_2} C(j_1,\tilde{\jmath}_2,j_3)\,=\,
B_{j_3} C(j_1,j_2,\tilde{\jmath}_3)
\ee
has been used at right. Making use of the identities (\ref{spinids}) and the action (\ref{oneunitsf}) of the one-unit spectral flow operator, the above normalizations can be verified by computing
\be
\amp{3mm}{\hat{\mc{V}}^{\mp 1}_{j_2}(m_2)\mc{V}^{\pm 1}_{j_1}(m_1)}\,=\,\omega_{0}
\amp{3mm}{\hat{\mc{V}}_{k/2}(\mp \txfc{1}{2}k)\mc{V}_{j_2}(m_2|1)\mc{V}^{\pm 1}_{j_1}(m_1)}\,=\,
\amp{3mm}{\hat{\mc{V}}_{j_2}(m_2)\mc{V}_{j_1}(m_1)}
\ee
where the corresponding limit of the expression (\ref{onesf}) has been compared to the two-point functions (\ref{glb2pt},\,\ref{sl2pt}). The factor \(F(j_1,j_2,j_3)\) may be determined more explicitly by factorizing the amplitude (\ref{sfoneampx}) using the \(H_3\) OPE
\be
\Phi_{j'}(x'|z)\Phi_{j_1}(x_1)\,=\,
\int\! d^2 x\!\int_{\mathbb{R}_{+}}\!\!ds
\amp{3mm}{\hat{\Phi}_{1-j}(x)\Phi_{j'}(x'|z)\Phi_{j_1}(x_1)}\left[\Phi_{j}(x)+\ldots\right]
\ee
where \(j=1/2+is\,\). Recognizing that the integrand in the OPE is invariant under \(j\goto 1-j\,\), the integral may be extended over the entire real \(s\) axis. The amplitude (\ref{sfoneampx}) is factorized in the \(z\goto 0\,\) limit while incorporating the result (\ref{sfonecnfblk}). The integrand vanishes along the original contour as well as for all poles which cross the real \(s\) axis except for the poles at \(j=\tilde{\jmath}_1\,\) and \(j=1-\tilde{\jmath}_1\,\), which lead to identical contributions. We find the result (\ref{sffunc}) given
\be
\oint_{\tilde{\jmath}_1}\!ds\,C(j,k/2,j_1)\,=\,
B_{j_1}B_{\tilde{\jmath}_1}/B_{0}\,=\,\pi^{-1}\,\nu^{1-k}\,b^{-2}\gamma(-b^2)
\ee
where the contour encircles the pole at \(j=\tilde{\jmath}_1\,\). This follows under analytic continuation of \(C(j,j',j_1)\) from \(j'=1/2+is'\in\mc{S}^{+}_{\sst \mc{C}}\,\) to \(j'=k/2\,\). Equivalently, the identity appearing in (\ref{sffunc2}) may be used in conjuction with (\ref{cfuncdlt}). This computation also shows that the factorization of the amplitude (\ref{sfoneampx}) involves only the single intermediate representation with primary field \(\Phi_{\tilde{\jmath}_1}(x)\,\).

The three-point function (\ref{onesf}) is consistent with the construction described by Ribault~\cite{Ribault_0507} which maps Liouville amplitudes to spectral flowed primary correlators in the \(H_3\) model. This is an extension of the work of Teschner and Ribault~\cite{RibaultTeschner_0502} where each unit of spectral flow violation in an \(H_3\) \(n\)-point function leads to the removal of one of the \(n-2\) degenerate Liouville primary field insertions in the mapping associated with the unflowed correlator (\ref{lvtoh3}). Thus for the \(n=3\) and \(\txsum_{i=1}^{3}w_{i}=\pm 1\,\), we expect the amplitude (\ref{onesf}) to be proportional to a three-point Liouville correlator without degenerate insertions. This is demonstrated by verifying the relation\footnote{This is equivalent to the determination of the constant \(c_k\) in~\cite{Ribault_0507}\,.}
\be
\frac{\pi\omega_{0}}{(k-1)}\,
\frac{F(j_1,j_2,j_3)}{\gamma(\hat{\jmath}-k/2)}\,=\,b^{-1}\,\nu^{-k/2}\,\frac{\Gamma(1+b^2)}{\Gamma(1-b^2)}\,C_{\sst L}(k/2-j_{p})
\ee
Here \(C_{\sst L}(a_{p})=C_{a_3a_2a_1}\) is the Liouville three-point function (\ref{lvcfunc}).

\section{The \texorpdfstring{$SL(2,R)$}{} fusion rules\label{sec_5}}

In this section the fusion rules of the respective spectral flow modules treated in subsection \ref{subsec3ptred}, which follow directly from the selection rules of subsection \ref{select_sec}, will be refined to describe the dependence of the OPE coefficients on the \(j\) quantum number. Here the term OPE coefficient is intended to refer to the expansion coefficients of products of elements of the \(SL(2,R)\) spectrum in terms of elements of this spectrum, which is presumed to form a complete set. As described in section \ref{sec_6}, a definition of such an expansion via the analytic continuation of factorized \(H_3\) amplitudes is elusive at best. In particular, the introduction of intermediate spectral flowed modules prior to A.C. from \(H_3\) to \(SL(2,R)\) amplitudes appears to be superfluous. This redundancy is associated with sets of alternative factorizations which are closely related to the equivalences between amplitudes of fixed total spectral flow described in subsection~\ref{sfequiv_sec}\,. In addition, the corresponding discrete intermediate states that arise via A.C. do not in general lie in the spectrum of normalizable state of the \(SL(2,R)\) model. This represents an apparent failure of the closure of an \(SL(2,R)\) OPE defined via A.C. of correlators factorized on the \(H_3\) spectrum. Before investigating these questions in detail, it is useful to provisionally introduce a definition, which was first proposed in \cite{Ribault_0507} and later investigated in \cite{BaronNunez_0810}\cite{IguriNunez_0908}, of the \(SL(2,R)\) OPE which incorporates at the outset both spectral flow conserving and non-conserving OPE coefficients in the \(H_3\) model.

A condensed notation will be briefly employed for vertex operators of definite affine weight by defining \(\mc{V}_{a}^{w}=\mc{V}_{j}^{w}(m)\,\), with \(J^{3}_{0}\) conjugation defined by \(\mc{V}_{a}^{w}\goto \mc{V}_{\bar{a}}^{-w}=\mc{V}_{j}^{-w}(-m)\,\). The two point function (\ref{glb2pt}\,,\,\ref{sl2pt}) of primary states is of the form
\be
\amp{3mm}{\hat{\mc{V}}_{a_{2}}^{w_{2}}\mc{V}_{a_{1}}^{w_{1}}}\,=\,G_{a_1}\,\delta_{w_1+w_2}\,\delta(a_1-\bar{a}_2)
\ee
where \(G_{a}=G_{\bar{a}}\,\). Using the notation of subsection \ref{sfequiv_sec}, this leads to an OPE of the form\footnote{Here only holomorphic dependence appears so that\((-1)^{m_1 w_2}\) denotes the sign \((-1)^{\ell_1 w_2}\,\), where \(\ell_1=m_1-\bar{m}_1\in\mathbb{Z}\,\).}
\be\label{OPEsum0}
\mc{V}_{a_2}^{w_2}(z_2)\,\mc{V}_{a_1}^{w_1}(z_1)\,=\,(-1)^{m_1w_2}\sum_{\hat{w}} \int da_3\,G^{-1}_{a_3}
\amp{3mm}{\hat{\mc{V}}_{\bar{a}_{3}}\mc{V}_{a_{2}}(1)\mc{V}^{-\hat{w}}_{a_{1}}}
z_{21}^{H_{21}}\left(\mc{V}_{a_3}^{w_3+\hat{w}}(z_1)+\ldots\right)
\ee
where we have used (\ref{3pfequiv}), and have defined\footnote{It has been found convenient both here and below to define \(w_3=w_1+w_2\,\), rather than to have \(w_3\) denote the spectral flow sector of the intermediate states with quantum number \(j_3\,\).} \(w_3=w_1+w_2\,\) while fixing the range of the sum to \(-1\leq \hat{w}\leq 1\,\).  Implicitly, the integral over \(a_3\) includes all of the distinct primary states in \(\mc{C}_{w_3+\hat{w}}\) and \(\mc{D}^{\pm}_{w_3+\hat{w}}\,\). More explicitly\footnote{The integral \(\int_{\mc{S}^{+}_{\sst \mc{C}}}dj\) is to be computed as \(\int_{0}^{\infty}\!ds\) for \(j=1/2+is\,\). Thus, we replace \(dj\) with \(ds=-idj\,\).} for some function \(f(a)=f_{j}(m)\,\),
\be
\int da\,f(a)\,=\,
\int_{\mc{S}^{+}_{\sst \mc{C}}}\!dj\int_{\sst -\frac{1}{2}}^{\sst \frac{1}{2}}\!d\alpha\;
\sum_{p\in\mathbb{Z}}f_{j}(\alpha+p) \ +\,
\sum_{\sigma=\pm 1}\,\int_{\mc{S}_{\sst \mc{D}}}\!dj\sum_{n=0}^{\infty}
f_{j}(\sigma(j+n))
\ee
Here we have ignored anti-holomorphic dependence since, as discussed below, when computing the fusion rules for flowed affine modules it is sufficient to consider\footnote{These states also have zero worldsheet spin (\(H-\bar{H}=0\)) in all spectral flow sectors. Using (\ref{mmbarint}), for \(m=\bar{m}\) we may write \(\int_{\sst -\frac{1}{2}}^{\sst \frac{1}{2}}\!d\alpha\;
\sum_{p\in\mathbb{Z}}f_{j}(\alpha+p) =\int_{\mathbb{R}}dm\,f_{j}(m)=2\pi^2\!\int d^2m\,\delta_{\ell}\,f_{j}(m)\,\).} sums over flowed primaries with \(\ell=m-\bar{m}=0\,\). Note that in the case of the discrete affine representations with identification \(\hat{\mc{D}}^{\pm}_{w}=\hat{\mc{D}}^{\mp}_{w\pm 1}\,\), the spectral flow quantum numbers appearing here are associated with the corresponding respective flowed primary states \(\mc{D}^{\pm}_{w}\oplus \mc{D}^{\mp}_{w\pm 1}\,\). Defining \(\hat{m}=m_1+m_2-\hat{w}k/2\,\), we first write
\bbb\label{sym3ptcof}
\amp{3mm}{\hat{\mc{V}}_{\bar{a}_{3}}\mc{V}_{a_{2}}(1)\mc{V}^{-\hat{w}}_{a_{1}}} &=&
\amp{3mm}{\hat{\mc{V}}_{j_{3}}(-m_3)\mc{V}_{j_{2}}(m_2|1)\mc{V}^{-\hat{w}}_{j_{1}}(m_1)}\nonumber\\[0.1cm]&=&
\pi^{-2}\,\delta^2(\hat{m}-m_3)\,(-1)^{-\ell_2\hat{w}}\,A^{\hat{w}}(a_{1},a_{2},\bar{a}_{3})
\eee
where \(A^{\hat{w}}(a_{p})=A^{\hat{w}}(j_{p}|m_1,m_2)\) corresponds to the 
symmetric coefficients in (\ref{glb3pt},\,\ref{onesf}). Then for \(\ell_1+\ell_2=0\,\) we have the OPE
\bbb\label{OPEsum}
\lefteqn{\mc{V}_{j_2}^{w_2}(m_2|1)\,\mc{V}_{j_1}^{w_1}(m_1)\,=\,
(-1)^{\ell_1 w_2}\sum_{\hat{w}}\,(-1)^{-\ell_2\hat{w}}\left[\mvsp{0.5cm}\right.
\ \int_{\mc{S}^{+}_{\sst \mc{C}}}\!dj_{3}\,A^{\hat{w}}(1-j_{3})
\left(\mc{V}_{j_3}^{w_3+\hat{w}}(\hat{m})+\ldots\right)}\hspace{1.5cm}\nonumber\\[0.2cm]
&&+\int_{\mc{S}_{\sst \mc{D}}}\!dj_{3}\sum_{n_{3}=0}^{\infty}A^{\hat{w}}(j_{3})\,
\frac{2(2j_3-1)^2}{D_{j_{3}}(n_3)\,B_{j_{3}}}
\,\delta(j_3+n_{3}-|\hat{m}|)
\left(\mc{V}_{j_3}^{w_3+\hat{w}}(\hat{m})+\ldots\right)
\hspace{0.2cm}\left.\mvsp{0.5cm}\right]\hspace{0.5cm}
\eee
where we have abbreviated \(A^{\hat{w}}(j_{3})=
A^{\hat{w}}(j_{1},j_{2},j_{3}|m_{1},m_{2})\,\). 

It is important to recognize that the above construction assumes that three-point amplitudes may be factorized on the \(SL(2,R)\) spectrum, and makes direct use of the form of the discrete two-point function (\ref{dsc2pf}) as well as the domain \(\mc{S}_{\sst \mc{D}}\) of the discrete spectrum. In particular, the expansions (\ref{OPEsum0},\,\ref{OPEsum}) essentially amount to an ansatz for the OPE, and are not taken to be derived through the analytic continuation of both flow conserving and non-conserving OPE coefficients\footnote{States in spectral flow sectors are non-normalizable in the \(H_3\) model, and thus do not appear in the  factorization of \(H_3\) amplitudes. However, the corresponding \(|w|=1\) three-point amplitude (\ref{onesf}) is well-defined in the \(H_3\) model, as is implied from its derivation \cite{Ribault_0507} in terms of three-point primary amplitudes in the Liouville CFT.} in the \(H_3\) model as in \cite{BaronNunez_0810}. As will be discussed in detail in subsection \ref{fussum}, this analytic continuation does not entirely restrict the summation of discrete fields appearing in the OPE to the domain \(\mc{S}_{\sst \mc{D}}\,\). Of course, depending on the domain of the product fields, many contributions from flowed affine modules appearing in (\ref{OPEsum}) vanish due to the selection rules described above. There is also a clear redundancy in the form of the OPE (\ref{OPEsum}) in the case of the discrete contributions since the flowed primary states in \(\mc{D}^{\mp}_{\tilde{\jmath}}(w_3\pm 1)\,\), where \(\tilde{\jmath}=k/2-j\,\) and \(w_3=w_1+w_2\,\), appear as descendant contributions in the OPE coefficients involving the \(\mc{D}^{\pm}_{j}(w_3)\,\) primaries. While the corresponding affine modules are identified, due to \(J_{0}^{3}\) conservation, only one of the flowed primaries in \(\mc{D}^{\pm}_{j}(w_3)\oplus\mc{D}^{\mp}_{\tilde{\jmath}}(w_3\pm 1)\) can appear in the OPE. It is thus helpful when considering OPE coeffcients involving flowed primary fields to include both terms in the OPE, choosing the representation of the operator as a flowed primary rather than the corresponding flowed descendant. The single state in the intersection \(\mc{B}^{\mp}_{\tilde{\jmath}}(w_3\pm 1)=\mc{B}^{\pm}_{j}(w_3)\,\) should appear only once, with (\ref{weylprimeq2}) and the normalization factor in the OPE ensuring that the result will be independent of the choice of either \(\mc{V}^{w_3}_{j}(\pm j)\) or \(\mc{V}^{w_3\pm 1}_{\tilde{\jmath}}(\mp \tilde{\jmath})\). Note that, due to the vanishing of the corresponding three-point functions, the primary states in \(\mc{D}^{\mp}_{\tilde{\jmath}}(w_3\pm 2)\,\) do not appear, and non-vanishing descendants with identical associated \(J^{3}_{0}\) values appear only once in the descendant contribution arising from the \(\mc{D}^{\pm}_{j}(w_3\pm 1)\,\) primaries. Each of the respective terms in the fusion rules of spectral elements of the flowed affine modules which are permitted by the selection rules will now be considered, with the aim of determining which \(j\) quantum numbers\footnote{Except where relevant, the \(\alpha\) quantum number of the irreducible representations \(\hat{\mc{C}}^{\alpha}_{j}(w)\) will be ignored.} appear with non-vanishing coefficients. It will be shown that all such coefficients are non-singular for elements of the spectrum, and that for the fusion rules (\ref{opeslct1}\,,\,\ref{opeslct2}\,,\,\ref{opeslct3}) each term has non-vanishing coefficients for some range of the respective \(j\) quantum numbers. Thus, as far as the indexing of representations by the spectral flow quantum number \(w\) is concerned, the selection rules are equivalent to the fusion rules; there are no spectral flow representations allowed by the selection rules which vanish\footnote{This statement will have to be reexamined below when discussing the computation of discrete string amplitudes following analytic continuation of \(H_3\) correlators as described in \cite{MaldacenaOoguri_0111}.} due to unrelated features of the form of the OPEs coefficients.

\subsection{Fusion rules for \texorpdfstring{$\hat{\mc{C}}_{w_2}\times\hat{\mc{C}}_{w_1}$}{}}

The fusion rule (\ref{opeslct3})
\be
\hat{\mc{C}}_{w_2}\times\hat{\mc{C}}_{w_1}\,\goto\,
\hat{\mc{C}}_{w_3}\oplus\hat{\mc{C}}_{w_3+1}\oplus\hat{\mc{C}}_{w_3-1}\oplus
\hat{\mc{D}}^{+}_{w_3}\oplus\hat{\mc{D}}^{-}_{w_3}
\ee
involves the following reduction of a general three-point descendant correlator
\be
\hat{\mc{D}}^{\mp}_{j_3}(-w_3)\otimes\hat{\mc{C}}_{j_2}(w_2)\otimes\hat{\mc{C}}_{j_1}(w_1)\,\Rightarrow\,
\mc{B}^{\mp}_{j_3}\otimes\mc{C}_{j_2}\otimes\mc{C}_{j_1}\label{ope1_1}
\ee
The OPE coefficients\footnote{As mentioned in subsection~\ref{subsec3ptred} above, the multiplicity two associated with the continuous representation \(\hat{\mc{C}}_{w_3}\) will be suppressed until the discussion appearing in section \ref{fussum}\,.} associated with \(\hat{\mc{C}}_{j_3}(w_3)\) and \(\hat{\mc{C}}_{j_3}(w_3\pm 1)\) will not be examined here since they are manifestly well-defined and will have some non-vanishing elements for all \(j_3\in \mc{S}^{+}_{\sst \mc{C}}\,\). Making use of \(\mc{B}^{\pm}_{j}=\mc{B}^{\mp}_{\tilde{\jmath}}(\pm 1)\,\) with \(\tilde{\jmath}=k/2-j\,\), it is somewhat simpler in the case of the discrete representations \(\hat{\mc{D}}^{\pm}_{j_3}(w_3)\) to consider the general spectral flow non-conserving amplitudes\footnote{Note that here we are making use of the fact that both the factor \(\omega_j\) in (\ref{weylprimeq}) and the factors multiplying the three-point function in (\ref{OPEsum}) are non-vanishing and non-singular for \(j_3\in\mc{S}_{\sst \mc{D}}\,\).} appearing in the discrete OPE coefficient in (\ref{OPEsum}) which involve elements of \(\mc{B}^{\pm}_{\tilde{\jmath}_3}(\mp 1)\otimes\mc{C}_{j_2}\otimes\mc{C}_{j_1}\,\). Since these are \(J_{0}^{3}\) conjugate representations, from (\ref{onesf}) we compute\footnote{For the purpose of computing the fusion rules, the fields appearing in the reduced primary three-point correlators considered here and below may be taken to satisfy \(\ell=m-\bar{m}=0\,\), and thus have zero worldsheet spin in all spectral flow sectors. The single exception, which we are not treating in detail, is the spectral flow conserving case \(\mc{C}_{j_3}\otimes\mc{C}_{j_2}\otimes\mc{C}_{j_1}\), in which \(\ell=0\) can only be imposed on one of the fields.}
\bbb\label{ope_1_cf}
\amp{3mm}{\hat{\mc{V}}^{-1}_{\tilde{\jmath}_{3}}(\tilde{\jmath}_3)\mc{V}_{j_{2}}(m_2|1)\mc{V}_{j_{1}}(m_1)}&=&
\pi^{-2}\,\delta^2(m_1+m_2-j_3)\;\frac{\pi\omega_{0}}{(k-1)}\,
\frac{B_{\tilde{\jmath}_3}}{(2\tilde{\jmath}_{3}-1)}
\nonumber\\[0.2cm]&&
\frac{C(j_1,j_2,j_3)}{\gamma(j_1+j_2-j_3)}\;\prod_{i\in\{1,2\}}\;
\frac{\Gamma(2j_{i}-1)}{\Gamma(1-2j_{i})}\frac{\Gamma(1-j_{i}-m_{i})}{\Gamma(j_{i}+m_{i})}
\eee
Given that none of the poles (\ref{gpoles}) in \(G(2j-1)\) or \(G(j)\) appear for \(j\in\mc{S}_{\sst \mc{D}}\,\), it is evident that for every \(j_3\in\mc{S}_{\sst \mc{D}}\,\) there exist nonzero amplitudes (all of which are non-singular) for \(j_1,j_2\in\mc{S}^{+}_{\sst \mc{C}}\,\). Of course, for fixed \(\pm(m_1+m_2)\in\mc{S}_{\sst \mc{D}}\) only \(\mc{B}^{\pm}_{j_3}\,\), not \(\mc{B}^{\mp}_{j_3}\,\), appears in the fusion rule for \(\mc{C}_{j_2}\times\mc{C}_{j_1}\,\). However, the above analysis is focused on the fusion of the affine modules \(\hat{\mc{C}}_{j_2}(w_2)\times\hat{\mc{C}}_{j_1}(w_1)\,\) for each pair \(j_1,j_2\in\mc{S}^{+}_{\sst \mc{C}}\,\). As follows trivially from the fact that the primary representations \(\mc{C}_{j}\) are \(J_{0}^{3}\) self-conjugate, for all \(j_3\in\mc{S}_{\sst \mc{D}}\,\) both of the modules \(\hat{\mc{D}}^{\pm}_{j_3}(w_3)\) make an appearance. Thus, with \(w_3=w_2+w_1\,\), we may write
\bbb\label{csqrfus}
\hat{\mc{C}}_{j_2}(w_2)\times\hat{\mc{C}}_{j_1}(w_1)&\goto&
\int_{\mc{S}^{+}_{\sst\mc{C}}}\!dj_3\,\left(\,\hat{\mc{C}}_{j_3}(w_3)
\oplus\hat{\mc{C}}_{j_3}(w_3+1)\oplus\hat{\mc{C}}_{j_3}(w_3-1)\,\right)
\nonumber\\[0.2cm]&&\oplus\int_{\mc{S}_{\mc{D}}}\!dj_3\,
\left(\,\hat{\mc{D}}^{+}_{j_3}(w_3)\oplus\hat{\mc{D}}^{-}_{j_3}(w_3)\,\right)
\eee

\subsection{Fusion rules for \texorpdfstring{$\hat{\mc{C}}_{w_2}\times\hat{\mc{D}}^{\pm}_{w_1}$}{}}

For the fusion rule (\ref{opeslct2})
\be
\hat{\mc{C}}_{w_2}\times\hat{\mc{D}}^{\pm}_{w_1}\,\goto\,
\hat{\mc{D}}^{\pm}_{w_3}\oplus\hat{\mc{C}}_{w_3}\oplus\hat{\mc{C}}_{w_3\pm 1}
\ee
the respective correlators are reduced as follows
\bbb
\hat{\mc{D}}^{\mp}_{j_3}(-w_3)\otimes\hat{\mc{C}}_{j_2}(w_2)\otimes\hat{\mc{D}}^{\pm}_{j_1}(w_1)\!&\Rightarrow&\!
\mc{B}^{\mp}_{j_3}\otimes\mc{C}_{j_2}\otimes\mc{B}^{\pm}_{j_1}\label{ope2_1}\\[0.2cm]
\hat{\mc{C}}_{j_3}(-w_3)\otimes\hat{\mc{C}}_{j_2}(w_2)\otimes\hat{\mc{D}}^{\pm}_{j_1}(w_1)\!&\Rightarrow&\!
\mc{C}_{j_3}\otimes\mc{C}_{j_2}\otimes\mc{B}^{\pm}_{j_1}\label{ope2_2}\\[0.2cm]
\hat{\mc{C}}_{j_3}(-w_3\mp 1)\otimes\hat{\mc{C}}_{j_2}(w_2)\otimes\hat{\mc{D}}^{\pm}_{j_1}(w_1)\!&\Rightarrow&\!
\mc{C}_{j_3}(\mp 1)\otimes\mc{C}_{j_2}\otimes\mc{B}^{\pm}_{j_1}\Rightarrow
\mc{C}_{j_3}\otimes\mc{C}_{j_2}\otimes\mc{B}^{\mp}_{\tilde{\jmath}_1}\label{ope2_3}
\eee
The cases (\ref{ope2_2},\,\ref{ope2_3}) are essentially equivalent for the present purposes to (\ref{ope1_1}). Making use of the \(J_{0}^{3}\) conjugation symmetry, and the fact that if \(j\in\mc{S}_{\sst \mc{D}}\,\) then \(\tilde{\jmath}=k/2-j\in\mc{S}_{\sst \mc{D}}\,\), it is sufficient to compute the correlator
\be
\amp{3mm}{\hat{\mc{V}}_{j_{3}}(-m_3)\mc{V}_{j_{2}}(m_2|1)\mc{V}^{-1}_{\tilde{\jmath}_{1}}(\tilde{\jmath}_1)}
\ee
This leads to a result equivalent to (\ref{ope_1_cf}) so that, for a given \(j_1=m_2-m_3\in\mc{S}_{\sst \mc{D}}\) and \(j_2\in\mc{S}^{+}_{\sst \mc{C}}\,\), for all \(j_3\in\mc{S}^{+}_{\sst \mc{C}}\,\) there are non-zero OPE coefficients, all of which are non-singular. The case (\ref{ope2_1}) requires the computation of
\bbb\label{ope_2_cf}
\amp{3mm}{\hat{\mc{V}}^{-1}_{\tilde{\jmath}_{3}}(\tilde{\jmath}_3)\mc{V}_{j_{2}}(m_2|1)\mc{V}_{j_{1}}(j_1)}&=&
\pi^{-2}\,\delta^2(j_1+m_2-j_3)\;\frac{\pi\omega_{0}}{(k-1)}\,
\frac{B_{\tilde{\jmath}_3}}{(2\tilde{\jmath}_{3}-1)}\,\frac{1}{(2j_{1}-1)}
\nonumber\\[0.2cm]&&
\frac{C(j_1,j_2,j_3)}{\gamma(j_1+j_2-j_3)}\;
\frac{\Gamma(2j_{2}-1)}{\Gamma(1-2j_{2})}\frac{\Gamma(1-j_{2}-m_{2})}{\Gamma(j_{2}+m_{2})}
\eee
Again, for \(j_1\in\mc{S}_{\sst \mc{D}}\,\) and \(j_2\in\mc{S}^{+}_{\sst \mc{C}}\,\), for all \(j_3\in\mc{S}_{\sst \mc{D}}\) there are non-zero OPE coefficients with \(m_2=j_3-j_1\), all of which are non-singular. Thus we have the fusion rule
\be\label{condscfus}
\hat{\mc{C}}_{j_2}(w_2)\times\hat{\mc{D}}^{\pm}_{j_1}(w_1)\,\goto\,
\int_{\mc{S}_{\mc{D}}}\!dj_3\;\hat{\mc{D}}^{\pm}_{j_3}(w_3)\;\oplus
\int_{\mc{S}^{+}_{\sst\mc{C}}}\!dj_3\,\left(\,\hat{\mc{C}}_{j_3}(w_3)
\oplus\hat{\mc{C}}_{j_3}(w_3\pm 1)\,\right)
\ee

\subsection{Fusion rules for \texorpdfstring{$\hat{\mc{D}}^{-}_{w_2}\times\hat{\mc{D}}^{+}_{w_1}$}{}}

It is convenient in the case of the OPE of discrete states to fix \(\pm w\geq 0\) for \(\hat{\mc{D}}^{\pm}_{w}\) for \(w_1\) and \(w_2\) and to treat the fusion rules (\ref{opeslct1}) and (\ref{opeslct4}) as distinct. For the fusion rule (\ref{opeslct1}) 
\be\label{opeslct1_2}
\hat{\mc{D}}^{-}_{w_2}\times\hat{\mc{D}}^{+}_{w_1}\,\goto\,
\hat{\mc{D}}^{+}_{w_3}\oplus\hat{\mc{D}}^{-}_{w_3}\oplus\hat{\mc{C}}_{w_3}
\ee
we have the following reductions\footnote{Using the \(J_{0}^{3}\) conjugation symmetry, and the arguments appearing in (\ref{subsec3ptred}), it may be seen that the result (\ref{Dcube1}) is independent of the sign of \(w_3=w_1+w_2\,\).} of OPE coefficients involving \(\hat{\mc{C}}_{w_3}\) and \(\hat{\mc{D}}^{\pm}_{w_3}\) 
\bbb
\hat{\mc{C}}_{j_3}(-w_3)\otimes\hat{\mc{D}}^{-}_{j_2}(w_2)\otimes\hat{\mc{D}}^{+}_{j_1}(w_1)\!&\Rightarrow&\!
\mc{C}_{j_3}\otimes\mc{B}^{-}_{j_2}\otimes\mc{B}^{+}_{j_1}\label{DsqrC1}\\[0.2cm]
\hat{\mc{D}}^{\mp}_{j_3}(-w_3)\otimes\hat{\mc{D}}^{-}_{j_2}(w_2)\otimes\hat{\mc{D}}^{+}_{j_1}(w_1)\!&\Rightarrow&\!
\mc{D}^{\mp}_{j_3}\otimes\mc{B}^{-}_{j_2}\otimes\mc{B}^{+}_{j_1}\label{Dcube1}
\eee
The case (\ref{DsqrC1}) involves computing the following correlator for an element of  
\(\mc{C}_{j_3}\otimes\mc{B}^{+}_{\tilde{\jmath}_2}(-1)\otimes\mc{B}^{+}_{j_1}\)
\be\label{DsqrCamp1}
\amp{3mm}{\hat{\mc{V}}_{j_{3}}(-m_3)\mc{V}^{-1}_{\tilde{\jmath}_{2}}(\tilde{\jmath}_{2}|1)\mc{V}_{j_{1}}(j_1)}
\ee
This is equivalent to (\ref{ope_2_cf}), and thus for \(j_1,j_2\in\mc{S}_{\sst \mc{D}}\,\), for all \(j_3\in\mc{S}^{+}_{\sst \mc{C}}\,\) there are non-zero OPE coefficients with \(m_3=j_1-j_2\,\), all of which are non-singular. Note that the irreducible continuous representations \(\hat{\mc{C}}^{\alpha_{3}}_{j_3}(w_3)\) which appear have \(\alpha_3=m_3-[m_3]\,\), where \([m_3]\) is \(m_3=j_1-j_2\) rounded to the nearest integer. For the case (\ref{Dcube1}), making use of the conjugation symmetry, we consider\footnote{This choice minimizes the required use of identities of the function \(G(j)\) in order to arrive at a non-singular expression for the corresponding correlators.} elements of \(\mc{D}^{+}_{j_3}\otimes\mc{B}^{+}_{j_2}\otimes\mc{B}^{+}_{\tilde{\jmath}_1}(-1)\,\) for the \(\hat{\mc{D}}^{+}_{w_3}\) term in (\ref{opeslct1_2}). For the \(\hat{\mc{D}}^{-}_{w_3}\) term we consider elements of \(\mc{D}^{+}_{j_3}\otimes\mc{B}^{+}_{\tilde{\jmath}_2}(-1)\otimes\mc{B}^{+}_{j_1}\). Since in (\ref{onesf}) these are related by \(j_1\rightarrow\tilde{\jmath}_{2}\) and \(j_2\rightarrow\tilde{\jmath}_{1}\,\), it is sufficient to calculate the correlator 
\bbb\label{DcubeCamp1}
\amp{3mm}{\hat{\mc{V}}_{j_{3}}(j_3+n_3)\mc{V}_{j_{2}}(j_{2}|1)\mc{V}^{-1}_{\tilde{\jmath}_{1}}(\tilde{\jmath}_1)}
&=&\pi^{-2}\,\delta^2(j_3+n_3+j_2-j_1)\,\frac{\pi\omega_{0}}{(k-1)}\,\frac{B_{\tilde{\jmath}_1}}{(2\tilde{\jmath}_{1}-1)}\,
\frac{1}{(2j_{2}-1)}\nonumber\\[0.2cm]
&&\frac{C(j_1,j_2,j_3)}{\gamma(j_3+j_2-j_1)}\,
\frac{(-1)^{n_3}}{(2j_{3}-1)}\,\frac{\Gamma^2(2j_{3})}{\Gamma^2(2j_{3}+n_{3})}
\eee
Given the form  (\ref{cfunc}) of \(C(j_1,j_2,j_3)\), and making use of the identity (\ref{gdefs2}), since none of the poles in \(G(2j-1)\) appear for \(j\in\mc{S}_{\sst \mc{D}}\,\), it may be seen that there are no zeros in this amplitude except for those enforced by the delta function. Poles in this amplitude could only appear in the product
\be
G(j_1+j_2+j_3-1)\,G(1-j_3+j_1-j_2)\,G(j_1+j_2-j_3)\,G(j_1+j_3-j_2)
\ee
Substituting \(j_3=j_1-j_2-n_3\,\), this may be written as
\be\label{DcubeCamp1_2}
G(1+n_3)\,G(2\tilde{\jmath}_1+n_3)\,G(2j_2+n_3)\,G(2j_3+n_3)
\ee
where (\ref{gdefs1}) has been used. It may be seen that for \(j_3\in\mc{S}_{\sst \mc{D}}\,\), which requires \(j_1-j_2\geq 1/2\,\) and imposes \(n_3\leq j_1-j_2-1/2\,\), no poles appear in (\ref{DcubeCamp1_2}). Note that values of \(j_3\) do not extend to the upper bound of \(\mc{S}_{\sst \mc{D}}\,\) since \(j_1-j_2\leq (k-2)/2<(k-1)/2\,\). Considering both terms in (\ref{Dcube1}), and defining \(j_{+}=j_1-j_2=-j_{-}\,\), we have\footnote{As above, the notation \([j_{\pm}]\) denotes \(j_{\pm}\) rounded to the nearest integer. The condition \(n_3\leq j_{\pm}-1/2\) is equivalent to fixing \([j_{\pm}]-1\) as the upper bound on the sum over \(n_3\,\).}
\be\label{discgenope}
\hat{\mc{D}}^{-}_{j_2}(w_2)\times\hat{\mc{D}}^{+}_{j_1}(w_1)\,\goto\,
\int_{\mc{S}^{+}_{\sst\mc{C}}}\!dj_3\;\hat{\mc{C}}^{\alpha_3}_{j_3}(w_3)\;\oplus
\sum_{\sigma=\pm}\sum_{n_3=0}^{[j_{\sigma}]-1}\int_{\mc{S}_{\mc{D}}}\!dj_3\,
\delta(j_3+n_3-j_{\sigma})\,\hat{\mc{D}}^{\sigma}_{j_{3}}(w_3) 
\ee
where for the continuous representations \(\alpha_3=j_{+}-[j_{+}]\,\). Note that the condition \(j_{\pm}\geq 1/2\) requires one of the discrete contributions to vanish. It is important to emphasize in this regard that, while \(w_1\geq 0\) and \(w_2\leq 0\,\), the condition \(\pm w\geq 0\) for \(\hat{\mc{D}}^{\pm}_{w}\) has not been imposed on \(w_3\,\). For \(w_{3}=w_1+w_2\neq 0\,\) it is perhaps more illuminating to use the conjugation symmetry to set \(w_3>0\,\) and write \(\hat{\mc{D}}^{-}_{j_{3}}(w_3)=\hat{\mc{D}}^{+}_{\tilde{\jmath}_{3}}(w_3-1)\,\). This leads to 
\bbb\label{discmpope}
\hat{\mc{D}}^{-}_{j_2}(w_2)\times\hat{\mc{D}}^{+}_{j_1}(w_1)&\goto&
\int_{\mc{S}^{+}_{\sst\mc{C}}}\!dj_3\;\hat{\mc{C}}^{\alpha_3}_{j_3}(w_3)\;\oplus
\sum_{n_3=0}^{[j_{+}]-1}\int_{\mc{S}_{\mc{D}}}\!dj_3\,\delta(j_3+n_3-j_{+})\,\hat{\mc{D}}^{+}_{j_{3}}(w_3)\nonumber\\[0.2cm]
&&\oplus\sum_{n_3=0}^{[j_{-}]-1}\int_{\mc{S}_{\mc{D}}}\!dj_3\,\delta(k/2-j_3+n_3-j_{-})\,\hat{\mc{D}}^{+}_{j_{3}}(w_3-1)
\eee
In the last term we have \(j_3=k/2+j_1-j_2+n_3\,\), and thus \(j_3\in\mc{S}_{\sst \mc{D}}\,\)  is equivalent to \(k/2+j_1-j_2+n_{3}\leq (k-1)/2\,\). As above, this requires \(j_{-}=j_2-j_1\geq 1/2\,\) and \(n_3\leq j_{-}-1/2\,\). Note that for \(-1/2\leq j_1-j_2\leq 1/2\) only the continuous representations \(\hat{\mc{C}}_{w_3}\) appear. It may be seen that the OPE coefficients involving discrete states vanish as \(j_{\pm}\goto 1/2\,\), a result which is consistent with the emergence of a continuum of states in \(\hat{\mc{C}}_{w_3}\) with finite OPE coefficients. The basic picture here is very similar to that which is treated in the following subsection and illustrated in Figure~\ref{fig4}\,.

\subsection{Fusion rules for \texorpdfstring{$\hat{\mc{D}}^{\pm}_{w_2}\times\hat{\mc{D}}^{\pm}_{w_1}$}{}}

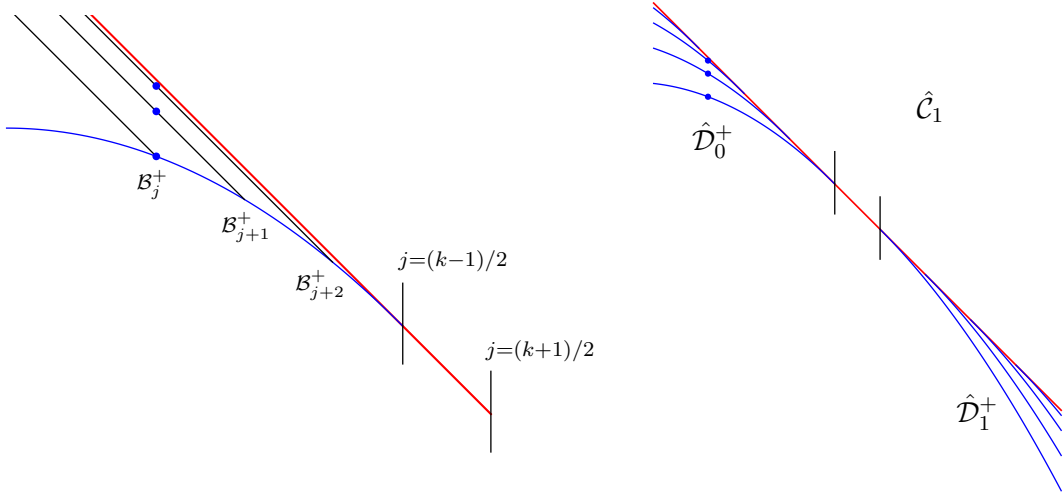
\begin{figure}[ht]
\begin{center}
\begin{picture}(426,200)
\includegraphics{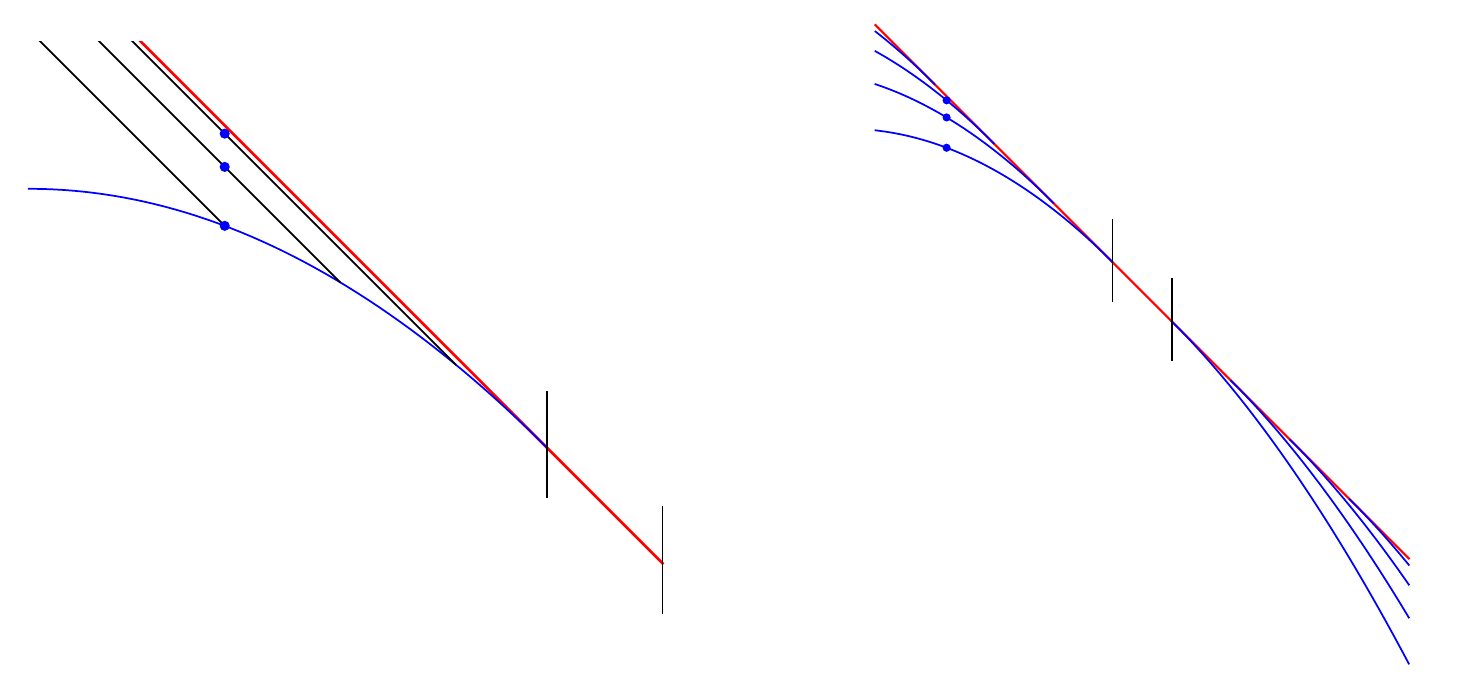}
\put(-271,94){\({\scriptstyle j=(k-1)/2}\)}
\put(-238,60){\({\scriptstyle j=(k+1)/2}\)}
\put(-370,123){\({\scriptstyle \mc{B}^{+}_{j}}\)}
\put(-338,107){\({\scriptstyle \mc{B}^{+}_{j+1}}\)}
\put(-310,85){\({\scriptstyle \mc{B}^{+}_{j+2}}\)}
\put(-160,138){\({\hat{\mc{D}}^{+}_{0}}\)}
\put(-60,35){\({\hat{\mc{D}}^{+}_{1}}\)}
\put(-75,150){\({\hat{\mc{C}}_{1}}\)}
\end{picture}
\caption{The figure at left above shows primary states in the discrete affine representations \(\hat{\mc{D}}^{+}_{0}=\hat{\mc{D}}^{-}_{1}\) which arise from fusion rule of \(\mc{B}^{+}_{j_1}(0)\times\mc{B}^{+}_{j_2}(0)\,\). The \textcolor{wtmblue}{blue} curve is the set of states \(\mc{B}^{+}_{0}=\mc{B}^{-}_{1}\,\) which are the intersections of the flowed primaries \(\mc{D}^{+}_{0}\) and \(\mc{D}^{-}_{1}\,\).  With a particular choice of \(j=j_1+j_2\,\), a finite number of discrete primary states (shown as \textcolor{wtmblue}{blue} dots) in \(\mc{D}^{-}_{\tilde{\jmath}-n}(1)\) (represented as black line segments), where \(n\in\mathbb{Z}_{\geq 0}\) and \(\tilde{\jmath}=k/2-j\), appear. The lowest of these states lies in \(\mc{B}^{+}_{j}(0)=\mc{B}^{-}_{\tilde{\jmath}}(1)\,\). The choices \(k=11\) and \(j=11/5\) have been made here, leading to a total of \(n_{max}+1=[\,\tilde{\jmath}\,]=3\) discrete states. The \textcolor{wtmred}{red} line is the upper bound on these discrete states where states in the continuum representations \(\hat{\mc{C}}_{1}\) begin. It may be seen that no states in \(\hat{\mc{D}}^{+}_{0}\) appear for \(j>(k-1)/2\). The same discrete states (also for \(k=11\)) are shown in the figure at right, which includes the affine representations \(\hat{\mc{D}}^{+}_{1}=\hat{\mc{D}}^{-}_{2}\) which also appear for \(j>(k+1)/2\). In this case the discrete primary states are elements of the representations \(\mc{D}^{+}_{-\tilde{\jmath}-n}(1)\,\). The \textcolor{wtmblue}{blue} curves at right are indexed by \(n\in\mathbb{Z}_{\geq 0}\) and give the locations of the discrete primary states for a particular value of \(j\). It is apparent that there are no discrete states for \((k-1)/2<j<(k+1)/2\,\).}
\label{fig4}
\end{center}
\end{figure}

The fusion rule (\ref{opeslct4})
\be
\hat{\mc{D}}^{\pm}_{w_2}\times\hat{\mc{D}}^{\pm}_{w_1} \,\goto\,
\hat{\mc{D}}^{\pm}_{w_3}\oplus\hat{\mc{D}}^{\pm}_{w_3\pm 1}
\oplus\hat{\mc{C}}_{w_3\pm 1}
\ee
requires the reduced correlators
\bbb
\hat{\mc{C}}_{j_3}(-w_3\mp 1)\otimes\hat{\mc{D}}^{\pm}_{j_2}(w_2)\otimes\hat{\mc{D}}^{\pm}_{j_1}(w_1)\!&\Rightarrow&\!
\mc{C}_{j_3}(\mp 1)\otimes\mc{B}^{\pm}_{j_2}\otimes\mc{B}^{\pm}_{j_1}\Rightarrow
\mc{C}_{j_3}\otimes\mc{B}^{\mp}_{\tilde{\jmath}_2}\otimes\mc{B}^{\pm}_{j_1}\label{DsqrC2}\\[0.2cm]
\hat{\mc{D}}^{\mp}_{j_3}(-w_3)\otimes\hat{\mc{D}}^{\pm}_{j_2}(w_2)\otimes\hat{\mc{D}}^{\pm}_{j_1}(w_1)\!&\Rightarrow&\!
\mc{D}^{\pm}_{\tilde{\jmath}_3}(\mp 1)\otimes\mc{B}^{\pm}_{j_2}\otimes\mc{B}^{\pm}_{j_1}\Rightarrow
\mc{D}^{\pm}_{\tilde{\jmath}_3}\otimes\mc{B}^{\mp}_{\tilde{\jmath}_2}\otimes\mc{B}^{\pm}_{j_1}\label{Dcube2}\\[0.2cm]
\hat{\mc{D}}^{\mp}_{j_3}(-w_3\mp 1)\otimes\hat{\mc{D}}^{\pm}_{j_2}(w_2)\otimes\hat{\mc{D}}^{\pm}_{j_1}(w_1)\!&\Rightarrow&\!
\mc{D}^{\mp}_{j_3}(\mp 1)\otimes\mc{B}^{\pm}_{j_2}\otimes\mc{B}^{\pm}_{j_1}\Rightarrow
\mc{D}^{\mp}_{j_3}\otimes\mc{B}^{\mp}_{\tilde{\jmath}_2}\otimes\mc{B}^{\pm}_{j_1}\label{Dcube3}
\eee
The case (\ref{DsqrC2}) involves the computation of 
\be\label{DsqrCamp2}
\amp{3mm}{\hat{\mc{V}}_{j_{3}}(-m_3)\mc{V}^{-1}_{j_{2}}(j_{2}|1)\mc{V}_{j_{1}}(j_1)}
\ee
which is equivalent to (\ref{DsqrCamp1}) under \(j_{2}\goto\tilde{\jmath}_{2}=k/2-j_{2}\,\). Thus for \(j_1,j_2\in\mc{S}_{\sst \mc{D}}\,\), for all \(j_3\in\mc{S}^{+}_{\sst \mc{C}}\,\) there are non-zero OPE coefficients\footnote{Again, the irreducible continuous representations \(\hat{\mc{C}}^{\alpha_{3}}_{j_3}(w_3)\) appearing in the OPE are restricted to \(\alpha_3=m_3-[m_3]\,\).} with \(m_3=j_1+j_2-k/2\,\), all of which are non-singular. For the cases (\ref{Dcube2}) and (\ref{Dcube3}) we consider elements of \(\mc{D}^{+}_{\tilde{\jmath}_3}\otimes\mc{B}^{+}_{j_2}(-1)\otimes\mc{B}^{+}_{j_1}\,\) and \(\mc{D}^{+}_{j_3}\otimes\mc{B}^{+}_{\tilde{\jmath}_2}(-1)\otimes\mc{B}^{+}_{\tilde{\jmath}_1}\,\), respectively. Since these are related by \(j_p\goto\tilde{\jmath}_p\,\), we consider (\ref{Dcube2}) and compute 
\bbb\label{DcubeCamp2}
\amp{3mm}{\hat{\mc{V}}_{\tilde{\jmath}_{3}}(\tilde{\jmath}_3+n_3)\mc{V}^{-1}_{j_{2}}(j_{2}|1)\mc{V}_{j_{1}}(j_1)}
&=&\pi^{-2}\,\delta^2(j_1+j_2+n_3-j_3)\,\frac{\pi\omega_{0}}{(k-1)}\,\frac{1}{(2j_{1}-1)}\,
\frac{1}{(2j_{2}-1)}\nonumber\\[0.2cm]
&&\frac{C(j_1,j_2,j_3)}{\gamma(j_1+j_2-j_3)}\,
\frac{B_{\tilde{\jmath}_{3}}}{(2\tilde{\jmath}_{3}-1)}\,\frac{(-1)^{n_3}\,\Gamma^2(2\tilde{\jmath}_{3})}{\Gamma^2(2\tilde{\jmath}_{3}+n_{3})}
\eee
which, of course, is just (\ref{DcubeCamp1}) with \(j_3\goto\tilde{\jmath}_{3}\) and \(j_1\goto\tilde{\jmath}_{1}\) after using (\ref{sffunc2}). As for (\ref{DcubeCamp1}), the amplitude (\ref{DcubeCamp2}) has no zeros except for those enforced by the delta function, while poles could only appear in the product
\be
G(j_1+j_2+j_3-1)\,G(j_3+j_2-j_1)\,G(1-j_1-j_2+j_3)\,G(j_1+j_3-j_2)
\ee
Substituting \(j_3=j_1+j_2+n_3\,\), this may be written as
\be\label{DcubeCamp2_2}
G(1+n_3)\,G(2j_1+n_3)\,G(2j_2+n_3)\,G(2\tilde{\jmath}_3+n_3)
\ee
It may be shown that for \(j_3\in\mc{S}_{\sst \mc{D}}\,\), which requires \(j_1+j_2\leq (k-1)/2\,\) and imposes \(n_3\leq (k-1)/2-j_1-j_2\,\), no poles appear in (\ref{DcubeCamp2_2}). Similarly, taking \(j_p\goto\tilde{\jmath}_p\,\), for (\ref{Dcube3}) the delta function fixes \(j_3=j_1+j_2-k/2-n_3\,\), which requires \(j_1+j_2\geq (k+1)/2\) and imposes \(n_3\leq j_1+j_2-(k+1)/2\,\); the corresponding amplitude is similarly well-defined. Defining \(g_{+}=j_1+j_2-k/2=-g_{-}\) the respective (mutually exclusive) conditions may be written as \(g_{\pm}\geq 1/2\) with \(n_3\leq g_{\pm}-1/2\,\). This leads to a fusion rule of the form
\bbb\label{discppope}
\hat{\mc{D}}^{\pm}_{j_2}(w_2)\times\hat{\mc{D}}^{\pm}_{j_1}(w_1)&\goto&
\int_{\mc{S}_{\sst\mc{C}}}\!dj_3\;\hat{\mc{C}}^{\alpha_3}_{j_3}(w_3\pm 1)\;\oplus
\sum_{n_3=0}^{[g_{-}]-1}\int_{\mc{S}_{\mc{D}}}\!dj_3\,\delta(k/2-j_3+n_3-g_{-})\,\hat{\mc{D}}^{\pm}_{j_{3}}(w_3)\nonumber\\[0.2cm]
&&\oplus\sum_{n_3=0}^{[g_{+}]-1}\int_{\mc{S}_{\mc{D}}}\!dj_3\,\delta(j_3+n_3-g_{+})\,\hat{\mc{D}}^{\pm}_{j_{3}}(w_3\pm 1)
\eee
where for the continuous representations \(\pm\alpha_3=g_{+}-[g_{+}]\,\). Note that for \((k-1)/2\leq j_1+j_2\leq (k+1)/2\) only continuous representations appear. It may be seen that the fusion rule (\ref{discmpope}) is identical to (\ref{discppope}) using 
\(\hat{\mc{D}}^{-}_{j_2}(w_2)=\hat{\mc{D}}^{+}_{\tilde{\jmath}_2}(w_2-1)\,\), so that (\ref{discgenope}) is the most general expression for the fusion rules of discrete flowed affine representations without imposing the condition \(\pm w\geq 0\) for \(\hat{\mc{D}}^{\pm}_{w}\,\). As illustrated\footnote{As above, the condensed notation \(\mc{D}^{\pm}_{w}\) denotes the set \(\mc{D}^{\pm}_{j}(w)\) for all \(j\in\mc{S}_{\mc{D}}\,\).} in Figure~\ref{fig4}\,, the somewhat intricate relations between the various terms in the fusion rules given here depend crucially on the range of the discrete spectrum \(\mc{S}_{\sst \mc{D}}\,\).

\section{The factorization of the four-point function\label{sec_6}}

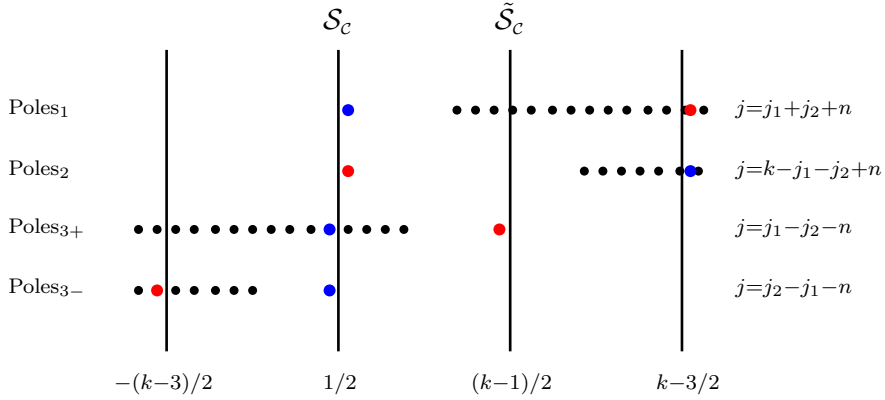
\begin{figure}[h]
\begin{center}
\begin{picture}(225,150)
\includegraphics{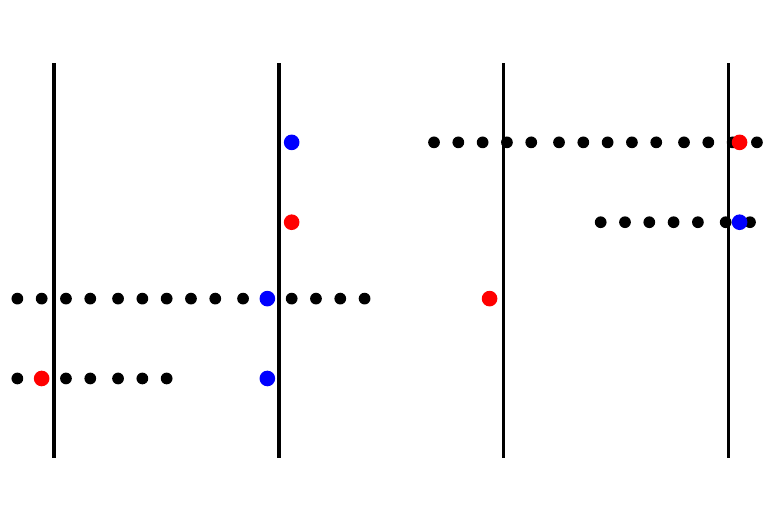}
\put(-151,4){\(\st 1/2\)}
\put(-151,140){\(\mc{S}_{\sst \mc{C}}\,\)}
\put(-230,4){\(\st -(k-3)/2\)}
\put(-95,4){\(\st (k-1)/2\)}
\put(-86,140){\(\tilde{\mc{S}}_{\sst \mc{C}}\,\)}
\put(-25,4){\(\st k-3/2\)}
\put(5,85){\(\st j=k-j_1-j_2+n\)}
\put(5,108){\(\st j=j_1+j_2+n\)}
\put(5,63){\(\st j=j_1-j_2-n\)}
\put(5,40){\(\st j=j_2-j_1-n\)}
\put(-270,85){\(\st \rm Poles_{2}\)}
\put(-270,108){\(\st \rm Poles_{1}\)}
\put(-270,63){\(\st \rm Poles_{3+}\)}
\put(-270,40){\(\st \rm Poles_{3-}\)}
\end{picture}
\caption{The figure above shows the locations of the sets of poles of the function \(C(j,j_1,j_2)\) in the complex \(j\) plane which can appear in the region between the contours \(j\in\mc{S}_{\sst\mc{C}} = 1/2+i\mathbb{R}\,\) and \(j\in\tilde{\mc{S}}_{\sst\mc{C}}=(k-1)/2+i\mathbb{R}\,\) for \(j_1,j_2 \in \mc{S}_{\sst\mc{D}}=(1/2\,,(k-1)/2)\subset\mathbb{R}\,\). For clarity the (real) poles have been separated along the imaginary axis, and the value \(k=20\) has been chosen. The \textcolor{wtmblue}{blue} dots show the locations of the leading (\(n=0\)) poles in both the case of minimum \(j_1+j_2=1\) for \(\rm Poles_{1}\) and \(\rm Poles_{2}\,\) and in the case of minimum \(j_1-j_2=0\) for \(\rm Poles_{3+}\) and \(\rm Poles_{3-}\,\). Note that none of the sets of poles cross the contour \(\mc{S}_{\sst\mc{C}}\) in the case \(j_1,j_2\in\mc{S}_{\sst\mc{C}}\,\). The \textcolor{wtmred}{red} dots show the locations of the leading (\(n=0\)) poles in both the case of maximum \(j_1+j_2=(k-1)\) for \(\rm Poles_{1}\) and \(\rm Poles_{2}\,\) and in the case of maximum \(j_1-j_2=(k-2)/2\) for \(\rm Poles_{3+}\) and \(\rm Poles_{3-}\,\). Note that none of the sets of poles cross the contour \(\tilde{\mc{S}}_{\sst\mc{C}}\) in the case \(j_1,j_2\in\tilde{\mc{S}}_{\sst\mc{C}}\,\). Also shown (in black) are the sets of poles in the case \(j_1+j_2=k/3<(k-1)/2\) and \(j_1-j_2=k/5\,\).}
\label{fig5}
\end{center}
\end{figure}

The three-point function of the \(H_3\) model in the \(\Phi_{j}(x)\) basis was derived~\cite{Teschner_9712} from null equations satisfied by the four-point function with a single insertion of fields in degenerate affine modules. An analysis of the corresponding zero-mode quantum mechanics and the form of the two-point function led to an assumed spectrum of normalizable states given by \(j\in\mc{S}^{+}_{\sst \mc{C}}=1/2+i\mathbb{R}_{\geq 0}\,\). Together these defined the OPE and an associated factorization of the four-point function~\cite{Teschner_9906} of the form\footnote{Here, as above, the worldsheet \(z\) argument in \(\Phi_{j}(x|z)\) is suppressed in \(\Phi_{j}(x)=\Phi_{j}(x|0)\,\), and the \(r\)-frame fields \(\Phi^{r}_{j}(r|z)\) are defined as in (\ref{rdef}) for \(r=1/x\,\).}
\be\label{H34pf}
\amp{3mm}{\hat{\Phi}^{r}_{j_4}(0)\Phi_{j_3}(1|1)\Phi_{j_2}(x|z)\Phi_{j_1}(0)}\,=\,
\int_{\mc{S}_{\sst \mc{C}}}\!\!dj\,\mc{C}(j)\;|\mc{F}_{j}(x|z)|^2
\ee
where the \(j_p\) dependence is suppressed in \(\mc{F}_{j}(x|z)\,\) and \(\mc{C}(j)\,=\,B_{j}^{-1}C(j_4,j_3,j)\,C(j,j_2,j_1)\,\) has been introduced. Note that this integral extends over \(j\in\mc{S}_{\sst \mc{C}}=1/2+i\mathbb{R}\,\), rather than the spectrum \(\mc{S}^{+}_{\sst \mc{C}}\), and thus the four-point function of the \(H_3\) model does not permit a chiral factorization in the conventional sense. This may also be seen from the fact that there are two independent (hypergeometric) solutions to the Knizhnik-Zamolodchikov equation at lowest order in \(z\,\) which are related by the reflection symmetry \(j\goto 1-j\,\) as required by monodromy invariance. Thus we have 
\be\label{4pfOz}
\mc{F}_{j}(x|z)=z^{h(j)-h_2-h_1}\left(x^{j-j_2-j_1}\,F_{j}(x)+\mc{O}(z)\right)
\ee
where \(F_{j}(x)={_{2}{\rm F}_{1}}(a,b;c;x)\) for
\be
a=j+j_2-j_1\hspace{1cm}b=j+j_3-j_4\hspace{1cm}c=2j
\ee
These solutions correspond to the contribution of primary fields to the intermediate states, from which descendant contributions follow uniquely~\cite{Teschner_9906}\cite{MaldacenaOoguri_0111}. The domain of validity of the expression (\ref{H34pf}) is given by
\bbb\label{jdom4pf}
|{\rm Re}(j_1-j_2)|<1/2\ \ &,&\hspace{0.3cm}|{\rm Re}(1-j_1-j_2)|<1/2\nonumber\\[0.2cm]
|{\rm Re}(j_4-j_3)|<1/2\ \ &,&\hspace{0.3cm}|{\rm Re}(1-j_4-j_3)|<1/2
\eee
This is clearly just two copies of the domain of validity of the OPE (\ref{H3OPEPhi}), outside of which (\ref{H34pf}) possesses a well-defined analytic continuation to \(j_{p}\in\mathbb{C}\,\) as poles in \(\mc{C}(j)\) cross the contour of integration and produce discrete contributions. The poles in \(\mc{C}(j)\) follow from the poles (\ref{gpoles}) in \(G(j)\) given by
\be
j=-f_{nm}\spwd{2cm}{and}j=(k-1)+f_{nm}
\ee
where 
\be
f_{nm}\,=\,n+m\,b^{-2}
\ee
From (\ref{cfunc}) the poles in \(C(j,j_1,j_2)\) are given by
\bbb
\pm(j-1/2)&=&1/2+f_{nm}+(j_1+j_2-1)\\[1mm]
\pm(j-1/2)&=&1/2+f_{nm}+(\tilde{\jmath}_1+\tilde{\jmath}_2-1)\\[1mm]
\pm((k-1)/2-j)&=&1/2+f_{nm}+(j_1+\tilde{\jmath}_2-1)\\[1mm]
\pm((k-1)/2-j)&=&1/2+f_{nm}+(\tilde{\jmath}_1+j_2-1)
\eee
In what follows some subset of \(j_p\) will be analytically continued from the domain (\ref{jdom4pf}) to 
the values \(\mc{S}_{\sst \mc{D}}\) of the discrete representations. In all cases for which \(j_p\) correspond to elements (continuous or discrete) of the spectrum only the following four sets of poles may appear in the region between the contours \(\mc{S}_{\sst \mc{C}}=1/2+i\mathbb{R}\) and \(\tilde{\mc{S}}_{\sst \mc{C}}=(k-1)/2+i\mathbb{R}\,\) at the boundaries of \(\mc{S}_{\sst \mc{D}}\). These are
\bbb\label{Cjpoles}
{\rm Poles}_{1\hspace*{6.6pt}}\hspace{0.5cm}&&j\,=\,j_1+j_2+n\\
{\rm Poles}_{2\hspace*{6.6pt}}\hspace{0.5cm}&&j\,=\,k-j_2-j_1+n\\
{\rm Poles}_{3+}\hspace{0.5cm}&&j\,=\,j_1-j_2-n\\
{\rm Poles}_{3-}\hspace{0.5cm}&&j\,=\,j_2-j_1-n
\eee
It will become relevant below that if \(j\in\mc{S}_{\sst \mc{C}}\) then \(\tilde{\jmath}=k/2-j\in\tilde{\mc{S}}_{\sst \mc{C}}\,\), and that while \(\mc{S}_{\sst \mc{C}}\) is preserved under \(j\goto 1-j\,\), \(\tilde{\mc{S}}_{\sst \mc{C}}\,\) is preserved under \(j\goto k-1-j\,\). As illustrated in figure~\ref{fig5}, under the A.C. considered \({\rm Poles}_{3\pm}\) may cross \(\mc{S}_{\sst \mc{C}}\) but not \(\tilde{\mc{S}}_{\sst \mc{C}}\,\), while the converse is true for \({\rm Poles}_{1}\) and \({\rm Poles}_{2}\,\).

\vspace{3mm}

In what follows it is helpful to define the condensed notation 
\be
\phi_{(j_2, j_1)}(x|z)=\Phi_{j_2}(x|z)\Phi_{j_1}(0)
\ee
and
\be
\hat{\phi}_{(j_4, j_3)}=\hat{\Phi}^{r}_{j_4}(0)\Phi_{j_3}(1|1)
\ee
Introducing the projection operator onto unflowed affine primaries
\be\label{pi0phi}
\Pi_{0}=\int_{\mc{S}^{+}_{\sst\mc{C}}}\!dj\!\int \!d^2 x\;
\Phi_{j}(x)\!\left.\right\rangle\left\langle\right.\!\hat{\Phi}_{1-j}(x)
\ee
the factorization of (\ref{H34pf}) for \(z\ll 1\) may be seen to follow from
\bbb
\amp{3mm}{\hat{\phi}_{(j_4, j_3)}\,\Pi_{0}\,\phi_{(j_2, j_1)}(x|z)}&=&
\int_{\mc{S}^{+}_{\sst \mc{C}}}\!\!dj\!\int\!d^2y\,
\amp{3mm}{\hat{\phi}_{(j_4, j_3)}\Phi_{j}(y)}\amp{3mm}{\hat{\Phi}_{1-j}(y)\phi_{(j_2, j_1)}(x|z)}
\nonumber\\&=&\int_{\mc{S}_{\sst \mc{C}}}\!\!dj\,
\mc{C}(j)\,z^{h(j)-h_2-h_1}\,x^{j-j_2-j_1}\left|F_j(x)\right|^2
\eee
where anti-holomorphic powers of \(z\) and \(x\) are suppressed. Given
\be\label{lamda1}
\lambda_{j}\,\equiv\,\mc{C}(1-j)/\mc{C}(j)\,=\,f(a,b,c)
\ee
where \(f(a,b,c)\) is defined in (\ref{hypfnc}), this follows from
\bbb
\lefteqn{\int\!d^2y\,\amp{3mm}{\hat{\phi}_{(j_4, j_3)}\Phi_{j}(y)}
\amp{3mm}{\hat{\Phi}_{1-j}(y)\phi_{(j_2, j_1)}(x|z)}\,=\,}\nonumber\\
&&\mc{C}(j)\,\abs{10pt}{z^{h(j)-h_2-h_1}\,x^{j-j_2-j_1}}^2\,
\left(\left|F_{j}(x)\right|^2+\lambda_{j}\left|x^{1-2j} F_{1-j}(x)\right|^2\right)
\eee
The KZ equation also implies that the \(H_3\) chiral blocks have a singularity of the form \((z-x)^{\delta}\,\), where \(\delta=k-\sum_{p}^{4}j_{p}\,\). This behavior, which is shown explicitly in (\ref{sfonecnfblk}), may be elucidated by an indentification between the four-point chiral block \(\mc{F}_{j}(x|z)\) of the \(H_3\) model and a five-point conformal block of Liouville theory with a degenerate operator insertion at worldsheet position \(x\,\). This (FZ) relation, which follows from a similar relation discovered by Fateev and Zamolodchikov between the four-point chiral block in the \(SU(2)\) WZNW model and a five-point conformal block in the minimal model, was the foundation of a proof of crossing symmetry~\cite{Teschner_0108} in the \(H_3\) model. The FZ relation was further investigated in~\cite{Ponsot_0204}, where it was found that the fusion matrices of the \(H_3\) model may be written as a sum of two fusion matrices in Liouville theory. This result was somewhat suggestive of a more straightforward and fundamental identification, described in~\cite{RibaultTeschner_0502} utilizing the \(\varphi_{j}(\mu)\,\) basis (\ref{phitolv}), between \(n\)-point primary correlators in the \(H_3\) model and \((2n-2)\)-point primary correlators with \(n-2\) degenerate field insertions in the Liouville CFT. The analytic continuation of two-point and three-point primary correlators of unflowed representations from the \(H_3\) model to the \(SL(2,R)\) CFT in the \(\mc{V}_{j}(m)\) basis was treated in~\cite{Satoh_0109}. The description of the associated fusion rules following from the OPE coefficients of primary fields was found to be in complete agreement with \(SL(2,R)\) representation theory. Making use of the symmetric coefficients of the unflowed three-point function defined in (\ref{sym3ptcof}), the \(H_3\) four-point function in the basis \(\mc{V}_{j}(m)\,\) of definite affine weight may be factorized as \cite{MincesNunez_0701}\cite{BaronNunez_0810}\cite{IguriNunez_0908}
\bbb\label{m4pffact}
\lefteqn{\amp{3mm}{\hat{\mc{V}}_{j_4}(m_4)\mc{V}_{j_3}(m_3|1)\mc{V}_{j_2}(m_2|z)\mc{V}_{j_1}(m_1)}\,=\,
\pi^{-2}\,\delta^2(\,\txsum_{p} m_p\,)}\hspace{1cm}\nonumber\\
&&\times\,\txfc{1}{2}\!\int_{\mc{S}_{\sst \mc{C}}}\!\!dj\,R^{-1}_{j}(m)\,
A^{0}(j_4,j_3,j|m_4,m_3)\,A^{0}(j,j_2,j_1|m_2,m_1)\;|\mc{F}^{\sst (0,0)}_{j}(m|z)|^2
\eee
where \(m=m_1+m_2=-m_3-m_4\,\). The \(j_p\) and \(m_p\) dependence is suppressed in \(\mc{F}^{\sst (0,0)}_{j}(m|z)\,\), where the superscript denotes the total spectral flow of the respective OPE coefficients. This factorization shares the domains of validity (\ref{jdom4pf}) and (\ref{h3glbdom}), outside of which discrete contributions are produced.

\subsection{The appearance of flowed intermediate states\label{flowedIM}}

As will be discussed in more detail below, an extensive investigation of correlators involving spectral flowed states, including the computation of the \(|w|=1\) three-point correlator, was carried out in~\cite{MaldacenaOoguri_0111}. Here the focus was on string amplitudes, employing normalizations\footnote{The associated boundary two-point function follows from the definition of the string two-point function, which is gauge-fixed by dividing by the singular factor \(V_{\rm conf}\) described in (\ref{vconf}).} relevant for describing correlation functions in the boundary CFT, and no attempt was made to directly describe a Lorentzian \(SL(2,R)\) CFT with \(AdS_3\) as the target space. In particular, the treatment of the string four-point function in~\cite{MaldacenaOoguri_0111} involved unflowed primary fields in the \(\Phi_{j}(x)\) basis factorized on the \(H_3\) spectrum in the \(x\ll 1\,\) limit. Integrating over the string modulus \(z\,\), four-point string amplitudes of unflowed discrete states were defined through analytic continuation from the \(H_3\) model in the \(\Phi_{j}(x)\) basis, without attempting to first compute the Lorentzian \(SL(2,R)\) four-point function in the \(\mc{V}_{j}(m)\) basis.  It was shown that for \(j_{p}\in\mc{S}_{\sst\mc{D}}\) the monodromy invariant solutions to the KZ equation expressed in terms of \(x\) and \(u=z/x\) lead to an alternative factorization of (\ref{H34pf}) given by
\be\label{H34pfMO}
\amp{3mm}{\hat{\phi}_{(j_4, j_3)}\phi_{(j_2, j_1)}(x|z)}\,=\,
\int_{\tilde{\mc{S}}_{\sst \mc{C}}}\!\!dj\,\mc{C}(j)\;|\mc{G}_{j}(x|u)|^2\,+\,{\rm Discrete}
\ee
where we have
\be\label{4pfOx}
\mc{G}_{j}(x|u)=\mc{F}_{j}(x|u x)=x^{j-j_2-j_1+h(j)-h_2-h_1}\left(u^{h(j)-h_2-h_1}\,G_{j}(u)+\mc{O}(x)\right)
\ee 
and where \(G_{j}(u)={_{2}{\rm F}_{1}}(\alpha,\beta;\gamma;u)\,\) for 
\be
\alpha=j_1+j_2-j\hspace{1cm}\beta=j_3+j_4-j\hspace{1cm}\gamma=k-2j
\ee
Note the appearance of the shifted contour \(\tilde{\mc{S}}_{\sst \mc{C}}=(k-1)/1+i\mathbb{R}\,\). In this context it is useful to note that, in analogy to (\ref{lamda1}),
\be
\tilde{\lambda}_{j}\,\equiv\,\mc{C}(k-1-j)/\mc{C}(j)\,=\,f(\alpha,\beta,\gamma)
\ee
Following integration over the string modulus \(z=ux\) the associated (on-shell) string amplitudes exhibit \(x\) behavior which identifies the intermediate states with flowed \(w=1\) affine modules. This identification, which will emerge in a similar guise below, makes use of the relation
\be\label{hjhtld}
h(j)+j\,=\,h(k-1-j)+(k-1-j)\,=\,h(\tilde{\jmath})+k/4
\ee
where\ \(h(j)=j(1-j)/(k-2)\,\) and \(\tilde{\jmath}=k/2-j\,\). It will now be shown that the result (\ref{H34pfMO}) for \(x\ll 1\,\) may be seen to follow from the projection onto \(w=1\) flowed affine primary states. First, it may be demonstrated that
\bbb\label{onezero3pf}
\amp{3mm}{\hat{\phi}_{(j_4, j_3)}\mc{V}^{+1}_{j}(m)}&=&
f_{j_4}\amp{3mm}{\hat{\phi}_{(\tilde{\jmath}_4, j_3)}\mc{V}_{j}(-m)}
\nonumber\\&=&-f_{j_4}\,\frac{C(j,j_3,\tilde{\jmath}_4)}{\gamma(j+j_3-\tilde{\jmath}_4)}\,
\frac{\Gamma(2j-1)}{\Gamma(1-2j)}\,\frac{g(j_{3}-m-\tilde{\jmath}_4)}{g(j- m)}
\eee
Here \(g(m)=\Gamma(m)/\Gamma(1-\bar{m})\,\) has been introduced which satisfies \(g(m)=(-1)^{L}g(\bar{m})\,\) for \(m-\bar{m}=L\in\mathbb{Z}\,\), and for \(m=\bar{m}\) we have \(g(m)=\gamma(m)\,\). This is equivalent to
\bbb
\amp{3mm}{\hat{\phi}_{(j_3,j_2)}\Phi^{+1}_{j}(x)}&=&\lim_{x_3\goto 0}\,x_3^{2j_3}
\amp{3mm}{\hat{\Phi}_{j_3}(x_3)\Phi_{j_2}(1|1)\Phi^{+1}_{j}(x)}\\
&=&f_{0}\lim_{z\goto 0}\,
z^{j}\amp{3mm}{\hat{\phi}_{(j_3,j_2)}\Phi_{j}(xz|z)\,\Phi_{k/2}(0)}\\
&=&f_{j_3}\,C(\tilde{\jmath}_3,j_2,j)\ (1-x^{-1})^{\tilde{\jmath}_3-j_2-j}\,x^{-2 j}
\eee
Also introduced above is
\be
f_{j}\,=\,f^{-1}_{\tilde{\jmath}}\,=\,
\omega_{j}\frac{(1-2j)}{(1-2\tilde{\jmath})}\,=\,
\nu^{k/2-1}\,b^2\,\pi\,B_{j}
\ee
Furthermore, suppressing anti-holomorphic variables
\be\label{phiflwrel}
\amp{3mm}{\hat{\mc{V}}^{-1}_{1-j}(-m)\phi_{(j_2, j_1)}(x|z)}\,=\,
z^{j_2}\,x^{-2j_2}\,f_{j_1}\amp{3mm}{\hat{\mc{V}}_{1-j}(m)\phi_{(j_2, \tilde{\jmath}_1)}(z/x|z)}
\ee 
Consider the following projection operators built from flowed primaries of definite affine weight
\be
\Pi_{w}=\txfc{1}{2}\!\int_{\mc{S}_{\sst\mc{C}}}\!dj\,
\pi^2\!\!\int\!d^2 m\;
\mc{V}_{j}^{w}(m)\!\left.\right\rangle\left\langle\right.\!\hat{\mc{V}}^{-w}_{1-j}(-m)
\ee
where \(\Pi_{0}\) is equivalent to (\ref{pi0phi}). The \(x\ll 1\) limit of the factorization of the unflowed four-point function may be seen to be equivalent to the projection onto the subspace of intermediate \(w=1\) spectral flowed affine primaries defined through the following analytic continuation. For \(j_p\in\mc{S}_{\sst\mc{C}}\,\), and using \(\tilde{\mc{C}}(j)\equiv B_{j}^{-1}C(\tilde{\jmath}_4,j_3,j)\,C(j,j_2,\tilde{\jmath}_1)=f_{\tilde{\jmath}_1}f_{\tilde{\jmath}_4}\,\mc{C}(\tilde{\jmath})\,\),
\bbb\label{proj1}
\lefteqn{
\amp{3mm}{\hat{\phi}_{(j_4, j_3)}\,\Pi_{1}\,\phi_{(j_2, j_1)}(x|z)}\,}\nonumber\\
&&\,=\,\underset{\tilde{\jmath}_4\goto j_4}{\st AC}\,
\underset{\tilde{\jmath}_1\goto j_1}{\st AC}\,
\txfc{1}{2}\!\int_{\mc{S}_{\sst \mc{C}}}\!\!dj\,
\pi^2\!\int\!d^2m\,\amp{3mm}{\hat{\phi}_{(\tilde{\jmath}_4, j_3)}\mc{V}^{+1}_{j}(m)}
\amp{3mm}{\hat{\mc{V}}^{-1}_{1-j}(-m)\phi_{(j_2, \tilde{\jmath}_1)}(x|z)}\nonumber\\[3mm]
&&\,=\,z^{j_2}\,x^{-2j_2}\,\underset{\tilde{\jmath}_4\goto j_4}{\st AC}\,
\underset{\tilde{\jmath}_1\goto j_1}{\st AC}\,
\txfc{1}{2}\!\int_{\mc{S}_{\sst \mc{C}}}\!\!dj
\!\int\!d^2y\,f_{\tilde{\jmath}_1}f_{\tilde{\jmath}_4}\amp{3mm}{\hat{\phi}_{(j_4, j_3)}\Phi_{j}(y)}
\amp{3mm}{\hat{\Phi}_{1-j}(y)\phi_{(j_2, j_1)}(z/x|z)}\nonumber\\[3mm]
&&\,=\,z^{j_2}\,x^{-2j_2}\!\int_{\tilde{\mc{S}}_{\sst \mc{C}}}\!\!dj\,
\mc{C}(\tilde{\jmath})\,z^{h(\tilde{\jmath})+\tilde{\jmath}-k/4-h_2-h(\tilde{\jmath}_1)}\,
u^{k/2-\tilde{\jmath}-j_2-\tilde{\jmath}_1}
\left|G_{\tilde{\jmath}}(u)\right|^2\nonumber\\[2mm]&&\,=\,\int_{\mc{S}_{\sst \mc{C}}}\!\!dj\,
\mc{C}(j)\, z^{h(j)-h_2-h_1}\,x^{j-j_2-j_1}\left|G_{j}(u)\right|^2
\eee
where in the third line the relation (\ref{phiflwrel}) has been used, followed by the insertion of \(\Pi_0\) in the form (\ref{pi0phi}). In the fourth line we have made use of the fact that, in the process of the A.C. \(j_1,j_4\goto \tilde{\jmath}_1,\tilde{\jmath}_4\,\), \({\rm Poles_{3\pm}}\) are avoided and no \({\rm Poles_1}\) or \({\rm Poles_2}\) are encountered if  the contour is also shifted from \(\mc{S}_{\sst \mc{C}}\) to \(\tilde{\mc{S}}_{\sst \mc{C}}\,\). In the final line (\ref{hjhtld}) has been used, and the change of variables \(j\goto \tilde{\jmath}=k/2-j\) has been made. Note that this final expression involves intermediate states \(\mc{V}^{+1}_{\tilde{\jmath}}(m)\) with \(\tilde{\jmath}=k/2-j\in\tilde{\mc{S}}_{\sst \mc{C}}\,\), and that no poles cross the contour for \(j_p\in\mc{S}_{\sst\mc{C}}\,\). The result (\ref{H34pfMO}) appearing in~\cite{MaldacenaOoguri_0111} may be seen to follow from shifting the contour in (\ref{proj1}) back to \(\tilde{\mc{S}}_{\sst \mc{C}}\) and implementing the A.C. to \(j_p\in\mc{S}_{\sst\mc{D}}\,\). It may be seen that (\ref{proj1}) is equivalent to 
\be\label{projcnv}
\amp{3mm}{\hat{\phi}_{(j_4, j_3)}\,\Pi_{1}\,\phi_{(j_2, j_1)}(x|z)}\,=\,
z^{j_2}\,x^{-2j_2}\,f_{j_1}f_{j_4}\,
\underset{j_4\goto \tilde{\jmath}_4}{\st AC}\,
\underset{j_1\goto \tilde{\jmath}_1}{\st AC}\,
\amp{3mm}{\hat{\phi}_{(j_4, j_3)}\,\Pi_{0}\,\phi_{(j_2, j_1)}(u|z)}
\ee
That the analytic continuation implemented here is correct will be shown from the corresponding \(w=1\) OPE coefficients for two unflowed operators in the \(\mc{V}_{j}(m)\) basis which are calculated below.

\subsection{Fusion rules via analytic continuation of extended \texorpdfstring{$H_3$}{} OPE\label{acfus}}

A somewhat different route is taken here than in \cite{BaronNunez_0810} to derive the fusion rules via analytic continuation from the \(H_3\) model. As long as one of the product fields in the OPE is taken to be in a discrete representation, the discussion in section~\ref{sec_4} shows that it is sufficient to examine the transformation of (\ref{proj1}) to include a single unflowed operator in the \(\mc{V}_{j}(m)\) basis. Thus we compute the projection onto \(w=1\) intermediate states of
\be
\amp{3mm}{\hat{\phi}_{(j_4, j_3)}\mc{V}_{j_2}(m_2|z)\,\Phi_{j_1}(0)}\,=\,
\mc{R}^{-1}_{j_2}R_{j_2}(m_2)\,\pi^{-1}\!\int d^{2}x\,x^{j_2-m_2-1}\,
\amp{3mm}{\hat{\phi}_{(j_4, j_3)}\phi_{(j_2, j_1)}(x|z)}
\ee
which, with \(u=z/x\,\) and \(j_p\in\mc{S}_{\sst\mc{C}}\,\), is computed as
\bbb\label{proj2}
\lefteqn{
\amp{3mm}{\hat{\phi}_{(j_4, j_3)}\,\Pi_{1}\,\mc{V}_{j_2}(m_2|z)\,\Phi_{j_1}(0)}}
\nonumber\\&&\,=\,\mc{R}^{-1}_{j_2}R_{j_2}(m_2)\,\pi^{-1}\!\!\int d^{2}x\,x^{j_2-m_2-1}
\underset{\tilde{\jmath}_4\goto j_4}{\st AC}\,
\underset{\tilde{\jmath}_1\goto j_1}{\st AC}
\int_{\tilde{\mc{S}}_{\sst \mc{C}}}\!\!dj\,\tilde{\mc{C}}(j)\,
z^{h(j)-h_2-h(\tilde{\jmath}_1)}\,x^{j-j_2-\tilde{\jmath}_1}
\left|\right.\hspace{-3pt}\tilde{G}_{j}(u)\hspace{-3pt}\left.\right|^2\hspace{1cm}
\eee
where \(\tilde{G}_{j}(u)={_{2}{\rm F}_{1}}(\tilde{\alpha},\tilde{\beta};\gamma;u)\,\) and 
\be
\tilde{\alpha}=\tilde{\jmath}_1+j_2-j\hspace{1cm}\tilde{\beta}=\tilde{\jmath}_4+j_3-j\hspace{1cm}\gamma=k-2j
\ee
Now,
\be
\tilde{\mc{C}}(k-1-j)/\tilde{\mc{C}}(j)=f(\tilde{\alpha},\tilde{\beta},\gamma)=\lambda_{\tilde{\jmath}}
\ee
And,
\be
\tilde{G}_{k-1-j}(u)\,=\,{_{2}{\rm F}_{1}}(1+\tilde{\alpha}-\gamma,1+\tilde{\beta}-\gamma;2-\gamma;u)
\ee
We compute the \(x\) integral by writing (\ref{proj1}) as a function of \(z\) and \(u=z/x\) in order to integrate over \(u\) in (\ref{proj2}) using \(x=z/u\,\). Following this the projection onto \(w=0\) intermediate affine primaries will be considered. In the computation of string amplitudes~\cite{MaldacenaOoguri_0111}\cite{AharonyKomargodski_0711}, the integral over the string modulus \(z\) may be computed first by writing (\ref{proj1}) as a function of \(x\) and \(u=z/x\) in order to integrate over \(u\) using \(z=ux\,\). Thus in the first case we consider
\be
\txfc{1}{2}\!\int_{\tilde{\mc{S}}_{\sst \mc{C}}}\!\!dj\,\tilde{\mc{C}}(j)\,
z^{h(j)+j-h_2-j_2-h(\tilde{\jmath}_1)-\tilde{\jmath}_1}\,u^{j_2+\tilde{\jmath}_1-j}
\left(\left|\right.\hspace{-3pt}\tilde{G}_{j}(u)\hspace{-3pt}\left.\right|^2+
\lambda_{\tilde{\jmath}}\left|\right.\hspace{-3pt}u^{1-\gamma}\,
\tilde{G}_{k-1-j}(u)\hspace{-3pt}\left.\right|^2\right)
\ee
while in the case of string amplitudes it is also useful to examine
\be
\txfc{1}{2}\!\int_{\tilde{\mc{S}}_{\sst \mc{C}}}\!\!dj\,\tilde{\mc{C}}(j)\,
x^{h(j)+j-h_2-j_2-h(\tilde{\jmath}_1)-\tilde{\jmath}_1}\,u^{h(j)-h_2-h(\tilde{\jmath}_1)}
\left(\left|\right.\hspace{-3pt}\tilde{G}_{j}(u)\hspace{-3pt}\left.\right|^2+
\lambda_{\tilde{\jmath}}\left|\right.\hspace{-3pt}u^{1-\gamma}\,
\tilde{G}_{k-1-j}(u)\hspace{-3pt}\left.\right|^2\right)
\ee
Making use of (\ref{hypint}) and (\ref{hjhtld}) the amplitude (\ref{proj2}) is given by
\bbb
\lefteqn{\amp{3mm}{\hat{\phi}_{(j_4, j_3)}\,\Pi_{1}\,\mc{V}_{j_2}(m_2|z)\,\Phi_{j_1}(0)}}
\hspace{2cm}\nonumber\\[2mm]&=&\mc{R}^{-1}_{j_2}R_{j_2}(m_2)\,
\underset{\tilde{\jmath}_4\goto j_4}{\st AC}\,
\underset{\tilde{\jmath}_1\goto j_1}{\st AC}\,
\txfc{1}{2}\!\int_{\tilde{\mc{S}}_{\sst \mc{C}}}\!\!dj\,\tilde{\mc{C}}(j)\,
z^{h(j)+j-h_2-m_2-h(\tilde{\jmath}_1)-\tilde{\jmath}_1}\nonumber\\&&
\hspace{1cm}\pi^{-1}\!\!\int d^{2}u\,u^{m_2+\tilde{\jmath}_1-j-1}
\left(\left|\right.\hspace{-3pt}\tilde{G}_{j}(u)\hspace{-3pt}\left.\right|^2+
\lambda_{\tilde{\jmath}}\left|\right.\hspace{-3pt}u^{1-\gamma}\,
\tilde{G}_{k-1-j}(u)\hspace{-3pt}\left.\right|^2\right)
\nonumber\\[2mm]&=&
f_{j_1}f_{j_4}\,\frac{\gamma(2j_2)}{(1-2j_2)}\,\frac{g(j_3+j_4-m_2-j_1)}{g(j_2+\bar{m}_2)}
\nonumber\\&&\label{weq1proj}\underset{\tilde{\jmath}_4\goto j_4}{\st AC}\,
\underset{\tilde{\jmath}_1\goto j_1}{\st AC}\,
\txfc{1}{2}\!\int_{\mc{S}_{\sst \mc{C}}}\!\!dj\,\mc{C}_{\sst S}(j)\,
z^{h(j)+k/4-m_2-\tilde{\jmath}_1-h(\tilde{\jmath}_1)-h_2}\,
\frac{g(j+m_2+\tilde{\jmath}_1-k/2)}{g(j-\bar{m}_2-\tilde{\jmath}_1+k/2)}\hspace{1cm}
\eee
Here we have introduced the symmetric function
\be
\mc{C}_{\sst S}(j)\,=\,\mc{C}_{\sst S}(1-j)\,=\,\frac{\mc{C}(j)\,\gamma(2j)}{\gamma(j+j_2-j_1)\gamma(j+j_3-j_4)}
\ee
From (\ref{cfunc}) and (\ref{gdefs2}) it may be seen that \(C(j,j_1,j_2)/\gamma(j+j_2-j_1)\) has the following poles
\bbb
\pm(j-1/2)&=&1/2+f_{nm}+(j_1+j_2-1)\\[1mm]
\pm(j-1/2)&=&1/2+f_{nm}+(\tilde{\jmath}_1+\tilde{\jmath}_2-1)\\[1mm]
\pm(j-1/2)&=&1/2+f_{nm}+j_1-j_2\\[1mm]
\pm(j-1/2)&=&(k-1)/2+f_{nm}+(\tilde{\jmath}_1+j_2-1)
\eee
The corresponding sets of poles which appear in the region between the contours \(\mc{S}_{\sst \mc{C}}\) and \(\tilde{\mc{S}}_{\sst \mc{C}}\) for \(j_p\) in the \(SL(2,R)\) spectrum differ from those given in (\ref{Cjpoles}). The sets \({\rm Poles}_{1}\) and \({\rm Poles}_{2}\) which can cross \(\tilde{\mc{S}}_{\sst \mc{C}}\) still appear. However, of the poles which can cross \(\mc{S}_{\sst \mc{C}}\,\), while \({\rm Poles}_{3-}\) appear, no \({\rm Poles}_{3+}\) at \(j=j_1-j_2-n\,\) appear. Rather new poles at \(1-j=j_2-j_1-n\,\) are introduced which are \(j\goto 1-j\,\) reflections of \({\rm Poles}_{3-}\,\).

Note that no sets of \(j\)-dependent or \(m\)-dependent poles in (\ref{weq1proj}) cross \(\mc{S}_{\sst \mc{C}}\) for \(m_2\in i\mathbb{R}\,\). Furthermore, under the A.C. \(\tilde{\jmath}_1\goto j_1\) and \(\tilde{\jmath}_4\goto j_4\) for \(j_p\in\mc{S}_{\sst \mc{C}}\,\) no \(j\)-dependent poles cross the contour. However, introducing the global \(J_{0}^{3}\) eigenvalue \(M=m+k/2=m_2+j_1\,\), and defining \(2m^{\pm}=E\pm |L|\,\) for \(E=m+\bar{m}\) and \(L=m-\bar{m}\in\mathbb{Z}\,\), \(m\)-dependent poles appear unless
\be
{\rm Re}(m^{+})={\rm Max}({\rm Re}(m),{\rm Re}(\bar{m}))>-1/2
\ee
If this condition is satisfied then
\bbb\label{weq1m4pf}
\lefteqn{\amp{3mm}{\hat{\phi}_{(j_4, j_3)}\,\Pi_{1}\,\mc{V}_{j_2}(m_2|z)\,\Phi_{j_1}(0)}}
\nonumber\\[2mm]&=&f_{j_1}f_{j_4}\,\frac{\gamma(2j_2)}{(1-2j_2)}\,
\frac{{g(j_3+j_4-m_2-j_1)}}{g(j_2+\bar{m}_2)}\,
\txfc{1}{2}\!\int_{\mc{S}_{\sst \mc{C}}}\!\!dj\,\tilde{\mc{C}}_{\sst S}(j)\,
z^{h(j)-m-k/4-h_1-h_2}\,\frac{g(j+m)}{g(j-\bar{m})}
\eee
where we have defined 
\be
\tilde{\mc{C}}_{\sst S}(j)\,=\,
\underset{j_4\goto \tilde{\jmath}_4}{\st AC}\,
\underset{j_1\goto \tilde{\jmath}_1}{\st AC}\,\mc{C}_{\sst S}(j)\,=\,\frac{\tilde{\mc{C}}(j)\,
\gamma(2j)}{\gamma(j+j_2-\tilde{\jmath}_1)\gamma(j+j_3-\tilde{\jmath}_4)}
\ee
The corresponding intermediate states in this expression are \(w=1\) affine primaries \(\mc{V}^{+1}_{j}(m)\,\) with conformal dimension \(H=h(j)-m-k/4\,\), and the amplitude is defined through A.C. from \({\rm Re}(m^{+})>-1/2\,\) and \(j_p\in\mc{S}_{\sst\mc{C}}\,\). As shown below, under further A.C. there are discrete contributions from \(j\)-dependent poles which are duplicated under \(j\goto 1-j\,\), thus canceling the factor of \(1/2\) in front of the integral. These cross \(\mc{S}_{\sst \mc{C}}\) for \(j=j_2-\tilde{\jmath}_1-n\) corresponding to \(\mc{D}^{+}_{j}(1)\,\) with \(1/2<j<k/2-1\) for \((k+1)/2<j_1+j_2<(k-1)\,\). For \({\rm Re}(m^{+})<-1/2\,\) there are duplicated \(m\)-dependent poles for \(j=-m^{+}-n\) crossing \(\mc{S}_{\sst \mc{C}}\) corresponding to \(\mc{D}^{-}_{j}(1)\,\). The associated \(w=1\) discrete contributions to the OPE via A.C. from the \(H_3\) model for \(j_1\in\mc{S}_{\sst \mc{D}}\,\) (so that \(\Phi_{j_1}(0)=(1-2j_1)\,\mc{V}_{j_1}(j_1)\in\mc{B}^{+}_{j_1}(0)\,\)) are now considered for comparison with the calculation of the fusion rules for elements of the \(SL(2,R)\) spectrum given in section~\ref{sec_5}\,.

\(\mc{C}_{0}\times\mc{B}^{+}_{0}\,\goto\,\mc{D}^{-}_{1}\)\\[1mm]
Consider \(m_2\in \mathbb{R}\) with \(j_2\in\mc{S}_{\sst \mc{C}}\,\) so that \(\mc{V}_{j_2}(m_2)\in\mc{C}_{j_2}(0)\,\). There are no \(j\)-dependent poles for \(j=j_1+j_2-k/2-n\) which cross \(\mc{S}_{\sst \mc{C}}\,\). There are \(m\)-dependent poles for \(m=m_2+j_1-k/2\,\) which cross \(\mc{S}_{\sst \mc{C}}\,\) for \(j=-m^{+}-n\,\) with \({\rm Re}(m^{+})<-1/2\,\). These correspond to states \(\mc{V}^{+1}_{j}(m)\in\mc{D}^{-}_{j}(1)\,\) with OPE coefficients
\bbb\label{opecof1}
\amp{3mm}{\hat{\phi}_{(j_4, j_3)}\,\Pi_{1}\,\mc{V}_{j_2}(m_2|z)\,\Phi_{j_1}(0)}
&=&\sum_{n=0}^{[-m^{+}]-1}f_{j_1}\amp{3mm}{\hat{\phi}_{(j_4, j_3)}\mc{V}^{+1}_{j_n}(m)}\,
z^{H_{n}-h_1-h_2}\,B_{j_n}^{-1}\nonumber\\&& \gamma(2j_2)\,\frac{(1-2j_n)}{(1-2j_2)}\,
\frac{C(j_{n},j_2,\tilde{\jmath}_1)}{\gamma(j_{n}+j_2-\tilde{\jmath}_1)}
\oint_{j_{n}}\!\!dj\,\frac{g(j+\bar{m})}{g(j_2+\bar{m}_2)}\hspace{1cm}
\eee
where \(j_n=-m^{+}-n\,\) and \(H_{n}=h(j_n)-m-k/4\,\).   The range of the sum is imposed by \({\rm Re}(j)>1/2\,\) which fixes \(0\leq n\leq [-m^{+}]-1\,\), where \([-m^{+}]\) is \(-m^{+}\) rounded to the nearest integer. The above OPE coefficient may be shown to be finite using
\be\label{mmbartilde}
\oint_{j_{n}}\!\!dj\,\frac{g(j+\bar{m})}{g(j'+\bar{m})}\,=\,
\oint_{j_{n}}\!\!dj\,\frac{g(j+m)}{g(j'+m)}\,=\,
\frac{2\pi\,(-1)^n\,\Gamma(1-j'-m^{-})}{\Gamma(1+n)\Gamma(1+n+|L|)\Gamma(j'+m^{+})}
\ee
where \(j'=j_2+m_2-m=j_2+\tilde{\jmath}_1\,\). It should be noted that while \(j_n>1/2\) there is no truncation of the intermediate states to \(j_n<(k-1)/2\,\). Of course this represents an apparent failure of the closure of the OPE on elements of the \(SL(2,R)\) spectrum.

\(\mc{D}^{+}_{0}\times\mc{B}^{+}_{0}\,\goto\,\mc{D}^{-}_{1}\)\\[1mm] 
Consider \(m_2=j_2+n_2\) with \(j_1+j_2<(k-1)/2\,\) so that \(\mc{V}_{j_2}(m_2)\in\mc{D}^{+}_{j_2}(0)\,\) for \(j_2\in\mc{S}_{\sst \mc{D}}\,\), and there are no \(j\)-dependent poles. There are \(m\)-dependent poles for \(m=j_2+n_2-\tilde{\jmath}_1\,\) which cross \(\mc{S}_{\sst \mc{C}}\,\) for \(j=-m^{+}-n\,\) with \({\rm Re}(m^{+})<-1/2\,\). Here \(m^{+}=j_2+n^{+}_2-\tilde{\jmath}_1\,\) where \(n^{+}_2={\rm Max}(n_2,\bar{n}_2)\,\). These correspond to states \(\mc{V}^{+1}_{j}(m)\in\mc{D}^{-}_{j}(1)\,\) with OPE coefficients which are identical in form to those in (\ref{opecof1}) with \(j_n=-m^{+}-n=\tilde{\jmath}_1-j_2-n^{+}_2-n\,\) and \(H_{n}=h(j_n)-m-k/4\,\). This may be seen to be finite from (\ref{mmbartilde}) with \(j'=j_2+\tilde{\jmath}_1\,\), and because the poles in \(C(j_{n},j_2,\tilde{\jmath}_1)\,\) cancel against the zeros in \(\gamma(j_{n}+j_2-\tilde{\jmath}_1)\,\) (equivalently \(\mc{C}_{S}(j)\) has no poles for \(j=j_1-j_2-\mathbb{Z}_{\geq 0}\,\)). It may be seen that all intermediate states appearing here satisfy \(j_n\in\mc{S}_{\sst \mc{D}}\,\).

\(\mc{D}^{-}_{0}\times\mc{B}^{+}_{0}\,\goto\,\mc{D}^{-}_{1}\)\\[1mm]
Now consider \(m_2=-j_2-n_2\) so that \(\mc{V}_{j_2}(m_2)\in\mc{D}^{-}_{j_2}(0)\,\). Due to the factor \(\Gamma(j_2+\bar{m}_2)\) in (\ref{weq1m4pf}) the amplitude vanishes for the \(j\)-dependent poles \(j=j_1+j_2-k/2-n\) which cross \(\mc{S}_{\sst \mc{C}}\) for \(j_1+j_2>(k+1)/2\,\). There are also no contributions from the original contour corresponding to states in \(\mc{C}_{j}(1)\,\). There are \(m\)-dependent poles for \(m=-j_2-n_2-\tilde{\jmath}_1\,\) which cross \(\mc{S}_{\sst \mc{C}}\,\) for \(j=-m^{+}-n\,\) with \({\rm Re}(m^{+})<-1/2\,\). Here \(m^{+}=-j_2-n^{-}_2-\tilde{\jmath}_1\,\) where \(n^{-}_2={\rm Min}(n_2,\bar{n}_2)\,\). These correspond to states \(\mc{V}^{+1}_{j}(m)\in\mc{D}^{-}_{j}(1)\,\) with OPE coefficients which are identical in form to those in (\ref{opecof1}) with \(j_n=-m^{+}-n=\tilde{\jmath}_1+j_2+n^{-}_2-n\,\) and \(H_{n}=h(j_n)-m-k/4\,\). In this case the factor (\ref{mmbartilde}) with \(j'=j_2+\tilde{\jmath}_1\,\) takes the form 
\be
\frac{2\pi\,(-1)^n\,\Gamma(1-j'-m^{-})}{\Gamma(1+n)\Gamma(1+n+|L|)\Gamma(j'+m^{+})}\,=\,
\frac{2\pi\,(-1)^n\,\Gamma(1+n^{+}_2)}{\Gamma(1+n)\Gamma(1+n+|L|)\Gamma(-n^{-}_2)}
\ee
where \(|L|=n^{+}_2-n^{-}_2\,\). The zeros which appear here are canceled by poles from \(C(j_{n},j_2,\tilde{\jmath}_1)\)
for \(j_n=\tilde{\jmath}_1+j_2+\mathbb{Z}_{\geq 0}\,\). This imposes \(n^{-}_2-n\geq 0\,\) and leads to finite OPE coefficients. Here again, while \(j_n>1/2\) there is no truncation of the intermediate states to \(j_n<(k-1)/2\,\).

\(\mc{D}^{+}_{0}\times\mc{B}^{+}_{0}\,\goto\,\mc{D}^{+}_{1}\)\\[1mm]
Consider \(m_2=j_2+n_2\) with \(j_1+j_2>(k+1)/2\,\) so that \(\mc{V}_{j_2}(m_2)\in\mc{D}^{+}_{j_2}(0)\,\), and there are no \(m\)-dependent poles which cross \(\mc{S}_{\sst \mc{C}}\,\). The \(j\)-dependent poles for \(j_n=j_1+j_2-k/2-n>1/2\) which cross \(\mc{S}_{\sst \mc{C}}\) correspond to states \(\mc{V}^{+1}_{j}(m)\in\mc{D}^{+}_{j}(1)\,\) with \(m=m_2+j_1-k/2=j_n+n+n_2\,\). The following OPE coefficients\footnote{Note that the three-point functions corresponding to (\ref{onezero3pf}) which appear in this expression actually vanish. This is irrelevant in terms of the computation of the OPE coefficients which determine the fusion rules via analytic continuation, and would not occur in general if all states were in the \(\mc{V}_{j}(m)\) basis.} may be seen to be finite 
\bbb\label{opecof2}
\amp{3mm}{\hat{\phi}_{(j_4, j_3)}\,\Pi_{1}\,\mc{V}_{j_2}(m_2|z)\,\Phi_{j_1}(0)}
&=&\sum_{n=0}^{[j_2-\tilde{\jmath}_1]-1}\amp{3mm}{\hat{\phi}_{(j_4, j_3)}\mc{V}^{+1}_{j_n}(j_n+n+n_2)}
z^{H_{n}-h_1-h_2}\,f_{j_1}\,B_{j_{n}}^{-1}\nonumber\\&& \frac{\gamma(2j_2)\,
g(2j_n+n+\bar{n}_2)}{g(2j_2+\bar{n}_2)\,\gamma(2j_n+n)}\,\frac{(1-2j_n)}{(1-2j_2)}\,
\oint_{j_{n}}\!\!dj\,C(j,j_2,\tilde{\jmath}_1)\hspace{1cm}
\eee
where \(H_n=h(j_n)-j_n-n-n_2-k/4\,\). All intermediate states appearing here satisfy \(j_n\in\mc{S}_{\sst \mc{D}}\,\).

We would also like to project onto \(w=0\) primary states. Using (\ref{hypint}), for \(j_p\in\mc{S}_{\sst\mc{C}}\,\)
\bbb\label{weq0m4pf}
\lefteqn{\amp{3mm}{\hat{\phi}_{(j_4, j_3)}\,\Pi_{0}\,\mc{V}_{j_2}(m_2|z)\,\Phi_{j_1}(0)}}\hspace{2cm}\nonumber \\&=&
\mc{R}^{-1}_{j_2}R_{j_2}(m_2)\,
\txfc{1}{2}\!\int_{\mc{S}_{\sst \mc{C}}}\!\!dj\,\mc{C}(j)\,z^{h(j)-h_2-h_1}\nonumber\\&&
\pi^{-1}\!\!\int d^{2}x\,x^{j-m_2-j_1-1}
\left(\left|\right.\hspace{-3pt}F_{j}(x)\hspace{-3pt}\left.\right|^2+
\lambda_{j}\left|\right.\hspace{-3pt}x^{1-c}\,
F_{1-j}(x)\hspace{-3pt}\left.\right|^2\right)\nonumber\\[2mm]&=&
\frac{\gamma(2j_2)}{(1-2j_2)}\,\frac{g(j_3-j_4+m_2+j_1)}{g(j_2-\bar{m}_2)}\,
\txfc{1}{2}\!\int_{\mc{S}_{\sst \mc{C}}}\!\!dj\,\mc{C}_{\sst S}(j)\,z^{h(j)-h_2-h_1}\,
\frac{g(j-m)}{g(j+\bar{m})}\hspace{1cm}
\eee
As above, we introduce the global \(J_{0}^{3}\) eigenvalue \(M=m=m_2+j_1\,\), and define \(2m^{\pm}=E\pm |L|\,\) for \(E=m+\bar{m}\) and \(L=m-\bar{m}\in\mathbb{Z}\,\). Note that no sets of \(j\)-dependent or \(m\)-dependent poles in (\ref{weq0m4pf}) cross \(\mc{S}_{\sst \mc{C}}\) for \(m_2\in i\mathbb{R}\,\) and \(j_p\in\mc{S}_{\sst\mc{C}}\,\). More generally, \(m\)-dependent poles appear unless
\be
{\rm Re}(m^{-})={\rm Min}({\rm Re}(m),{\rm Re}(\bar{m}))<1/2
\ee
The corresponding intermediate states in this expression are unflowed primaries \(\mc{V}_{j}(m)\,\), and the amplitude is defined through A.C. from \({\rm Re}(m^{-})<1/2\,\) and \(j_p\in\mc{S}_{\sst\mc{C}}\,\). As shown below, there are discrete contributions from duplicated \(j\)-dependent poles crossing \(\mc{S}_{\sst \mc{C}}\) for \(j=j_2-j_1-n\) (\({\rm Poles}_{3-}\)) corresponding to \(\mc{D}^{-}_{j}(0)\,\) with \(1/2<j<k/2-1\,\) for \(1/2<j_2-j_1<k/2-1\,\). For \({\rm Re}(m^{-})>1/2\,\) there are duplicated poles for \(j=m^{-}-n\) crossing \(\mc{S}_{\sst \mc{C}}\) corresponding to \(\mc{D}^{+}_{j}(0)\,\). The associated discrete \(w=0\) contributions to the OPE via A.C. from the \(H_3\) model for \(j_1\in\mc{S}_{\sst \mc{D}}\,\) are now considered for comparison with the results of section~\ref{sec_5}\,.

\(\mc{C}_{0}\times\mc{B}^{+}_{0}\,\goto\,\mc{D}^{+}_{0}\)\\[1mm]
Consider \(m_2\in \mathbb{R}\) with \(j_2\in\mc{S}_{\sst \mc{C}}\,\) so that \(\mc{V}_{j_2}(m_2)\in\mc{C}_{j_2}(0)\,\). There are no \(j\)-dependent poles for \(j=j_2-j_1-n\) which cross \(\mc{S}_{\sst \mc{C}}\,\). There are \(m\)-dependent poles for \(m=m_2+j_1\,\) which cross \(\mc{S}_{\sst \mc{C}}\,\) for \(j=m^{-}-n\,\) with \({\rm Re}(m^{-})>1/2\,\). These correspond to states \(\mc{V}_{j}(m)\in\mc{D}^{+}_{j}\,\) with OPE coefficients
\bbb\label{opecof3}
\amp{3mm}{\hat{\phi}_{(j_4, j_3)}\,\Pi_{0}\,\mc{V}_{j_2}(m_2|z)\,\Phi_{j_1}}
&=&\sum_{n=0}^{[m^{-}]-1}\amp{3mm}{\hat{\phi}_{(j_4, j_3)}\mc{V}_{j_n}(m)}\,
z^{H_{n}-h_1-h_2}\,B_{j_n}^{-1}\nonumber\\&& \gamma(2j_2)\,\frac{(1-2j_n)}{(1-2j_2)}\,
\frac{C(j_{n},j_2,j_1)}{\gamma(j_{n}+j_2-j_1)}
\oint_{j_{n}}\!\!dj\,\frac{g(j-\bar{m})}{g(j_2-\bar{m}_2)}\hspace{1cm}
\eee
where \(j_{n}=m^{-}-n\,\) and \(H_{n}=h(j_n)\,\). The range of the sum is imposed by \({\rm Re}(j)>1/2\,\) which fixes \(0\leq n\leq [m^{-}]-1\,\). Note that, due to the relation
\bbb\label{onezero4pf}
\amp{3mm}{\hat{\phi}_{(j_4, j_3)}\,\Pi_{1}\,\mc{V}_{j_2}(m_2|z)\,\Phi_{j_1}}&=&
\frac{\gamma(2j_2)}{(1-2j_2)}\,
\frac{\Gamma(1-j_2-m_2)}{\Gamma(j_2+\bar{m}_2)}\,\frac{\Gamma(j_3+j_4-m_2-j_1)}
{\Gamma(1-j_3-j_4+\bar{m}_2+j_1)}\nonumber\\&& f_{j_1}\,f_{j_4}\,
\txfc{1}{2}\!\int_{\mc{S}_{\sst \mc{C}}}\!\!dj\,\tilde{\mc{C}}_{S}(j)\,
z^{h(j)-m_2-j_1+k/4-h_1-h_2}\,
\nonumber\\&&\frac{\Gamma(j+m_2+j_1-k/2)\,\Gamma(1-j+m_2+j_1-k/2)}
{\Gamma(1-j-\bar{m}_2-j_1+k/2)\,\Gamma(j-\bar{m}_2-j_1+k/2)}
\nonumber\\[2mm]&=& z^{-m_2}\,f_{j_1}\,f_{j_4}
\amp{3mm}{\hat{\phi}_{(\tilde{\jmath}_4, j_3)}\,\Pi_{0}\,\mc{V}_{j_2}(-m_2|z)\Phi_{\tilde{\jmath}_1}(0)}
\eee
which is valid without discrete contributions for \(m_2>(k-1)/2\,\) prior to analytic continuation, this OPE coefficient is mapped to (\ref{opecof1}) (divided by \(z^{-m_2}\,f_{j_1}\,f_{j_4}\)) given (\ref{onezero3pf}) if the identifications \(m_2\goto -m_2\,\), \(j_1\goto\tilde{\jmath}_1\,\), and \(j_4\goto\tilde{\jmath}_4\,\) are made. This relation, which follows directly from (\ref{projcnv}), and which implies \(m\goto -m\,\), allows the above arguments for the \(w=1\) case to be followed closely. Thus, (\ref{opecof3}) may be shown to be finite using (\ref{mmbartilde}) with \(m_{\pm}\goto -m_{\pm}\) and \(j'=j_2+j_1\,\). In addition, as for (\ref{opecof1}), while \(j_n>1/2\) there is no truncation of the intermediate states to \(j_n<(k-1)/2\,\).

\(\mc{D}^{-}_{0}\times\mc{B}^{+}_{0}\,\goto\,\mc{D}^{+}_{0}\)\\[1mm]
Using (\ref{onezero4pf}) this follows directly from the case \(\mc{D}^{+}_{0}\times\mc{B}^{+}_{0}\,\goto\,\mc{D}^{-}_{1}\) treated above. Here \(m_2=-j_2-n_2\) with \(j_1-j_2>1/2\,\) so that \(\mc{V}_{j_2}(m_2)\in\mc{D}^{-}_{j_2}(0)\,\), and there are no \(j\)-dependent poles. There are \(m\)-dependent poles for \(m=j_1-j_2-n_2\,\) which cross \(\mc{S}_{\sst \mc{C}}\,\) for \(j=m^{-}-n\,\) with \({\rm Re}(m^{-})>1/2\,\). Here \(m^{-}=j_1-j_2-n^{+}_2\,\) where \(n^{+}_2={\rm Max}(n_2,\bar{n}_2)\,\). These correspond to states \(\mc{V}_{j}(m)\in\mc{D}^{+}_{j}(0)\,\) with OPE coefficients which are identical in form to those in (\ref{opecof3}) with \(j_n=m^{-}-n=j_1-j_2-n^{+}_2-n\,\) and \(H_{n}=h(j_n)\,\). As for \(\mc{D}^{+}_{0}\times\mc{B}^{+}_{0}\,\goto\,\mc{D}^{-}_{1}\,\), all OPE coefficients are finite and all intermediate states appearing here satisfy \(j_n\in\mc{S}_{\sst \mc{D}}\,\).

\pagebreak

\(\mc{D}^{+}_{0}\times\mc{B}^{+}_{0}\,\goto\,\mc{D}^{+}_{0}\)\\[1mm]
Again, using (\ref{onezero4pf}) this follows directly from the case \(\mc{D}^{-}_{0}\times\mc{B}^{+}_{0}\,\goto\,\mc{D}^{-}_{1}\) treated above. Here \(m_2=j_2+n_2\) so that \(\mc{V}_{j_2}(m_2)\in\mc{D}^{+}_{j_2}(0)\,\). Due to the factor \(\Gamma(j_2-\bar{m}_2)\) in (\ref{weq0m4pf}) the amplitude vanishes for the \(j\)-dependent poles \(j=j_2-j_1-n\) which cross \(\mc{S}_{\sst \mc{C}}\) for \(j_2-j_1>1/2\,\). There are also no contributions from the original contour corresponding to states in \(\mc{C}_{j}(0)\,\). There are \(m\)-dependent poles for \(m=j_1+j_2+n_2\,\) which cross \(\mc{S}_{\sst \mc{C}}\,\) for \(j=m^{-}-n\,\) with \({\rm Re}(m^{-})>1/2\,\). Here \(m^{-}=j_1+j_2+n^{-}_2\,\) where \(n^{-}_2={\rm Min}(n_2,\bar{n}_2)\,\). These correspond to states \(\mc{V}_{j}(m)\in\mc{D}^{+}_{j}(0)\,\) with OPE coefficients which are identical in form to those in (\ref{opecof3}) with \(j_n=m^{-}-n=j_1+j_2+n^{-}_2-n\,\) and \(H_{n}=h(j_n)\,\). As for \(\mc{D}^{-}_{0}\times\mc{B}^{+}_{0}\,\goto\,\mc{D}^{-}_{1}\), all OPE coefficients are finite and the condition \(n^{-}_2-n\geq 0\,\) is imposed. Also, while \(j_n>1/2\) there is no truncation of the intermediate states to \(j_n<(k-1)/2\,\).

\(\mc{D}^{-}_{0}\times\mc{B}^{+}_{0}\,\goto\,\mc{D}^{-}_{0}\)\\[1mm]
Using (\ref{onezero4pf}) this case follows from the case \(\mc{D}^{+}_{0}\times\mc{B}^{+}_{0}\,\goto\,\mc{D}^{+}_{1}\). Here \(m_2=-j_2-n_2\) with \(j_2-j_1>1/2\,\) so that \(\mc{V}_{j_2}(m_2)\in\mc{D}^{-}_{j_2}(0)\,\), and there are no \(m\)-dependent poles which cross \(\mc{S}_{\sst \mc{C}}\,\). The \(j\)-dependent poles for \(j_n=j_2-j_1-n>1/2\) which cross \(\mc{S}_{\sst \mc{C}}\) correspond to states \(\mc{V}_{j}(m)\in\mc{D}^{-}_{j}(0)\,\) with \(m=j_1-j_2-n_2=-j_n-n-n_2\,\) and \(H_n=h(j_n)\,\). As is consistent with the identification with \(\mc{D}^{+}_{0}\times\mc{B}^{+}_{0}\,\goto\,\mc{D}^{+}_{1}\,\), all intermediate states appearing here satisfy \(j_n\in\mc{S}_{\sst \mc{D}}\,\), with finite OPE coefficients given by
\bbb\label{opecof4}
\amp{3mm}{\hat{\phi}_{(j_4, j_3)}\,\Pi_{0}\,\mc{V}_{j_2}(m_2|z)\,\Phi_{j_1}}
&=&\sum_{n=0}^{[j_2-j_1]-1}\amp{3mm}{\hat{\phi}_{(j_4, j_3)}\mc{V}_{j_n}(-j_n-n-n_2)}\,
z^{H_{n}-h_1-h_2}\,B_{j_n}^{-1}\nonumber\\&& 
\frac{\gamma(2j_2)\,g(2j_n+n+\bar{n}_2)}{g(2j_2+\bar{n}_2)\,\gamma(2j_n+n)}
\frac{(1-2j_n)}{(1-2j_2)}\oint_{j_{n}}\!\!dj\,C(j,j_2,j_1)\hspace{1cm}
\eee

\subsection{Summary of the fusion rules \label{fussum}}

The fusion rules derived in the previous subsection may be seen to be identical\footnote{The vertex operator normalizations appearing in these notes, which impose \(\mc{V}_{0}(0)=\mathbbm{1}\,\) and lead to a finite discrete two-point function coefficient, differ from those of \cite{BaronNunez_0810}.} to those which appear in \cite{BaronNunez_0810} through analytic continuation of the OPE of the \(H_3\) CFT extended to include (non-normalizable) spectral flowed continuous representations
\be\label{acflowope}
\mc{V}_{j_2}^{w_2}(m_2|1)\,\mc{V}_{j_1}^{w_1}(m_1)\,=\,
(-1)^{\ell_1 w_2}\sum_{|w|\leq 1}\,(-1)^{-\ell_2 w}
\int_{\mc{S}^{+}_{\sst \mc{C}}}\!dj_{3}\,A^{w}(1-j_{3}|-\hat{m})
\left(\mc{V}_{j_3}^{w_3+w}(\hat{m})+\ldots\right)
\ee
Here \(\mc{V}_{j_1}(m_1)\) and \(\mc{V}_{j_2}(m_2)\,\) are elements of the \(H_3\) spectrum with \(j_p\in\mc{S}^{+}_{\sst\mc{C}}\) and \(m_p+\bar{m}_p\in i\mathbb{R}\,\), and the conventions here are as in (\ref{OPEsum}). Except for the absence of a direct imposition of the closure of the \(SL(2,R)\) spectrum, this leads to an expression identical to that in (\ref{OPEsum}), and thus the above comments concerning the patent redundancy of the discrete contributions apply here as well. For intermediate states which are elements of the spectrum, the fusion rules\footnote{For another derivation of the \(SL(2,R)\) fusion rules see~\cite{Ribault_1406}.} following from (\ref{acflowope}) are also in complete agreement with those computed in section~\ref{sec_5}, which are restated below in (\ref{cntfus}), (\ref{mixfus}), (\ref{mpope}), and (\ref{ppope}). Using \(\hat{\mc{D}}^{\pm}_{j_3}(w_3)=\hat{\mc{D}}^{\mp}_{\tilde{\jmath}_3}(w_3\pm 1)\,\), and disregarding subtleties which arise at the boundaries of \(\mc{S}_{\sst \mc{D}}\) for intermediate states in discrete representations, it should be noted that the fusion rule (\ref{mpope}) can be computed via analytic continuation of the \(w=0\,\) OPE coefficient alone, while the fusion rule (\ref{ppope}) can be computed via analytic continuation of the \(w=\pm 1\,\) OPE coefficient alone. In contrast, the fusion rule (\ref{cntfus}) requires all three \(|w|\leq 1\) OPE coefficients, while the fusion rule (\ref{mixfus}) requires two OPE coefficients with \(0\leq \pm w \leq 1\,\). Depending on the boundary (\(j=1/2\,\) or \(\,j=(k-1)/2\)) of \(\mc{S}_{\sst \mc{D}}\) being considered, and depending on the representations of the respective product and intermediate states, there are essentially three cases which appear in the results of the analytic continuation of (\ref{acflowope}) that are summarized below. The first case is when the intermediate states are constrained by the affine algebra (though charge conservation or selection rules), as described in the treatment of section~\ref{sec_5}, to fall in the spectrum of normalizable states due to the domain of the product states. The second case is when the analytic continuation procedure itself restricts intermediate states to the spectrum. The third case is when product states in the spectrum produce intermediate discrete states outside of \(\mc{S}_{\sst \mc{D}}\,\), thus representing an apparent failure of the OPE defined by (\ref{acflowope}) to close on the spectrum of the \(SL(2,R)\) model. 

It will be assumed here, as in \cite{BaronNunez_0810}, that the non-vanishing OPE coefficients involving primary fields which appear in the previous subsection extend to the respective flowed affine modules appearing in (\ref{acflowope}), thus permitting comparison with the \(SL(2,R)\) fusion rules of section~\ref{sec_5}. These fusion rules are assembled again here, with the definition \(w_3=w_1+w_2\,\), and with detailed ranges of the \(j_3\) quantum number appearing only where relevant. For the fusion rule of two continuous states we have
\be\label{cntfus}
\hat{\mc{C}}_{w_2}\times\hat{\mc{C}}_{w_1}\,\goto\,
2\,\hat{\mc{C}}_{w_3}\oplus\hat{\mc{C}}_{w_3+1}\oplus\hat{\mc{C}}_{w_3-1}\oplus
\hat{\mc{D}}^{+}_{w_3}\oplus\hat{\mc{D}}^{-}_{w_3}
\ee
The factor of \(2\) appearing in the flow-conserving contribution of the continuous representations corresponds to two independent intermediate states for fixed \(j_3\) in the \(\hat{\mc{C}}_{j_3}(w_3)\,\) affine module. This follows from the fact that the associated three-point function of primary fields~(\ref{glbint}) is a sum of two linearly independent functions of \(j_1\) and \(j_2\,\). When at least one of these fields is in a discrete representation there is only a single independent contribution which exhibits a chirally factorized form (either \(D_{12} C^{12}\bar{C}^{12}\) or \(D_{21} C^{21}\bar{C}^{21}\) depending on the discrete modules). In the representation theory of \(SL(2,R)\), this is seen in the doubling of the contribution of the continuous representations to the Clebsch-Gordon coefficients of the tensor product of two continuous representations. More generally, the identification of \(|w|=1\) three-point \(H_3\) correlators with three-point correlators in Liouville~\cite{Ribault_0507} makes it apparent why all multiplicities in the fusion rules described here, with the exception of \(\hat{\mc{C}}_{w_2}\times\hat{\mc{C}}_{w_1}\goto 2\,\hat{\mc{C}}_{w_3}\,\), are trivial. As discussed above~(\ref{redreps}), all three point correlators in the \(SL(2,R)\) model may ultimately be reduced to those involving either three unflowed continuous primaries or to those involving two unflowed primaries and a single \(|w|=1\) primary. The former case leads to a fusion coefficient of multiplicity two due to the appearance of two intermediate Virasoro representations in the factorization of the degenerate Liouville four-point amplitude, while the latter cases should be expected to have multiplicity one due to the single Liouville three-point amplitude. Since the relevant ranges of the \(j_3\) quantum number for the discrete representations \(\hat{\mc{D}}^{\pm}_{w_3}\) under analytic continuation from the \(H_3\) model are essentially the same as those for the following case of \(\hat{\mc{C}}_{w_2}\times\hat{\mc{D}}^{\pm}_{w_1}\,\), the corresponding OPE of primary states was omitted in the preceding subsection. However, the analytic continuation implemented in \cite{BaronNunez_0810} leads to the bound \(j_3\geq 1/2\) for the non-vanishing unflowed OPE coefficients containing the primary representations \(\mc{D}^{\pm}_{j_3}(w_3)\,\), and the complementary bound \(j_3\leq (k-1)/2\,\), or \(\tilde{\jmath}_3=k/2-j_3\geq 1/2\,\), for the non-vanishing \(|w|=1\) flowed OPE coefficients containing the primary representations \(\mc{D}^{\mp}_{\tilde{\jmath}_3}(w_3\pm 1)\,\).

For one continuous and one discrete state we have
\be\label{mixfus}
\hat{\mc{C}}_{w_2}\times\hat{\mc{D}}^{\pm}_{w_1}\,\goto\,
\hat{\mc{D}}^{\pm}_{w_3}\oplus\hat{\mc{C}}_{w_3}\oplus\hat{\mc{C}}_{w_3\pm 1}
\ee
The fusion rules of the preceding subsection impose constraints through analytic continuation on the \(j_3\) quantum number for the discrete representations \(\hat{\mc{D}}^{\pm}_{j_3}(w_3)=\hat{\mc{D}}^{\mp}_{\tilde{\jmath}_3}(w_3\pm 1)\,\). In particular, in the OPE of primary fields given by \(\mc{C}_{0}\times\mc{B}^{+}_{0}\,\goto\,\mc{D}^{+}_{0}\) the intermediate primary representation \(\mc{D}^{+}_{j_3}(0)\) is constrained at the lower boundary of \(\mc{S}_{\sst \mc{D}}\) to satisfy \(j_3>1/2\,\), but is unconstrained at the upper boundary. Similarly, in the OPE of primary fields given by \(\mc{C}_{0}\times\mc{B}^{+}_{0}\,\goto\,\mc{D}^{-}_{1}\) the corresponding intermediate flowed primary representation \(\mc{D}^{-}_{\tilde{\jmath}_3}(1)\) is constrained to satisfy \(\tilde{\jmath}_3>1/2\,\), but is not constrained to satisfy \(\tilde{\jmath}_3<(k-1)/2\,\). As emphasized in \cite{BaronNunez_0810}, these cases are complementary since \(\tilde{\jmath}=k/2-j\,\), and the overlap in the two cases is \(\mc{S}_{\sst \mc{D}}\,\). However, note again that both the \(w=0\) and \(w=\pm 1\) OPE coefficients in (\ref{acflowope}) are required to produce (\ref{mixfus}), and that neither of the respective expansions exhibit closure on the \(SL(2,R)\) spectrum. It will be argued below that these properties may be related due to the equivalence of alternative factorizations required by a definition of \(SL(2,R)\) amplitudes through analytic continuation from the \(H_3\) model.

For the fusion rules of discrete representations we take \(\pm w\geq 0\) for \(\hat{\mc{D}}^{\pm}_{w}\,\), and treat the below fusion rules (\ref{mpope}) and (\ref{ppope}) as distinct despite the equivalence as expressed in (\ref{discgenope}).  Defining, \(j=|j_1-j_2|\) and \(\sigma={\rm sign}(j_1-j_2)\,\), in section~\ref{sec_5} the following result was found under the assumption that the OPE closes on the \(SL(2,R)\) spectrum
\be\label{mpope}
\hat{\mc{D}}^{-}_{j_2}(w_2)\times\hat{\mc{D}}^{+}_{j_1}(w_1)\,\goto\,
\hat{\mc{C}}_{w_3}\;\oplus\sum_{n=0}^{[\,j\,]-1}\hat{\mc{D}}^{\sigma}_{j-n}(w_3) 
\ee
Here it was necessary to impose the condition \(j_3=j-n\in\mc{S}_{\sst\mc{D}}\) at the lower bound of \(\mc{S}_{\sst\mc{D}}\) (\(j_3\geq 1/2\)) through the upper bound\footnote{Again, here \([j]\) is \(j\) rounded to the nearest integer.} of the sum, which requires \(j\geq 1/2\,\). The condition \(j_3\leq (k-1)/2\,\) at the upper bound of \(\mc{S}_{\sst\mc{D}}\) is satisfied automatically if \(j_1,j_2\in\mc{S}_{\sst\mc{D}}\,\). Now consider the constraints imposed in the preceding subsection though analytic continuation from the \(H_3\) model. For the OPE of primary fields for \(\mc{D}^{-}_{0}\times\mc{B}^{+}_{0}\,\goto\,\mc{D}^{+}_{0}\,\) given above \(j_1-j_2>1/2\) and \(j_3>1/2\) are imposed by the analytic continuation procedure. Conversely, for the OPE coefficients in the case \(\mc{D}^{-}_{0}\times\mc{B}^{+}_{0}\,\goto\,\mc{D}^{-}_{0}\) analytic continuation imposes \(j_2-j_1>1/2\) and \(j_3>1/2\,\). Thus the non-vanishing \(w=0\) OPE coefficients exhibit closure on the \(SL(2,R)\) spectrum in the case (\ref{mpope}). Perhaps not coincidentally, since the unflowed continuous module \(\hat{\mc{C}}_{w_3}\) is also produced by (\ref{weq0m4pf}), the \(w=0\) OPE coefficients also produce all terms in (\ref{mpope}). This is to be contrasted with the \(w=1\) OPE coefficients appearing in \(\mc{D}^{-}_{0}\times\mc{B}^{+}_{0}\,\goto\,\mc{D}^{-}_{1}\,\). The intermediate states \(\mc{V}^{+1}_{\tilde{\jmath}_3}(m)\in\mc{D}^{-}_{\tilde{\jmath}_3}(1)\,\) are not constrained to satisfy \(\tilde{\jmath}_3<(k-1)/2\,\). Thus \(j_3=k/2-\tilde{\jmath}_3\) violates the lower bound on \(\mc{S}_{\sst\mc{D}}\,\) and the \(w=1\) OPE coefficients fail to exhibit closure on the \(SL(2,R)\) spectrum. The \(w=-1\) OPE coefficients defined through analytic continuation may be seen to behave similarly in the case \(\mc{D}^{-}_{0}\times\mc{B}^{+}_{0}\,\goto\,\mc{D}^{+}_{-1}\,\). This behavior should be seen in the context of the fact that \(\hat{\mc{C}}_{w_3}\) is not produced by the \(w=\pm 1\) OPE coefficients, so that these two contributions, unlike the \(w=0\) case, do not produce all of the representations appearing in (\ref{mpope}) either separately or together.

Finally, defining \(j=j_1+j_2\,\) with \(\tilde{\jmath}=k/2-j\,\), and again assuming as in section~\ref{sec_5} that the OPE closes on the \(SL(2,R)\) spectrum we have 
\be\label{ppope}
\hat{\mc{D}}^{\pm}_{j_2}(w_2)\times\hat{\mc{D}}^{\pm}_{j_1}(w_1)\,\goto\,
\hat{\mc{C}}_{w_3\pm 1}\;\oplus
\sum_{n=0}^{[\,\tilde{\jmath}\,]-1}\hat{\mc{D}}^{\pm}_{j+n}(w_3)\,
\oplus\sum_{n=0}^{[-\tilde{\jmath}\,]-1}\hat{\mc{D}}^{\pm}_{-\tilde{\jmath}-n}(w_3\pm 1)
\ee
In the flow conserving discrete sum the condition \(j_3=j+n\in\mc{S}_{\sst\mc{D}}\) is satisfied automatically at the lower bound of \(\mc{S}_{\sst\mc{D}}\) if \(j_1,j_2\in\mc{S}_{\sst\mc{D}}\,\). However, it was necessary in section~\ref{sec_5} to impose \(j_3\leq (k-1)/2\,\) at the upper bound of \(\mc{S}_{\sst\mc{D}}\) through the upper bound of the sum, which requires \(j\leq (k-1)/2\,\). In the flow non-conserving discrete sum the condition \(j_3=-\tilde{\jmath}-n\in\mc{S}_{\sst\mc{D}}\) is satisfied automatically at the upper bound of \(\mc{S}_{\sst\mc{D}}\) if \(j_1,j_2\in\mc{S}_{\sst\mc{D}}\,\), and was imposed in section~\ref{sec_5} at the lower bound of \(\mc{S}_{\sst\mc{D}}\) by the upper bound of the sum, which requires \(j\geq (k+1)/2\,\). Since these cases are clearly mutually exclusive, a single term must appear in (\ref{ppope}) for a given value of \(j\), as follows from the equivalence of (\ref{mpope}) and (\ref{ppope}) for the fusion rules of the affine modules. The OPE of primary fields treated in the previous subsection in the case \(\mc{D}^{+}_{0}\times\mc{B}^{+}_{0}\,\goto\,\mc{D}^{+}_{0}\) shows that the upper bound \(j_3<(k-1)/2\,\) in \(\mc{D}^{+}_{j_3}(0)\subset \hat{\mc{D}}^{+}_{j_3}(0)\,\) does not follow though analytic continuation from the \(H_3\) model using the \(w=0\) OPE coefficients. Conversely, in the case \(\mc{D}^{+}_{0}\times\mc{B}^{+}_{0}\,\goto\,\mc{D}^{-}_{1}\) the upper bound \(j_3<(k-1)/2\,\) in \(\mc{D}^{-}_{\tilde{\jmath}_3}(1)\subset \hat{\mc{D}}^{+}_{j_3}(0)\,\) follows though analytic continuation from the \(H_3\) model using the \(w=1\) OPE coefficients. Similarly, in the case \(\mc{D}^{+}_{0}\times\mc{B}^{+}_{0}\,\goto\,\mc{D}^{+}_{1}\,\) the lower bound \(j_3>1/2\) in \(\mc{D}^{+}_{j_3}(1)\subset \hat{\mc{D}}^{+}_{j_3}(1)\,\) for non-vanishing OPE coefficients follows though analytic continuation. Again we see that when the OPE coefficients for intermediate states in a particular spectral flow sector, in this case for \(w=1\), close on the \(SL(2,R)\) spectrum, all modules in the corresponding OPE (\ref{ppope}) are produced. When all of the modules in the OPE do not appear in the corresponding intermediate states, as in the \(w=0\) case here, the OPE coefficients produced via analytic continuation fail to close on the spectrum.

As has been observed in \cite{BaronNunez_0810}, it is clear that the closure of the OPE to \(j_3\in\mc{S}_{\sst\mc{D}}\,\) for intermediate discrete representation does not follow directly from the definition of the \(SL(2,R)\) CFT through analytic continuation from the \(H_3\) model. Furthermore, despite expectations \cite{BaronNunez_0810}\cite{IguriNunez_0908} expressed in the literature, it is not clear that any natural mechanism (imposed by crossing symmetry, the KZ equation, etc.) exists to avoid discrete intermediate states outside of the \(SL(2,R)\) spectrum. Any such mechanism would appear to violate the analytic properties on which much of the theory is constructed. As discussed in the following subsection, the redundancy of (\ref{acflowope}) for discrete intermediate contributions is part of a much more significant set of conjectured equivalences between the flowed and unflowed contributions to the factorization of general \(SL(2,R)\) correlators via analytic continuation from the \(H_3\) model. It will be argued that these equivalences, which are intimately connected to the equivalences between correlators of fixed total spectal flow discussed in subsection~\ref{sfequiv_sec}, obviate the need for a construction such as (\ref{acflowope}). However, a consistent picture has not emerged due the failure of intermediate states for certain choices of equivalent factorizations to lie in the spectrum of the \(SL(2,R)\) CFT.

It is worthwhile pointing out that, in contrast to the fusion rules of the \(SL(2,R)\) CFT described here and in \cite{BaronNunez_0810}, the discrete flowed representation \(\hat{\mc{D}}^{+}_{1}\,\), which appears in the \(w=1\) spectral flow non-conserving term \(\,\hat{\mc{D}}^{+}_{0}\times\hat{\mc{D}}^{+}_{0}\goto\hat{\mc{D}}^{+}_{1}\oplus\,\hat{\mc{C}}_{1}\,\) in the fusion rule (\ref{opeslct4}), does not make a contribution to intermediate states of the discrete string four-point amplitude computed in~\cite{MaldacenaOoguri_0111}. In particular, the string four-point amplitude of unflowed discrete states apparently fails to factorize unless  \(j_1+j_2\leq (k+1)/2\,\) and \(j_3+j_4\leq (k+1)/2\,\). This constraint is consistent with the bound in~\cite{MaldacenaOoguri_0111} which imposes \(\sum_{p=1}^{n}j_p<(k+n-3)\,\) for well-defined \(n\)-point amplitudes of discrete unflowed representations. This bound is satisfied for \(n=3\) due to the fusion rules for discrete states discussed above. However, for \(n=4\) these fusion rules imply that intermediate states in \(\hat{\mc{C}}_{1}\) and \(\hat{\mc{D}}^{+}_{1}\,\) should appear outside of this bound. Specifically, as pictured in Figure~\ref{fig4}\,, non-vanishing coefficients in the \(SL(2,R)\) OPE considered above lead to the emergence of intermediate states in the \(\hat{\mc{D}}^{+}_{1}\) representations precisely when the above constraint is violated. It is argued in~\cite{MaldacenaOoguri_0111} that the corresponding string correlators factorize on three-point amplitudes which vanish for states in \(\hat{\mc{D}}^{+}_{1}\,\). The normalization of these amplitudes differ from those of the \(SL(2,R)\) CFT since they incorporate the boundary two-point function, equivalently the string two-point function normalized by dividing by the divergent residual \(SL(2,C)\) subgroup volume (\ref{vconf}), rather than the worldsheet two-point function. The distinction between the factorizations of \(SL(2,R)\) correlators and string amplitudes~\cite{MaldacenaOoguri_0111}\cite{CardonaKirsch_1007}\cite{KirschWirtz_1106} has been the subject of some investigation \cite{BaronNunez_0810}, as has the nature of both local and non-local contributions from the worldsheet OPE to boundary correlators~\cite{AharonyKomargodski_0711}. However, a comprehensive understanding of these issues for general string amplitudes has been elusive, and certainly warrants further investigation.

\subsection{Alternative factorizations \label{altfacts}}

A definition of an OPE in the \(SL(2,R)\) CFT though analytic continuation of (\ref{acflowope}) was originally suggested in \cite{Ribault_0507}, which extended the results of~\cite{RibaultTeschner_0502} to include the spectral flowed representations~\cite{MaldacenaOoguri_0001}\cite{MaldacenaOoguriSon_0005}\cite{MaldacenaOoguri_0111} appearing as non-normalizable fields in the \(H_3\) model. This work described the construction of \(n\)-point primary correlators in the \(H_3\) model with \(w\) units of total spectral flow in terms of \((2n-2-|w|)\)-point primary correlators with \(n-2-|w|\) degenerate field insertions in the Liouville CFT. As discussed in \cite{BaronNunez_0810}, it should be evident that the definition (\ref{acflowope}), being an extension of the \(H_3\) OPE, is difficult to reconcile with a definition of the correlators of the \(SL(2,R)\) CFT which arises via analytic continuation from the corresponding correlators in the \(H_3\) model. A further assertion appears in~\cite{Ribault_0507} which points to a possible resolution of the problem of defining an OPE for the \(SL(2,R)\) model. This states that correlators in the \(H_3\) model permit a factorization using either the \(w=0\) or the \(w=\pm 1\) OPE depending on the appropriate selection rules. If more than one OPE is permitted then the corresponding factorizations are taken to be equivalent. A variant of this conjecture as applied to spectral flow conserving amplitudes in the \(SL(2,R)\) model appears in~\cite{BaronNunez_0810}, with the implication that some form of non-trivial equivalence also applies to intermediate states in spectral flow non-conserving amplitudes. More specifically, as assumed in~\cite{MaldacenaOoguri_0111} for the string four-point function, the spectral flow conserving four-point function of the \(SL(2,R)\) model, which is equivalent to the unflowed four-point function,  is taken to arise though analytic continuation from that of the \(H_3\) model factorized exclusively on unflowed normalizable states as in (\ref{m4pffact}). This might seem like a questionable assertion since, as outlined in the preceding subsection, the \(H_3\) OPE (\ref{H3OPEglb}) produces only unflowed intermediate representations when the product fields are analytically continued to unflowed normalizable states in the \(SL(2,R)\) model. Of course, if the assertion is correct, then the more elaborate definition of the \(SL(2,R)\) CFT following from (\ref{acflowope}) exhibits an apparent redundancy that is much more extensive than that which applies to intermediate discrete states due to \(\hat{\mc{D}}^{\pm}_{j_3}(w_3)=\hat{\mc{D}}^{\mp}_{\tilde{\jmath}_3}(w_3\pm 1)\,\). A further complication involves the general failure, depending on the product states and the spectral flow sector \(w\) of the intermediate states, of the OPE defined in (\ref{acflowope}) to close on the \(SL(2,R)\) spectrum. This raises the question as to whether the apparently redundant contributions are genuinely equivalent.

The general picture that emerges in the preceding paragraph is one of seeming inconsistencies which appear to be compounded in the case of spectral flow non-conserving correlators. As part of an effort to clarify this picture, an outline of equivalent factorizations of flow conserving and non-conserving amplitudes in the \(SL(2,R)\) CFT will be given below. As is true of the computation of (\ref{onesf}), this will utilize only the \(H_3\) OPE, equivalently the \(H_3\) factorization ansatz on unflowed normalizable states, and will generally involve the appearance, rather than the prior introduction as in (\ref{acflowope}), of both the \(w=0\) and \(|w|=1\) OPE coefficients. It might be argued that all \(|w|\leq 1\) OPE coefficients are required to enforce consistency with the flow non-conserving three-point function, since the unflowed OPE coefficient manifestly does not include flowed states. However, it should be kept in mind that the \(|w|=1\) three-point function itself arises from an \(H_3\) four-point function factorized on unflowed states. As discussed below, it may be conjectured that all \(SL(2,R)\) correlators may be factorized in this way, with spectral flow non-conserving OPE coefficients arising from the introduction of spectral flow operators \(\mc{V}_{k/2}(\pm k/2)\,\) in \(H_3\) correlators, and appearing as governed by the specific nature of the analytic continuation, but without being required to be inserted at the outset of the factorization procedure. In the case of the unflowed four-point amplitude (\ref{m4pffact}), an alternative factorization on \(|w|=1\) OPE coefficients may be formulated from the equivalent flow conserving amplitude
\bbb\label{m4pffactsfcns}
\lefteqn{\amp{3mm}{\hat{\mc{V}}^{\pm 1}_{j_4}(m_4)\mc{V}_{j_3}(m_3|1)\mc{V}_{j_2}(m_2|z)\mc{V}^{\mp 1}_{j_1}(m_1)}\,=\,
\pi^{-2}\,\delta^2(\,\txsum_{p} m_p\,)}\hspace{1cm}\nonumber\\
&&\times\,\txfc{1}{2}\!\int_{\mc{S}_{\sst \mc{C}}}\!\!dj\,R^{-1}_{j}(m)\,
A^{\pm 1}(j_4,j_3,j|m_4,m_3)\,A^{\mp 1}(j,j_2,j_1|m_2,m_1)\;|\mc{F}^{{\sst (\pm 1,\mp 1)}}_{j}(m|z)|^2
\eee
where \(m=m_1+m_2\mp k/2=-m_3-m_4\mp k/2\,\). The descendant contributions contained in the chiral block are taken to arise from the terms suppressed in the resolution of the identity on the \(H_3\) spectrum
\be\label{h3csos}
\mathbbm{1}\,=\,P_{0}\,=\,\pi^2\!\int \!d^2 m\int_{\mc{S}^{+}_{\sst\mc{C}}}dj\;
\mc{V}_{j}(m)\!\left.\right\rangle\left\langle\right.\!\hat{\mc{V}}_{1-j}(-m)\;+\ldots
\ee
Here we have treated the flow conserving \(SL(2,R)\) amplitude as a six-point function, including the insertion of two spectral flow operators, in the \(H_3\) model factorized on unflowed states prior to the analytic continuation which defines the spectral elements of \(SL(2,R)\,\). This is conjectured to be a valid procedure since the associated four-point amplitudes produced prior to analytic continuation are those which define the \(|w|=1\) spectral flow non-conserving three point amplitudes. Equivalently, it is effectively assumed that, following the analytic continuation which defines the non-normalizable \(H_3\) states \(\mc{V}^{\mp 1}_{j_1}(m_1)\) and \(\mc{V}^{\pm 1}_{j_4}(m_4)\,\), prior to further analytic continuation the state \(\mc{V}_{j_2}(m_2|z)\mc{V}^{\mp 1}_{j_1}(m_1)\!\left.\right\rangle\) can be expanded in the complete set (\ref{h3csos}) for the purpose of contraction with \(\left\langle\right.\!\hat{\mc{V}}^{\pm 1}_{j_4}(m_4)\mc{V}_{j_3}(m_3|1)\,\). Given \(m=m_1+m_2\mp k/2\,\), this factorization is taken to share the domains of validity (\ref{jdom4pf}) and (\ref{h3glbdom}), outside of which discrete contributions are produced via analytic continuation. Thus, while the conjectured equivalence of (\ref{m4pffact}) and (\ref{m4pffactsfcns}) is consistent with the identity
\be\label{flwid}
\amp{3mm}{\hat{\mc{V}}^{\pm 1}_{j_4}(m_4)\mc{V}_{j_3}(m_3|1)\mc{V}_{j_2}(m_2|z)\mc{V}^{\mp 1}_{j_1}(m_1)}\,=\,
z^{\pm m_2}\amp{3mm}{\hat{\mc{V}}_{j_4}(m_4)\mc{V}_{j_3}(m_3|1)\mc{V}_{j_2}(m_2|z)\mc{V}_{j_1}(m_1)}
\ee
the corresponding factorizations are defined via analytic continuation from distinct domains. Two related identities involving projection onto flowed and unflowed primary states are given by
\bbb\label{factproj1}
\lefteqn{\amp{3mm}{\hat{\mc{V}}^{\pm 1}_{j_4}(m_4)\mc{V}_{j_3}(m_3|1)\,\Pi_{\mp 1}\,
\mc{V}_{j_2}(m_2|z)\mc{V}^{\mp 1}_{j_1}(m_1)}}\hspace{4cm}\nonumber\\&=&
z^{\pm m_2}\amp{3mm}{\hat{\mc{V}}_{j_4}(m_4)\mc{V}_{j_3}(m_3|1)\,\Pi_{0}\,\mc{V}_{j_2}(m_2|z)\mc{V}_{j_1}(m_1)}
\eee
and,
\bbb\label{factproj2}
\lefteqn{\amp{3mm}{\hat{\mc{V}}^{\pm 1}_{j_4}(m_4)\mc{V}_{j_3}(m_3|1)\,\Pi_{0}\,
\mc{V}_{j_2}(m_2|z)\mc{V}^{\mp 1}_{j_1}(m_1)}}\hspace{4cm}\nonumber\\&=&
z^{\pm m_2}\amp{3mm}{\hat{\mc{V}}_{j_4}(m_4)\mc{V}_{j_3}(m_3|1)\,
\Pi_{\pm 1}\,\mc{V}_{j_2}(m_2|z)\mc{V}_{j_1}(m_1)}
\eee
These follow, respectively, from 
\be
\amp{3mm}{\hat{\mc{V}}^{\pm 1}_{j}(-m)\mc{V}_{j_2}(m_2|z)\mc{V}^{\mp 1}_{j_1}(m_1)}
\,=\,z^{\pm m_2}\amp{3mm}{\hat{\mc{V}}_{j}(-m)\mc{V}_{j_2}(m_2|z)\mc{V}_{j_1}(m_1)}
\ee
and
\be
\amp{3mm}{\hat{\mc{V}}^{\pm 1}_{j}(-m)\mc{V}_{j_2}(m_2|z)\mc{V}_{j_1}(m_1)}
\,=\,z^{\pm m_2}\amp{3mm}{\hat{\mc{V}}_{j}(-m)\mc{V}_{j_2}(m_2|z)\mc{V}^{\pm 1}_{j_1}(m_1)}
\ee
Note that both sides of (\ref{factproj1}) and (\ref{factproj2}) are defined via analytic continuation from identical domains, though the corresponding domains of the two equations are distinct. These follow from (\ref{h3glbdom}) with \(m=m_1+m_2\) for (\ref{factproj1}) and \(m=m_1+m_2\pm k/2\) for (\ref{factproj2}), corresponding to the domains of (\ref{m4pffact}) and (\ref{m4pffactsfcns}), respectively.

If one assumes that the identity (\ref{factproj2}) extends to factorization on the corresponding complete sets of states \(P_w\,\), where \(P_{0}\) is the identity of the \(H_3\) CFT given by (\ref{h3csos}), and
\be\label{pplsmns}
P_{\pm}=\pi^2\int \!d^2 m\int_{\mc{S}^{+}_{\sst\mc{C}}}dj\;
\mc{V}_{j}^{\pm 1}(m)\!\left.\right\rangle\left\langle\right.\!\hat{\mc{V}}^{\mp 1}_{1-j}(-m)\;+\ldots
\ee
then validity of (\ref{m4pffact}) and (\ref{m4pffactsfcns}), together with the identity (\ref{flwid}), produces an equivalent factorization of (\ref{m4pffact}) expressed as
\bbb\label{factequiv}
\amp{3mm}{\hat{\mc{V}}_{j_4}(m_4)\mc{V}_{j_3}(m_3|1)\mc{V}_{j_2}(m_2|z)\mc{V}_{j_1}(m_1)}&=&
\amp{3mm}{\hat{\mc{V}}_{j_4}(m_4)\mc{V}_{j_3}(m_3|1)\,P_{0}\,\mc{V}_{j_2}(m_2|z)\mc{V}_{j_1}(m_1)}\nonumber\\&=&
\amp{3mm}{\hat{\mc{V}}_{j_4}(m_4)\mc{V}_{j_3}(m_3|1)\,P_{1}\,\mc{V}_{j_2}(m_2|z)\mc{V}_{j_1}(m_1)}
\eee
A number of comments are in order here. First, as for the factorizations on \(P_0\) appearing in (\ref{m4pffact}) and (\ref{m4pffactsfcns}), the domains from which the analytic continuations are taken differ between the \(P_0\) and \(P_1\) factorizations in (\ref{factequiv}). This has consequences in terms of the respective flowed and unflowed discrete intermediate representations which are expected to appear. In particular, for a set of affine modules which are equivalent given \(\hat{\mc{D}}^{\pm}_{0}=\hat{\mc{D}}^{\mp}_{\pm 1}\,\), both of the specific intermediate modules \(\hat{\mc{D}}^{\pm}_{j}(0)\) and \(\hat{\mc{D}}^{\mp}_{\tilde{\jmath}}(\pm 1)\) will not appear if one lies outside the spectrum of states of the \(SL(2,R)\,\) CFT. This is equivalent to the discussion of the closure of the OPE following from analytic continuation as discussed in subsection~\ref{fussum}\,. It should also be noted that while the relation (\ref{factequiv}) may be implied by (\ref{m4pffact}) and (\ref{m4pffactsfcns}), if the validity of these expressions is accepted, then there does not appear to be a need to introduce expansions such as (\ref{pplsmns}) on non-normalizable states in the \(H_3\) model in order to define either flowed or unflowed \(SL(2,R)\) correlators.
In contrast, the definition implicit in the analytic continuation of the OPE (\ref{acflowope}) suggests that the left side of (\ref{factequiv}) should be computed via summation over contributions involving the insertion of the \(H_3\)  identity (\ref{h3csos}) as well as the two projectors \(P_{\pm}\,\) in (\ref{pplsmns}).

In the discussions appearing in \cite{BaronNunez_0810}\cite{IguriNunez_0908} the validity of the alternative factorization (\ref{factequiv}) is taken to be implied from the assumed validity of the factorization of unflowed correlators on the \(H_3\) spectrum as in (\ref{m4pffact}). This is a weaker conjecture than assuming the validity of (\ref{m4pffactsfcns}), and applies in the case of the restricted set of amplitudes
\bbb\label{hwflwid}
\lefteqn{\amp{3mm}{\hat{\mc{V}}_{j_4}(-j_4)\mc{V}_{j_3}(m_3|1)\mc{V}_{j_2}(m_2|z)\mc{V}_{j_1}(j_1)}}
\hspace{4cm}\nonumber\\&=&
z^{-m_2}\amp{3mm}{\hat{\mc{V}}^{+ 1}_{j_4}(-j_4)\mc{V}_{j_3}(m_3|1)
\mc{V}_{j_2}(m_2|z)\mc{V}^{-1}_{j_1}(j_1)}\nonumber\\ &=&
z^{-m_2}\,\omega_{j_1}\,\omega_{j_4}
\amp{3mm}{\hat{\mc{V}}_{\tilde{\jmath}_4}(\tilde{\jmath}_4)\mc{V}_{j_3}(m_3|1)
\mc{V}_{j_2}(m_2|z)\mc{V}_{\tilde{\jmath}_1}(-\tilde{\jmath}_1)}
\eee
where use has been made of the identification (\ref{weylprimeq}). In this case the identities (\ref{factproj1}) and (\ref{factproj2}) both reduce to
\bbb\label{hwfctprj}
\lefteqn{\amp{3mm}{\hat{\mc{V}}_{j_4}(-j_4)\mc{V}_{j_3}(m_3|1)\,\Pi_1\,\mc{V}_{j_2}(m_2|z)\mc{V}_{j_1}(j_1)}}
\hspace{4cm}\nonumber\\&=&
\omega_{j_1}\,\omega_{j_4}\amp{3mm}{\hat{\mc{V}}^{-1}_{\tilde{\jmath}_4}(\tilde{\jmath}_4)
\mc{V}_{j_3}(m_3|1)\,\Pi_1\,\mc{V}_{j_2}(m_2|z)
\mc{V}^{+1}_{\tilde{\jmath}_1}(-\tilde{\jmath}_1)}\nonumber\\ &=&
z^{-m_2}\,\omega_{j_1}\,\omega_{j_4}
\amp{3mm}{\hat{\mc{V}}_{\tilde{\jmath}_4}(\tilde{\jmath}_4)\mc{V}_{j_3}(m_3|1)
\,\Pi_0\,\mc{V}_{j_2}(m_2|z)\mc{V}_{\tilde{\jmath}_1}(-\tilde{\jmath}_1)}
\eee
Thus, assuming that both the unflowed correlators appearing in (\ref{hwflwid}) can both be factorized on \(P_0\) leads to the statement (\ref{factequiv}) of equivalent factorizations in the case of these amplitudes. It is important to note that (\ref{factequiv}) does not directly follow from this argument, though this is strongly suspected \cite{BaronNunez_0810} due to the assumed identification of the analytic continuation of the unflowed \(H_3\) four-point correlator and the unflowed \(SL(2,R)\) four-point correlator. Of course, the same issues related to the initial domains of the analytic continuations, and the related closure of the OPE, apply here as well. Given the equivalences outlined here, there might appear to be a small distinction between a direct analytic continuation from the \(H_3\) model and the explicit introduction of the flow non-conserving terms in (\ref{acflowope}). However, at least for spectral flow conserving amplitudes, these equivalent terms lead to redundant contributions beyond those immediately apparent for discrete states as outlined in section~\ref{sec_5}, and will be taken here to be entirely superfluous in terms of the process of factorization of \(SL(2,R)\) correlators. This conclusion is reinforced by the results of \cite{IguriNunez_0908}, which computes the flow-conserving \(SL(2,R)\) four-point function via a well-justified free-field calculation following on the results of \cite{IguriNunez_0705}. While the free-field methods avoid the introduction of the \(H_3\) factorization ansatz, they are rigorously defined only for (discrete) amplitudes which respect \(1-\sum_{p}j_p\in \mathbb{Z}_{\geq 0}\,\). In this context, flow conserving four-point amplitudes are expressed in the semi-classical limit \cite{IguriNunez_0908} alternatively in terms of either flow conserving or flow non-conserving OPE coefficients. It is further argued that general flow-conserving four-point amplitudes may be computed via analytic continuation, thereby strengthening the arguments appearing in \cite{BaronNunez_0810}. It would be helpful to develop an understanding of these results beyond the semi-classical approximation, including the details of the associated descendant contributions.

In the case of the spectral flow non-conserving \(w=1\) four point function, beginning with a five-point function in the \(H_3\) model, it is expected that the following factorization is valid
\bbb\label{weq1fact}
\lefteqn{\amp{3mm}{\hat{\mc{V}}^{1}_{j_4}(m_4)\mc{V}_{j_3}(m_3|1)\mc{V}_{j_2}(m_2|z)\mc{V}_{j_1}(m_1)}\,=\,
\pi^{-2}\,\delta^2(\,\txsum_{p} m_p\,)}\hspace{1cm}\nonumber\\
&&\times\,\txfc{1}{2}\!\int_{\mc{S}_{\sst \mc{C}}}\!\!dj\,R^{-1}_{j}(m)\,
A^{1}(j_4,j_3,j|m_4,m_3)\,A^{0}(j,j_2,j_1|m_2,m_1)\;|\mc{F}^{{\sst (1,0)}}_{j}(m|z)|^2
\eee
where \(m=m_1+m_2=-m_3-m_4-k/2\,\). This rests on the same assumptions as the alternative factorization (\ref{m4pffactsfcns}). If the prescription implicit in (\ref{acflowope}) is used a single additional term is expected here involving the intermediate flowed states given by the sum \(P_{-}\) in (\ref{pplsmns}). However, given the equivalence between the \(w=0\) three-point amplitudes \(\amp{3mm}{\hat{\mc{V}}^{1}_{j_4}(m_4)\mc{V}_{j_3}(m_3|1)\mc{V}_{j}^{-1}(m)}\) and \(\amp{3mm}{\hat{\mc{V}}_{j_4}(m_4)\mc{V}_{j_3}(m_3|1)\mc{V}_{j}(m)}\), the resulting expression is essentially redundant and reflects the form of the factorization of the equivalent \(w=1\) four-point function 
\be
\amp{3mm}{\hat{\mc{V}}_{j_4}(m_4)\mc{V}_{j_3}(m_3|1)\mc{V}_{j_2}(m_2|z)\mc{V}^{1}_{j_1}(m_1)}
\ee
on unflowed \(H_3\) intermediate states. Finally, consider the maximally flow non-conserving \(w=2\) four-point function. Distributing the spectral flow quantum numbers to take advantage of the above factorization assumption leads to
\bbb\label{weq2fact}
\lefteqn{\amp{3mm}{\hat{\mc{V}}^{1}_{j_4}(m_4)\mc{V}_{j_3}(m_3|1)\mc{V}_{j_2}(m_2|z)\mc{V}^{1}_{j_1}(m_1)}\,=\,
\pi^{-2}\,\delta^2(\,\txsum_{p} m_p\,)}\hspace{1cm}\nonumber\\
&&\times\,\txfc{1}{2}\!\int_{\mc{S}_{\sst \mc{C}}}\!\!dj\,R^{-1}_{j}(m)\,
A^{1}(j_4,j_3,j|m_4,m_3)\,A^{1}(j,j_2,j_1|m_2,m_1)\;|\mc{F}^{{\sst (1,1)}}_{j}(m|z)|^2
\eee
where \(m=m_1+m_2+k/2=-m_3-m_4-k/2\,\). It may be seen that the selection rules constrain the OPE prescription (\ref{acflowope}) to also produce this single non-vanishing contribution. Each of the above factorizations may be understood in terms of the factorization of equivalent amplitudes in Liouville theory. In the maximally flow non-conserving case (\ref{weq2fact}) this leads to a factorized Liouville four point amplitude, with associated three-point functions corresponding to \(w=1\) three-point amplitudes in the \(H_3\) model prior to analytic continuation to \(SL(2,R)\) amplitudes. It should be noted that each of the respective factorizations (\ref{m4pffact}), (\ref{m4pffactsfcns}), (\ref{weq1fact}), and (\ref{weq2fact}) correspond to amplitudes in the \(H_3\) model involving normalizable intermediate states, and require further analytic continuation, in general with the appearance of discrete contributions, for the computation of correlators in the \(SL(2,R)\) CFT.

The presumed equivalence of alternative factorizations, together with the associated complimentary domains of the OPE under analytic continuation described in subsection \ref{fussum} was argued in \cite{BaronNunez_0810} to imply the existence of a presently unknown (presumably non-analytic) mechanism which is required to impose the closure of the OPE on the spectrum of states. This argument is partially based on an analysis of the primary states appearing in the following equivalent fusion rules
\be\label{bpp1}
\mc{B}^{+}_{j_2}(-1)\times\mc{B}^{+}_{j_1}(0)\,=\,
\mc{B}^{-}_{\tilde{\jmath}_2}(0)\times\mc{B}^{+}_{j_1}(0)\goto\mc{C}_{0}\;\oplus
\sum_{n=0}^{[\,\tilde{\jmath}\,]-1}\mc{D}^{-}_{\tilde{\jmath}-n}(0)\;\oplus
\sum_{n=0}^{[-\tilde{\jmath}\,]-1}\mc{D}^{+}_{-\tilde{\jmath}-n}(0)\,+\,\ldots
\ee
where we have defined \(j=j_1+j_2\,\), and made use of the selection rules of subsection \ref{select_sec} and the fusion rules  (\ref{mpope}) and (\ref{ppope}). The assertion made in \cite{BaronNunez_0810} is that, despite the operator identity \(\mc{V}_{j}(\pm j)=\omega_{j}\mc{V}_{\tilde{\jmath}}(\mp \tilde{\jmath})\,\), the representations appearing in (\ref{bpp1}) for the respective flowed and unflowed OPE coefficients via analytic continuation do not coincide. Of course the representation content in these expressions would be identical if the upper bounds on the sums, which enforce closure on the \(SL(2,R)\) spectrum, were respected under analytic continuation. This is certainly the case for the flowed (\(w=1\)) and unflowed (\(w=0\)) OPE coefficients for \(\mc{B}^{+}_{j_2}(-1)\) and  \(\mc{B}^{-}_{\tilde{\jmath}_2}(0)\), respectively. However, since \(\mc{B}^{+}_{j_2}(-1)=\mc{B}^{-}_{\tilde{\jmath}_2}(0)\) is taken as an operator identity in the \(SL(2,R)\) CFT, if (\ref{bpp1}) were not a precise identity even when the OPE does not close on the spectrum this would represent a significant pathology in the \(SL(2,R)\) CFT as defined via analytic continuation from the \(H_3\) model. That this is not the case can be seen from an examination of the particular case \(j>(k-1)/2\,\) for which the \(\mc{D}^{-}_{\tilde{\jmath}-n}(0)\) terms do not appear in the fusion rules involving spectral elements. However, identical contributions in \(\mc{B}^{+}_{j+n}(-1)\in\mc{D}^{-}_{\tilde{\jmath}-n}(0)\) appear outside the bound \(j<(k-1)/2\) for both (not just the former as stated in \cite{BaronNunez_0810}) the unflowed (\(w=0\)) and flowed (\(w=-1\)) OPE coefficients for \(\mc{B}^{+}_{j_2}(-1)\) and \(\mc{B}^{-}_{\tilde{\jmath}_2}(0)\), respectively. This follows from the \(\mc{D}^{+}_{0}\times\mc{B}^{+}_{0}\,\goto\,\mc{D}^{+}_{0}\) and (under conjugation) \(\mc{D}^{-}_{0}\times\mc{B}^{+}_{0}\,\goto\,\mc{D}^{-}_{1}\) OPE coefficients considered in subsection \ref{acfus} above.

\section{Concluding remarks \label{sec_7}}

The purpose of these notes has been to elucidate issues in the \(SL(2,R)\) CFT that remain unresolved in the literature, and to establish consistent notation and techniques for further efforts in this direction. The vertex operator normalization for states of definite affine weight used in these notes has led to well-defined OPE coefficients for all representations permitted by the selection rules, and a confirmation of fusion rules~\cite{BaronNunez_0810} previously derived from an examination of poles in analytically continued OPE coefficients from the \(H_3\) coset model. The alternative factorizations described in section~\ref{sec_6} above, which depend crucially on the selection rules and the equivalence of correlators of fixed total spectral flow, suggest the existence of a consistent picture of the \(SL(2,R)\) CFT under analytic continuation from the \(H_3\) model. In particular, it appears unnecessary to introduce an extended \(H_3\) OPE as in (\ref{acflowope}). Furthermore, give the equivalences between correlators of fixed total spectral flow outlined in subsection~\ref{sfequiv_sec}\,, the introduction of flowed intermediate states, as opposed to \(|w|=1\) OPE coefficients, does not appear to be required for the computation of arbitrary \(SL(2,R)\) correlation functions. However, regardless of the validity of these arguments, the analytically continued factorized expressions for \(SL(2,R)\) correlators generically include discrete intermediate contributions outside the spectrum of normalizable states, with no apparent natural mechanism for their truncation. While it might be hoped that use of the various equivalent factorizations, perhaps accompanied by judicious contour choices, could eliminate these contributions, it is difficult to avoid the conclusion that the failure of closure of the \(SL(2,R)\) OPE is an inevitable consequence of an attempt to define it via analytic continuation from the \(H_3\) model. If this is correct then it might be thought that the failure of closure represents a fundamental pathology of an associated definition of the \(SL(2,R)\) CFT, at least as expressed in the \(\mc{V}^{w}_{j}(m)\) basis. This does not necessarily imply a consequent related pathology in string theory on \(AdS_3\,\), which could be defined via analytic continuation \cite{MaldacenaOoguri_0111} entirely in terms of boundary amplitudes via worldsheet integration of \(H_3\) correlators in the \(\Phi_{j}(x)\) basis. This is consistent with the fact that the failure of closure is entirely due to \(m\)-dependent rather than \(j\)-dependent poles in the expressions analyzed in subsection~\ref{acfus}\,. It may also be that the failure of closure is required due to the fact that alternative factorizations defined via analytic continuation from the \(H_3\) model, which are conjectured to produce identical analytic expressions for correlation functions, are generally distinct in terms of the representations which appear in the OPE coefficients. As mentioned in subsection~\ref{fussum}\,, for choices of factorization for which this representation content is identical to that appearing in the \(SL(2,R)\) fusion rules, the OPE closes on the spectrum of states of the \(SL(2,R)\) CFT. Thus, for instance, the states for \(j>(k-1)/2\) in the unflowed OPE coefficients for \(\mc{D}^{+}_{j_2}(0)\times\mc{D}^{+}_{j_1}(0)\goto\mc{D}^{+}_{j}(0)\) may be conjectured to appear precisely to provide the additional spectral content which appears in the \(w=1\) case. In this context, it is tempting to view the additional intermediate states outside of the spectrum as somewhat similar to those which appear for correlators involving normalizable states in the \(H_3\) model if the countour is deformed from \(\mc{S}^{+}_{\sst\mc{C}}\) to enclose poles corresponding to intermediate non-normalizable fields. The difference for the \(SL(2,R)\) CFT is that there is apparently no choice of contour or factorization which leads to the closure of the OPE in the most general case. To resolve the issue of closure a better understanding of how discrete descendant states make an appearance in the analytically continued four-point function seems required. It may be hoped that the mapping between the \(H_3\) and Liouville CFTs described in~\cite{RibaultTeschner_0502}\cite{Ribault_0507} could be useful in this regard. More broadly, a proof of crossing symmetry and a greater understanding of \(SL(2,R)\) string amplitudes, particularly spectral flow non-conserving correlators and non-local contributions~\cite{AharonyKomargodski_0711} on the worldsheet, seems desirable. Insight gained from such investigations may also shed further light on the nature of the boundary CFT, particularly on questions~\cite{KutasovSeiberg_9903} related to the variable central extensions of the corresponding boundary affine algebras.  Given the extensive, and largely successful, efforts which have gone into investigating the \(SL(2,R)\) CFT, these issues might appear to be a relatively minor loose ends. However, as has been repeatedly evident in the now long history of this model, their resolution may illuminate aspects of worldsheet string models with both non-trivial time-dependence and relevance to a holographic description of spacetime physics, as well as the general structure of non-rational conformal field theories, in ways that cannot easily be anticipated.

\section*{Acknowledgements}

The author would like to thank Sylvain Ribault for helpful suggestions and extensive comments on these notes. Thanks also to Carmen Nunez for insights into some of the problems addressed above.

\appendix

\section*{Appendix :}

\section{Some preliminaries for \texorpdfstring{$AdS_3$}{} and \texorpdfstring{$H_{3}$}{}\label{sec_A}}

The string action in global coordinates on \(AdS_3\) takes the form
\be
S[t,\rho,\varphi]\,=\,\frac{k}{2\pi}\int\!d^2z\,
(-\cosh^2\!\rho\,\partial t\bar{\partial}t+\partial\rho\bar{\partial}\rho
+\sinh^2\!\rho\,(\partial\varphi\bar{\partial}\varphi+\partial t\bar{\partial}\varphi-\partial\varphi\bar{\partial}t))
\ee
The corresponding metric is given by
\be\label{glbmet}
G\,=\,k\left(-\cosh^2\!\rho\,dt^2+d\rho^2+\sinh^2\!\rho\,d\varphi^2\right)
\ee
and the \(B\) field is
\be
B\,=\,k\sinh^2\!\rho\,dt\wedge d\varphi
\ee
The metric describes a symmetric space of constant negative curvature with \(R_{ab}=-(2/k)\,g_{ab}\,\), and volume form given by \(\varepsilon=-\txfc{1}{2}dB\,\). Note that the above action is defined on a Euclidean worldsheet and has an imaginary part coming from the \(B\) field. Under continuation to a Lorentzian worldsheet, with \(\sigma^{2}\goto i\sigma^{0}\,\), where \(z=\sigma^{1}+i\sigma^{2}\,\) and \(d^2z=2d\sigma^{1}d\sigma^{2}\,\), the action is real. Choosing \(x^{\pm}=t\mp\varphi\,\), the following vector fields generate isometries on \(AdS_3\)
\bbb
K^{1}_{\pm} &= & -\txfc{1}{2}\sin(x^{\pm})\,\partial_{\rho}-\cos(x^{\pm})\,
(f_{+}\,\partial_{\pm}+f_{-}\,\partial_{\mp})\\
K^{2}_{\pm} &= & -\txfc{1}{2}\cos(x^{\pm})\,\partial_{\rho}+\sin(x^{\pm})\,
(f_{+}\,\partial_{\pm}+f_{-}\,\partial_{\mp})\\
K^{3}_{\pm} &= & -\partial_{\pm}
\eee
where, \(f_{\pm}\,=\,\txfc{1}{2}(\tanh\rho\pm\tanh^{-1}\!\rho)\,\). 
It may be shown that \(\com{K^{a}_{+}}{K^{b}_{-}}=0\,\) and, suppressing identical subscripts,
\be
\com{K^{1}}{K^{2}}=K^{3}\hspace{1.2cm}
\com{K^{2}}{K^{3}}=-K^{1}\hspace{1.2cm}
\com{K^{3}}{K^{1}}=-K^{2}
\ee
These generators comprise the \(SO(2,1)\) algebra, with the Killing vector fields \(K^{a}_{\pm}\) satisfying the \(SO(2,2)=SO(2,1)\times SO(2,1)\) algebra. The \(SL(2,R)\) group manifold may be seen to be the double cover of \(SO(2,1)=SL(2,R)/\mathbb{Z}_{2}\,\), which corresponds to the Poincaire patch of \(SL(2,R)\,\). \(AdS_3\) may be seen to be the universal cover of \(SL(2,R)\), which may be constructed through the identification \(t\sim t+2\pi\,\). Defining the generators \(J^{a}=-iK^{a}\,\), and taking \(J^{\pm}=J^{1}\pm iJ^2\,\) we find the algebra
\be\label{sl2alg}
\com{J^{3}}{J^{\pm}}=\pm J^{\pm}\hspace{2cm}\com{J^{+}}{J^{-}}=-2J^{3}
\ee
The associated Killing fields are \(J^{3}_{\pm}=i\partial_{\pm}\,\), and for \(\sigma=\pm 1\,\),
\be
J^{\sigma}_{\pm}\,=\,e^{-i \sigma x^{\pm}}
\left(f_{+}\,i\partial_{\pm}+f_{-}\,i\partial_{\mp}-\txfc{1}{2}\sigma\partial_{\rho}\right)
\ee
The continuation to \(H_{3}\) is given by \(t\goto -i\tau\) and leads to
\be
S[\tau,\rho,\varphi]\,=\,\frac{k}{2\pi}\int\!d^2z\,
(\partial\tau\bar{\partial}\tau+\partial\rho\bar{\partial}\rho
+\sinh^2\!\rho\,(\partial\varphi-i\partial\tau)(\bar{\partial}\varphi+i\bar{\partial}\tau))
\ee
which is manifestly real and positive. Defining \(w=\varphi+i\tau\,\), under this continuation the above Killing vector fields take the form
\bbb
K^{1}_{+} \, =&\! -\bar{K}^{1}_{-} \!&= \; \txfc{1}{2}\sin(w)\,\partial_{\rho}+\cos(w)\,
(f_{+}\,\partial_{w}-f_{-}\,\partial_{\bar{w}})\\
K^{2}_{+} \, =&\! \bar{K}^{2}_{-} \!&= \; -\txfc{1}{2}\cos(w)\,\partial_{\rho}+\sin(w)\,
(f_{+}\,\partial_{w}-f_{-}\,\partial_{\bar{w}})\\
K^{3}_{+} \, =&\! -\bar{K}^{3}_{-} \!&= \; \partial_{w}
\eee
where \(\bar{K}^{a}_{\pm}=(K^{a}_{\pm})^*\,\) and \(\bar{w}=(w)^*=\varphi-i\tau\,\). We may then form the real vector fields
\bbb
R^{1} = &  \!iK^{1}_{+}+iK^{1}_{-}\hspace{1cm} & B^{1}=K^{1}_{+}-K^{1}_{-} \\
R^{2} = &  \!iK^{2}_{+}-iK^{2}_{-}\hspace{1cm} & B^{2}=K^{2}_{+}+K^{2}_{-} \\
R^{3} = &  \!K^{3}_{+}-K^{3}_{-}\hspace{1cm} & B^{3}=-iK^{3}_{+}-iK^{3}_{-}
\eee
where \(R^{a}\) and \(B^{a}\) comprise the rotation and boost generators, respectively, for \(SO(3,1)\,\). Thus,
\be
\exp\left(\theta_{a}R^{a}+\beta_{a}B^{a}\right)\in SO(3,1)
\spwd{1cm}{for}\theta_{a},\beta_a\in\mathbb{R}
\ee
As above, suppressing subscripts, we may also define the generators \(J^{a}=-iK^{a}\,\), and \(J^{\pm}=J^{1}\pm iJ^2\,\), which satisfy the algebra (\ref{sl2alg}). It may be seen that \(J^{3}\equiv J^{3}_{+}=\bar{J}^{3}_{-}\,\) and \(J^{\pm}\equiv J^{\pm}_{+}=\bar{J}^{\pm}_{-}\,\). The associated Killing fields are \(J^{3}=-i\partial_{w}\,\) and
\be
J^{\pm}\,=\,-e^{\pm i w}
\left(f_{+}\,i\partial_{w}-f_{-}\,i\partial_{\bar{w}}\pm\txfc{1}{2}\,\partial_{\rho}\right)
\ee
Defining \(c\cdot J=c_{3}J^{3}+c_{+}J^{+}+c_{-}J^{-}\,\) with \(c_{\alpha}\in\mathbb{C}\,\), an element of \(SO(3,1)=SL(2,C)/\mathbb{Z}_{2}\,\) is constructed as follows
\be
\exp\left(c\cdot J+\bar{c}\cdot\bar{J}\,\right)\in SO(3,1)
\spwd{1cm}{for}c_{\alpha}\in\mathbb{C}
\ee
Another commonly used coordinate system on \(H_3\) is given by the transformation
\bbb
\gamma & = & e^{\tau-i\varphi}\,\tanh\rho\nonumber\\[1mm]
\phi & = & -\tau+\ln(\cosh\rho)
\eee
The corresponding massless fields are
\be
G\,=\,k\,(\,d\phi^2+e^{2\phi}d\gamma d\bar{\gamma}\,)
\ee
and, defining \(\tilde{B}=B-ik\,d\,(\ln(\cosh\rho)\,d\varphi)\,\), which is \(B\) up to an exact form, we have 
\be
\tilde{B}\,=\,\txfc{1}{2}\,k\,e^{2\phi}\,d\bar{\gamma}\wedge d\gamma
\ee
This leads to the action
\be
S[\phi,\gamma,\bar{\gamma}]\,=\,\frac{k}{2\pi}\int\!d^2z\,
(\partial\phi\bar{\partial}\phi+e^{2\phi}\partial\bar{\gamma}\bar{\partial}\gamma)
\ee
In the \((\phi,\gamma,\bar{\gamma})\,\) system, we have
\be
J^{+}\,=\,-\partial_{\gamma}\hspace{1cm} J^{3}\,=\,\txfc{1}{2}\partial_{\phi}-\gamma\partial_{\gamma}
\hspace{1cm} J^{-}\,=\,\gamma\partial_{\phi}-\gamma^2\partial_{\gamma}+e^{-2\phi}\partial_{\bar{\gamma}}
\ee
Note that the \((\gamma,\bar{\gamma},\phi)\)  coordinates cover all of \(\mc{H}_{3}\) and are defined through a continuation from global \(AdS_3\,\); there is an inequivalent continuation using these coordinates from \(\mc{H}_{3}\) to the Poincare patch of \(SL(2,R)\,\).

\section{Wavefunctions and norms on \texorpdfstring{$AdS_3$}{} and \texorpdfstring{$H_{3}$}{}\label{sec_B}}

\subsection{Unitary representations of \texorpdfstring{$SL(2,R)$}{}}

States in the zero-mode quantum mechanics of strings on \(AdS_3\) are constructed out of unitary representations of the isometry group \(SL(2,R)\times SL(2,R)\,\). What follows is a short review of the unitary representations\footnote{Note that the representations of the universal cover of \(SL(2,R)\,\), which is the \(AdS_{3}\) spacetime, are considered here. Thus the parameter \(j\) has continuous, rather than half-integral, eigenvalues in the discrete representations.} of \(SL(2,R)\)~\cite{DixonPeskinLykken}. Consider a state \(\ket{j,m}\) in a unitary representation of \(SL(2,R)\) which satisfies \(J^{3}\ket{j,m}=m\ket{j,m}\) and \(C\ket{j,m}=j(1-j)\ket{j,m}\,\), where the Casimir operator is given by
\be
C\,=\,-J^{3}J^{3}+\txfc{1}{2}\left(J^{+}J^{-}+J^{-}J^{+}\right)
\ee
Since \(C=C^{\dagger}\) we may use the (redundant) parameterization \(j\in\mathbb{R}\,\) or \(\,2j-1\in i\,\mathbb{R}\,\). Assuming that \(\pm m>0\) and \(\brkt{j,m}{j,m}>0\,\), then the following inequality is satisfied
\be
\parallel\!J^{\pm}\ket{j,m}\!\parallel^2\,=\,\bra{j,m}J^{\mp}J^{\pm}\ket{j,m}\,=\,
\bra{j,m}J^{\pm}J^{\mp}\pm 2J^3\ket{j,m}\,\geq\,\pm 2m\brkt{j,m}{j,m}>0
\ee
Thus if \(\ket{j,m}\) for \(\pm m>0\) is normalizable then all states \(\ket{j,m\pm n}\) with \(n\in\mathbb{Z}_{\geq 0}\) are also normalizable. Thus repeated application of \(J^{\mp}\) to \(\ket{j,m}\) for \(\pm m>0\) will either produce an infinite number of normalizable states or there will be a state \(\ket{j,\pm l}\) with \(l>0\) for which \(J^{\mp}\ket{j,\pm l}=0\,\). In the latter case we have
\be\label{casbnd}
j(1-j)\ket{j,\pm l}\,=\,
\left(-J^{3}J^{3}\pm J^{3}+J^{\pm}J^{\mp}\right)\ket{j,\pm l}\,=\,l(1-l)\ket{j,\pm l}
\ee
Since \(l\in\mathbb{R}_{> 0}\,\), we must have \(j\in\mathbb{R}\,\). For \(j\geq 1\) there is the solution \(l=j\,\), for \(j\leq 0\) there is the solution \(l=1-j\,\), and for \(0<j<1\) there are the two solutions \(l=j\,\) and \(l=1-j\,\). It is conventional\footnote{Note that these representations are indexed by the lowest \(J^{3}\) eigenvalue (\(\pm j\) for \(\mc{D}^{\pm}_{j}\,\) with \(j>0\,\)), rather than the value \(j(1-j)\) of the Casimir.} to impose the restriction \(l=j>0\,\) and denote these (discrete) representations by \(\mc{D}^{\pm}_{j}\,\). In the case of a non-terminating set of states, taking \(-1/2\leq\alpha<1/2\) to be the \(J^{3}\) eigenvalue of smallest absolute value, consider the state \(\ket{j,\alpha}\,\). For \(\pm \alpha\geq 0\) we have
\be\label{contcond}
\parallel\!J^{\mp}\ket{j,\alpha}\!\parallel^2\,=\,
\bra{j,\alpha}J^{\pm}J^{\mp}\ket{j,\alpha}\,=\,
\bra{j,\alpha}\left(C+J^{3}J^{3}\mp J^3\right)\ket{j,\alpha}>0
\ee
Thus, \(j(1-j)>|\alpha|(1-|\alpha|)\,\). For \(\,2j-1\in i\,\mathbb{R}\,\) this condition is satisfied, and the corresponding (continuous) representations are denoted by \(\mc{C}^{\alpha}_{j}\,\). For \(\,j\in\mathbb{R}\,\) (\ref{contcond}) leads to \(|j-1/2|<1/2-|\alpha|\,\), and the corresponding (exceptional) representations are denoted by \(\mc{E}^{\alpha}_{j}\,\). There is also the identity representation consisting of the single state \(\ket{0,0}\,\).	
	
\subsection{Wavefunctions on \texorpdfstring{$AdS_3$}{}}

Depending on the norm that is chosen, only a subset of the above representations will appear in a scalar field quantization on \(AdS_{3}\)~\cite{Teschnermss_9712}\cite{BalasubramanianKrausLawrence_9805}\cite{Hikida_0403}\cite{Ribault_0912}. In the string zero-mode quantum mechanics, we expect the corresponding Hilbert space to be a direct sum of tensor products of two identical representations\footnote{As in the treatment of the \(SL(2,R)\) CFT, a condensed notation is adopted here which suppresses the left/right doubling of diagonal representations. Thus, for example, \(\mc{D}^{\sigma}_{j}\otimes\mc{D}^{\sigma}_{j}\goto\mc{D}^{\sigma}_{j}\,\), \((m_{+},m_{-})\goto m\,\), \((x^{+},x^{-})\goto x\,\), and etc.} of \(SL(2,R)\,\), with a norm on states inherited through analytic continuation from the theory on \(H_{3}\,\). Using the coordinates \(x^{\pm}=t\mp\varphi\,\) from section~\ref{sec_A} above, and defining \(z=\cosh^{-2}\!\rho\,\), the generators take the form \(J^{3}_{\pm}=i\partial_{\pm}\,\), and for \(\sigma=\pm 1\,\),
\be
J^{\sigma}_{\pm}\,=\,\txfc{1}{2}\,(1-z)^{-1/2}\,e^{-i\sigma x^{\pm}}\,
\left((2-z)\,i\partial_{\pm}-z\,i\partial_{\mp}+2\sigma z(1-z)\,\partial_{z}\right)
\ee
The wavefunction corresponding to the basis element \(\ket{j,m}\equiv\ket{\mc{V}_{j}(m)}\,\) is given by\footnote{It should be noted that the notation used in this appendix to denote the zero-mode solutions, in this case \(\mc{V}_{j}(m)\,\), is identical to that used in the main body of the paper to denote the corresponding CFT primary vertex operators. The relevant normalizations have been chosen so that quantities such as the reflection coefficients \(R_{j}(m)\,\)(\ref{qmrefcof}) coincide with the \(k\goto\infty\) limit of the corresponding CFT expressions. Note that this limit is to be distinguished from the 2+1 dimensional flat-space limit, which also requires the scaling of both the charges and the  vertex operators.}
\be\label{pardef}
\mc{V}_{j}(m|x,z)\,=\,e^{-i(m_{+}x^{+}+m_{-}x^{-})}\,\mc{U}_{j}(m|z)
\ee
Consider first the representation \(\mc{D}^{\sigma}_{j}\,\), with terminating state which satisfies \(J^{-\sigma}_{\pm}\ket{j,\sigma j}=0\,\), leading to the conventionally normalized solution \(\mc{U}_{j}(\sigma j)=z^{j}\,\). 
The Hilbert space \(\mc{D}^{\sigma}_{j}\) is then constructed from
\be\label{discstates}
\ket{j,m}\,=\,\frac{\Gamma(2j)}{\Gamma(2j+n_{+})}\frac{\Gamma(2j)}{\Gamma(2j+n_{-})}\,
\left(J^{\sigma}_{+}\right)^{n_{+}}\left(J^{\sigma}_{-}\right)^{n_{-}}\ket{j,\sigma j}
\ee
where \(m_{\pm}=\sigma(j+n_{\pm})\,\) with \(n_{\pm}\in\mathbb{Z}_{\geq 0}\,\) and \(j\in\mathbb{R}_{>0}\,\). Suppressing identical subscripts, the normalization has been chosen to produce the relations
\bbb\label{glbalg1}
J^{3}\ket{j,m} & = & m\ket{j,m}\\ \label{glbalg2}
J^{\pm}\ket{j,m} & = & (m\pm j)\ket{j,m\pm1}
\eee
For the states (\ref{discstates}) with the action of the algebra (\ref{glbalg1}, \ref{glbalg2}), any norm on \(\mc{D}^{\sigma}_{j}\,\) must satisfy
\be\label{gennrm}
\brkt{j,m}{j',m'}\,=\,
\delta_{\sigma\sigma'}\,\delta^{2}_{nn'}\,
\frac{\Gamma(2j)\Gamma(1+n_{+})}{\Gamma(2j+n_{+})}\,
\frac{\Gamma(2j)\Gamma(1+n_{-})}{\Gamma(2j+n_{-})}\,\brkt{j,\sigma j}{j',\sigma j'}
\ee
More generally, we are looking for solutions to the Klein-Gordon equation
\be
\left(\nabla^2-\mu^2\right)\mc{V}_{j}(m)\,=\,0
\ee
of mass \(\mu^2=-4j(1-j)/k\in\mathbb{R}\,\), energy \(E=m_{+}+m_{-}\in\mathbb{R}\,\), and angular momentum \(\ell=m_{+}-m_{-}\in\mathbb{Z}\,\). The solutions which are regular at \(z=1\) (\(\rho=0\)) can be written in terms of hypergeometric functions
\be\label{parfnc}
\mc{U}_{j}(m)\,=\,\frac{\Gamma(A_{1-j}^{+})\Gamma(A_{1-j}^{-})}{\Gamma(1+|\ell|)\Gamma(1-2j)}\,(1-z)^{|\ell|/2}\,z^{j}\,
_{2}{\rm F}_{1}(A_{j}^{+},A_{j}^{-},1+|\ell|,1-z)
\ee
where \(A^{\pm}_{j}=j+(|\ell|\pm E)/2\,\), and \(\mc{U}_{j}(m)\) is invariant under \(m_{\pm}\goto-m_{\pm}\) and \(m_{\pm}\goto m_{\mp}\,\). The asymptotic behavior for \(z\goto 0\) (\(\rho\goto\infty\)) is given by 
\be\label{parasym}
\mc{U}_{j}(m)\,\sim\,z^{j}+R_{j}(m)\,z^{1-j}
\ee
The reflection coefficient is given by
\be\label{qmrefcof}
R_{j}(m)\,=\,R^{-1}_{1-j}(m)\,=\,\frac{\Gamma(2j-1)}{\Gamma(1-2j)}\,
\frac{\Gamma(1-j-m_{+})\,\Gamma(1-j+m_{-})}{\Gamma(j-m_{+})\,\Gamma(j+m_{-})}
\ee
with the wavefunctions (\ref{pardef}) satisfying
\be\label{refrel}
\mc{V}_{j}(m)\,=\,R_{j}(m)\,\mc{V}_{1-j}(m)
\ee 
It may seen that \(R_{j}(m)=0\) for elements of \(\mc{D}^{\sigma}_{j}\,\) with \(j\in\mathbb{R}_{>0}\,\), and that the solutions given by (\ref{parfnc}) are identical to those given by (\ref{discstates}) and take the form
\be
\mc{U}_{j}(m)\,=\,(1-z)^{(n_{+}-n_{-})/2}\,z^{j}\,
_{2}{\rm F}_{1}(-n_{-},2j+n_{+},2j,z)
\ee

\subsection{The \texorpdfstring{$\mc{L}^2$}{} and Klein-Gordon norms on \texorpdfstring{$AdS_3$}{}}

We would like to consider the following \(\mc{L}^2\) norm on \(AdS_{3}\,\)
\be\label{l2nrm}
\brktamp{3mm}{\mc{V}}{\mc{V}'}_{\sst\mc{L}^2}\,=\,
(\pi\sqrt{k})^{-3}\!\int\!d^3 x\sqrt{g}\;\bar{\mc{V}}\,\mc{V}'
\ee
where \(\bar{\mc{V}}=(\mc{V})^*\,\). It is the analogue of this norm which appears in the \(SL(2,R)\) CFT that arises through continuation of the Euclidean target space theory on \(H_{3}\,\). From (\ref{pardef}),
\be\label{l2nrm2}
\brktamp{3mm}{\mc{V}}{\mc{V}'}_{\sst\mc{L}^2}\,=\,
2\pi^{-1}\delta(E'-E)\,\delta_{\ell\ell'}\!
\int_{0}^{1}\!dz\,z^{-2}\,\bar{\mc{U}}\,\mc{U}'
\ee
As may be computed through analytic continuation of the Euclidean expression (\ref{qmpoin2pt}), it may be seen from (\ref{parasym}) that this integral converges in the delta function sense for \(\,2j-1\in i\mathbb{R}\,\), and thus \(\mc{C}^{\alpha}_{j}\) appears in the spectrum associated with the \(\mc{L}^2\) norm. It may similarly be seen that this integral does not converge for \(j\in\mathbb{R}_{>0}\) unless \(R_{j}(m)=0\) and \(j>1/2\,\). Thus neither the identity representation nor \(\mc{E}^{\alpha}_{j}\) appear in the \(\mc{L}^2\) spectrum.  Furthermore, with the restriction \(j,j'>1/2\,\), \(\mc{D}^{\sigma}_{j}\) appears with norm for the terminating state given by
\be\label{l2vnrm}
\brkt{\mc{V}_{j}(\sigma j)}{\mc{V}_{j'}(\sigma j')}_{\sst\mc{L}^2}\,=\,
\brkt{j,\sigma j}{j',\sigma j'}_{\sst\mc{L}^2}\,=\,\frac{\pi^{-1}}{(2j-1)}\,\delta(j-j')
\ee
which agrees with (\ref{dsc2pf}) in the \(k\goto 0\,\) limit. Note that this expression does not include a reflection term (\(j\goto 1-j\)) since the reflection relation (\ref{refrel}) degenerates (\(R_{j}(m)=0\)) for elements of \(\mc{D}^{\pm}_{j}\,\). It is helpful in this case to introduce the wavefunctions  
\be\label{wref}
\mc{W}_{j}(m)\,=\,\mc{W}_{1-j}(m)\,=\,\Omega_{1-j}(m)\mc{V}_{j}(m)
\ee
where,
\be
\Omega_{j}(m)\,=\,R_{j}(m)\,\Omega_{1-j}(m)\,=\,
\frac{\Gamma(2j-1)}{\Gamma(j-m_{+})\Gamma(j+m_{-})}
\ee
For elements of \(\mc{D}^{\sigma}_{j}\) with \(m_{\pm}=\sigma(j+n_{\pm})\,\) for \(j>1/2\) and \(n_{\pm}\in\mathbb{Z}_{\geq 0}\,\) this leads to the norm
\bbb\label{qmwnrm}
\brkt{\mc{W}_{j}(m)}{\mc{W}_{j'}(m')}&=&
\Omega_{1-j}(m)\,\Omega_{1-j'}(m')\,\brkt{\mc{V}_{j}(m)}{\mc{V}_{j'}(m')}\nonumber\\[0.1cm]
&=&\frac{\pi^{-1}}{(2j-1)}\,
\left(\frac{\Gamma(1+n_{+})\,\Gamma(2j+n_{+})}{\Gamma(1+n_{-})\,\Gamma(2j+n_{-})}\right)^{\sigma}\,
\!\!\delta_{\sigma\sigma'}\,\delta^{2}_{nn'}\,\delta(j-j')
\eee
which agrees with (\ref{discw2pf}) in the \(k\goto 0\,\) limit. Note that while the corresponding reflected wavefunction \(\mc{W}_{1-j}(m)=\mc{W}_{j}(m)\) is non-degenerate, the wavefunction \(\mc{W}_{j}(m)\) with \(m_{\pm}=\sigma(1-j+n_{\pm})\,\) for \(j>1/2\) and \(n_{\pm}\in\mathbb{Z}_{\geq 0}\,\) is non-normalizable.

We may also consider the Klein-Gordon norm on \(AdS_{3}\)
\be\label{kgnrm1}
\brktamp{3mm}{\mc{V}}{\mc{V}'}_{\sst \rm KG}\,=\,
\frac{i}{2\pi\sqrt{k}}\int_{\Sigma}d^2x\sqrt{g_{\sst\Sigma}}\;
n^{a}\left(\bar{\mc{V}}\partial_{a}\mc{V}'-\mc{V}'\partial_{a}\bar{\mc{V}}\right)
\ee
Here \(\Sigma\) is a spacelike surface for \(AdS_{3}\,\), which we choose to be a surface of constant \(t\,\), and \(n^a\) is the corresponding future-directed unit vector field normal to \(\Sigma\,\). From (\ref{pardef}),
\be\label{kgnrm2}
\brktamp{3mm}{\mc{V}}{\mc{V}'}_{\sst \rm KG}\,=\,
\txfc{1}{2}(E'+E)\,e^{-i(E'-E)t}\,\delta_{\ell\ell'}\!
\int_{0}^{1}\!dz\,z^{-1}\,\bar{\mc{U}}\,\mc{U}'
\ee
It may be seen that elements of \(\mc{C}^{\alpha}_{j}\) are (point rather than continuum) normalizable under this norm. This is also true of elements of \(\mc{E}^{\alpha}_{j}\,\). For \(\mc{D}^{\sigma}_{j}\) it may be seen from (\ref{parasym}) that this integral converges for all \(j>0\,\), with norm for the terminating state given by
\be
\brkt{j,\sigma j}{j,\sigma j}_{\sst\rm KG}\,=\,1
\ee
Thus, although \(\mc{C}^{\alpha}_{j}\) may be excluded since the associated mass is tachyonic (both in the sense of the associated bosonic string vertex operators and due to the violation of the Breitenlohner-Freedman bound on \(AdS_3\)), all unitary representations of \(SL(2,R)\) are normalizable under the Klein-Gordon norm. Note that here, unlike in the case (\ref{l2vnrm}) of the \(\mc{L}^2\) norm, \(j=j'\) must be imposed, and wavefunctions of different \(AdS_3\) mass are not orthogonal under the KG norm (\ref{kgnrm2}). Thus the KG norm is not suitable to serve as a metric on primary fields in the \(SL(2,R)\) CFT.

\subsection{Wavefunctions on \texorpdfstring{$H_{3}$}{}}

Solutions to the Klein-Gordon equation on \(\mc{H}_3\) in global coordinates arise via continuation of the solutions \(\mc{V}_{j}(m)\) given above (\ref{pardef}, \ref{parfnc}). Normalized solutions require the condition \(E\goto -i\omega\,\) with \(\omega\in\mathbb{R}\,\), so that \(e^{-iEt}=e^{i\omega\tau}\,\), as well as the condition \(j=1/2+is\,\) with \(s\in\mathbb{R}\,\). To conform to the CFT conventions introduced above, we will denote\footnote{Here \(2m=2m_{+}=-i\omega+\ell\,\) and \(2\bar{m}=2m_{-}=-i\omega-\ell\,\), so that \(\bar{m}=-(m)^*\,\). This corresponds to the fact that the Killing vector fields \(J^{3}\) and \(\bar{J}^{3}\) are anti-hermitian operators with respect to the \(\mc{L}^2\) norm (\ref{l2nrm}) on \(\mc{H}_3\,\).} the corresponding solutions on \(\mc{H}_3\) by \(\mc{V}_{j}(m)\,\). The operators \(J^{a}\) and their conjugates act on \(\mc{V}_{j}(m)\,\) as given by (\ref{glbalg1}, \ref{glbalg2}). For comparison with other solutions introduced below, we express the asymptotic (\(\phi\goto\infty\)) form of \(\mc{V}_{j}(m)\) in the (\(\phi,\gamma,\bar{\gamma}\)) coordinates as
\be
\mc{V}_{j}(m)\,\sim\,
\gamma^{-m}\,\bar{\gamma}^{-\bar{m}}\left((e^{2\phi}\gamma\bar{\gamma})^{-j}+
R_{j}(m)\,(e^{2\phi}\gamma\bar{\gamma})^{j-1}\right)
\ee
The wavefunction \(\mc{V}_{j}(m)\) satisfies the reflection relation \(\mc{V}_{j}(m)=R_{j}(m)\mc{V}_{1-j}(m)\) with coefficient \(R_{j}(m)\) which is the \(k\goto\infty\) limit of that (\ref{glbrefcof}) which applies to the corresponding CFT vertex operator.

Another basis of solutions to the Klein-Gordon equation on \(\mc{H}_3\) which is often utilized is 
\be\label{phifuncs}
\Phi_{j}(x)\,=\,(1-2j)\left(e^{-\phi}+e^{\phi}|\gamma-x|^2\right)^{-2j}
\ee
These functions satisfy \(J^a\Phi_{j}(x)=-D^{a} \Phi_{j}(x)\,\), and the complex conjugate relation, where \(D^a\) are the \(x\)-dependent operators given by (\ref{phigen}). They may be seen to have the asymptotic form
\be\label{phiasym}
\Phi_{j}(x)\,\sim\,-\,e^{-2(1-j)\phi}\,\pi\delta^2(\gamma-x)+
(1-2j)\,e^{-2j\phi}\,|\gamma-x|^{-4j}
\ee
As can be verified using (\ref{poinrefint}), the wavefunctions \(\Phi_{j}(x)\) satisfy a reflection relation which is given by the \(k\goto\infty\) limit of the reflection relation (\ref{phiref}) for the analogous primary fields of the \(H_{3}\) CFT. For \({\rm Re}(j)>1/2\) it may be seen from (\ref{phiasym}) that, given the volume factor \(e^{2\phi}\) which appears in (\ref{qmpoin2pt}), the divergence of the norm of \(\Phi_{j}(x)\) is concentrated around \(\gamma=x\,\), while for \({\rm Re}(j)<1/2\) the norm diverges for all \(\gamma\,\). The wavefunctions \(\Phi_{j}(x)\) are related to \(\mc{V}_{j}(m)\) via the transform (\ref{phitoglb2}). 

The Fourier transform of \(\Phi_{j}(x)\) leads to a third basis of solutions to the Klein-Gordon equation on \(\mc{H}_3\,\). The analogous basis (\ref{phitolv}) of primary vertex operators has proven useful in relating the \(H_{3}\) and Liouville CFTs. Given (\ref{kfuncfour}), this basis is given by
\be\label{phiwfunc}
\varphi_{j}(\mu)\,=\,
\frac{e^{\mu\gamma-\bar{\mu}\bar{\gamma}}}{\Gamma(1-2j)}\;2|\mu|e^{-\phi}\,
{\rm K}_{2j-1}(2|\mu|e^{-\phi})
\ee
These functions satisfy \(J^a\varphi_{j}(\mu)=-F^{a}\varphi_{j}(\mu)\,\), and the complex conjugate relation, where \(F^a\) are the \(\mu\)-dependent operators given by (\ref{fops}). The wavefunctions \(\varphi_{j}(\mu)\) have the asymptotic form
\be
\varphi_{j}(\mu)\,\sim\,
\frac{e^{\mu\gamma-\bar{\mu}\bar{\gamma}}}{\Gamma(1-2j)}\,
\left(|\mu|^{2j}e^{-2j\phi}\,\Gamma(1-2j)+|\mu|^{2(1-j)}e^{-2(1-j)\phi}\,\Gamma(2j-1)\right)
\ee
and satisfy the reflection relation \(\varphi_{j}(\mu)=R_{j}\,\varphi_{1-j}(\mu)\,\). The reflection coefficient \(R_{j}\) is the \(k\goto\infty\) limit of that  which appears in the CFT (\ref{h3lvref},\,\ref{h3ref}). In addition, the wavefunctions \(\varphi_{j}(\mu)\) are related to the modes \(\mc{V}_{j}(m)\) via the transform (\ref{h3lvtoglb}). In analogy with the CFT, the associated two-point function with \(j=1/2+is\,\) is defined as
\bbb\label{qmpoin2pt}
\vev{\hat{\varphi}_{j_2}(\mu_2)\varphi_{j_1}(\mu_1)} & = &
\brkt{\bar{\varphi}_{j_2}(\mu_2)}{\varphi_{j_1}(\mu_1)}_{\sst\mc{L}^2}\,=\,
\pi^{-3}\!\!\int\!d^2\gamma d\phi\,e^{2\phi}\,\varphi_{j_2}(\mu_2)\varphi_{j_1}(\mu_1)\nonumber\\
& = & |\mu_1|^2\,\delta^2(\mu_2+\mu_1)\left(\delta(s_2+s_1)+R_{j_1}\,\delta(s_2-s_1)\right)
\eee
Here we have used the orthogonality relation for \({\rm K}\) Bessel functions (\ref{bslKnrm}). It may be seen that this relation corresponds precisely to the \(k\goto\infty\) limit of the CFT two-point function (\ref{mu2pf}). Similarly, using the transforms (\ref{phitolv},\,\ref{h3lvtoglb}) it may be shown that the above forms for the wavefunctions \(\Phi_{j}(x)\) and \(\mc{V}_{j}(m)\)  lead to the \(k\goto\infty\) limit of the CFT two-point functions (\ref{h32pf},\,\ref{glb2pt}).

\section{Degenerate modules and the Knizhnik-Zamolodchikov equation \label{sec_C}}

Degenerate affine modules in the \(H_{3}\) CFT exist at \(h=j(1-j)/(k-2)\) for 
\be\label{refdeg}
2j_{r,s}-1\,=\,\pm\left(r+s(k-2)\right)
\ee
with \(r,s\in\mathbb{Z}_{>0}\,\). There are also degenerate modules for
\be\label{norefdeg}
2j_{m}-1\,=\,-m
\ee
with \(m\in\mathbb{Z}_{>0}\,\), values of \(j\) for which the reflection symmetry \(j\goto 1-j\,\) is not satisfied. Note that there are degenerate operators at \(j=0\) and \(j=k/2\) but none in the range \(0<2j-1<(k-2)\,\), which corresponds to the physical spectrum for discrete states \(\mc{S}_{\sst\mc{D}}\,\)(\ref{discspec}) of the \(SL(2,R)\) CFT. As may be seen semi-classically from the form of the wave functions (\ref{phifuncs}), the finite dimensional representations (\ref{norefdeg}) satisfy \((D^{+})^{m}\Phi_{(1-m)/2}=0\,\), where the representation of the algebra (\ref{phigen}) has been used. These representations may be seen to be the analogues of the half-integer representations of \(SU(2)\,\).

The Knizhnik-Zamolodchikov (KZ) equation follows from the Sugawara construction of the stress tensor (\ref{sugawara}). Considering the action of \(L_{-1}\) (\ref{viramodes2}) on a primary field \(\mc{O}\) we find
\be\label{sugmnsone}
\left[\,(k-2)L_{-1}-\left(-2J^{3}_{-1}J^{3}_{0}+J^{+}_{-1}J^{-}_{0}+J^{-}_{-1}J^{+}_{0}\right)\,\right]\cdot\mc{O}\,=\,0
\ee
Suppressing anti-holomorphic dependence, a correlator of \(n\) primary fields \(\Phi_{j_{r}}(x_r|z_r)\) satisfies
\be
(k-2)\frac{\partial\,}{\partial z_{q}}\amp{3mm}{{\textstyle \prod}_{r=1}^{n}\Phi_{j_{r}}(x_r|z_r)}\,=\,-2\sum_{p\neq q}(z_p-z_q)^{-1}\,\eta_{ab}\,D_{q}^{a}D_{p}^{b}\,\amp{3mm}{{\textstyle \prod}_{r=1}^{n}\Phi_{j_{r}}(x_r|z_r)}
\ee
where \(D_{p}^{a}\) to corresponds to (\ref{phigen}) with \((j,x)\goto(j_p,x_p)\,\).
More explicitly, the KZ equation reads,
\bbb\label{kzphi}
\lefteqn{(k-2)\frac{\partial\,}{\partial z_{q}}\amp{3mm}{{\textstyle \prod}_{r=1}^{n}\Phi_{j_{r}}(x_r|z_r)}\,=\,-\sum_{p\neq q}\frac{1}{(z_p-z_q)}
\left[(x_p-x_q)^2\frac{\partial^2\,}{\partial x_q\partial x_p}\right.}\hspace{2cm}\nonumber\\
&& \left.+2(x_p-x_q)\left(j_p\frac{\partial\,}{\partial x_q}-j_q\frac{\partial\,}{\partial x_p}\right)-2j_q\,j_p\right]\amp{3mm}{{\textstyle \prod}_{r=1}^{n}\Phi_{j_{r}}(x_r|z_r)}
\eee
Suppressing anti-holomorphic dependence, consider the lowest weight state \(\mc{V}_{k/2}(k/2)\,\) which is an element of the degenerate module \(j_{{\sst 1,1}}\) appearing in (\ref{refdeg}). We will assume\footnote{This is a non-trivial assumption since, while the norm of \(J^{-}_{-1}\cdot\mc{V}_{k/2}(k/2)\,\) vanishes given the \(SL(2,R)\) current algebra (\ref{modealg}), in a non-unitary theory it should not be expected that zero-norm states vanish identically. Nevertheless, as stressed in~\cite{GiveonKutasov_0106}, it does not appear that either the \(H_3\) or \(SL(2,R)\) CFT can be defined absent this assumption.} that the associated null descendant decouples from correlation functions, producing the operator equation
\be\label{nullglb}
J^{-}_{-1}\cdot\mc{V}_{k/2}(k/2)\,=\,0
\ee
Since \(J^{-}_{0}\cdot\mc{V}_{k/2}(k/2)=0\,\), in this case the KZ equation (\ref{sugmnsone}) reads
\be\label{kzglb}
\left(L_{-1}+J^{3}_{-1}\right)\cdot\mc{V}_{k/2}(k/2)\,=\,0
\ee
It is helpful to express (\ref{nullglb}) and (\ref{kzglb}) in the \(\Phi_{j}(x)\) basis, from which they may be derived by taking \(\mc{V}_{k/2}(k/2)=\Phi_{k/2}(0)/(1-k)\,\). In the case of the field \(\Phi_{k/2}(x)\,\), the null state condition may be derived as follows. Since  \(J_{0}^{+}\Phi_{j}(x)=\frac{\partial}{\partial x}\Phi_{j}(x)\,\) from (\ref{phigen}), for any operator \(\mc{O}\) we may define an operator translated with respect to the boundary coordinates \((x,\bar{x})\) as
\be
\mc{O}(x)\,=\,e^{J_{0}^{+}x}\,\mc{O}\,e^{-J_{0}^{+}x}
\ee
where anti-holomorphic dependence is suppressed. In particular we define
\bbb
J^{+}(x|z)&=& J^{+}(z)\\[0.1cm]
J^{3}(x|z)&=& J^{3}(z)-xJ^{+}(z)\\[0.1cm]
J^{-}(x|z)&=& J^{-}(z)-2xJ^{3}(z)+x^2J^{+}(z)
\eee
and introduce the modes expanded around \(z\,\)
\be
J^{a}_{n}(x)\,=\,\frac{1}{2\pi i}\oint dw\,(w-z)^{n}J^{a}(x|w)
\ee
The null state condition (\ref{nullglb}) then takes the form \(J^{-}_{-1}(x)\cdot\Phi_{k/2}(x)=0\,\), which follows from (\ref{phitohighglb}). For a correlator of \(\Phi_{k/2}(x|z)\,\) with \(n\) primary fields \(\Phi_{j_{r}}(x_r|z_r)\) this translates to
\be\label{nullphi}
\sum_{p=1}^{n}\,\frac{(x_p-x)}{(z_p-z)}
\left((x_p-x)\frac{\partial\,}{\partial x_{p}}+2j_{p}\right)
\amp{3mm}{\Phi_{k/2}(x|z){\textstyle \prod}_{r=1}^{n}\Phi_{j_{r}}(x_r|z_r)}\,=\,0
\ee
It follows from (\ref{vircaalg}) that \(L_{-1}(x)=L_{-1}\,\), and thus for \(\Phi_{k/2}(x)\,\) (\ref{kzglb}) reads
\be
\left(L_{-1}+J^{3}_{-1}(x)\right)\cdot\Phi_{k/2}(x)\,=\,0
\ee
Here we have used
\bbb
J^{+}_{0}(x)\cdot\Phi_{j}(x) &=& \frac{\partial\,}{\partial x}\Phi_{j}(x)\\[0.1cm]
J^{3}_{0}(x)\cdot\Phi_{j}(x) &=& j\,\Phi_{j}(x)\\[0.2cm]
J^{-}_{0}(x)\cdot\Phi_{j}(x) &=& 0
\eee
This leads to the KZ equation with \(\Phi_{k/2}(x)\,\) and \(n\) primary fields \(\Phi_{j_{r}}(x_r|z_r)\)
\be\label{kzphi2}
\ampbox{5mm}{\frac{\partial\,}{\partial z}-\sum_{p=1}^{n}\,(z_p-z)^{-1}
\left((x_p-x)\frac{\partial\,}{\partial x_{p}}+j_{p}\right)}
\amp{3mm}{\Phi_{k/2}(x|z)\,{\textstyle \prod}_{r=1}^{n}
\Phi_{j_{r}}(x_r|z_r)}\,=\,0
\ee
It may be seen that this equation follows from (\ref{kzphi}) and (\ref{nullphi}).

\section{Conventions for vertex operators \label{sec_D}}

In describing correlators of Virasoro descendants and primary operators of indefinite conformal weight or non-zero spin, it is helpful to employ a somewhat more flexible notation~\cite{Gaberdiel_9910} than is used in most expressions in this paper. In this convention, which suppresses anti-holomorphic dependence, the vertex operator associated with a general state \(\varphi\) is denoted by \(V(\varphi|z)\) with \(V(\varphi|0)\equiv V(\varphi)=\varphi\,\). Where no risk of confusion exists the more common notation \(V(\varphi|z)=\varphi(z)\,\) will be employed. The action of a general Mobius transformation \(z\goto\gamma(z)\) is given by 
\be
D_{\gamma}V(\varphi|z)D^{-1}_{\gamma}\,=\,V((\gamma')^{L_{0}}\,e^{gL_{1}}\varphi|\gamma(z))
\ee
where \(g(z)=\gamma''/(2\gamma')\,\). For some worldsheet operator \(\mc{O}\,\), this is consistent with the more common notation \(V(\mc{O}\varphi|z)=\mc{O}\cdot\varphi(z)\,\). For \(\gamma(z)=u=1/z\) we have \(u\)-frame operator 
\be\label{hattrans}
\hat{V}(\varphi|z)\,=\,D_{\gamma}V(\varphi|z)D^{-1}_{\gamma}
\,=\,V((-z^{-2})^{L_{0}}\,e^{-z^{-1}L_{1}}\varphi|z^{-1})\,=\,
V(e^{zL_{1}}\,(-z^{-2})^{L_{0}}\varphi|z^{-1})
\ee
where we have used
\be
\alpha^{L_{0}}e^{\beta L_{1}}=e^{\alpha^{-1}\beta L_{1}}\alpha^{L_{0}}
\ee
The symmetry of the Mobius-fixed two-point function
\be
\amp{3.5mm}{\hat{V}(\varphi_{2})V(\varphi_{1})}\,=\,
\amp{3.5mm}{\hat{\varphi}_{2}\,\varphi_{1}}\,=\,
\amp{3.5mm}{\hat{\varphi}_{1}\,\varphi_{2}}
\ee
permits it to be used as a metric on the space of states. 
Using Mobius transformations the two-point function can be written as
\bbb\label{twoptdesc}
\amp{3.5mm}{\varphi_{2}(z_2)\varphi_{1}(z_1)}&=&
\amp{3.5mm}{\hat{V}(e^{-L_{1}}(z_{21}^{-1})^{L_{0}}\varphi_{2})
V(e^{-L_{1}}(-z_{21}^{-1})^{L_{0}}\varphi_{1})}\\[0.1cm]
&=&\amp{3.5mm}{\hat{V}(e^{-z_{21}^{-1}L_{1}}\varphi_{2})
V(e^{-z_{21}L_{1}}(-z_{21}^{-2})^{L_{0}}\varphi_{1})}
\eee
For primary fields \(\phi_p\) of definite conformal dimension this implies \(h_1=h_2\,\) and 
\be
\amp{3.5mm}{\phi_{2}(z_2)\phi_{1}(z_1)}\,=\,
(-z_{21}^{-2})^{h_{1}}\amp{3.5mm}{\hat{\phi}_{2}\,\phi_{1}}
\ee

\section{Some useful relations \label{sec_E}}

Here is a convenient representation of the delta function which appears in the computation of the two-point function (\ref{h32pf})
\be\label{phidlt}
\lim_{\epsilon\goto  0}\,\epsilon\,x^{\epsilon-p-1}\bar{x}^{\epsilon-q-1}\,=\,
\frac{\pi}{p!q!}\,\partial^{p}\bar{\partial}^{q}\delta^2(x)\spwd{1cm}{for}p,q\in\mathbb{Z}_{\geq 0}
\ee
The following definition is useful to relate expressions involving \(\mc{V}_{j}(m)\) to those involving \(\Phi_{j}(x)\)
\be
\delta^2(x-y)\,=\,|y|^{-2}\int d^2m\,(y/x)^{-m}\,(\bar{y}/\bar{x})^{-\bar{m}}
\ee
For the primary fields of definite affine weight \(\mc{V}_{j}(m)\) we have the definitions \(E=-i\omega=m+\bar{m}\,\), \(\ell=m-\bar{m}\in\mathbb{Z}\,\). The definition of the expression \(\delta^2(m)\) appearing in (\ref{wfunc}) is then given by
\be\label{mdelta}
\delta^2(m)\,=\,(2\pi)^2\,\delta(\omega)\,\delta_{\ell}\,=\,
\int_{\mathbb{C}}d^2x\,x^{m-1}\,\bar{x}^{\bar{m}-1}
\ee
so that
\be\label{mmbarint}
\int d^2m\,=\frac{1}{(2\pi)^2}\sum_{\ell\in\mathbb{Z}}\int d\omega
\ee
For the Fourier transform in (\ref{phitolv}) we use the normalization
\be
\delta^2(\mu)\,=\,\pi^{-2}\!\!\int d^2x\,e^{\mu x-\bar{\mu}\bar{x}}
\ee

The following integral is required above in (\ref{h3ref})
\be\label{int1}
\pi^{-1}\int d^2x\,e^{\mu x-\bar{\mu}\bar{x}}\,|x|^{-4j}\,=\,
\frac{\Gamma(1-2j)}{\Gamma(2j)}\,|\mu|^{2(2j-1)}
\ee
this may be seen to follow from
\be\label{int3}
\pi^{-1}\!\!\int d^2 x\,
x^{-j+m}\,\bar{x}^{-j+\bar{m}}\,e^{x-\bar{x}}\,=\,\frac{\Gamma(1-j+\bar{m})}{\Gamma(j-m)}
\ee
which is required in (\ref{h3lvtoglb}) and elsewhere. Here is an important Dotsenko-Fateev integral which is required in (\ref{glbrefcof}) and throughout this paper
\be\label{int2}
\pi^{-1}\!\!\int d^2x\,
x^{\alpha-1}(1-x)^{\beta-1}\,\bar{x}^{\bar{\alpha}-1}(1-\bar{x})^{\bar{\beta}-1}\,=\,
\frac{\Gamma(\alpha)\Gamma(\beta)\Gamma(1-\bar{\alpha}-\bar{\beta})}
{\Gamma(1-\bar{\alpha})\Gamma(1-\bar{\beta})\Gamma(\alpha+\beta)}
\ee
This integral is defined for \(\alpha-\bar{\alpha}\in\mathbb{Z}\,\) and \(\beta-\bar{\beta}\in\mathbb{Z}\,\), and, using \(\Gamma(x)\Gamma(1-x)=\pi/\sin(\pi x)\,\), it may be shown to be invariant under \((\alpha,\bar{\alpha};\beta,\bar{\beta})\goto(\bar{\alpha},\alpha;\bar{\beta},\beta)\). 

The reflection relation (\ref{phiref}) in the \(\Phi_{j}(x)\) basis and the three-point function (\ref{mu3pf}) in the \(\varphi_{j}(\mu)\) basis require the integral
\bbb\label{louvint}
\lefteqn{\pi^{-1}\!\!\int d^2x\,|x|^{2(a-1)}|1-x|^{2(b-1)}|x-y|^{2(c-1)}}\hspace{3cm}\nonumber\\
&=& \frac{\gamma(b)\gamma(a+c-1)}{\gamma(a+b+c-1)}\,{_{2}\mc{F}_{1}}(1-c,2-a-b-c;2-a-c;y)\nonumber\\
& & + \;\frac{\gamma(a)\gamma(c)}{\gamma(a+c)}\,|y|^{2(a+c-1)}\,{_{2}\mc{F}_{1}}(1-b,a;a+c;y)
\eee
Where \(\gamma(x)=\Gamma(x)/\Gamma(1-x)\,\), and \({_{2}\mc{F}_{1}}(a,b;c;y)={_{2}{\rm F}_{1}}(a,b;c;y)\,{_{2}{\rm F}_{1}}(a,b;c;\bar{y})\,\). Here we have the representation
\be
\label{F21sum}
{_{2}{\rm F}_{1}}(a,b;c;1)\,=\,\sum_{n=0}^{\infty}
\frac{\mc{P}_{n}(a)\mc{P}_{n}(b)}{\mc{P}_{n}(c)\mc{P}_{n}(1)}
\ee
This sum is convergent for \({\rm Re}(c-a-b)>0\,\). Here \(\mc{P}_{n}(x)=\Gamma(x+n)/\Gamma(x)\,\) is the Pochhammer function. Here are some \({_{2}{\rm F}_{1}}\) identities
\bbb\label{hypid1}
\lefteqn{{_{2}{\rm F}_{1}}(a,b;c;z)\,=\,
(1-z)^{(c-a-b)}\;{_{2}{\rm F}_{1}}(c-a,c-b;c;z)}\hspace{1cm}\\[0.1cm]
&=& \ \frac{\Gamma(c)\Gamma(c-a-b)}{\Gamma(c-a)\Gamma(c-b)}\ 
{_{2}{\rm F}_{1}}(a,b;a+b+1-c;1-z)\nonumber \\
&& \!\!+\,\frac{\Gamma(c)\Gamma(a+b-c)}{\Gamma(a)\Gamma(b)}\,
(1-z)^{(c-a-b)}\,{_{2}{\rm F}_{1}}(c-a,c-b;1+c-a-b;1-z)
\eee

The three-point function (\ref{glb3pt}) in the \(\mc{V}_{j}(m)\) basis requires the integral
\bbb\label{glbint}
W(j_p|m_1,m_2) &=& \int d^2x\,d^2y\;x^{-j_{1}-m_{1}}\bar{x}^{-j_{1}-\bar{m}_{1}}\,
y^{-j_{2}-m_{2}}\bar{y}^{-j_{2}-\bar{m}_{2}} \nonumber \\
& & \hspace*{1cm}\times\;|1-x|^{2(\hat{\jmath}-2j_{2}-1)}\, 
|1-y|^{2(\hat{\jmath}-2j_{1}-1)}\,|x-y|^{2(\hat{\jmath}-2j_{3}-1)}\nonumber \\[3mm]
&=& D_{12}C^{12}\bar{C}^{12}+D_{21}C^{21}\bar{C}^{21}+
D_{3}\left(C^{12}\bar{C}^{21}+C^{21}\bar{C}^{12}\right)
\eee
where, defining \(s(x)=\sin(\pi x)\,\),
\bbb
D_{12}(j_1,j_2,j_3,m_1,m_2)\,=\,
\frac{s(j_2+m_2)s(j_3+j_1-j_2)}{s(j_1-m_1)s(j_2-m_2)s(j_3-m_1-m_2)} \hspace{4cm}\nonumber \\[1mm]
\hspace{1cm}\times\left(s(j_1+m_1)s(j_1-m_1)s(j_2+m_2)-s(j_2-m_2)s(j_2-j_3-m_1)s(j_2+j_3-m_1)\right)
\eee
with \(D_{21}=D_{12}(j_2,j_1,j_3,m_2,m_1)\,\), and 
\be
D_{3}(j_1,j_2,j_3,m_1,m_2)\,=\,
-\frac{s(\hat{\jmath}-2j_1)s(\hat{\jmath}-2j_2)s(j_1+m_1)s(j_2+m_2)s(j_1+j_2-m_1-m_2)}{s(j_1-m_1)s(j_2-m_2)s(j_3-m_1-m_2)} 
\ee
where \(\hat{\jmath}=j_1+j_2+j_3\,\). We also have
\bbb
C^{12}(j_1,j_2,j_3,m_1,m_2)\,=\,
\frac{\Gamma(\hat{\jmath}-1)\Gamma(1-j_3-m_1-m_2)}{\Gamma(j_3-m_1-m_2)}\hspace{4cm}\nonumber \\
\hspace{1cm}\times\,{_{3}{\rm G}_{2}}(j_3-m_1-m_2,\hat{\jmath}-2j_2,1-j_2-m_2\,;j_1-j_2-m_1-m_2+1,j_1+j_3-m_2)
\eee
with \(C^{21}=C^{12}(j_2,j_1,j_3,m_2,m_1)\,\), \(\bar{C}^{12}=C^{12}(j_1,j_2,j_3,\bar{m}_1,\bar{m}_2)\,\), and \(\bar{C}^{21}=C^{12}(j_2,j_1,j_3,\bar{m}_2,\bar{m}_1)\,\), where we have defined
\be
{_{3}{\rm G}_{2}}(a_1,a_2,a_3\,;b_1,b_2)\,=\,
\frac{\Gamma(a_1)\Gamma(a_2)\Gamma(a_3)}{\Gamma(b_1)\Gamma(b_2)}\ 
{_{3}{\rm F}_{2}}(a_1,a_2,a_3\,;b_1,b_2\,;1)
\ee
Here we have the representation
\be
\label{F32sum}
{_{3}{\rm F}_{2}}(a_1,a_2,a_3\,;b_1,b_2\,;1)\,=\,\sum_{n=0}^{\infty}
\frac{\mc{P}_{n}(a_1)\mc{P}_{n}(a_2)\mc{P}_{n}(a_3)}{\mc{P}_{n}(b_1)\mc{P}_{n}(b_2)\mc{P}_{n}(1)}
\ee
which is convergent for \({\rm Re}(\hat{b}-\hat{a})>0\,\), where \(\hat{a}=a_{1}+a_{2}+a_{3}\) and \(\hat{b}=b_{1}+b_{2}\,\). Here \(\mc{P}_{n}(x)=\Gamma(x+n)/\Gamma(x)\,\) is the Pochhammer function.
Here are some \({_{3}{\rm G}_{2}}\) identities
\bbb\label{gid1}
\lefteqn{{_{3}{\rm G}_{2}}(a_1,a_2,a_3\,;b_1,b_2)}\nonumber\\[0.2cm]
&=&\frac{\Gamma(a_{2})\Gamma(a_{3})}
{\Gamma(b_{1}-a_{1})\Gamma(b_{2}-a_{1})}\,
{_{3}{\rm G}_{2}}(b_{1}-a_{1},b_{2}-a_{1},\hat{b}-\hat{a}\,;
\hat{b}-\hat{a}+a_{2},\hat{b}-\hat{a}+a_{3})\\[0.2cm]
&=&\frac{\Gamma(a_{2})\Gamma(a_{3})\Gamma(\hat{b}-\hat{a})}
{\Gamma(b_{1}-a_{2})\Gamma(b_{2}-a_{1})\Gamma(b_{1}-a_{3})}
\,{_{3}{\rm G}_{2}}(a_{1},b_{1}-a_{2},b_{1}-a_{3}\,;
b_{1},\hat{b}-\hat{a}+a_{1})\\[0.2cm]
&=&\ \frac{s(b_{1}-a_{2})s(b_{2}-a_{2})}
{s(a_{1})s(a_{3}-a_{2})}\,
{_{3}{\rm G}_{2}}(a_{2},1+a_{2}-b_{1},1+a_{2}-b_{2}\,;
1+a_{2}-a_{3},1+a_{2}-a_{1})\nonumber\\[0.15cm]
&&\!\!+\,\frac{s(b_{1}-a_{3})s(b_{2}-a_{3})}
{s(a_{1})s(a_{2}-a_{3})}\,
{_{3}{\rm G}_{2}}(a_{3},1+a_{3}-b_{1},1+a_{3}-b_{2}\,;
1+a_{3}-a_{2},1+a_{3}-a_{1})
\eee

The three-point function in both Liouville theory and the \(H_3\) model involves the function \(G(j)\) which is defined in terms of  \(\Upsilon_{b}(x)\) by (\ref{upsdefs}). Defining \(Q=b+b^{-1}\,\), the function \(\Upsilon_{b}(x)\) has the integral representation
\be\label{upsdefs1}
\ln\Upsilon_{b}(x)\,=\,\int_{0}^{\infty}\frac{dt}{t}\left(e^{-2t}\,(Q/2-x)^2-\frac{\sinh^2[(Q/2-x)t]}{\sinh(bt)\sinh(t/b)}\right)
\ee
which converges for \(0<{\rm Re}(x)<Q\,\). Note that \(\Upsilon_{b}(x)\) satisfies
\bbb
\Upsilon_{b}(x)&=&\Upsilon_{b}(Q-x)\,=\,\Upsilon_{b^{-1}}(x)\label{upsdefs2}\\
\Upsilon_{b}(x+b)&=&\gamma(bx)\,b^{1-2bx}\,\Upsilon_{b}(x)\label{upsdefs3}\\
\Upsilon_{b}(x+b^{-1})&=&\gamma(b^{-1}x)\,b^{2b^{-1}x-1}\,\Upsilon_{b}(x)\label{upsdefs4}
\eee
Using these relations and (\ref{upsdefs1}) it may be shown that \(\Upsilon_{b}(x)\) is an entire function with zeros at 
\be\label{upsdefs5}
x=-nb-mb^{-1}\spwd{1cm}{and}x=(n+1)b+(m+1)b^{-1}\spwd{1cm}{for}n,m\in\mathbb{Z}_{\geq 0}
\ee

The following integral is employed in section~\ref{sec_6}
\bbb\label{hypint}
\lefteqn{\pi^{-1}\!\!\int d^{2}u\,u^{d-1}\,\bar{u}^{\bar{d}-1}
\left(\left|{_{2}{\rm F}_{1}}(a,b;c;u)\right|^2+f(a,b,c)\,
\left|u^{1-c}\,{_{2}{\rm F}_{1}}(1+a-c,1+b-c;2-c;u)\right|^2\right)}\hspace{3cm}
\nonumber\\&&\,=\,\frac{\gamma(c)}{\gamma(a)\gamma(b)}\,
\frac{\Gamma(d)\Gamma(1-c+d)}{\Gamma(1-\bar{d})\Gamma(c-\bar{d})}
\frac{\Gamma(a-\bar{d})\Gamma(b-\bar{d})}
{\Gamma(1-a+d)\Gamma(1-b+d)}\hspace{2.5cm}
\eee
where
\be\label{hypfnc}
f(a,b,c)\,=\,-\frac{\gamma^2(c)}{(1-c)^2}\,
\frac{\gamma(1-a)\,\gamma(1-b)}{\gamma(c-a)\,\gamma(c-b)}
\ee

The following integral may be used to verify the \(k\goto\infty\) limit of the reflection relation (\ref{phiref}) for the wavefunctions (\ref{phifuncs})
\be\label{poinrefint}
\left(1+|\alpha|^2\right)^{-2j}\,=\,(1-2j)\,\pi^{-1}\!\int d^2\beta\,
|\alpha-\beta|^{-4j}\left(1+|\beta|^2\right)^{-2(1-j)}
\ee
The Fourier transform of (\ref{poinrefint}) is given by
\be\label{kfuncfour}
\pi^{-1}\!\!\int d^2 x\,e^{\mu x-\bar{\mu}\bar{x}}\,\left(1+|x|^2\right)^{-2j}\,=\,
\frac{2|\mu|^{(2j-1)}}{\Gamma(2j)}\;
\,{\rm K}_{1-2j}(2|\mu|)
\ee
which is required in (\ref{phiwfunc}). The following orthogonality relation for the \(K\) Bessel functions for \(a,b\in\mathbb{R}\) is used in the computation of the two-point function (\ref{qmpoin2pt})
\be\label{bslKnrm}
\int_{0}^{\infty}dx\,x^{-1}\,K_{ia}(x)\,K_{ib}(x)\,=\,
\frac{\pi}{2a}\,\frac{\pi}{\sinh(\pi a)}\left(\delta(a+b)+\delta(a-b)\right)
\ee

\pagebreak

\end{document}